\documentclass[10pt,a4paper]{article}
\usepackage[T1]{fontenc}
\usepackage[utf8]{inputenc}
\usepackage{lmodern,microtype}
\usepackage[margin=20mm,headheight=14pt]{geometry}
\usepackage{amsmath,amssymb,mathtools}
\usepackage{booktabs,tabularx,array,enumitem}
\usepackage{fancyhdr}
\usepackage{graphicx}
\usepackage{pdfpages}
\usepackage{xcolor}
\usepackage{cite}
\usepackage[hidelinks]{hyperref}
\hypersetup{pdftitle={Quiver Lab: A Computing Tool for Quiver Gauge Theories},pdfauthor={Zhenghao Zhong},pdfsubject={A compact user guide and computational methods note for Quiver Lab 1.7.3}}
\let\originalnormalsize\normalsize
\renewcommand{\normalsize}{\originalnormalsize\fontsize{11}{13.2}\selectfont}
\setlist{nosep,leftmargin=1.5em,topsep=4pt}
\newcommand{\ui}[1]{\textsf{#1}}
\newcommand{\PE}{\operatorname{PE}}
\newcommand{\PL}{\operatorname{PL}}
\newcommand{\rank}{\operatorname{rank}}
\newcommand{\wt}{\operatorname{wt}}
\newcommand{\USp}{\mathrm{USp}}
\newcolumntype{Y}{>{\raggedright\arraybackslash}X}
\newcommand{\topic}[1]{\clearpage\section{#1}}
\begin{document}
\thispagestyle{plain}
\begin{center}
{\LARGE\bfseries Quiver Lab\\[3pt] A Computing Tool for Quiver Gauge Theories\par}
\vspace{10pt}
{\large Zhenghao Zhong\par}
\vspace{8pt}
{\small Shanghai Institute for Mathematics and Interdisciplinary Sciences (SIMIS),\\
Block A, International Innovation Plaza, No. 657 Songhu Road, Shanghai, China\par}
\vspace{6pt}
{\small\href{mailto:zhong@simis.cn}{\texttt{zhong@simis.cn}}\par}
\end{center}
\vspace{2pt}
\begin{abstract}
Quiver Lab is a browser-based workspace for drawing quivers and exploring the moduli spaces of three-dimensional $\mathcal N=4$ theories. It computes Higgs- and Coulomb-branch Hilbert series, plethystic logarithms, supported refinements, and decay and fission Hasse diagrams. This note explains the Light and Expert workflows and the main computational methods.
\end{abstract}
\vspace{18pt}
\begin{center}
\includegraphics[width=0.9\linewidth]{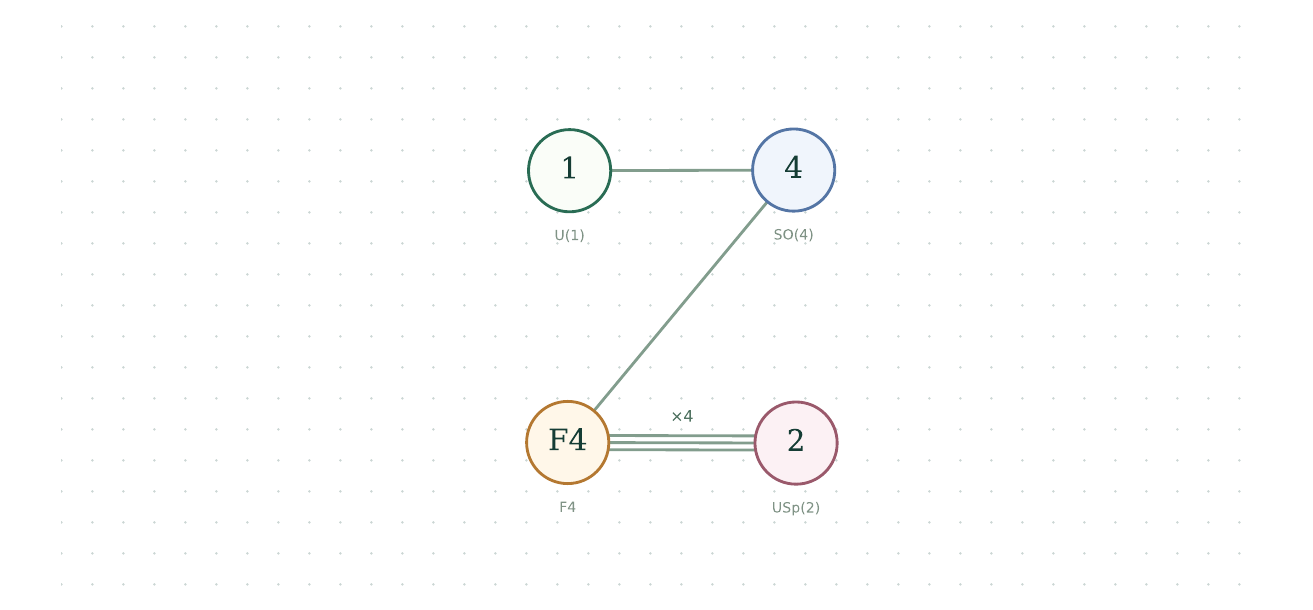}
\end{center}
\clearpage
\section{Introduction}
Substantial progress has been made in understanding the moduli spaces of three-dimensional $\mathcal N=4$ quiver gauge theories since Hilbert-series methods entered the subject. A Hilbert series counts gauge-invariant chiral operators by degree, equivalently holomorphic functions on the moduli space, and organises information about its dimension, symmetries, generators and relations~\cite{Plethystics}. For suitable Higgs branches, the Molien--Weyl integral projects matter fields onto gauge invariants while imposing the complex moment-map equations. For Coulomb branches, the monopole formula introduced in 2013 counts monopole operators dressed by residual gauge invariants~\cite{CHZ}.

Since around 2017, increasing attention has also been paid to Hasse diagrams~\cite{KP,QS,Hasse}. These describe how the moduli space decomposes into symplectic leaves, how their closures meet, and the transverse singularities connecting neighbouring strata. Physically, they reveal patterns of partial Higgsing along the moduli space. Despite this progress, the algorithms and their implementations remain dispersed (or non-existent), motivating a common computational workspace.

With the assistance of GPT-Astra, the author has realised a decade-long ambition to assemble these methods in one ecosystem: Quiver Lab. Its computational reach \textbf{far} exceeds the author's own capabilities with \ui{Mathematica}; some computations that used to take days now take seconds. There remains considerable scope to make it faster and \textbf{even more} efficient. The current release is a standalone HTML application running in a web browser; an installable software version is planned.

As a demonstration, the author collected and tested 1,639 Coulomb-branch quiver examples from 74 papers published since 2013. The database records 1,576 matches, 47 apparent literature typos and 16 other mismatches (we have verified that the errors lie in the literature). \hyperref[app:database]{Appendix~A} lists the quivers and Hilbert series, with notes on discrepancies. This collection also serves as a reusable Hilbert-series database, available in HTML form on \href{https://quiverlab.org/\#database}{the Quiver Lab webpage}, with quiver data exportable for import into Quiver Lab. Further databases are planned. The quivers in the database originate from Refs.~\cite{CHZ,db14020016,db14032384,db14036523,db14086835,db150806813,db150901294,db150905402,HK,db160402155,db160705728,db160908034,db170103825,db170603781,db170709784,db170905818,db171010247,KP,db171206604,db180101129,db180408857,db180601890,db180702521,db180703221,db180805226,db190110493,db190305822,db190506297,db190900667,db200205716,db200404082,db200409620,db200409738,db200616994,db200704667,db200705019,db200715600,db200801027,db201005889,db201009732,db201107033,db201212827,db201212852,db210112235,db210206190,db210305458,db210508064,db210700754,db210709077,db210907237,db210907493,db211006940,db211011365,db211100026,db211102831,db211104745,db211210825,db220301951,db220316550,db220605362,db220705764,db220807270,db220915016,db221002966,db221004921,db230102409,db230316939,DF,db240609626,db241112802,db241114531,db241215202,db250503875,db251014489}.

The author hopes that Quiver Lab will make the study of quiver gauge theories more approachable, more convenient and more fun for everyone.

\vspace{12pt}
\begin{center}
\textbf{Quiver Lab website}\\[4pt]
\href{https://quiverlab.org/}{\texttt{https://quiverlab.org/}}
\end{center}

\topic{Getting started: the editor and Light mode}
Open \href{https://quiverlab.org/}{Quiver Lab}. Choose an example, such as \ui{SQED~$\cdot$~2 flavours}, or start from \ui{Blank canvas}. Circles denote gauge groups and squares denote flavour groups. Choose \ui{Gauge node} or \ui{Flavour node}, then click the canvas; select an existing node to edit its group and size. Use \ui{Bifundamental} on two distinct nodes to add an ordinary edge. Select the edge to change its multiplicity or full/half-hypermultiplet assignment. Drag nodes to arrange the drawing; undo, redo and erase are available.

Light mode accepts $\mathrm U(n)$, $\mathrm{SU}(n)$, $\mathrm{SO}(N)$ and $\USp(2r)$ with ordinary bifundamental matter or half-hypermultiplets. The $\mathrm O(N)$ option, loops, trifundamentals, charged edges and non-simply-laced bonds belong to Expert mode. Drawing an allowed input does not ensure that every calculation has a supported method.

\textbf{A first calculation.}
\begin{enumerate}
\item For an unframed U quiver (one with no flavour nodes), \ui{Remove diagonal U(1)} removes the overall U(1) that does not act on bifundamental matter. Leave it off for the SQED example because it has a flavour node.
\item Click \ui{Compute geometry}. The Coulomb branch and Higgs branch Hilbert series are given in real time up to $t^{20}$.
\item For supported all-unitary inputs, inspect \ui{Hasse diagram} and \ui{Decay candidates}. The \ui{Decay} button runs that exploration separately.
\item Use \ui{Save quiver} for portable JSON, \ui{Export SVG} for the drawing, and \ui{Export results} for the calculation record. Browser autosave retains your workspace on this device.
\end{enumerate}
Light uses preset time and memory limits, with ten seconds per engine. \ui{Cancel} retains all completed coefficients, even if the calculation has not reached $t^{20}$. Switch to Expert mode to adjust the settings or view the plethystic logarithm.

\textbf{What the powers of $t$ mean.} A Hilbert series counts independent holomorphic operators by degree. Each complex scalar field $Q$ or $\widetilde Q$ in a hypermultiplet contributes $t$. A full free hypermultiplet contains both, so
\[
 H_{\mathrm{free\ hyper}}(t)=\PE[t+t]=\PE[2t]
 =\frac{1}{(1-t)^2}=1+2t+3t^2+\cdots.
\]
Here $\PE[t]=1/(1-t)$ counts polynomials in one free complex scalar; PE is the plethystic exponential, defined in Section~\ref{sec:higgs}. On the Coulomb branch, the complex scalar $\varphi$ in the vector multiplet contributes $t^2$, while a monopole of scaling dimension $\Delta$ contributes $t^{2\Delta}$. ``Through $t^N$'' means that all coefficients up to degree $N$ are included, with remainder $O(t^{N+1})$.

\topic{Expert mode: specifying and running a calculation}
Switching to \ui{Expert mode} keeps the current quiver and exposes the full editor, calculation settings and result tabs. Start with a modest finite order, inspect the method and diagnostics, and then increase the order or choose an all-orders attempt.

\subsection*{Matter and gauge-group data}
The editor adds $\mathrm O(N)$, finite $\mathbb Z_q$ factors and exceptional groups $G_2,F_4,E_6,E_7,E_8$. Each calculation reports whether it supports the chosen quiver.

To add adjoint, symmetric or antisymmetric matter, connect a gauge node to itself and choose the loop type in the inspector. To add a \ui{Trifundamental}, select that tool and click three distinct nodes. Select the diamond junction to edit its matter settings.

For a non-simply-laced bond, choose its direction and lacing in the edge inspector. Such a bond changes the Coulomb monopole formula; it is different from several copies of an ordinary hypermultiplet. A generic Higgs calculation is not available for these bonds.

If the theory requires a particular gauge-group quotient, set it under \ui{Global form \& quotients}. These choices determine the magnetic charges included in the Coulomb calculation. The main workflow below uses the standard settings.

\subsection*{Choosing a method}
\begin{tabularx}{\linewidth}{@{}p{0.30\linewidth}Y@{}}
\toprule
\textbf{Method} & \textbf{What it requests}\\
\midrule
\ui{Truncated series} & Exact coefficients through the selected degree.\\[4pt]
\ui{Exact rational} & A supported closed formula or exact resummation/integration; otherwise a finite-series fallback when time remains.\\[4pt]
\ui{Rational search $\cdot$ Coulomb} & An exact attempt followed by a bounded rational reconstruction from longer Coulomb prefixes.\\[4pt]
\ui{Rational search $\cdot$ Higgs} & The analogous Higgs search, with its own validity checks and reconstruction assumptions.\\
\bottomrule
\end{tabularx}

\ui{Compute geometry} runs the branches and eligible Hasse calculation. \ui{Higgs} and \ui{Coulomb} run one branch; \ui{Decay} is independent. Refined and direct-check calculations have their own buttons.

\subsection*{Time, order and memory}
The initial Expert settings are a truncated series through $t^6$, Normal work, ten seconds, one million sparse terms and $128$ MiB. Orders extend to $t^{80}$; time allowances range from three seconds to thirty minutes per engine. \ui{Large} and \ui{Patient} raise work limits; Patient raises a short default allowance to five minutes. Higgs sparse-term and memory limits remain separate. More resources do not make an unsupported representation supported or a nonpositive UV monopole grading convergent.

\topic{Coulomb Hilbert series: magnetic charges and dressing}
The monopole formula takes the form~\cite{CHZ}
\begin{equation}
 H_C(t,\boldsymbol z)=\sum_{m\in X_*(G)/W}
 \boldsymbol z^{J(m)}t^{2\Delta(m)}P_G(t;m),
 \qquad P_G(t;m)=\prod_a\frac{1}{1-t^{2d_a(m)}}.
\end{equation}
Here $X_*(G)$ is the chosen cocharacter lattice, $W$ the Weyl group, and $d_a(m)$ are the Casimir degrees of the residual gauge group centralising $m$. Refinement retains the topological charges $J(m)$; the ordinary branch tab sets their fugacities to one. The conformal dimension is:
\begin{equation}
 \Delta(m)=-\sum_{\alpha>0}|\alpha(m)|
 +\frac14\sum_{\rho\in\wt(M_{\mathbb C})}|\rho(m)|.
\end{equation}
Here $M_{\mathbb C}$ is the full holomorphic symplectic matter representation. A full hyper in $R$ contributes $\tfrac12\sum_{\rho\in R}|\rho(m)|$, while a half hyper contributes half of that. Trifundamentals use tensor-product weights of all three legs.

\subsection*{How the summation is accelerated}
The engines reuse residual dressing factors and use lower bounds on the monopole degree to discard charge sectors that cannot contribute through the requested order. Tree dynamic programming, graph minimum-cut bounds and magnetic-charge gluing avoid enumerating the full Cartesian product when applicable. In a tree, a subtree is summed once for each boundary charge and its truncated series is passed to the adjacent node. Graph elimination extends this idea to larger boundaries; its cost depends strongly on connectivity. These reorganisations preserve the lattice sum. Exact integer arithmetic is used for coefficients.

\topic{Higgs Hilbert series: integration and moment-map equations}\label{sec:higgs}
The Higgs branch is a hyperk\"ahler quotient, and its Hilbert series is given by the Molien--Weyl integral~\cite{Molien1897,Weyl1926,BHM2010}:
\begin{equation}
 H_H(t,\boldsymbol x)=\int_G d\mu_G(g)\,
 \PE\!\left[t\,\chi_{\mathrm{matter}}(g,\boldsymbol x)
       -t^2\chi_{\mathrm{adj}}(g)\right],
 \label{eq:higgs}
\end{equation}
where $\chi_{\mathrm{matter}}$ is the character of the full matter representation and $\chi_{\mathrm{adj}}$ is the adjoint character of the effective gauge group $G$. For $f(0,\boldsymbol x)=0$, the plethystic exponential is~\cite{BPS2006}
\begin{equation}
 \PE[f(t,\boldsymbol x)]
 =\exp\!\left(\sum_{k\geq1}\frac{f(t^k,\boldsymbol x^k)}{k}\right).
\end{equation}
In this definition, all fugacities, including the gauge variables in \eqref{eq:higgs}, are raised to the $k$th power. The positive term counts hypermultiplet coordinates; the negative term imposes the complex moment-map equations. Haar integration projects onto gauge invariants.

\subsection*{Finite Molien--Weyl evaluation}
The engine expands through the requested degree as sparse Laurent polynomials and extracts the gauge constant term with the Haar measure. It retains exact integer coefficients and discards terms that cannot affect the requested order. Local integration and character contraction reduce intermediate storage; simple unframed unitary trees have a character-contraction route that avoids a large explicit torus expansion.

For $\mathrm O(N)$ factors the integral includes both disconnected components~\cite{HK}. Several O nodes require all component combinations with their averaging weights; replacing O by SO would change the projection. Finite gauge factors use a modular projection. Full trifundamentals contribute the actual tensor-product character, not a triangle of bifundamental characters. Gauge-singlet ordinary hypers multiply supported finite unrefined Higgs output by $(1-t)^{-2h}$.

\subsection*{The role of the flatness check}
The subtraction in \eqref{eq:higgs} is valid as a Hilbert-series computation only when the required regularity is established. For supported ordinary unitary quivers, the software uses moment-map flatness criteria based on quiver dimension vectors and their decompositions~\cite{CB}.

\textbf{For quivers with sufficient matter to avoid incomplete Higgsing, the flatness check is generally expected to pass. For most such quivers, we strongly recommend turning off \ui{Check moment-map flatness} to speed up the computation.}

\subsection*{All-orders integration}
\ui{Exact rational} attempts to find a rational function valid to all orders in $t$. Depending on the quiver, it uses a known formula, sums a character expansion, or evaluates the Molien--Weyl integral by residues. If no supported method finishes within the available resources, the program falls back to a finite expansion when time permits.

\topic{Exact rational expressions and rational searches}
Finite coefficients, an all-orders evaluation and a fitted rational expression answer different questions. The result panel records which calculation succeeded; selecting \ui{Exact rational} does not by itself guarantee that an all-orders answer was found.

\subsection*{Exact evaluation}
For supported Coulomb inputs, magnetic charges decompose into nested threshold layers. A strict chain of these layers fixes the residual dressing structure. Arbitrarily repeated layers are summed by geometric series, schematically
\begin{equation}
 \sum_{r\geq1}t^{r d}=\frac{t^d}{1-t^d},\qquad d>0.
\end{equation}
\subsection*{Reconstruction from a finite series}
Both rational searches first attempt an exact method. Otherwise, we compute the Hilbert series to a finite order and use its coefficients to conjecture a rational function. We then check the conjecture against higher-order coefficients withheld from the fit. When resources permit, we compute further coefficients for additional checks.

The Coulomb search uses $D=2\rank(G)$ as the expected complex dimension. It tries denominators such as $(1-t^k)^D$, mixed-degree denominators and Pad\'e approximations. Candidate filters include the pole order at $t=1$, suitable cyclotomic denominator factors\footnote{The author would like to thank Rudolph Kalveks for discussions about this many years ago.}, and nonnegative integral predicted coefficients. By default, four coefficient positions are withheld; where feasible, another eight degrees are computed after fitting.

The Higgs search uses finite integral data and exact withheld-coefficient checks. With flatness checking off, it reconstructs the formal series $F(t)$. In either branch, successful finite checks do not prove an all-orders identity. The bounded search can also miss a rational function that exists.

\subsection*{Reading the outcome}
\begin{tabularx}{\linewidth}{@{}p{0.29\linewidth}Y@{}}
\toprule
\textbf{Result status} & \textbf{Meaning}\\
\midrule
Certified finite series & All coefficients through the stated completed degree are included within the engine's reported scope.\\[4pt]
Exact rational form & An all-orders identity follows from an implemented formula, integration or resummation, subject to its stated interpretation.\\[4pt]
Conjectured rational form & A candidate matches finite data and passes the listed tests; the uncomputed tail remains conjectural.\\[4pt]
Stopped / unsupported & A resource or scope limit was reached. Retain only the explicitly completed result, if one exists.\\
\bottomrule
\end{tabularx}

\topic{Plethystic logarithms, refinement and symmetry}
\subsection*{Plethystic logarithms}
In Expert mode, each branch can switch between \ui{Hilbert series} and \ui{Plethystic log}. For a series with constant coefficient one,
\begin{equation}
 \PL[H](t,\boldsymbol x)=\sum_{k\geq1}\frac{\mu(k)}{k}
       \log H(t^k,\boldsymbol x^k),
\end{equation}
with $\mu$ the M\"obius function. The software expands this expression with exact arithmetic through the available degree. For a complete intersection, a finite PL records generator degrees positively and relation degrees negatively. For a general ring, higher terms mix relations and syzygies. A truncated PL cannot prove that the full logarithm terminates or determine the equations of the variety.

\subsection*{Using the refined tab}
Open \ui{Refined Hilbert series}, choose $t^4$, $t^6$, $t^8$ or $t^{10}$, and compute either or both branches. The default is $t^4$. This tab always performs a finite expansion, independently of the main exact/search choice, while retaining the relevant resource and flatness settings. It currently requires $\mathrm U(n)$ gauge nodes and supported matter; SU, orthosymplectic, exceptional and finite gauge factors are outside this refinement interface. Additional representations and several advanced quotient/projection settings are also excluded explicitly.

\subsection*{Irreps and Dynkin Labels}
The refined Hilbert series is decomposed into characters of irreducible representations~\cite{HWG2014}. Dynkin labels identify each simple-factor representation and abelian charges are shown separately. If decomposition cannot be completed, the available weight expansion remains useful. The Lie algebra does not by itself determine the faithful global form; global-form claims require separate all-orders charge information.

\subsection*{Conjectured generators (BETA)}
For eligible certified refined results with total unitary gauge rank at most $15$, Quiver Lab names candidate generators from positive PL terms strictly before the first negative degree. It can display an adjoint as a traceless matrix $M^i{}_j$, fundamentals as vectors, and product-group representations as tensors with the corresponding indices.

These names describe representation content. The program does not derive relations such as $M^2=0$, and this heuristic need not find every generator. Positive terms after a negative degree may involve syzygies and are not promoted to a complete generator list.

\topic{Decay, fission and Hasse diagrams}
\subsection*{The published unitary rule}
The \ui{Hasse diagram} tab concerns the Coulomb branch of the drawn three-dimensional quiver. When that drawing is a magnetic quiver for another theory, it describes the latter theory's Higgs branch. The engine implements the supported unitary decay-and-fission construction~\cite{DF}: enumerate admissible connected good rank vectors, form allowed multisets of residual factors, impose their partial order, and remove transitive edges to obtain cover relations.

For framed inputs, Quiver Lab first uses an equivalent unframed quiver. For ordinary bifundamental matter, the local goodness condition is~\cite{DFLetter,DF}
\begin{equation}
 e_i(\boldsymbol k)=\sum_j a_{ij}k_j-2k_i\geq0,
\end{equation}
where $a_{ij}$ counts the edges between gauge nodes $i$ and $j$, and $k_i$ is the rank at node $i$. The overall diagonal $\mathrm U(1)$ is decoupled. The implementation also enforces the construction's additional exclusions and multiset conditions; balance alone is not the entire rule. Directed non-simply-laced bonds use the corresponding directed balance and weighted ungauging convention.

\subsection*{Reading the Hasse diagram and the Decays}
The Hasse diagram shows the symplectic leaves, ordered by inclusion of their closures. Higher-dimensional leaves appear above the lower-dimensional leaves in their closures. The dimensions shown are quaternionic. Decay and fission transitions are distinguished: solid lines indicate decay, and dotted lines indicate fission.

\ui{Decay candidates} shows the starting quiver and the next 1--10 steps of decay and fission, selected with the \ui{Layers} dropdown (two by default). Select a daughter quiver and use its \ui{Show next} button to explore further; \ui{Back to input} restores the starting view. The complete Hasse diagram remains available in \ui{Hasse diagram}. Its interpretation as a stratification into symplectic leaves follows the decay-and-fission conjecture.

\subsection*{Trifundamentals (BETA)}
For connected all-U gauge support with full trifundamentals and ordinary loopless full bifundamentals, \ui{Decay} runs a separate conjectural explorer. It scans residual ranks $0\leq k_i\leq N_i$, holds flavour ranks fixed, and requires at each surviving gauge node
\begin{equation}
 e_i(\boldsymbol k)=\sum_j a_{ij}k_j
   +\sum_{\{i,j,\ell\}}h_{ij\ell}k_jk_\ell-2k_i\geq0.
\end{equation}
Here the first sum includes flavour neighbours with their fixed sizes, and $h_{ij\ell}$ counts full trifundamentals. Rank-zero gauge nodes and their incident matter are removed. Every candidate has its entire ineffective continuous centre recomputed and removed, independently of the main Coulomb quotient toggle, and must pass the exact strictly-positive monopole-grading certificate. Local balance is insufficient.

The surviving ranks are ordered componentwise and reduced to covers; the interface uses the same 1--10 layer selector, with two layers shown by default. This is a conjectural rank-decay poset, not an established symplectic-leaf diagram. Labels give residual effective gauge rank, not leaf dimension. No trifundamental fission rule, transverse slices or complete geometric stratification are asserted.

\topic{A short example and reproducible checks}
Choose \ui{SQED $\cdot$ 2 flavours}: one $\mathrm U(1)$ gauge circle joined to a flavour square of size two by one full-hyper edge. No diagonal $\mathrm U(1)$ removal is needed. Both branches are the $A_1$ surface singularity, and the built-in engines give
\begin{equation}
 H_H(t)=H_C(t)=\frac{1+t^2}{(1-t^2)^2}
 =1+3t^2+5t^4+7t^6+\cdots,\qquad \PL=3t^2-t^4.
\end{equation}
Light mode displays the successive coefficients. In Expert mode, \ui{Exact rational} returns the fraction; refinement resolves the adjoint $[2]$ of $\mathfrak{su}(2)$ at degree two. In standard SU(2) character notation,
\begin{equation}
 H(t,x)=\sum_{n\geq0}[2n]_x\,t^{2n},\qquad
 \PL[H]=[2]_x\,t^2-[0]_x\,t^4.
\end{equation}
This reproduces the result of the monopole-formula paper~\cite{CHZ}.

The ordinary Hasse output has the open leaf and the origin, with an $A_1$ transition. For this note, the two branch expansions through $t^6$ were checked in the actual worker using finite, exact, refined and direct-evaluation paths; their coefficients agree.

\textbf{Direct verification.} The \ui{Don't believe me?} tab computes the Coulomb-branch Hilbert series perturbatively from the monopole formula, without the acceleration methods used in the main calculation. It performs a similar direct evaluation of the Higgs-branch Hilbert series. Skeptical users can compare these traditional calculations with the accelerated results.

\textbf{Keeping a result.} Export both the quiver JSON and result JSON. Record the software version, grading, global form/quotient, method, flatness setting and actual completed order.

\vspace{2\baselineskip}
\textbf{Acknowledgement.} The author is deeply indebted to Rudolph Kalveks for many years of discussion and optimisation in the art of Hilbert series computation. The rational Higgs-branch Hilbert series search in Quiver Lab, in particular, draws heavy inspiration from the \texttt{Nested\_Weyl} integrals developed by Rudolph. The author is also indebted to Antoine Bourget, whose Decay and Fission Mathematica file shows, for the first time, that Hasse diagrams of complicated quivers can be drawn in less than a second. GPT-Astra has been used in the development of Quiver Lab and in the preparation of this user manual.

\clearpage
\vspace*{\fill}
\noindent\begin{minipage}{\linewidth}
\centering
\textcolor{red}{Zhenghao: As generations before us have done,\\
we now pass the \textbf{Quiver Sword} into the hands of a new generation.}\par
\vspace{10pt}
\includegraphics[width=\linewidth,height=0.76\textheight,keepaspectratio]{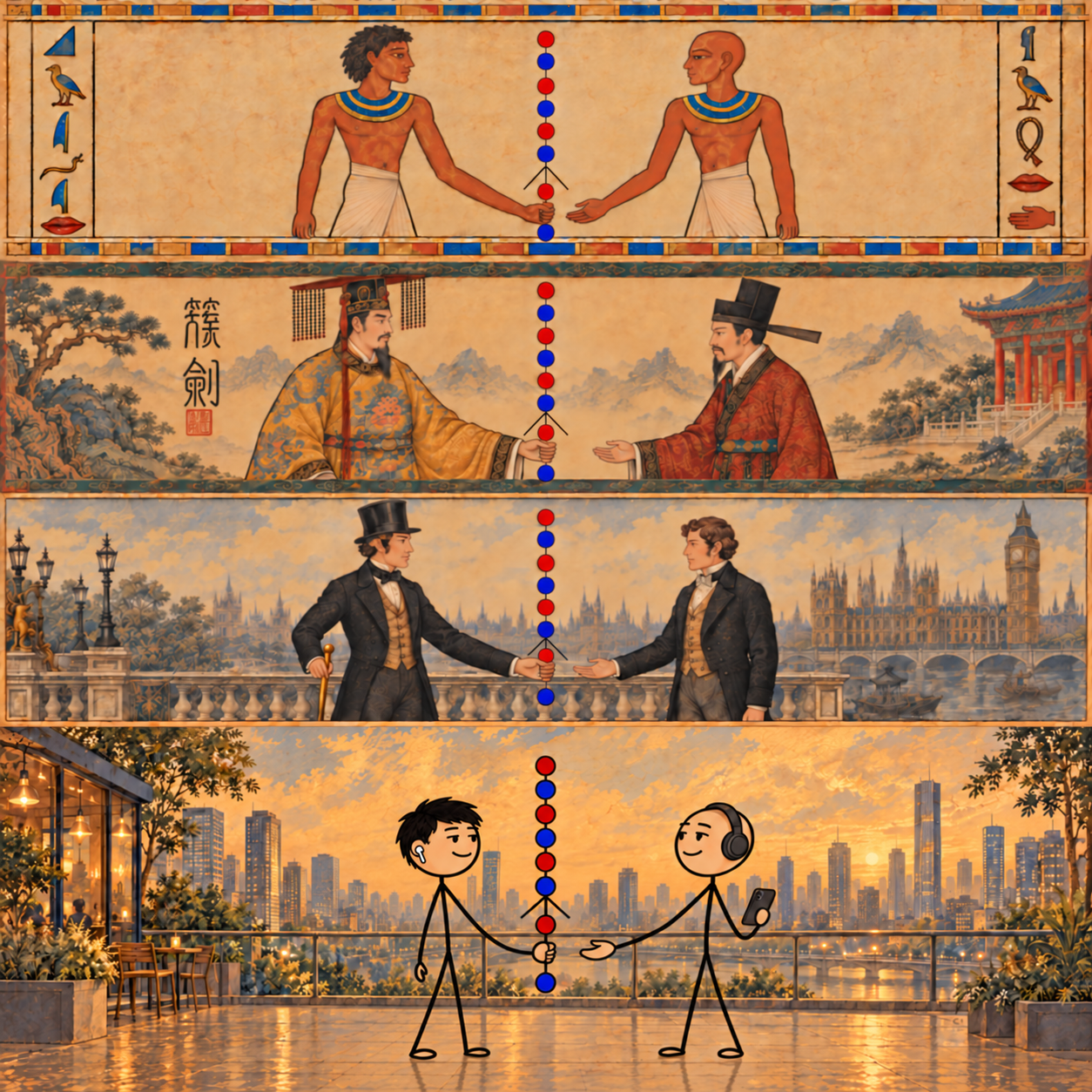}\par
\vspace{6pt}
{\small Generated by Astra. With the right ranks, the quiver gives $E_7$.\par}
\end{minipage}
\par\vspace*{\fill}

\clearpage
\section*{Appendix A: Quiver Database}
\label{app:database}
\addcontentsline{toc}{section}{Appendix A: Quiver Database}
This appendix collects 1,639 Coulomb-branch quiver examples from 74 papers published since 2013, listed in Refs.~\cite{CHZ,db14020016,db14032384,db14036523,db14086835,db150806813,db150901294,db150905402,HK,db160402155,db160705728,db160908034,db170103825,db170603781,db170709784,db170905818,db171010247,KP,db171206604,db180101129,db180408857,db180601890,db180702521,db180703221,db180805226,db190110493,db190305822,db190506297,db190900667,db200205716,db200404082,db200409620,db200409738,db200616994,db200704667,db200705019,db200715600,db200801027,db201005889,db201009732,db201107033,db201212827,db201212852,db210112235,db210206190,db210305458,db210508064,db210700754,db210709077,db210907237,db210907493,db211006940,db211011365,db211100026,db211102831,db211104745,db211210825,db220301951,db220316550,db220605362,db220705764,db220807270,db220915016,db221002966,db221004921,db230102409,db230316939,DF,db240609626,db241112802,db241114531,db241215202,db250503875,db251014489}. The tables pair each quiver with its Hilbert series, using the grading $t^{2\Delta}$, and give notes on the literature comparison where needed. The same collection is available through the \ui{Quiver Database} tab on the \href{https://quiverlab.org/\#database}{Quiver Lab webpage}.

The source papers contain explicit Hilbert series as well as results that are stated or conjectured without a full series calculation. We also compute these claimed cases with Quiver Lab, providing an explicit computational check. Whenever computationally reasonable, our calculations target an expansion through $t^{30}$, even when the literature gives a lower-order expansion. The achieved order is recorded for each calculation; agreement with the literature is assessed over the orders available in both series.

The database records 1,576 matched entries, 47 apparent typographical errors in the literature, and 16 other mismatches (we have verified that the errors lie in the literature). The accompanying notes discuss the typographical errors and mismatches.

\clearpage
\pagenumbering{arabic}
\renewcommand{\thepage}{A-\arabic{page}}


\section*{Supplementary material}

## 2101.12235 · Factorised 3d N=4 orthosymplectic quivers (89)

**0857 2101.12235-MQ11-N2__N2 / N = 2**

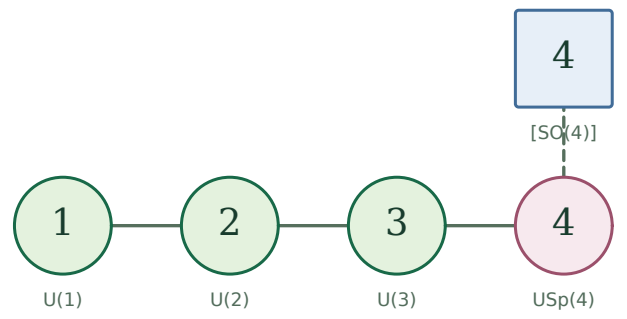


$1 + 30t^2 + 433t^4 + 4070t^6 + 28384t^8 + 158174t^{10} + O(t^{12})$

**0858 2101.12235-MQ11-N3__N3 / N = 3**

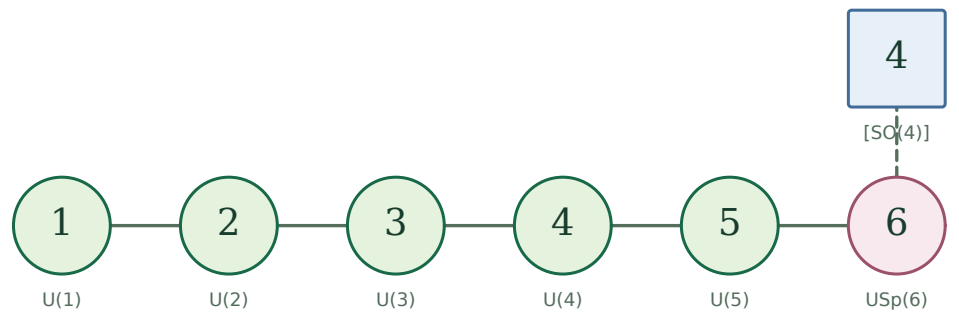


$1 + 70t^2 + 2413t^4 + 54670t^6 + 917244t^8 + 12178110t^{10} + O(t^{12})$

**0859 2101.12235-MQ13-I-N2__N2 / N = 2**

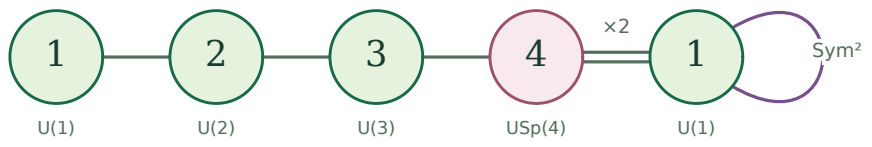


÷ $\mathbb{Z}_2$ diagonal

$1 + 31t^2 + 12t^3 + 464t^4 + 320t^5 + 4574t^6 + 4188t^7 + 33908t^8 + 36488t^9 + 203160t^{10} + O(t^{11})$

**0860 2101.12235-MQ13-I-N3__N3 / N = 3**

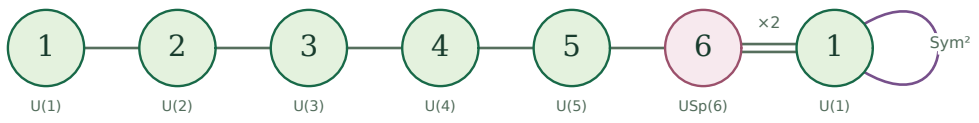


÷ $\mathbb{Z}_2$ diagonal

$1 + 71t^2 + 2524t^4 + 59674t^6 + 1053332t^8 + 14792806t^{10} + O(t^{12})$

**0861 2101.12235-MQ13-II-N2__N2 / N = 2**

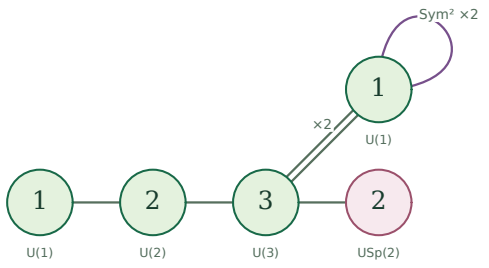


÷ $\mathbb{Z}_2$ diagonal

$1 + 31t^2 + 8t^3 + 444t^4 + 200t^5 + 4059t^6 + 2432t^7 + 27344t^8 + 19560t^9 + 147137t^{10} + O(t^{11})$

**0862 2101.12235-MQ13-II-N3__N3 / N = 3**

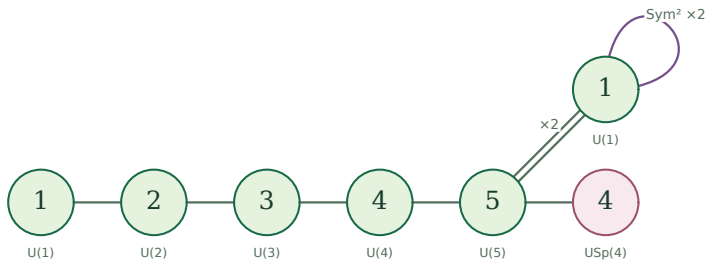


÷ $\mathbb{Z}_2$ diagonal

$1 + 71t^2 + 2514t^4 + 58827t^6 + 1020589t^8 + 13995278t^{10} + O(t^{12})$

| QUIVER | HILBERT SERIES |
|---|---|

**0863 2101.12235-MQ33-I-N1__N1** / N = 1

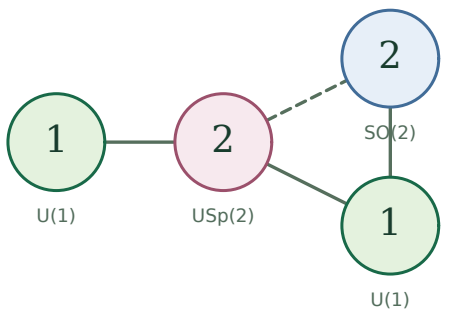


÷ $\mathbb{Z}_2$ diagonal

$1 + 16t^2 + 118t^4 + 560t^6 + 2003t^8 + 5888t^{10} + O(t^{12})$

**0864 2101.12235-MQ33-I-N2__N2** / N = 2

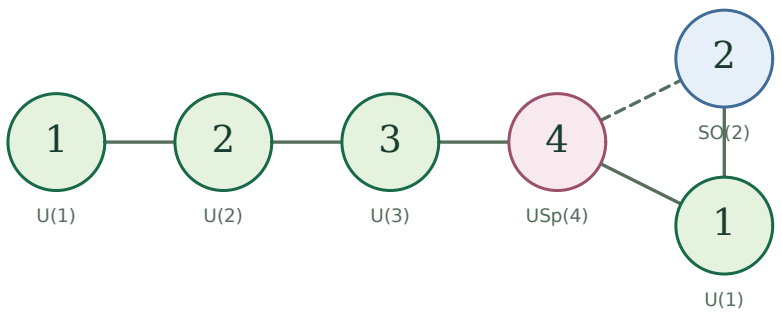


÷ $\mathbb{Z}_2$ diagonal

$1 + 32t^2 + 24t^3 + 496t^4 + 664t^5 + 5254t^6 + 9040t^7 + 43368t^8 + 82976t^9 + 294996t^{10} + O(t^{11})$

**0865 2101.12235-MQ33-II-N1__N1** / N = 1

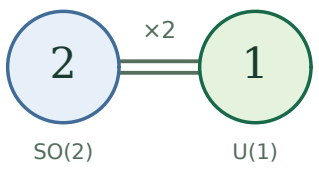


÷ $\mathbb{Z}_2$ diagonal

$1 + 6t^2 + 19t^4 + 44t^6 + 85t^8 + 146t^{10} + O(t^{12})$

**0866 2101.12235-MQ33-II-N2__N2** / N = 2

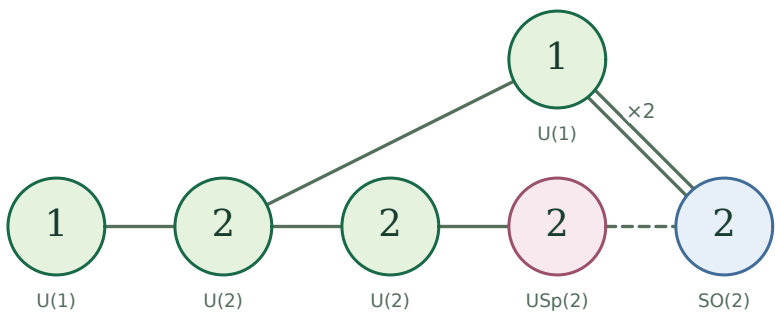


÷ $\mathbb{Z}_2$ diagonal

$1 + 32t^2 + 16t^3 + 456t^4 + 416t^5 + 4104t^6 + 4960t^7 + 27490t^8 + 38400t^9 + 148792t^{10} + O(t^{11})$

**0867 2101.12235-MQ33-III-N1__N1** / N = 1

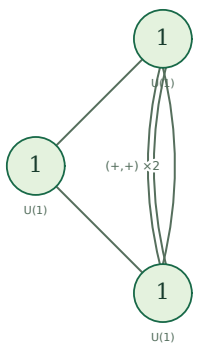

÷ $\mathbb{Z}_2$ diagonal

$1 + 11t^2 + 56t^4 + 192t^6 + 517t^8 + 1183t^{10} + O(t^{12})$

**0868 2101.12235-MQ33-III-N2__N2** / N = 2

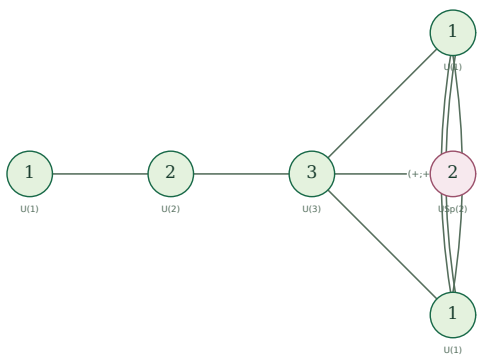

÷ $\mathbb{Z}_2$ diagonal

$1 + 32t^2 + 20t^3 + 476t^4 + 540t^5 + 4671t^6 + 6920t^7 + 34989t^8 + 58628t^9 + 214034t^{10} + O(t^{11})$

| QUIVER | HILBERT SERIES |
| --- | --- |

**0869 2101.12235-MQ3p3p-II-N2__N2** / N = 2

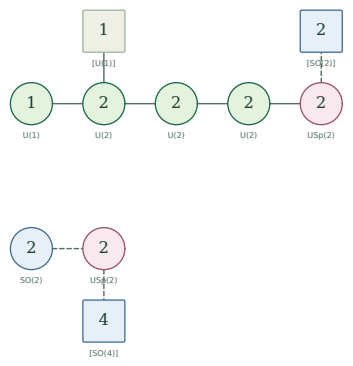

$1 + 54t^2 + 1283t^4 + 18412t^6 + 185691t^8 + 1438022t^{10} + O(t^{12})$

**0870 2101.12235-MQ3p3p-III-N1__N1** / N = 1

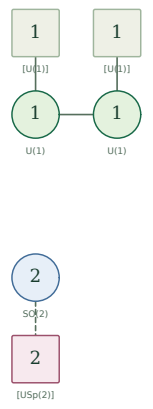

$1 + 11t^2 + 56t^4 + 192t^6 + 517t^8 + 1183t^{10} + O(t^{12})$

**0871 2101.12235-MQ3p3p-III-N2__N2** / N = 2

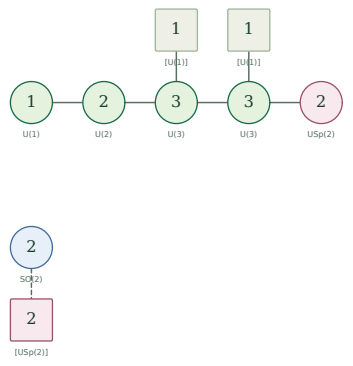

$1 + 51t^2 + 1200t^4 + 17824t^6 + 191099t^8 + 1596553t^{10} + O(t^{12})$

**0872 2101.12235-MQ3p4-I-N1__N1** / N = 1

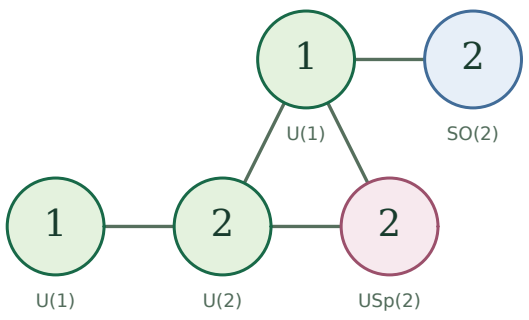


÷ $\mathbb{Z}_2$ diagonal

$1 + 32t^2 + 419t^4 + 3312t^6 + 18736t^8 + 83392t^{10} + O(t^{12})$

**0873 2101.12235-MQ3p4-I-N2__N2** / N = 2

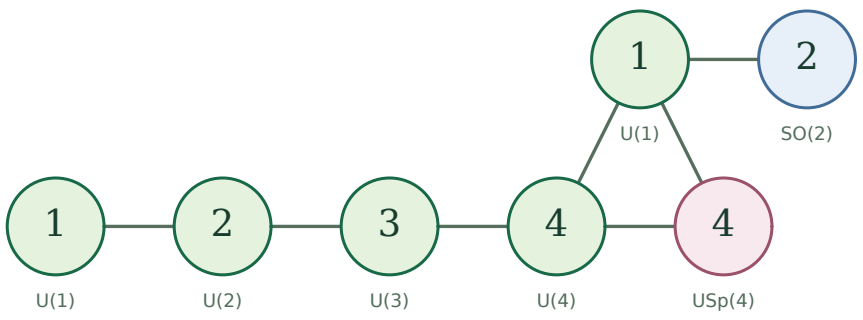


÷ $\mathbb{Z}_2$ diagonal

$1 + 52t^2 + 40t^3 + 1327t^4 + 1780t^5 + 22500t^6 + 39140t^7 + 286968t^8 + 570120t^9 + 2939292t^{10} + O(t^{11})$

**0874 2101.12235-MQ3p4-II-N1__N1** / N = 1

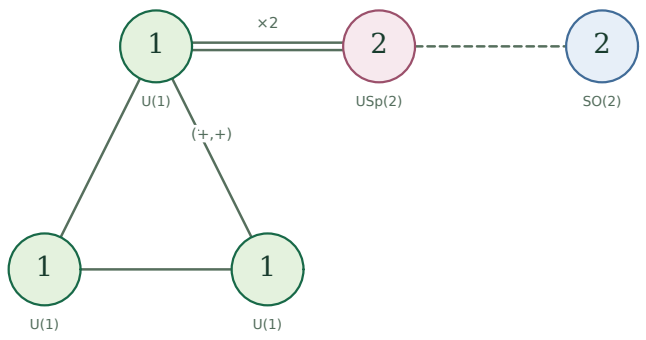


÷ $\mathbb{Z}_2$ diagonal

$1 + 27t^2 + 277t^4 + 1727t^6 + 7852t^8 + 28628t^{10} + O(t^{12})$

| QUIVER | HILBERT SERIES |
|---|---|

**0875 2101.12235-MQ3p4-II-N2__N2** / N = 2

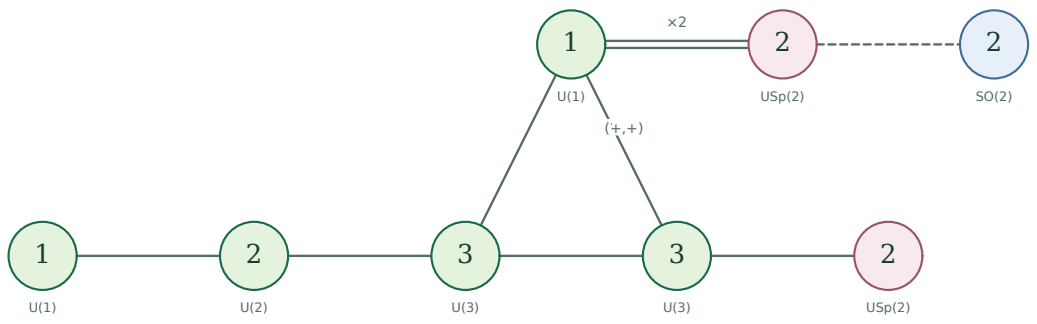


÷ $\mathbb{Z}_2$ diagonal

$1 + 55t^2 + 40t^3 + 1413t^4 + 1900t^5 + 23399t^6 + 41680t^7 + 287256t^8 + 585260t^9 + 2808576t^{10} + O(t^{11})$

**0876 2101.12235-MQ44-N1__N1** / N = 1

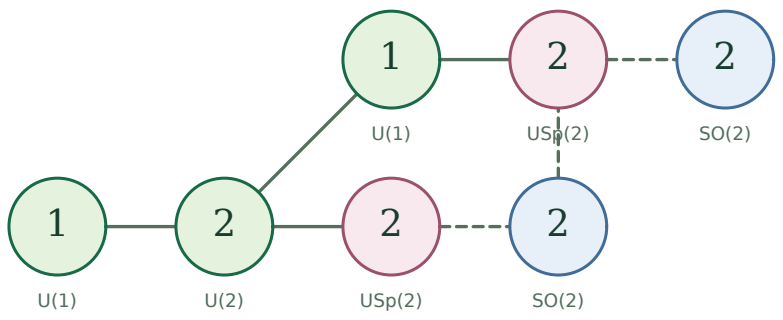


÷ $\mathbb{Z}_2$ diagonal

$1 + 48t^2 + 976t^4 + 11600t^6 + 95350t^8 + 598352t^{10} + O(t^{12})$

**0877 2101.12235-MQ44-N2__N2** / N = 2

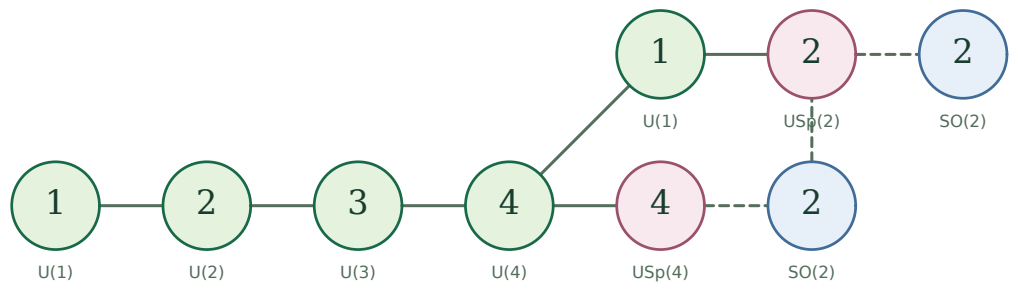


÷ $\mathbb{Z}_2$ diagonal

$1 + 56t^2 + 80t^3 + 1544t^4 + 3880t^5 + 30192t^6 + 93240t^7 + 468888t^8 + 1510680t^9 + 5964152t^{10} + O(t^{11})$

**0878 2101.12235-MQ55-N1__N1** / N = 1

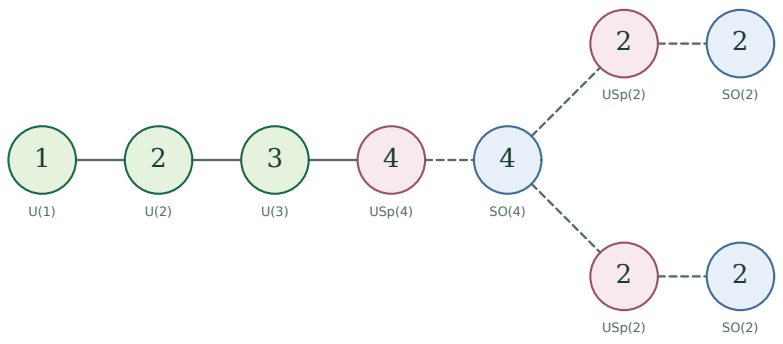


÷ $\mathbb{Z}_2$ diagonal

$1 + 90t^2 + 3565t^4 + 84588t^6 + 1386700t^8 + 17100048t^{10} + O(t^{12})$

**0879 2101.12235-MQ55-N2__N2** / N = 2

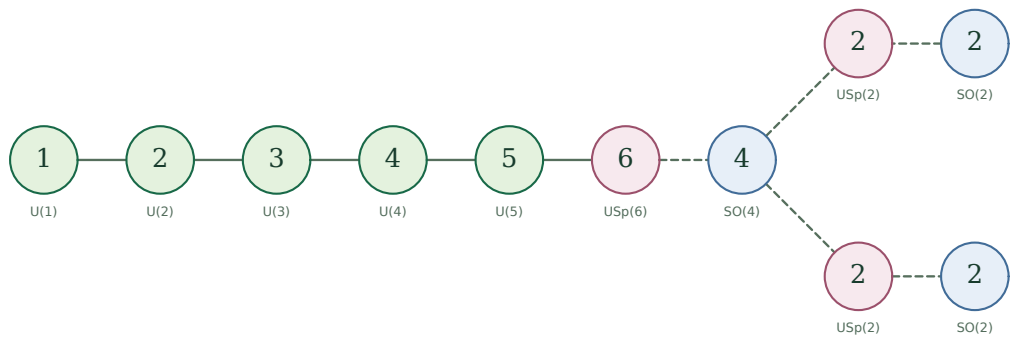


÷ $\mathbb{Z}_2$ diagonal

$1 + 82t^2 + 160t^3 + 3329t^4 + O(t^5)$

**0880 2101.12235-MQ5p5p-N1__N1** / N = 1

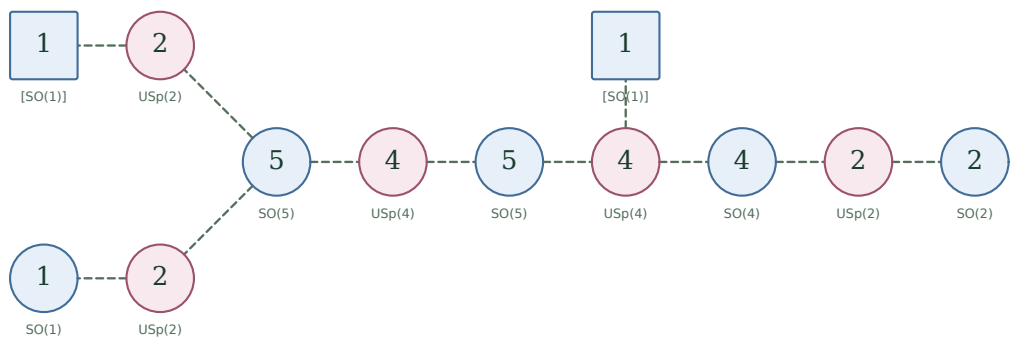


$1 + 90t^2 + 3565t^4 + 84588t^6 + 1386700t^8 + 17100048t^{10} + O(t^{12})$

| QUIVER | HILBERT SERIES |
|---|---|

**0881 2101.12235-MQ5p5p-N2__N2 / N = 2**

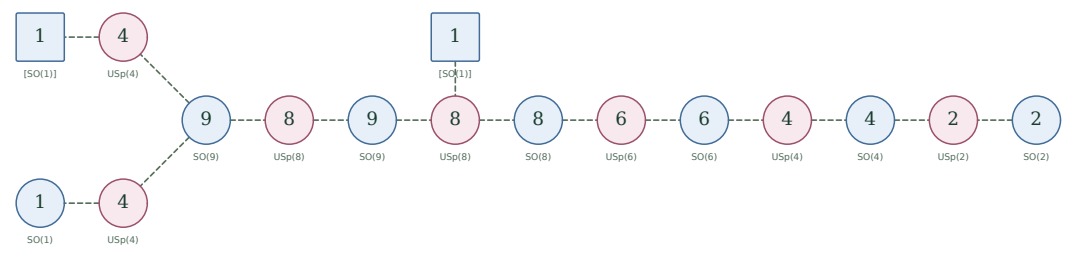

$1 + 182t^{2} + 16443t^{4} + 977366t^{6} + O(t^{8})$

**0882 2101.12235-MQ5p6-N1__N1 / N = 1**

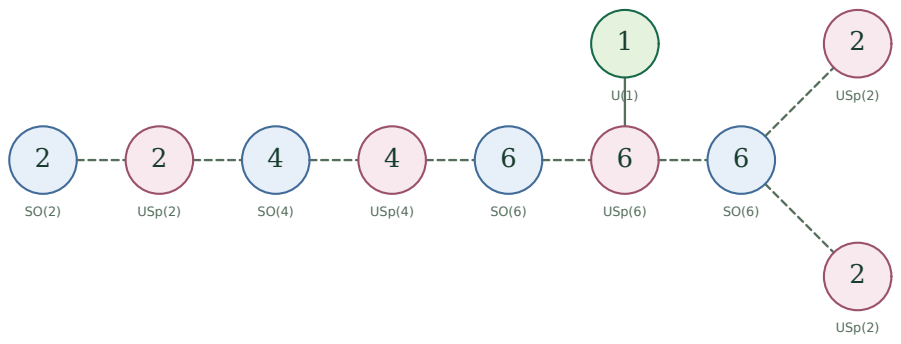


÷ ℤ₂ diagonal

$1 + 123t^{2} + 6710t^{4} + 220812t^{6} + 5027328t^{8} + 85856661t^{10} + O(t^{12})$

**0883 2101.12235-MQ5p6-N2__N2 / N = 2**

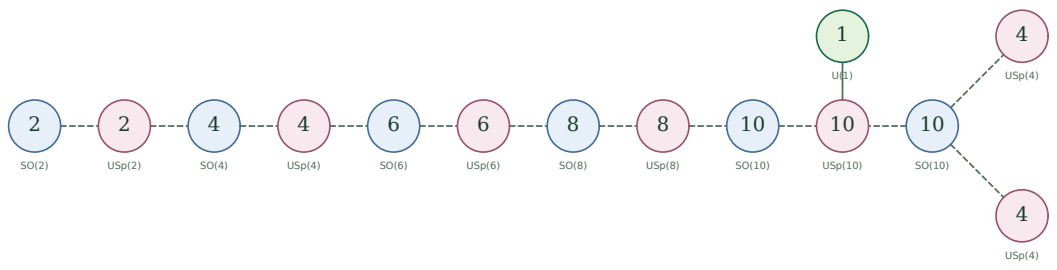

÷ ℤ₂ diagonal

$1 + 183t^{2} + 128t^{3} + 16626t^{4} + 21632t^{5} + 1000427t^{6} + O(t^{7})$

**0884 2101.12235-MQ77-N1__N1 / N = 1**

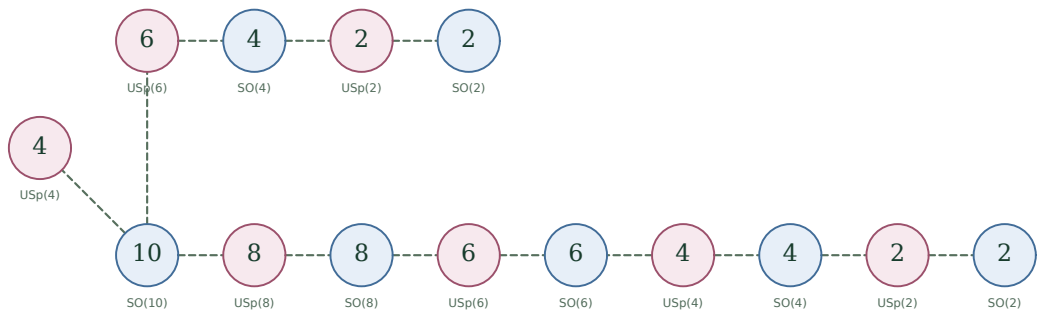


÷ ℤ₂ diagonal

$1 + 266t^{2} + 32431t^{4} + O(t^{6})$

**0885 2101.12235-unitary-5__N3 / N = 3**

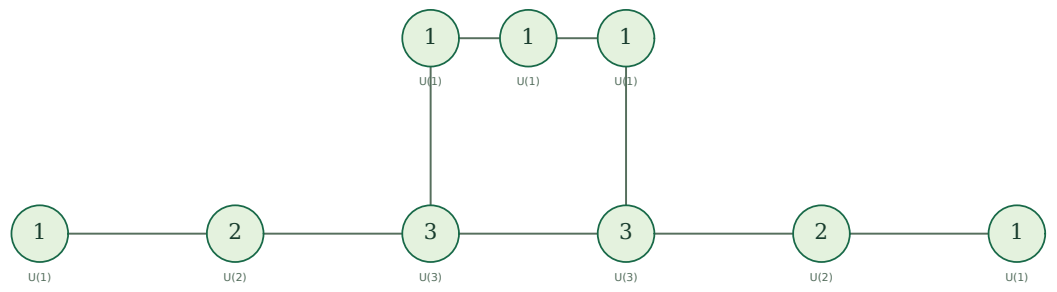


÷ U(1) diagonal

$1 + 52t^{2} + 1468t^{4} + 27964t^{6} + 394715t^{8} + 4374340t^{10} + 39669647t^{12} + 303747588t^{14} + 2012229795t^{16} + 11759705000t^{18} + 61589772213t^{20} + 292841477184t^{22} + 1277736950841t^{24} + 5162551809228t^{26} + 19464239364301t^{28} + 68931441556344t^{30} + O(t^{31})$

**0886 2101.12235-unitary-5__N4 / N = 4**

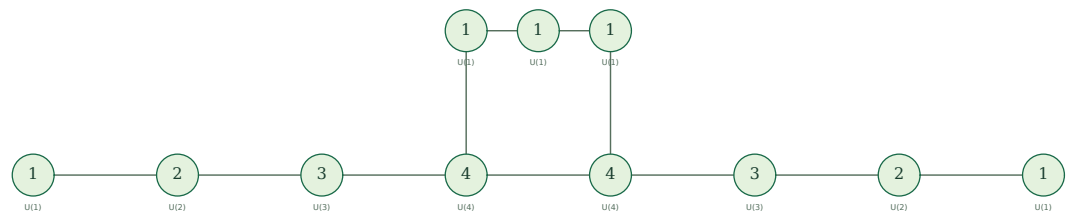

÷ U(1) diagonal

$1 + 84t^{2} + 3488t^{4} + 504t^{5} + 95532t^{6} + 34776t^{7} + 1943115t^{8} + 1192896t^{9} + 31366904t^{10} + 27160020t^{11} + 419622411t^{12} + 462380364t^{13} + 4796523216t^{14} + 6287377212t^{15} + 47914702998t^{16} + 71239038120t^{17} + 425516792768t^{18} + 692798689584t^{19} + 3404352021225t^{20} + 5911093971636t^{21} + 24796760557388t^{22} + 45003751117956t^{23} + 165850512015862t^{24} + 309870894520968t^{25} + 1025894306429580t^{26} + 1950721346008008t^{27} + 5904780206562910t^{28} + 11329223438385648t^{29} + 31793056618508040t^{30} + O(t^{31})$

| QUIVER | HILBERT SERIES |
| --- | --- |

**0887 2101.12235-unitary-5__N5** / N = 5

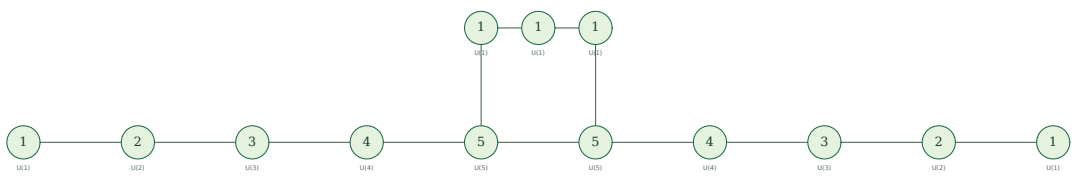

÷ U(1) diagonal

$1 + 124t^{2} + 7628t^{4} + 312340t^{6} + 9605569t^{8} + 236835648t^{10} + 4873920527t^{12} + 86009591492t^{14} + 1326956406498t^{16} + 18162112967300t^{18} + 223104841195359t^{20} + 2483216894019592t^{22} + 25244586651182441t^{24} + 236032827090670472t^{26} + 2041994485227512752t^{28} + 16433903210489251228t^{30} + O(t^{31})$

**0888 2101.12235-unitary-6__N2** / N = 2

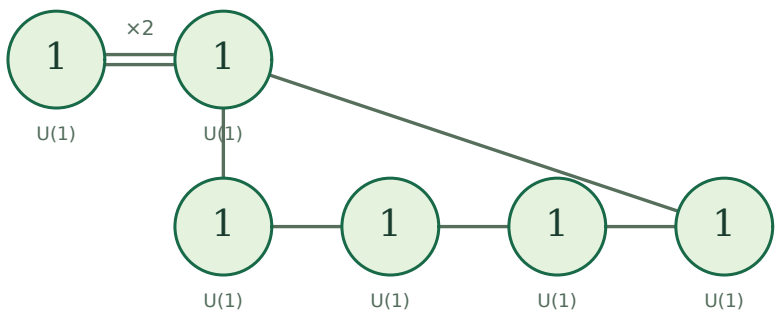


÷ U(1) diagonal

$1 + 27t^{2} + 277t^{4} + 1727t^{6} + 7852t^{8} + 28628t^{10} + 88604t^{12} + 241604t^{14} + 595529t^{16} + 1351779t^{18} + 2865005t^{20} + 5730231t^{22} + 10905856t^{24} + 19882656t^{26} + 34910656t^{28} + 59297632t^{30} + O(t^{31})$

**0889 2101.12235-unitary-6__N3** / N = 3

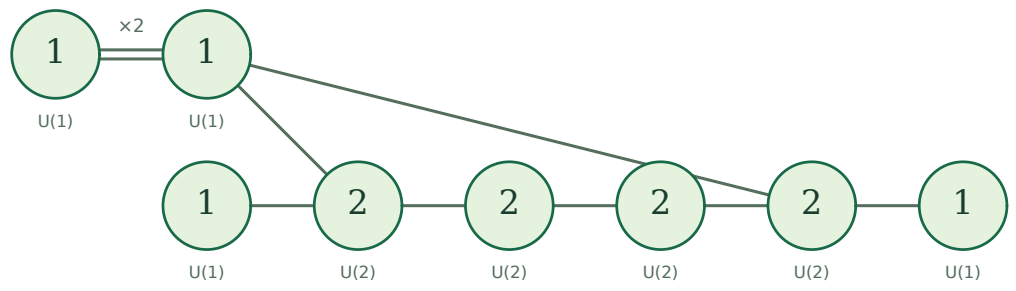


÷ U(1) diagonal

$1 + 51t^{2} + 1276t^{4} + 20140t^{6} + 226723t^{8} + 1966321t^{10} + 13848576t^{12} + 82252480t^{14} + 423729954t^{16} + 1934526374t^{18} + 7960512072t^{20} + 29926716680t^{22} + 103921771814t^{24} + 336377743394t^{26} + 1022626422464t^{28} + 2938730799360t^{30} + O(t^{31})$

**0890 2101.12235-unitary-6__N4** / N = 4

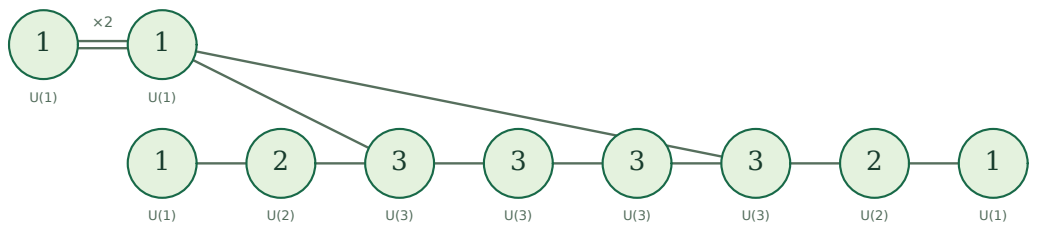

÷ U(1) diagonal

$1 + 83t^{2} + 3404t^{4} + 92044t^{6} + 1838763t^{8} + 28867913t^{10} + 370808440t^{12} + 4012633240t^{14} + 37417404869t^{16} + 306163841151t^{18} + 2231144846268t^{20} + 14661516846588t^{22} + 87791328524055t^{24} + 483304778052237t^{26} + 2465022737141952t^{28} + 11725762499536512t^{30} + O(t^{31})$

**0891 2101.12235-unitary-7__N3** / N = 3

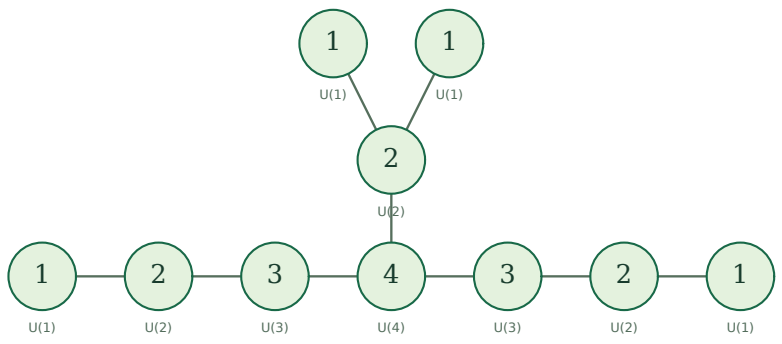

÷ U(1) diagonal

$1 + 69t^{2} + 2630t^{4} + 68469t^{6} + 1334326t^{8} + 20554428t^{10} + 259940921t^{12} + 2776383765t^{14} + 25605813759t^{16} + 207577975304t^{18} + 1500882475500t^{20} + 9797522082553t^{22} + 58337354026603t^{24} + 319623998025036t^{26} + 1623552008753296t^{28} + 7696020129255172t^{30} + O(t^{31})$

**0892 2101.12235-unitary-7__N4** / N = 4

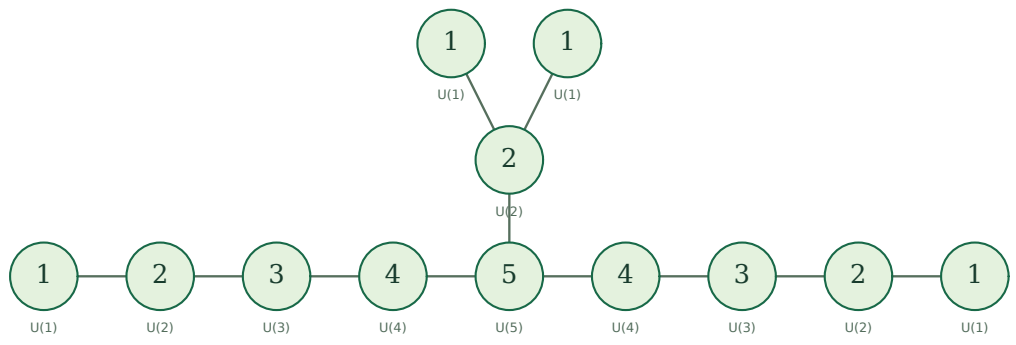

÷ U(1) diagonal

$1 + 105t^{2} + 5464t^{4} + 1008t^{5} + 187979t^{6} + 89040t^{7} + 4812375t^{8} + 3914112t^{9} + 98016707t^{10} + 114263040t^{11} + 1660106065t^{12} + 2494111232t^{13} + 24129832812t^{14} + 43467385760t^{15} + 308025068658t^{16} + 630859155632t^{17} + 3512485135171t^{18} + 7852968124112t^{19} + 36230687905618t^{20} + 85697321171888t^{21} + 341211871036039t^{22} + 833773842393264t^{23} + 2955122530979783t^{24} + 7329300323825712t^{25} + 23672958424599318t^{26} + 58840726993355600t^{27} + 176281723081721243t^{28} + 435234654352963120t^{29} + 1225627512295821054t^{30} + O(t^{31})$

| QUIVER | HILBERT SERIES |
|---|---|
| **0893 2101.12235-unitary-7__N5** / **N = 5** 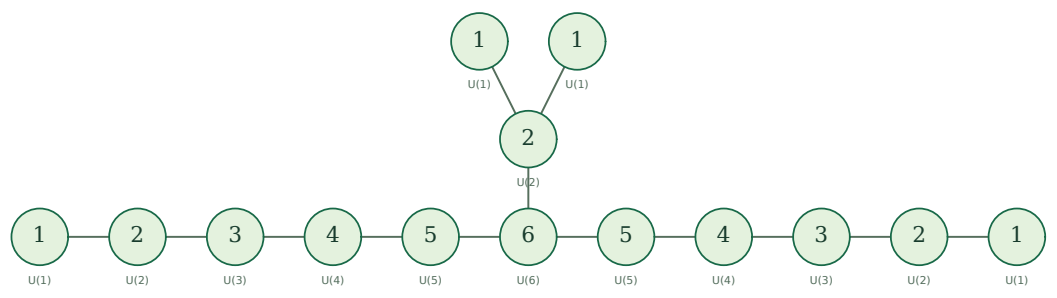 ÷ U(1) diagonal | $1 + 149t^{2} + 11030t^{4} + 544709t^{6} + 20259950t^{8} + 605953840t^{10} + 15171263329t^{12} + 326597764105t^{14} + 6161349770919t^{16} + 103325642106920t^{18} + 1557678804604059t^{20} + 21303594220799778t^{22} + 266355783168655301t^{24} + 3064514771999485985t^{26} + 32631982658729168551t^{28} + 323228321155447196184t^{30} + O(t^{31})$ |
| **0894 2101.12235-unitary-product-11__N1** / **N = 1** 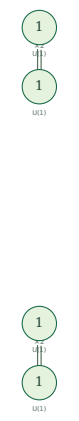 ÷ U(1) diagonal on {n0, n1} · ÷ U(1) diagonal on {n2, n3} | $1 + 6t^{2} + 19t^{4} + 44t^{6} + 85t^{8} + 146t^{10} + 231t^{12} + 344t^{14} + 489t^{16} + 670t^{18} + 891t^{20} + 1156t^{22} + 1469t^{24} + 1834t^{26} + 2255t^{28} + 2736t^{30} + O(t^{31})$ |
| **0895 2101.12235-unitary-product-11__N2** / **N = 2** 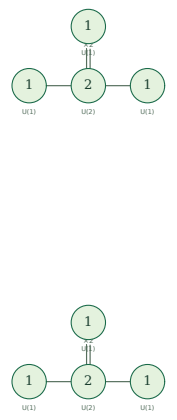 ÷ U(1) diagonal on {n0, n1, n2, n3} · ÷ U(1) diagonal on {n4, n5, n6, n7} | $1 + 30t^{2} + 433t^{4} + 4070t^{6} + 28384t^{8} + 158174t^{10} + 738949t^{12} + 2992502t^{14} + 10765339t^{16} + 35047556t^{18} + 104762398t^{20} + 290849396t^{22} + 756993136t^{24} + 1861233236t^{26} + 4350709870t^{28} + 9720660452t^{30} + O(t^{31})$ |
| **0896 2101.12235-unitary-product-11__N3** / **N = 3** 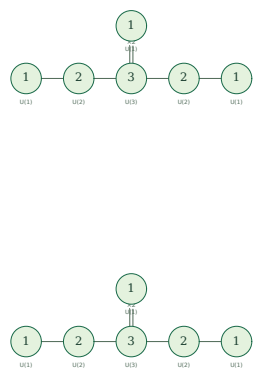 ÷ U(1) diagonal on {n0, n1, n2, n3, n4, n5} · ÷ U(1) diagonal on {n6, n7, n8, n9, n10, n11} | $1 + 70t^{2} + 2413t^{4} + 54670t^{6} + 917244t^{8} + 12178110t^{10} + 133538239t^{12} + 1246364230t^{14} + 10126813819t^{16} + 72895340310t^{18} + 471446143616t^{20} + 2771287831670t^{22} + 14949798696365t^{24} + 74617520274750t^{26} + 347007549260757t^{28} + 1512730593193880t^{30} + O(t^{31})$ |
| **0897 2101.12235-unitary-product-13-I__N1** / **N = 1** 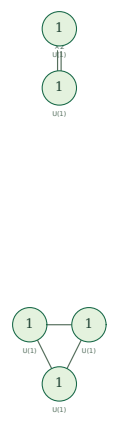 ÷ U(1) diagonal on {n0, n1} · ÷ U(1) diagonal on {n2, n3, n4} | $1 + 11t^{2} + 56t^{4} + 192t^{6} + 517t^{8} + 1183t^{10} + 2408t^{12} + 4488t^{14} + 7809t^{16} + 12859t^{18} + 20240t^{20} + 30680t^{22} + 45045t^{24} + 64351t^{26} + 89776t^{28} + 122672t^{30} + O(t^{31})$ |
| **0898 2101.12235-unitary-product-13-I__N2** / **N = 2** 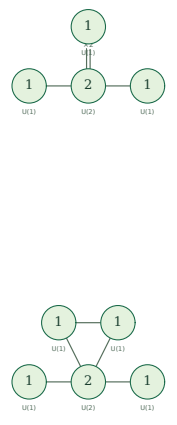 ÷ U(1) diagonal on {n0, n1, n2, n3} · ÷ U(1) diagonal on {n4, n5, n6, n7, n8} | $1 + 31t^{2} + 12t^{3} + 464t^{4} + 320t^{5} + 4574t^{6} + 4188t^{7} + 33908t^{8} + 36488t^{9} + 203160t^{10} + 240816t^{11} + 1028521t^{12} + 1293080t^{13} + 4534807t^{14} + 5908068t^{15} + 17801694t^{16} + 23682016t^{17} + 63278060t^{18} + 85138920t^{19} + 206406526t^{20} + 279117040t^{21} + 624524484t^{22} + 845264900t^{23} + 1768384710t^{24} + 2388921496t^{25} + 4720742482t^{26} + 6353700812t^{27} + 11955055046t^{28} + 16012016440t^{29} + 28873493708t^{30} + O(t^{31})$ |

| QUIVER | HILBERT SERIES |
|---|---|
| **0899 2101.12235-unitary-product-13-I__N3** / **N = 3** 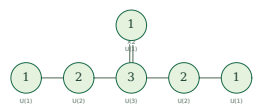 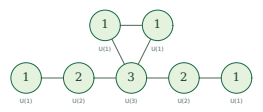 ÷ U(1) diagonal on {n0, n1, n2, n3, n4, n5} · ÷ U(1) diagonal on {n6, n7, n8, n9, n10, n11, n12} | $1 + 71t^{2} + 2524t^{4} + 59674t^{6} + 1053332t^{8} + 14792806t^{10} + 172161943t^{12} + 1708746337t^{14} + 14777430350t^{16} + 113239217496t^{18} + 779397113316t^{20} + 4872439062898t^{22} + 27928372167315t^{24} + 147957845069701t^{26} + 729496768886438t^{28} + 3367521382761122t^{30} + O(t^{31})$ |
| **0900 2101.12235-unitary-product-13-II__N1** / **N = 1** 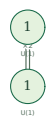 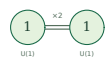 ÷ U(1) diagonal on {n0, n1} · ÷ U(1) diagonal on {n2, n3} | $1 + 6t^{2} + 19t^{4} + 44t^{6} + 85t^{8} + 146t^{10} + 231t^{12} + 344t^{14} + 489t^{16} + 670t^{18} + 891t^{20} + 1156t^{22} + 1469t^{24} + 1834t^{26} + 2255t^{28} + 2736t^{30} + O(t^{31})$ |
| **0901 2101.12235-unitary-product-13-II__N2** / **N = 2** 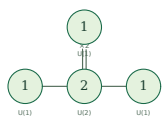 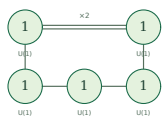 ÷ U(1) diagonal on {n0, n1, n2, n3} · ÷ U(1) diagonal on {n4, n5, n6, n7, n8} | $1 + 31t^{2} + 8t^{3} + 444t^{4} + 200t^{5} + 4059t^{6} + 2432t^{7} + 27344t^{8} + 19560t^{9} + 147137t^{10} + 118672t^{11} + 665274t^{12} + 584424t^{13} + 2615931t^{14} + 2446880t^{15} + 9168557t^{16} + 8989528t^{17} + 29174912t^{18} + 29648808t^{19} + 85486048t^{20} + 89300504t^{21} + 233239828t^{22} + 248895496t^{23} + 597902886t^{24} + 648688848t^{25} + 1450673094t^{26} + 1594307360t^{27} + 3351719494t^{28} + 3720677544t^{29} + 7412247664t^{30} + O(t^{31})$ |
| **0902 2101.12235-unitary-product-13-II__N3** / **N = 3** 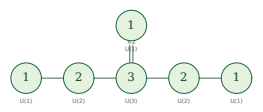 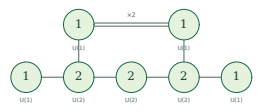 ÷ U(1) diagonal on {n0, n1, n2, n3, n4, n5} · ÷ U(1) diagonal on {n6, n7, n8, n9, n10, n11, n12} | $1 + 71t^{2} + 2514t^{4} + 58827t^{6} + 1020589t^{8} + 13995278t^{10} + 158139785t^{12} + 1516773826t^{14} + 12629107393t^{16} + 92911572756t^{18} + 612659101045t^{20} + 3663900927577t^{22} + 20069727446826t^{24} + 101545784143165t^{26} + 478010564347501t^{28} + 2106593894680572t^{30} + O(t^{31})$ |
| **0903 2101.12235-unitary-product-33-I__N1** / **N = 1** 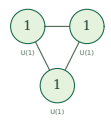 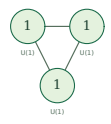 ÷ U(1) diagonal on {n0, n1, n2} · ÷ U(1) diagonal on {n3, n4, n5} | $1 + 16t^{2} + 118t^{4} + 560t^{6} + 2003t^{8} + 5888t^{10} + 14988t^{12} + 34176t^{14} + 71445t^{16} + 139216t^{18} + 255970t^{20} + 448240t^{22} + 752999t^{24} + 1220480t^{26} + 1917464t^{28} + 2931072t^{30} + O(t^{31})$ |
| **0904 2101.12235-unitary-product-33-I__N2** / **N = 2** 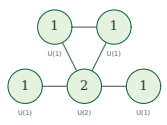 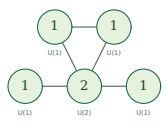 ÷ U(1) diagonal on {n0, n1, n2, n3, n4} · ÷ U(1) diagonal on {n5, n6, n7, n8, n9} | $1 + 32t^{2} + 24t^{3} + 496t^{4} + 664t^{5} + 5254t^{6} + 9040t^{7} + 43368t^{8} + 82976t^{9} + 294996t^{10} + 584208t^{11} + 1706109t^{12} + 3375600t^{13} + 8585020t^{14} + 16680824t^{15} + 38291570t^{16} + 72502424t^{17} + 153746176t^{18} + 282881192t^{19} + 562963652t^{20} + 1006221592t^{21} + 1900530760t^{22} + 3302962336t^{23} + 5970244704t^{24} + 10103882688t^{25} + 17588471380t^{26} + 29036155512t^{27} + 48919001110t^{28} + 78916158616t^{29} + 129187833552t^{30} + O(t^{31})$ |

| QUIVER | HILBERT SERIES |
|---|---|

**0905 2101.12235-unitary-product-33-I__N3** / **N = 3**

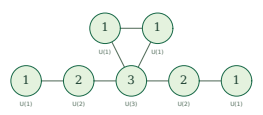

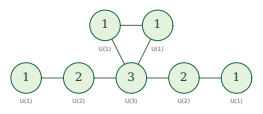

÷ U(1) diagonal on {n0, n1, n2, n3, n4, n5, n6} · ÷ U(1) diagonal on {n7, n8, n9, n10, n11, n12, n13}

$1 + 72t^{2} + 2636t^{4} + 64830t^{6} + 1198696t^{8} + 17719800t^{10} + 217795659t^{12} + 2287813036t^{14} + 20965716336t^{16} + 170340753610t^{18} + 1243128376480t^{20} + 8237653890004t^{22} + 50020924459445t^{24} + 280517851876564t^{26} + 1462761807896884t^{28} + 7134483341749692t^{30} + O(t^{31})$

**0906 2101.12235-unitary-product-33-II__N1** / **N = 1**





÷ U(1) diagonal on {n0, n1} · ÷ U(1) diagonal on {n2, n3}

$1 + 6t^{2} + 19t^{4} + 44t^{6} + 85t^{8} + 146t^{10} + 231t^{12} + 344t^{14} + 489t^{16} + 670t^{18} + 891t^{20} + 1156t^{22} + 1469t^{24} + 1834t^{26} + 2255t^{28} + 2736t^{30} + O(t^{31})$

**0907 2101.12235-unitary-product-33-II__N2** / **N = 2**





÷ U(1) diagonal on {n0, n1, n2, n3, n4} · ÷ U(1) diagonal on {n5, n6, n7, n8, n9}

$1 + 32t^{2} + 16t^{3} + 456t^{4} + 416t^{5} + 4104t^{6} + 4960t^{7} + 27490t^{8} + 38400t^{9} + 148792t^{10} + 223840t^{11} + 681924t^{12} + 1062432t^{13} + 2729368t^{14} + 4305376t^{15} + 9754099t^{16} + 15374976t^{17} + 31652168t^{18} + 49482864t^{19} + 94506536t^{20} + 145929968t^{21} + 262426376t^{22} + 399402368t^{23} + 683684780t^{24} + 1024710816t^{25} + 1683385784t^{26} + 2484337392t^{27} + 3941504900t^{28} + 5729237056t^{29} + 8821808128t^{30} + O(t^{31})$

**0908 2101.12235-unitary-product-33-II__N3** / **N = 3**

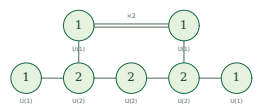

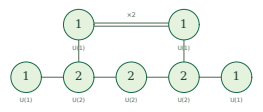

÷ U(1) diagonal on {n0, n1, n2, n3, n4, n5, n6} · ÷ U(1) diagonal on {n7, n8, n9, n10, n11, n12, n13}

$1 + 72t^{2} + 2616t^{4} + 63116t^{6} + 1130796t^{8} + 16005324t^{10} + 186338856t^{12} + 1837023558t^{14} + 15680401536t^{16} + 117956067528t^{18} + 793385308608t^{20} + 4829150282664t^{22} + 26870840918148t^{24} + 137868765191796t^{26} + 657126839723793t^{28} + 2928370036737216t^{30} + O(t^{31})$

**0909 2101.12235-unitary-product-33-III__N1** / **N = 1**





÷ U(1) diagonal on {n0, n1, n2} · ÷ U(1) diagonal on {n3, n4}

$1 + 11t^{2} + 56t^{4} + 192t^{6} + 517t^{8} + 1183t^{10} + 2408t^{12} + 4488t^{14} + 7809t^{16} + 12859t^{18} + 20240t^{20} + 30680t^{22} + 45045t^{24} + 64351t^{26} + 89776t^{28} + 122672t^{30} + O(t^{31})$

**0910 2101.12235-unitary-product-33-III__N2** / **N = 2**





÷ U(1) diagonal on {n0, n1, n2, n3, n4} · ÷ U(1) diagonal on {n5, n6, n7, n8, n9}

$1 + 32t^{2} + 20t^{3} + 476t^{4} + 540t^{5} + 4671t^{6} + 6920t^{7} + 34989t^{8} + 58628t^{9} + 214034t^{10} + 377288t^{11} + 1110572t^{12} + 1984004t^{13} + 5019157t^{14} + 8912136t^{15} + 20162660t^{16} + 35234092t^{17} + 73168310t^{18} + 125253944t^{19} + 243049474t^{20} + 406859436t^{21} + 747146380t^{22} + 1222750152t^{23} + 2144913646t^{24} + 3433865332t^{25} + 5794692256t^{26} + 9084138284t^{27} + 14827942722t^{28} + 22789371952t^{29} + 36137638716t^{30} + O(t^{31})$

| QUIVER | HILBERT SERIES |
|---|---|
| **0911 2101.12235-unitary-product-33-III__N3 / N = 3** 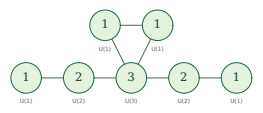 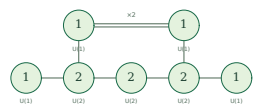 ÷ U(1) diagonal on {n0, n1, n2, n3, n4, n5, n6} · ÷ U(1) diagonal on {n7, n8, n9, n10, n11, n12, n13} | $1 + 72t^{2} + 2626t^{4} + 63973t^{6} + 1164696t^{8} + 16857592t^{10} + 201849673t^{12} + 2056666721t^{14} + 18216492934t^{16} + 142639584152t^{18} + 1001000389075t^{20} + 6367759568818t^{22} + 37073850330766t^{24} + 199178510559049t^{26} + 994464618529178t^{28} + 4642890295700722t^{30} + O(t^{31})$ |
| **0912 2101.12235-unitary-product-3p3p-I__N1 / N = 1** 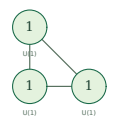 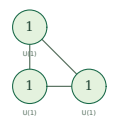 ÷ U(1) diagonal on {n0, n1, n2} · ÷ U(1) diagonal on {n3, n4, n5} | $1 + 16t^{2} + 118t^{4} + 560t^{6} + 2003t^{8} + 5888t^{10} + 14988t^{12} + 34176t^{14} + 71445t^{16} + 139216t^{18} + 255970t^{20} + 448240t^{22} + 752999t^{24} + 1220480t^{26} + 1917464t^{28} + 2931072t^{30} + O(t^{31})$ |
| **0913 2101.12235-unitary-product-3p3p-I__N2 / N = 2** 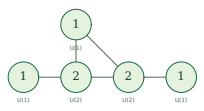 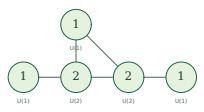 ÷ U(1) diagonal on {n0, n1, n2, n3, n4} · ÷ U(1) diagonal on {n5, n6, n7, n8, n9} | $1 + 48t^{2} + 1126t^{4} + 17248t^{6} + 194729t^{8} + 1735152t^{10} + 12762768t^{12} + 80014352t^{14} + 438014218t^{16} + 2133466416t^{18} + 9386562252t^{20} + 37764895984t^{22} + 140361844202t^{24} + 486061067984t^{26} + 1579603241040t^{28} + 4847285726000t^{30} + O(t^{31})$ |
| **0914 2101.12235-unitary-product-3p3p-I__N3 / N = 3** 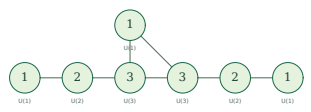 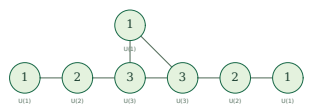 ÷ U(1) diagonal on {n0, n1, n2, n3, n4, n5, n6} · ÷ U(1) diagonal on {n7, n8, n9, n10, n11, n12, n13} | $1 + 96t^{2} + 4558t^{4} + 142784t^{6} + 3321959t^{8} + 61268224t^{10} + 933799130t^{12} + 12106951264t^{14} + 136429375817t^{16} + 1358626096896t^{18} + 12117127987468t^{20} + 97850777101888t^{22} + 722075758450356t^{24} + 4907432483630528t^{26} + 30925260625737556t^{28} + 181769112625065728t^{30} + O(t^{31})$ |
| **0915 2101.12235-unitary-product-3p3p-II__N1 / N = 1** 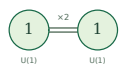 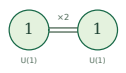 ÷ U(1) diagonal on {n0, n1} · ÷ U(1) diagonal on {n2, n3} | $1 + 6t^{2} + 19t^{4} + 44t^{6} + 85t^{8} + 146t^{10} + 231t^{12} + 344t^{14} + 489t^{16} + 670t^{18} + 891t^{20} + 1156t^{22} + 1469t^{24} + 1834t^{26} + 2255t^{28} + 2736t^{30} + O(t^{31})$ |
| **0916 2101.12235-unitary-product-3p3p-II__N2 / N = 2** 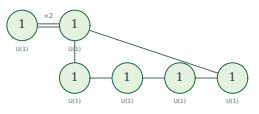 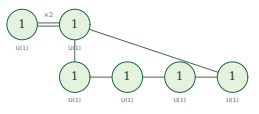 ÷ U(1) diagonal on {n0, n1, n2, n3, n4, n5} · ÷ U(1) diagonal on {n6, n7, n8, n9, n10, n11} | $1 + 54t^{2} + 1283t^{4} + 18412t^{6} + 185691t^{8} + 1438022t^{10} + 9055657t^{12} + 48248544t^{14} + 223859306t^{16} + 924323068t^{18} + 3454148958t^{20} + 11839273304t^{22} + 37623636478t^{24} + 111839064252t^{26} + 313272039626t^{28} + 832024292704t^{30} + O(t^{31})$ |

| QUIVER | HILBERT SERIES |
|---|---|

**0917 2101.12235-unitary-product-3p3p-II__N3 / N = 3**

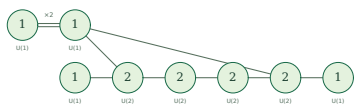

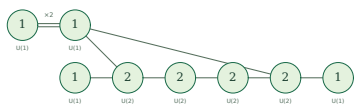

÷ U(1) diagonal on {n0, n1, n2, n3, n4, n5, n6, n7} · ÷ U(1) diagonal on {n8, n9, n10, n11, n12, n13, n14, n15}

$1 + 102t^{2} + 5153t^{4} + 170432t^{6} + 4135902t^{8} + 78455668t^{10} + 1212478590t^{12} + 15727513344t^{14} + 175185507429t^{16} + 1706438870462t^{18} + 14753731119237t^{20} + 114635129182144t^{22} + 808889585198872t^{24} + 5230036084319568t^{26} + 31226057726752472t^{28} + 173317134576291648t^{30} + O(t^{31})$

**0918 2101.12235-unitary-product-3p3p-III__N1 / N = 1**





÷ U(1) diagonal on {n0, n1, n2} · ÷ U(1) diagonal on {n3, n4}

$1 + 11t^{2} + 56t^{4} + 192t^{6} + 517t^{8} + 1183t^{10} + 2408t^{12} + 4488t^{14} + 7809t^{16} + 12859t^{18} + 20240t^{20} + 30680t^{22} + 45045t^{24} + 64351t^{26} + 89776t^{28} + 122672t^{30} + O(t^{31})$

**0919 2101.12235-unitary-product-3p3p-III__N2 / N = 2**

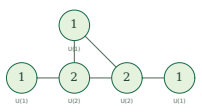

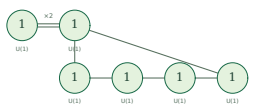

÷ U(1) diagonal on {n0, n1, n2, n3, n4} · ÷ U(1) diagonal on {n5, n6, n7, n8, n9, n10}

$1 + 51t^{2} + 1200t^{4} + 17824t^{6} + 191099t^{8} + 1596553t^{10} + 10928128t^{12} + 63496000t^{14} + 321538122t^{16} + 1448135822t^{18} + 5894196656t^{20} + 21961971056t^{22} + 75705165550t^{24} + 243547089194t^{26} + 736595463744t^{28} + 2107484597760t^{30} + O(t^{31})$

**0920 2101.12235-unitary-product-3p3p-III__N3 / N = 3**

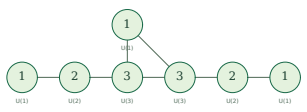

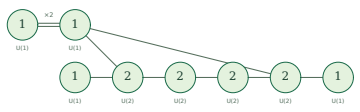

÷ U(1) diagonal on {n0, n1, n2, n3, n4, n5, n6} · ÷ U(1) diagonal on {n7, n8, n9, n10, n11, n12, n13, n14}

$1 + 99t^{2} + 4851t^{4} + 156161t^{6} + 3709298t^{8} + 69345142t^{10} + 1063817097t^{12} + 13793519067t^{14} + 154555552459t^{16} + 1522838402409t^{18} + 13381015869764t^{20} + 106081112581020t^{22} + 766210090844612t^{24} + 5084434043538780t^{26} + 31221253447313740t^{28} + 178523941124487732t^{30} + O(t^{31})$

**0921 2101.12235-unitary-product-3p4-I__N1 / N = 1**





÷ U(1) diagonal on {n0, n1, n2} · ÷ U(1) diagonal on {n3, n4, n5, n6, n7}

$1 + 32t^{2} + 419t^{4} + 3312t^{6} + 18736t^{8} + 83392t^{10} + 309784t^{12} + 999088t^{14} + 2876029t^{16} + 7540832t^{18} + 18285423t^{20} + 41492256t^{22} + 88928632t^{24} + 181371776t^{26} + 354155288t^{28} + 665421344t^{30} + O(t^{31})$

| QUIVER | HILBERT SERIES |
| --- | --- |
| **0922 2101.12235-unitary-product-3p4-I__N2 / N = 2** 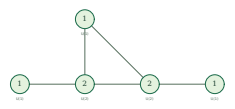 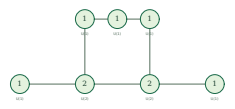 ÷ U(1) diagonal on {n0, n1, n2, n3, n4} · ÷ U(1) diagonal on {n5, n6, n7, n8, n9, n10, n11} | $1 + 52t^{2} + 40t^{3} + 1327t^{4} + 1780t^{5} + 22500t^{6} + 39140t^{7} + 286968t^{8} + 570120t^{9} + 2939292t^{10} + 6219780t^{11} + 25149772t^{12} + 54427500t^{13} + 184632748t^{14} + 399129252t^{15} + 1186476430t^{16} + 2528281844t^{17} + 6781825932t^{18} + 14144770824t^{19} + 34942081882t^{20} + 71087212100t^{21} + 164114415116t^{22} + 325256247036t^{23} + 709423391650t^{24} + 1369626796584t^{25} + 2845791683160t^{26} + 5355532817344t^{27} + 10669165448190t^{28} + 19592075881796t^{29} + 37615896370132t^{30} + O(t^{31})$ |
| **0923 2101.12235-unitary-product-3p4-I__N3 / N = 3** 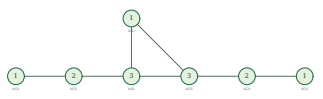 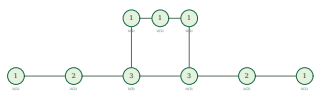 ÷ U(1) diagonal on {n0, n1, n2, n3, n4, n5, n6} · ÷ U(1) diagonal on {n7, n8, n9, n10, n11, n12, n13, n14, n15} | $1 + 100t^{2} + 5091t^{4} + 174328t^{6} + 4486522t^{8} + 92150964t^{10} + 1569197296t^{12} + 22753478260t^{14} + 286628991661t^{16} + 3186709241424t^{18} + 31672336938989t^{20} + 284424387582236t^{22} + 2328796422631965t^{24} + 17520755213517036t^{26} + 121946378651345077t^{28} + 789875048426435428t^{30} + O(t^{31})$ |
| **0924 2101.12235-unitary-product-3p4-II__N1 / N = 1** 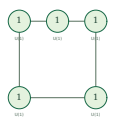 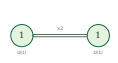 ÷ U(1) diagonal on {n0, n1, n2, n3, n4} · ÷ U(1) diagonal on {n5, n6} | $1 + 27t^{2} + 277t^{4} + 1727t^{6} + 7852t^{8} + 28628t^{10} + 88604t^{12} + 241604t^{14} + 595529t^{16} + 1351779t^{18} + 2865005t^{20} + 5730231t^{22} + 10905856t^{24} + 19882656t^{26} + 34910656t^{28} + 59297632t^{30} + O(t^{31})$ |
| **0925 2101.12235-unitary-product-3p4-II__N2 / N = 2** 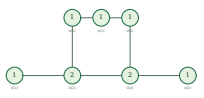 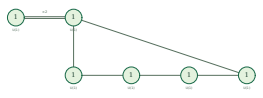 ÷ U(1) diagonal on {n0, n1, n2, n3, n4, n5, n6} · ÷ U(1) diagonal on {n7, n8, n9, n10, n11, n12} | $1 + 55t^{2} + 40t^{3} + 1413t^{4} + 1900t^{5} + 23399t^{6} + 41680t^{7} + 287256t^{8} + 585260t^{9} + 2808576t^{10} + 6050060t^{11} + 22796460t^{12} + 49736520t^{13} + 157972756t^{14} + 341201712t^{15} + 954794766t^{16} + 2018084800t^{17} + 5121161834t^{18} + 10535980556t^{19} + 24728438718t^{20} + 49423176668t^{21} + 108800033450t^{22} + 211220544232t^{23} + 440644312586t^{24} + 831632942132t^{25} + 1657044146570t^{26} + 3044273893320t^{27} + 5828846955590t^{28} + 10439911449732t^{29} + 19301943315222t^{30} + O(t^{31})$ |
| **0926 2101.12235-unitary-product-3p4-II__N3 / N = 3** 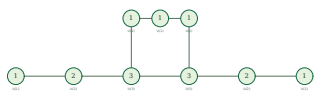 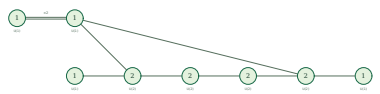 ÷ U(1) diagonal on {n0, n1, n2, n3, n4, n5, n6, n7, n8} · ÷ U(1) diagonal on {n9, n10, n11, n12, n13, n14, n15, n16} | $1 + 103t^{2} + 5396t^{4} + 189324t^{6} + 4968050t^{8} + 103508306t^{10} + 1778538919t^{12} + 25887137157t^{14} + 325728771826t^{16} + 3600788622634t^{18} + 35438734830426t^{20} + 314013045942310t^{22} + 2528989938938934t^{24} + 18666100530956306t^{26} + 127170226009877266t^{28} + 804787292535672554t^{30} + O(t^{31})$ |
| **0927 2101.12235-unitary-product-44__N1 / N = 1** 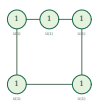 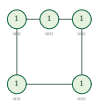 ÷ U(1) diagonal on {n0, n1, n2, n3, n4} · ÷ U(1) diagonal on {n5, n6, n7, n8, n9} | $1 + 48t^{2} + 976t^{4} + 11600t^{6} + 95350t^{8} + 598352t^{10} + 3053296t^{12} + 13224752t^{14} + 50129875t^{16} + 170108000t^{18} + 525728128t^{20} + 1499862944t^{22} + 3992550028t^{24} + 10002922400t^{26} + 23756262400t^{28} + 53799799904t^{30} + O(t^{31})$ |

| QUIVER | HILBERT SERIES |
|---|---|

**0928 2101.12235-unitary-product-44__N2 / N = 2**

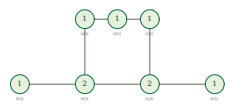

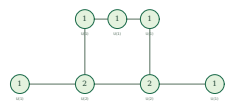

÷ U(1) diagonal on {n0, n1, n2, n3, n4, n5, n6} · ÷ U(1) diagonal on {n7, n8, n9, n10, n11, n12, n13}

$1 + 56t^{2} + 80t^{3} + 1544t^{4} + 3880t^{5} + 30192t^{6} + 93240t^{7} + 468888t^{8} + 1510680t^{9} + 5964152t^{10} + 18665600t^{11} + 63205328t^{12} + 187321360t^{13} + 568498816t^{14} + 1586088264t^{15} + 4421468514t^{16} + 11630846104t^{17} + 30242307864t^{18} + 75321029016t^{19} + 184613300080t^{20} + 437421691648t^{21} + 1018526793520t^{22} + 2306557003152t^{23} + 5133278694420t^{24} + 11157963050608t^{25} + 23850659407280t^{26} + 49949609133392t^{27} + 102966150340240t^{28} + 208457686611304t^{29} + 415829300101768t^{30} + O(t^{31})$

**0929 2101.12235-unitary-product-44__N3 / N = 3**

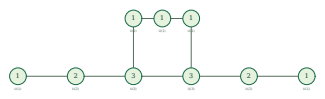

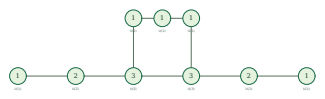

÷ U(1) diagonal on {n0, n1, n2, n3, n4, n5, n6, n7, n8} · ÷ U(1) diagonal on {n9, n10, n11, n12, n13, n14, n15, n16, n17}

$1 + 104t^{2} + 5640t^{4} + 208600t^{6} + 5852710t^{8} + 131901344t^{10} + 2475139190t^{12} + 39651821224t^{14} + 552532311079t^{16} + 6796473470664t^{18} + 74693350511020t^{20} + 740902006830080t^{22} + 6691121248341485t^{24} + 55434522771040280t^{26} + 424119506024895306t^{28} + 3014214309587561728t^{30} + O(t^{31})$

**0930 2101.12235-unitary-product-55__N1 / N = 1**





÷ U(1) diagonal on {n0, n1, n2, n3, n4, n5} · ÷ U(1) diagonal on {n6, n7, n8, n9, n10, n11}

$1 + 90t^{2} + 3565t^{4} + 84588t^{6} + 1386700t^{8} + 17100048t^{10} + 167857416t^{12} + 1365541740t^{14} + 9486469554t^{16} + 57589566980t^{18} + 311107661634t^{20} + 1517583491228t^{22} + 6764929124760t^{24} + 27832549734992t^{26} + 106571838573900t^{28} + 382476746226204t^{30} + O(t^{31})$

**0931 2101.12235-unitary-product-55__N2 / N = 2**

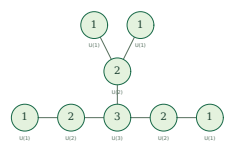

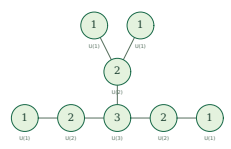

÷ U(1) diagonal on {n0, n1, n2, n3, n4, n5, n6, n7} · ÷ U(1) diagonal on {n8, n9, n10, n11, n12, n13, n14, n15}

$1 + 82t^{2} + 160t^{3} + 3329t^{4} + 11680t^{5} + 98836t^{6} + 423776t^{7} + 2405524t^{8} + 10483680t^{9} + 48619712t^{10} + 200381888t^{11} + 820182244t^{12} + 3140635520t^{13} + 11715364684t^{14} + 41720848544t^{15} + 144167357556t^{16} + 480204640160t^{17} + 1553154598704t^{18} + 4869260886496t^{19} + 14853317240416t^{20} + 44084090762912t^{21} + 127594599784704t^{22} + 360348815245696t^{23} + 994568390589610t^{24} + 2684682220685312t^{25} + 7095953462298856t^{26} + 18378981712871584t^{27} + 46691478344202890t^{28} + 116432464726364896t^{29} + 285217877407464462t^{30} + O(t^{31})$

**0932 2101.12235-unitary-product-55__N3 / N = 3**

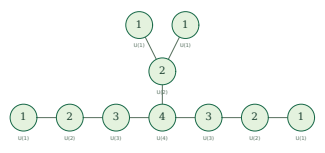

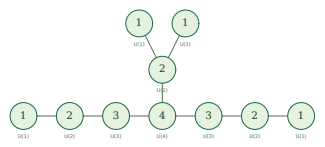

÷ U(1) diagonal on {n0, n1, n2, n3, n4, n5, n6, n7, n8, n9} · ÷ U(1) diagonal on {n10, n11, n12, n13, n14, n15, n16, n17, n18, n19}

$1 + 138t^{2} + 10021t^{4} + 499878t^{6} + 19034274t^{8} + 585392784t^{10} + 15062951627t^{12} + 332260839696t^{14} + 6396749967288t^{16} + 109000942084204t^{18} + 1662702915042538t^{20} + 22920061877544244t^{22} + 287824936264788621t^{24} + 3315723208815584108t^{26} + 35254793325910693581t^{28} + 347854154871551847960t^{30} + O(t^{31})$

| QUIVER | HILBERT SERIES |
| --- | --- |
| **0933 2101.12235-unitary-product-5p5p__N1 / N = 1** 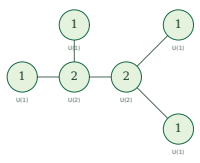 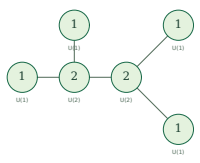 ÷ U(1) diagonal on {n0, n1, n2, n3, n4, n5} · ÷ U(1) diagonal on {n6, n7, n8, n9, n10, n11} | $1 + 90t^{2} + 3565t^{4} + 84588t^{6} + 1386700t^{8} + 17100048t^{10} + 167857416t^{12} + 1365541740t^{14} + 9486469554t^{16} + 57589566980t^{18} + 311107661634t^{20} + 1517583491228t^{22} + 6764929124760t^{24} + 27832549734992t^{26} + 106571838573900t^{28} + 382476746226204t^{30} + O(t^{31})$ |
| **0934 2101.12235-unitary-product-5p5p__N2 / N = 2** 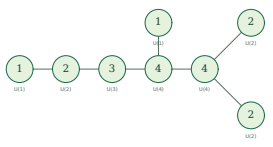 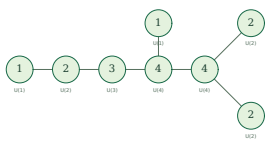 ÷ U(1) diagonal on {n0, n1, n2, n3, n4, n5, n6, n7} · ÷ U(1) diagonal on {n8, n9, n10, n11, n12, n13, n14, n15} | $1 + 182t^{2} + 16443t^{4} + 977366t^{6} + 42848547t^{8} + 1475064682t^{10} + 41500186778t^{12} + 981444419592t^{14} + 19926320712081t^{16} + 353137745216620t^{18} + 5537049500191090t^{20} + 77680846683111490t^{22} + 984456112010725004t^{24} + 11363058478646759304t^{26} + 120314917511278823448t^{28} + 1175985452991944995288t^{30} + O(t^{31})$ |
| **0935 2101.12235-unitary-product-5p5p__N3 / N = 3** 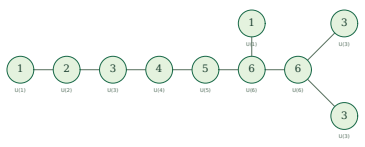 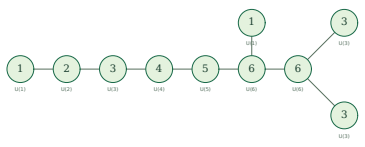 ÷ U(1) diagonal on {n0, n1, n2, n3, n4, n5, n6, n7, n8, n9} · ÷ U(1) diagonal on {n10, n11, n12, n13, n14, n15, n16, n17, n18, n19} | $1 + 306t^{2} + 46629t^{4} + 4718044t^{6} + 356537448t^{8} + 21456973062t^{10} + 1070800645180t^{12} + 45560073634990t^{14} + 1686498157265604t^{16} + 55157853453884154t^{18} + 1613331163533772153t^{20} + 42619234809241136992t^{22} + 1025159634735626072229t^{24} + 22608272477706608805688t^{26} + 459827558582923897135340t^{28} + 8669663524521570749214588t^{30} + O(t^{31})$ |
| **0936 2101.12235-unitary-product-5p6__N1 / N = 1** 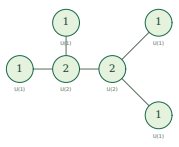 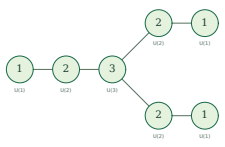 ÷ U(1) diagonal on {n0, n1, n2, n3, n4, n5} · ÷ U(1) diagonal on {n6, n7, n8, n9, n10, n11, n12} | $1 + 123t^{2} + 6710t^{4} + 220812t^{6} + 5027328t^{8} + 85856661t^{10} + 1160906385t^{12} + 12919480684t^{14} + 121836107745t^{16} + 995927436569t^{18} + 7185225712236t^{20} + 46428889066092t^{22} + 271980778978818t^{24} + 1459155335653719t^{26} + 7231291138577817t^{28} + 33348754937791440t^{30} + O(t^{31})$ |
| **0937 2101.12235-unitary-product-5p6__N2 / N = 2** 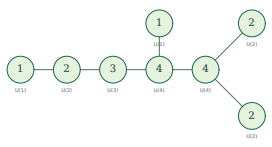 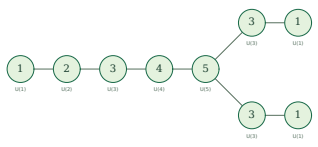 ÷ U(1) diagonal on {n0, n1, n2, n3, n4, n5, n6, n7} · ÷ U(1) diagonal on {n8, n9, n10, n11, n12, n13, n14, n15, n16} | $1 + 183t^{2} + 128t^{3} + 16626t^{4} + 21632t^{5} + 1000427t^{6} + 1815296t^{7} + 44866419t^{8} + 100543872t^{9} + 1599857948t^{10} + 4128989696t^{11} + 47241732159t^{12} + 134052513792t^{13} + 1187644477615t^{14} + 3584945182080t^{15} + 25931548491019t^{16} + 81273381997312t^{17} + 499209765899717t^{18} + 1595661433838080t^{19} + 8573910292940615t^{20} + 27582651744094592t^{21} + 132645907276973100t^{22} + 425370215657202688t^{23} + 1863541593477943823t^{24} + 5915986181358373376t^{25} + 23940878934177503972t^{26} + 74874664568479450112t^{27} + 282978273155790598403t^{28} + 868992599593680948352t^{29} + 3094155700520577209535t^{30} + O(t^{31})$ |

| QUIVER | HILBERT SERIES |
|---|---|

**0938 2101.12235-unitary-product-5p6__N3 / N = 3**

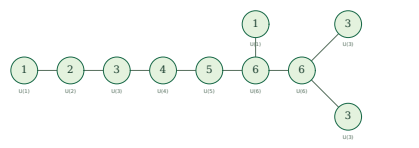

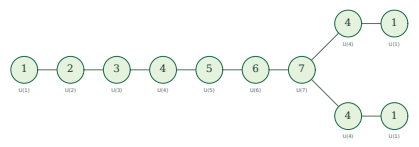

÷ U(1) diagonal on {n0, n1, n2, n3, n4, n5, n6, n7, n8, n9} · ÷ U(1) diagonal on {n10, n11, n12, n13, n14, n15, n16, n17, n18, n19, n20}

$1 + 307t^{2} + 47448t^{4} + 4912948t^{6} + 382692982t^{8} + 23879613754t^{10} + 1241496712968t^{12} + 55242998572232t^{14} + 2145381950985488t^{16} + 73803621307318672t^{18} + 2275498755357607893t^{20} + 63477136597338616313t^{22} + 1614777185598732535538t^{24} + 37709191971317819474815t^{26} + 813016853396166226368849t^{28} + 16264030547468809240751292t^{30} + O(t^{31})$

**0939 2101.12235-unitary-product-66__N1 / N = 1**

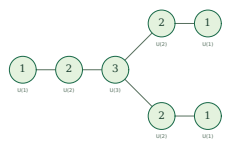

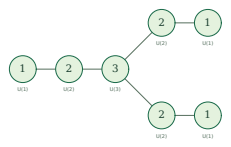

÷ U(1) diagonal on {n0, n1, n2, n3, n4, n5, n6} · ÷ U(1) diagonal on {n7, n8, n9, n10, n11, n12, n13}

$1 + 156t^{2} + 10944t^{4} + 466596t^{6} + 13807080t^{8} + 306524790t^{10} + 5377829028t^{12} + 77405354312t^{14} + 940221343776t^{16} + 9853246779220t^{18} + 90680857312617t^{20} + 743597747463636t^{22} + 5499029802140784t^{24} + 37048975552593240t^{26} + 229389602023982920t^{28} + 1314982419051078540t^{30} + O(t^{31})$

**0940 2101.12235-unitary-product-66__N2 / N = 2**

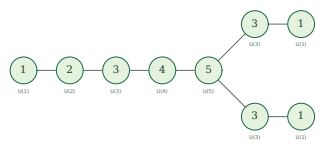

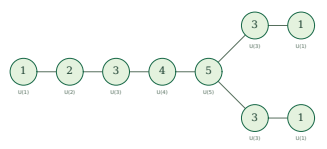

÷ U(1) diagonal on {n0, n1, n2, n3, n4, n5, n6, n7, n8} · ÷ U(1) diagonal on {n9, n10, n11, n12, n13, n14, n15, n16, n17}

$1 + 184t^{2} + 256t^{3} + 16810t^{4} + 43520t^{5} + 1040056t^{6} + 3674112t^{7} + 49469875t^{8} + 206409216t^{9} + 1925805206t^{10} + 8703438848t^{11} + 63418779630t^{12} + 294056655104t^{13} + 1801033108182t^{14} + 8290154332416t^{15} + 44696692490453t^{16} + 200425546653440t^{17} + 979719926063866t^{18} + 4237436679956480t^{19} + 19144937799760589t^{20} + 79508499488723712t^{21} + 336351068256306004t^{22} + 1339393653875000320t^{23} + 5353557330835115870t^{24} + 20448861172149924864t^{25} + 77734446082834081484t^{26} + 285171347538469251072t^{27} + 1036157097293424805820t^{28} + 3657040896809078287616t^{29} + 12750665336573546537520t^{30} + O(t^{31})$

**0941 2101.12235-unitary-product-66__N3 / N = 3**

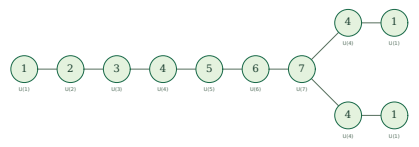

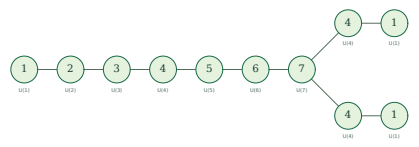

÷ U(1) diagonal on {n0, n1, n2, n3, n4, n5, n6, n7, n8, n9, n10} · ÷ U(1) diagonal on {n11, n12, n13, n14, n15, n16, n17, n18, n19, n20, n21}

$1 + 308t^{2} + 48268t^{4} + 5109184t^{6} + 409454864t^{8} + 26421447906t^{10} + 1426480824340t^{12} + 66144082915864t^{14} + 2684517226599308t^{16} + 96748084561243286t^{18} + 3131392050137233289t^{20} + 91861504908160265684t^{22} + 2461117166463944695488t^{24} + 60607358750628705380270t^{26} + 1379471216109334974035532t^{28} + 29159906662488205397527364t^{30} + O(t^{31})$

**0942 2101.12235-unitary-product-77__N1 / N = 1**

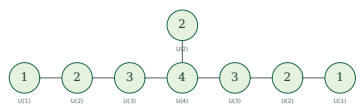

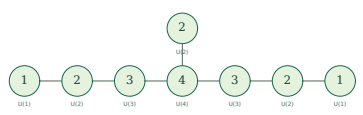

÷ U(1) diagonal on {n0, n1, n2, n3, n4, n5, n6, n7} · ÷ U(1) diagonal on {n8, n9, n10, n11, n12, n13, n14, n15}

$1 + 266t^{2} + 32431t^{4} + 2437890t^{6} + 128299853t^{8} + 5085554028t^{10} + 159353197186t^{12} + 4086829420626t^{14} + 88074412819494t^{16} + 1628326501845402t^{18} + 26262027726102820t^{20} + 374634842317768270t^{22} + 4782070777992608385t^{24} + 55161495417837540740t^{26} + 579903325545777980196t^{28} + 5597305572994396231404t^{30} + O(t^{31})$

**Paper typo:** The negative HWG relation is missing the square on the spinor highest-weight fugacity μ_(2N+4). Restoring the square, already present in the cited source, gives 128297273 rather than 128299853 at $t^8$.

| QUIVER | HILBERT SERIES |
|---|---|

**0943 2101.12235-unitary-product-77__N2 / N = 2**

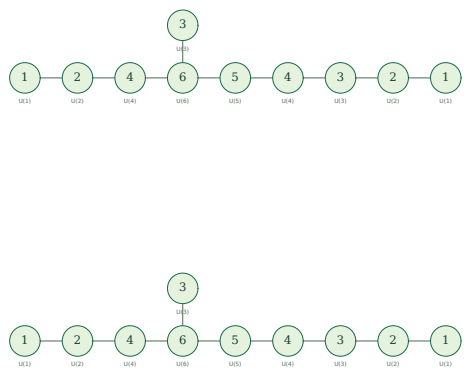

÷ U(1) diagonal on {n0, n1, n2, n3, n4, n5, n6, n7, n8, n9} · ÷ U(1) diagonal on {n10, n11, n12, n13, n14, n15, n16, n17, n18, n19}

$1 + 246t^{2} + 512t^{3} + 30109t^{4} + 117760t^{5} + 2549000t^{6} + 13477888t^{7} + 170617424t^{8} + 1034931200t^{9} + 9557541070t^{10} + 60338314240t^{11} + 458387567850t^{12} + 2853399523560t^{13} + 19040787759464t^{14} + 113850326071908t^{15} + 691532804574288t^{16} + 3930933694335116t^{17} + 22167853515446246t^{18} + 119509857102878228t^{19} + 633046617493715185t^{20} + 3240550672531134500t^{21} + 16244108637769631662t^{22} + 79155925406598050924t^{23} + 377478354767463811374t^{24} + 1756102143657503592292t^{25} + 7999155231663513634542t^{26} + 35630487305689002341240t^{27} + 155537044930037155837632t^{28} + 665117914152878287434828t^{29} + 2790273156505751713888346t^{30} + O(t^{31})$

**Paper typo:** The negative HWG relation is missing the square on the spinor highest-weight fugacity μ_(2N+4). Restoring the square, already present in the cited source, gives 9557503228 rather than 9557541070 at t¹⁰.

**0944 2101.12235-unitary-product-77__N3 / N = 3**

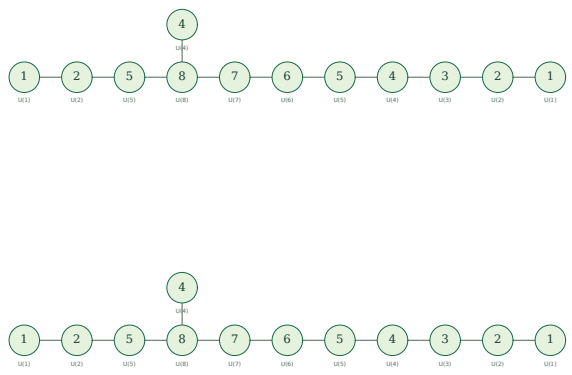

÷ U(1) diagonal on {n0, n1, n2, n3, n4, n5, n6, n7, n8, n9, n10, n11} · ÷ U(1) diagonal on {n12, n13, n14, n15, n16, n17, n18, n19, n20, n21, n22, n23}

$1 + 386t^{2} + 76319t^{4} + 10247902t^{6} + 1046685523t^{8} + 86428143088t^{10} + 5992688067924t^{12} + 358029677867722t^{14} + 18778166306642169t^{16} + 876926029244395996t^{18} + 36869324156724995565t^{20} + 1408148161149556289676t^{22} + 49218208138806799111498t^{24} + 1584182066967625797471492t^{26} + 47206898113583917327680594t^{28} + 1308402686150128923609353976t^{30} + O(t^{31})$

**0945 2101.12235-two-U2-4F**

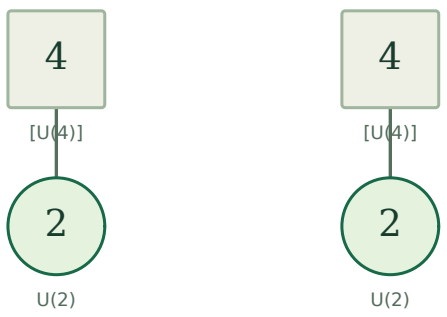


$1 + 6t^{2} + 27t^{4} + 90t^{6} + 259t^{8} + 648t^{10} + 1478t^{12} + 3096t^{14} + 6085t^{16} + 11286t^{18} + 19985t^{20} + 33930t^{22} + 55623t^{24} + 88320t^{26} + 136460t^{28} + 205632t^{30} + O(t^{31})$

## 2102.06190 · Branes, Quivers, and the Affine Grassmannian (39)

**0946 2102.06190-ac2**

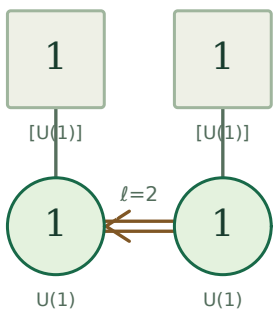


$1 + 4t^{2} + 6t^{3} + 9t^{4} + 16t^{5} + 26t^{6} + 30t^{7} + 49t^{8} + 62t^{9} + 78t^{10} + 102t^{11} + 131t^{12} + 150t^{13} + 194t^{14} + 228t^{15} + 267t^{16} + 318t^{17} + 376t^{18} + 420t^{19} + 499t^{20} + 564t^{21} + 636t^{22} + 724t^{23} + 821t^{24} + 900t^{25} + 1024t^{26} + 1130t^{27} + 1245t^{28} + 1380t^{29} + 1526t^{30} + O(t^{31})$

**0947 2102.06190-ac3**

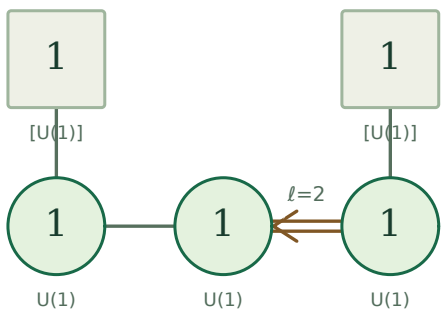


$1 + 9t^{2} + 12t^{3} + 36t^{4} + 60t^{5} + 130t^{6} + 180t^{7} + 351t^{8} + 476t^{9} + 777t^{10} + 1056t^{11} + 1594t^{12} + 2052t^{13} + 2976t^{14} + 3752t^{15} + 5127t^{16} + 6408t^{17} + 8463t^{18} + 10308t^{19} + 13332t^{20} + 16016t^{21} + 20124t^{22} + 23988t^{23} + 29577t^{24} + 34728t^{25} + 42273t^{26} + 49168t^{27} + 58848t^{28} + 68028t^{29} + 80454t^{30} + O(t^{31})$

**0948 2102.06190-ac4**

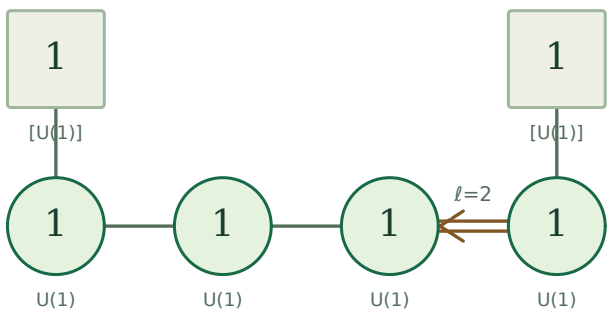


$1 + 16t^{2} + 20t^{3} + 100t^{4} + 160t^{5} + 470t^{6} + 700t^{7} + 1673t^{8} + 2408t^{9} + 4816t^{10} + 6840t^{11} + 12186t^{12} + 16740t^{13} + 27710t^{14} + 37092t^{15} + 57585t^{16} + 75732t^{17} + 111918t^{18} + 144248t^{19} + 205486t^{20} + 260360t^{21} + 358966t^{22} + 448788t^{23} + 602243t^{24} + 742852t^{25} + 975350t^{26} + 1188892t^{27} + 1530498t^{28} + 1847208t^{29} + 2337828t^{30} + O(t^{31})$

| QUIVER | HILBERT SERIES |
| --- | --- |

**0949 2102.06190-ac5**

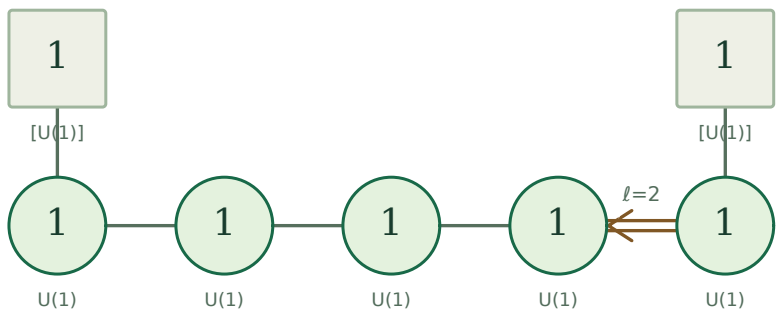


$1 + 25t^{2} + 30t^{3} + 225t^{4} + 350t^{5} + 1365t^{6} + 2100t^{7} + 6160t^{8} + 9240t^{9} + 22176t^{10} + 32700t^{11} + 68190t^{12} + 98010t^{13} + 185350t^{14} + 259952t^{15} + 455235t^{16} + 625690t^{17} + 1030835t^{18} + 1389570t^{19} + 2181501t^{20} + 2889070t^{21} + 4356585t^{22} + 5681600t^{23} + 8282010t^{24} + 10648902t^{25} + 15086830t^{26} + 19149560t^{27} + 26469300t^{28} + 33212200t^{29} + 44931128t^{30} + O(t^{31})$

**0950 2102.06190-acn__n6 / n = 6**

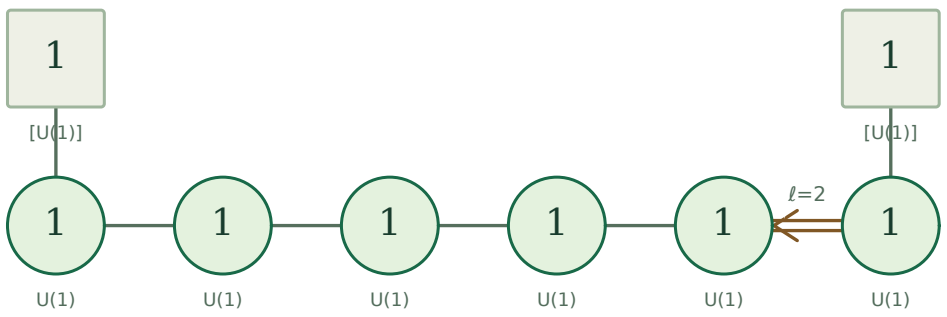


$1 + 36t^{2} + 42t^{3} + 441t^{4} + 672t^{5} + 3388t^{6} + 5292t^{7} + 18900t^{8} + 29148t^{9} + 82908t^{10} + 125928t^{11} + 304722t^{12} + 453222t^{13} + 975612t^{14} + 1419418t^{15} + 2791503t^{16} + 3980340t^{17} + 7284368t^{18} + 10191090t^{19} + 17599113t^{20} + 24190056t^{21} + 39813984t^{22} + 53853144t^{23} + 85119034t^{24} + 113449194t^{25} + 173250684t^{26} + 227805326t^{27} + 337717248t^{28} + 438595290t^{29} + 633612960t^{30} + O(t^{31})$

**0951 2102.06190-acn__n7 / n = 7**

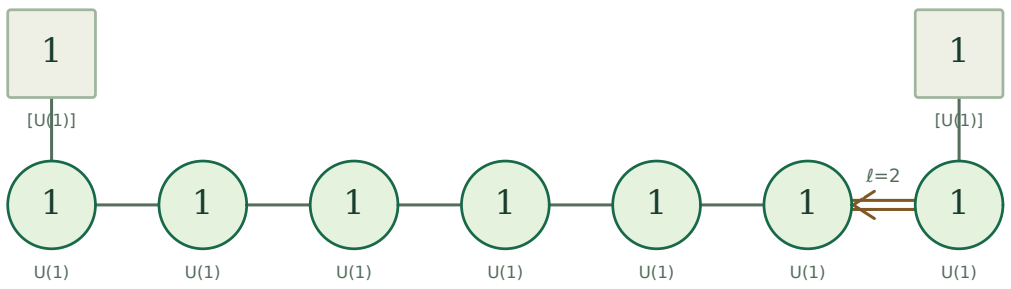


$1 + 49t^{2} + 56t^{3} + 784t^{4} + 1176t^{5} + 7476t^{6} + 11760t^{7} + 50568t^{8} + 79464t^{9} + 265188t^{10} + 412104t^{11} + 1148070t^{12} + 1753752t^{13} + 4275986t^{14} + 6406400t^{15} + 14091077t^{16} + 20713784t^{17} + 41965105t^{18} + 60571056t^{19} + 114779224t^{20} + 162825728t^{21} + 291901288t^{22} + 407476944t^{23} + 697107474t^{24} + 958682088t^{25} + 1575910350t^{26} + 2137377032t^{27} + 3394418210t^{28} + 4545043328t^{29} + 7004239298t^{30} + O(t^{31})$

**0952 2102.06190-acn__n8 / n = 8**

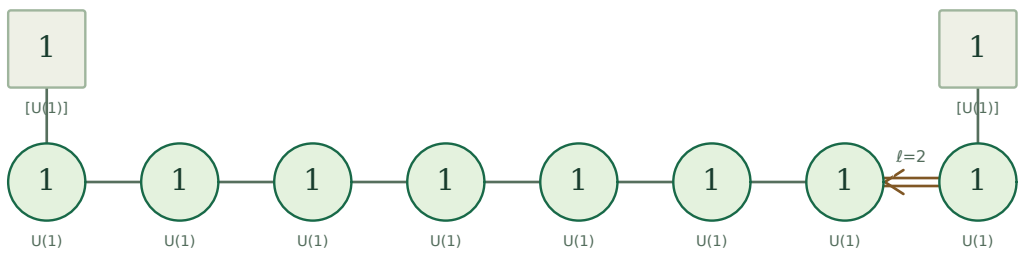


$1 + 64t^{2} + 72t^{3} + 1296t^{4} + 1920t^{5} + 15060t^{6} + 23760t^{7} + 121572t^{8} + 193512t^{9} + 750816t^{10} + 1187472t^{11} + 3781206t^{12} + 5899608t^{13} + 16208764t^{14} + 24869416t^{15} + 60930441t^{16} + 91869024t^{17} + 205357672t^{18} + 304332912t^{19} + 631161040t^{20} + 919902096t^{21} + 1792424376t^{22} + 2571462432t^{23} + 4753101978t^{24} + 6718309560t^{25} + 11870132580t^{26} + 16545685184t^{27} + 28113832638t^{28} + 38680213088t^{29} + 63519467748t^{30} + O(t^{31})$

**0953 2102.06190-G2-twoframes**

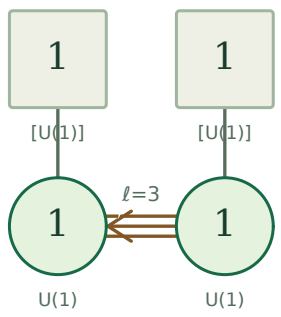


$1 + 4t^{2} + 17t^{4} + 36t^{6} + 75t^{8} + 124t^{10} + 203t^{12} + 296t^{14} + 429t^{16} + 580t^{18} + 781t^{20} + 1004t^{22} + 1287t^{24} + 1596t^{26} + 1975t^{28} + 2384t^{30} + O(t^{31})$

**0954 2102.06190-G2-oneframe**

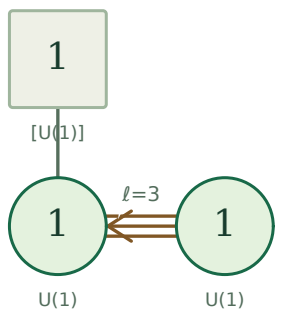


$1 + 4t^{2} + 8t^{3} + 9t^{4} + 20t^{5} + 30t^{6} + 36t^{7} + 57t^{8} + 76t^{9} + 90t^{10} + 124t^{11} + 155t^{12} + 180t^{13} + 230t^{14} + 276t^{15} + 315t^{16} + 384t^{17} + 448t^{18} + 504t^{19} + 595t^{20} + 680t^{21} + 756t^{22} + 872t^{23} + 981t^{24} + 1080t^{25} + 1224t^{26} + 1360t^{27} + 1485t^{28} + 1660t^{29} + 1826t^{30} + O(t^{31})$

**0955 2102.06190-h-nk__n2_k4 / n = 2 · k = 4**

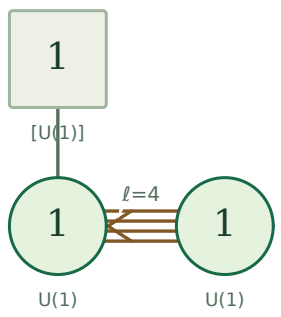


$1 + 4t^{2} + 19t^{4} + 40t^{6} + 85t^{8} + 140t^{10} + 231t^{12} + 336t^{14} + 489t^{16} + 660t^{18} + 891t^{20} + 1144t^{22} + 1469t^{24} + 1820t^{26} + 2255t^{28} + 2720t^{30} + O(t^{31})$

| QUIVER | HILBERT SERIES |
|---|---|
| **0956 2102.06190-h-nk__n2_k5** / n = 2 · k = 5 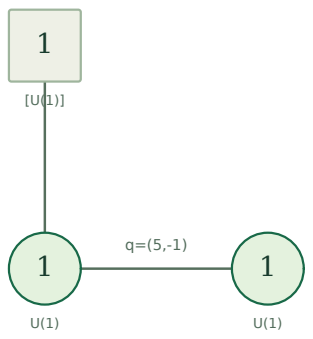 | $1 + 4t^{2} + 9t^{4} + 12t^{5} + 16t^{6} + 28t^{7} + 25t^{8} + 48t^{9} + 58t^{10} + 72t^{11} + 97t^{12} + 100t^{13} + 142t^{14} + 164t^{15} + 193t^{16} + 236t^{17} + 250t^{18} + 316t^{19} + 355t^{20} + 404t^{21} + 470t^{22} + 500t^{23} + 595t^{24} + 656t^{25} + 730t^{26} + 824t^{27} + 875t^{28} + 1004t^{29} + 1092t^{30} + O(t^{31})$ |
| **0957 2102.06190-h-nk__n2_k6** / n = 2 · k = 6 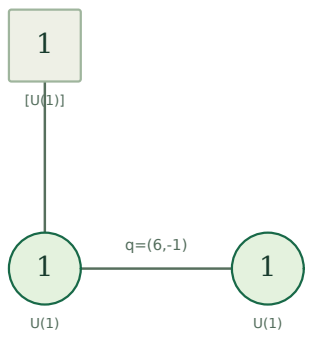 | $1 + 4t^{2} + 9t^{4} + 30t^{6} + 57t^{8} + 90t^{10} + 155t^{12} + 230t^{14} + 315t^{16} + 448t^{18} + 595t^{20} + 756t^{22} + 981t^{24} + 1224t^{26} + 1485t^{28} + 1826t^{30} + O(t^{31})$ |
| **0958 2102.06190-h-nk__n3_k1** / n = 3 · k = 1 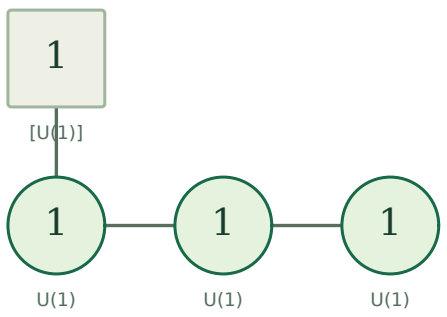 | $1 + 6t + 21t^{2} + 56t^{3} + 126t^{4} + 252t^{5} + 462t^{6} + 792t^{7} + 1287t^{8} + 2002t^{9} + 3003t^{10} + 4368t^{11} + 6188t^{12} + 8568t^{13} + 11628t^{14} + 15504t^{15} + 20349t^{16} + 26334t^{17} + 33649t^{18} + 42504t^{19} + 53130t^{20} + 65780t^{21} + 80730t^{22} + 98280t^{23} + 118755t^{24} + 142506t^{25} + 169911t^{26} + 201376t^{27} + 237336t^{28} + 278256t^{29} + 324632t^{30} + O(t^{31})$ |
| **0959 2102.06190-h-nk__n3_k3** / n = 3 · k = 3 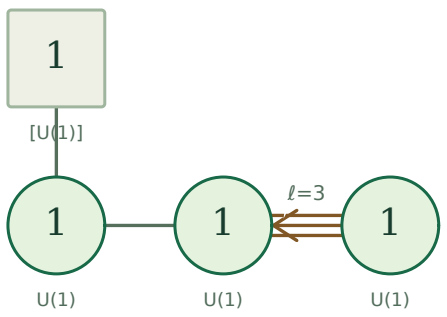 | $1 + 9t^{2} + 20t^{3} + 36t^{4} + 90t^{5} + 156t^{6} + 252t^{7} + 441t^{8} + 670t^{9} + 981t^{10} + 1476t^{11} + 2066t^{12} + 2826t^{13} + 3906t^{14} + 5172t^{15} + 6741t^{16} + 8820t^{17} + 11221t^{18} + 14112t^{19} + 17766t^{20} + 21932t^{21} + 26838t^{22} + 32832t^{23} + 39591t^{24} + 47412t^{25} + 56727t^{26} + 67132t^{27} + 79002t^{28} + 92862t^{29} + 108218t^{30} + O(t^{31})$ |
| **0960 2102.06190-h-nk__n3_k4** / n = 3 · k = 4 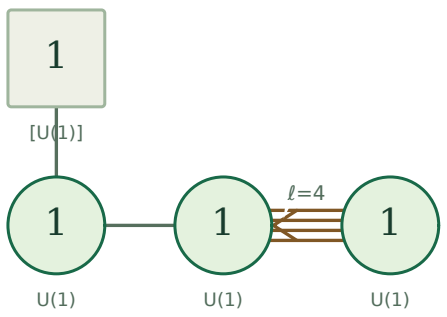 | $1 + 9t^{2} + 66t^{4} + 226t^{6} + 651t^{8} + 1491t^{10} + 3108t^{12} + 5796t^{14} + 10197t^{16} + 16797t^{18} + 26598t^{20} + 40326t^{22} + 59423t^{24} + 84903t^{26} + 118728t^{28} + 162248t^{30} + O(t^{31})$ |
| **0961 2102.06190-h-nk__n4_k1** / n = 4 · k = 1 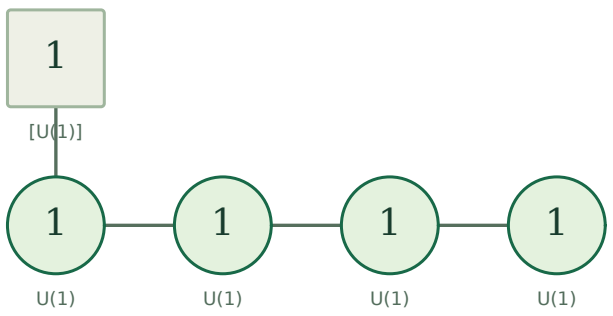 | $1 + 8t + 36t^{2} + 120t^{3} + 330t^{4} + 792t^{5} + 1716t^{6} + 3432t^{7} + 6435t^{8} + 11440t^{9} + 19448t^{10} + 31824t^{11} + 50388t^{12} + 77520t^{13} + 116280t^{14} + 170544t^{15} + 245157t^{16} + 346104t^{17} + 480700t^{18} + 657800t^{19} + 888030t^{20} + 1184040t^{21} + 1560780t^{22} + 2035800t^{23} + 2629575t^{24} + 3365856t^{25} + 4272048t^{26} + 5379616t^{27} + 6724520t^{28} + 8347680t^{29} + 10295472t^{30} + O(t^{31})$ |
| **0962 2102.06190-h-nk__n4_k3** / n = 4 · k = 3 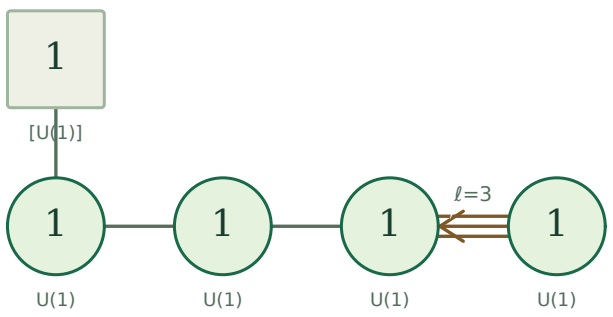 | $1 + 16t^{2} + 40t^{3} + 100t^{4} + 280t^{5} + 568t^{6} + 1120t^{7} + 2185t^{8} + 3800t^{9} + 6436t^{10} + 10688t^{11} + 16766t^{12} + 25760t^{13} + 38900t^{14} + 56792t^{15} + 81593t^{16} + 115592t^{17} + 160140t^{18} + 219080t^{19} + 296346t^{20} + 394536t^{21} + 519996t^{22} + 679080t^{23} + 876315t^{24} + 1121592t^{25} + 1424676t^{26} + 1792912t^{27} + 2241030t^{28} + 2783440t^{29} + 3431428t^{30} + O(t^{31})$ |

**QUIVER** | **HILBERT SERIES**

**0963 2102.06190-h-nk__n4_k4** / n = 4 · k = 4

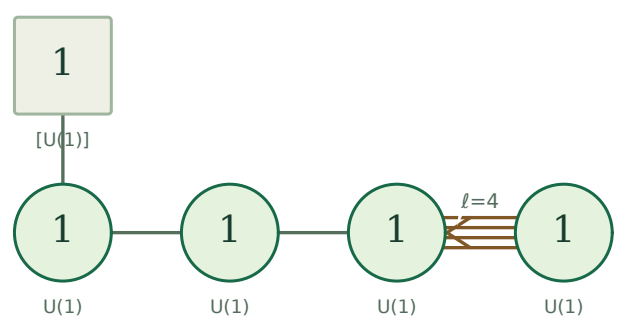


$1 + 16t^{2} + 170t^{4} + 848t^{6} + 3235t^{8} + 9696t^{10} + 25236t^{12} + 58080t^{14} + 122661t^{16} + 240240t^{18} + 444158t^{20} + 780208t^{22} + 1315015t^{24} + 2135744t^{26} + 3362600t^{28} + 5147328t^{30} + O(t^{31})$

**0964 2102.06190-hbar-nk__n2_k4** / n = 2 · k = 4

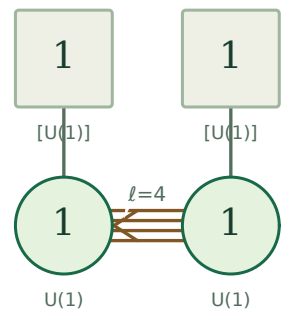


$1 + 4t^{2} + 9t^{4} + 10t^{5} + 16t^{6} + 24t^{7} + 25t^{8} + 42t^{9} + 54t^{10} + 64t^{11} + 89t^{12} + 90t^{13} + 130t^{14} + 146t^{15} + 177t^{16} + 210t^{17} + 230t^{18} + 282t^{19} + 323t^{20} + 362t^{21} + 426t^{22} + 450t^{23} + 539t^{24} + 588t^{25} + 662t^{26} + 738t^{27} + 795t^{28} + 900t^{29} + 988t^{30} + O(t^{31})$

**0965 2102.06190-hbar-nk__n2_k5** / n = 2 · k = 5

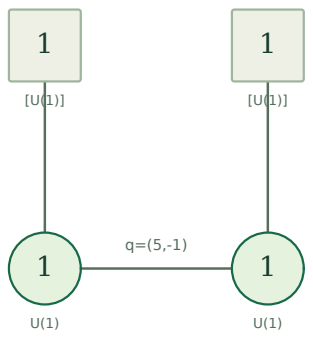


$1 + 4t^{2} + 9t^{4} + 28t^{6} + 53t^{8} + 84t^{10} + 143t^{12} + 212t^{14} + 291t^{16} + 412t^{18} + 547t^{20} + 696t^{22} + 901t^{24} + 1124t^{26} + 1365t^{28} + 1676t^{30} + O(t^{31})$

**0966 2102.06190-hbar-nk__n2_k6** / n = 2 · k = 6

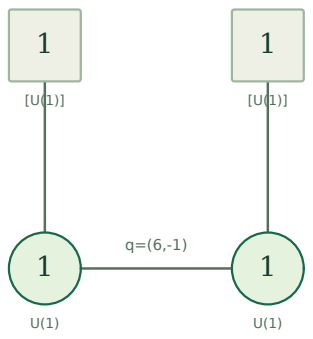


$1 + 4t^{2} + 9t^{4} + 16t^{6} + 14t^{7} + 25t^{8} + 32t^{9} + 36t^{10} + 54t^{11} + 49t^{12} + 80t^{13} + 90t^{14} + 110t^{15} + 137t^{16} + 144t^{17} + 190t^{18} + 182t^{19} + 249t^{20} + 262t^{21} + 314t^{22} + 350t^{23} + 385t^{24} + 446t^{25} + 462t^{26} + 550t^{27} + 595t^{28} + 662t^{29} + 738t^{30} + O(t^{31})$

**0967 2102.06190-hbar-nk__n3_k3** / n = 3 · k = 3

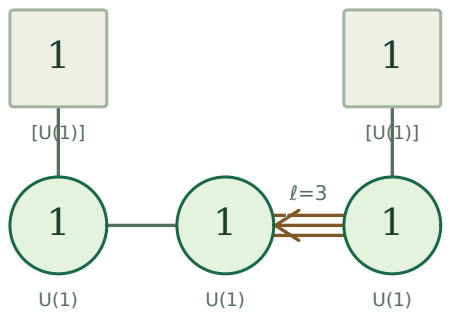


$1 + 9t^{2} + 56t^{4} + 190t^{6} + 533t^{8} + 1217t^{10} + 2514t^{12} + 4682t^{14} + 8203t^{16} + 13503t^{18} + 21334t^{20} + 32332t^{22} + 47579t^{24} + 67963t^{26} + 94956t^{28} + 129740t^{30} + O(t^{31})$

**0968 2102.06190-hbar-nk__n3_k4** / n = 3 · k = 4

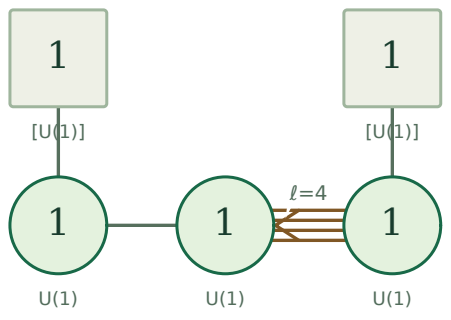


$1 + 9t^{2} + 36t^{4} + 30t^{5} + 100t^{6} + 126t^{7} + 225t^{8} + 336t^{9} + 531t^{10} + 720t^{11} + 1114t^{12} + 1350t^{13} + 2088t^{14} + 2492t^{15} + 3585t^{16} + 4326t^{17} + 5755t^{18} + 7056t^{19} + 9072t^{20} + 10910t^{21} + 13830t^{22} + 16140t^{23} + 20353t^{24} + 23484t^{25} + 28995t^{26} + 33374t^{27} + 40140t^{28} + 46278t^{29} + 54852t^{30} + O(t^{31})$

**0969 2102.06190-hbar-nk__n3_k5** / n = 3 · k = 5

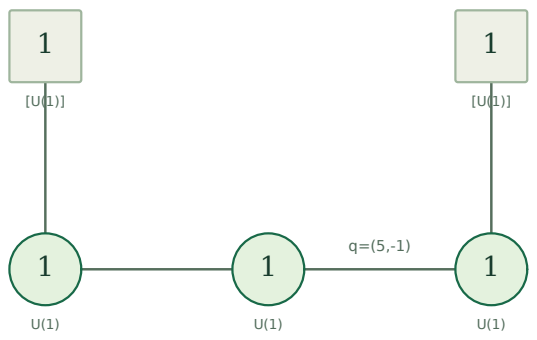


$1 + 9t^{2} + 36t^{4} + 142t^{6} + 393t^{8} + 873t^{10} + 1816t^{12} + 3414t^{14} + 5889t^{16} + 9765t^{18} + 15426t^{20} + 23298t^{22} + 34311t^{24} + 49107t^{26} + 68382t^{28} + 93588t^{30} + O(t^{31})$

| QUIVER | HILBERT SERIES |
| --- | --- |

**0970 2102.06190-hbar-nk__n4_k3** / n = 4 · k = 3

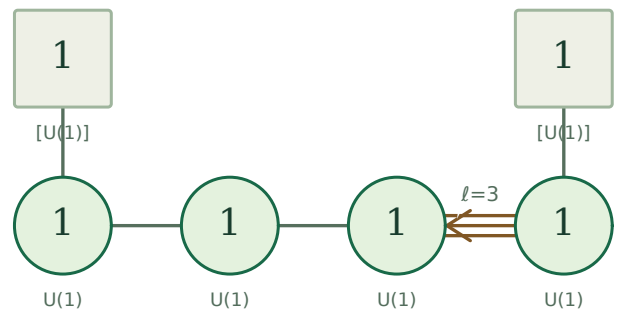


$1 + 16t^{2} + 140t^{4} + 680t^{6} + 2513t^{8} + 7456t^{10} + 19196t^{12} + 43968t^{14} + 92395t^{16} + 180488t^{18} + 332788t^{20} + 583648t^{22} + 982123t^{24} + 1593440t^{26} + 2506136t^{28} + 3833568t^{30} + O(t^{31})$

**0971 2102.06190-hbar-nk__n4_k4** / n = 4 · k = 4

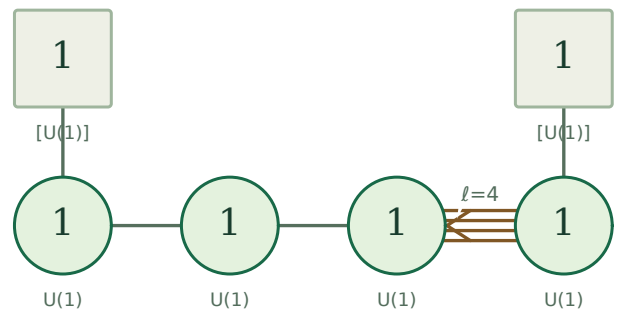


$1 + 16t^{2} + 100t^{4} + 70t^{5} + 400t^{6} + 448t^{7} + 1225t^{8} + 1680t^{9} + 3466t^{10} + 4800t^{11} + 8816t^{12} + 11550t^{13} + 20120t^{14} + 25550t^{15} + 41785t^{16} + 52528t^{17} + 80250t^{18} + 100960t^{19} + 146454t^{20} + 182790t^{21} + 255856t^{22} + 314230t^{23} + 429465t^{24} + 520182t^{25} + 694970t^{26} + 833680t^{27} + 1087970t^{28} + 1297390t^{29} + 1659154t^{30} + O(t^{31})$

**0972 2102.06190-hbar-nk__n4_k5** / n = 4 · k = 5

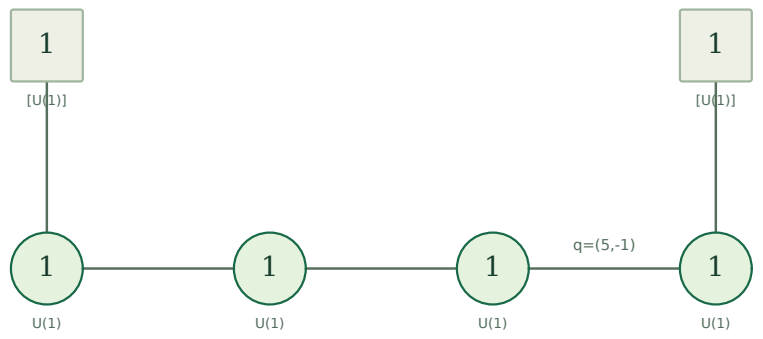


$1 + 16t^{2} + 100t^{4} + 512t^{6} + 1897t^{8} + 5536t^{10} + 14228t^{12} + 32712t^{14} + 68357t^{16} + 133584t^{18} + 246348t^{20} + 431488t^{22} + 725759t^{24} + 1177944t^{26} + 1851084t^{28} + 2831440t^{30} + O(t^{31})$

**0973 2102.06190-C2-l0-3**

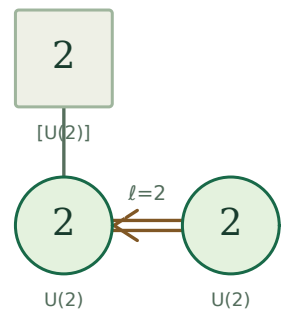


$1 + 10t^{2} + 65t^{4} + O(t^{5})$

**0974 2102.06190-C2-l0-4**

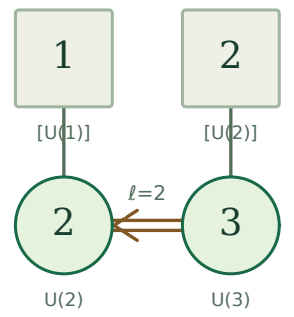


$1 + 10t^{2} + 65t^{4} + O(t^{5})$

**0975 2102.06190-C2-l0-5**

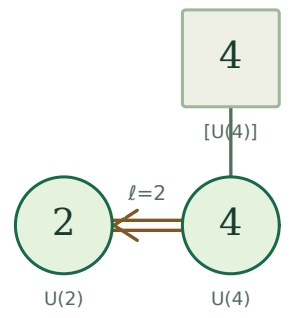


$1 + 10t^{2} + 65t^{4} + O(t^{5})$

**0976 2102.06190-C2-l0-6**

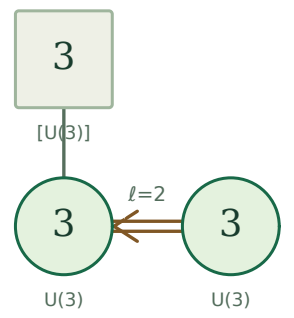


$1 + 10t^{2} + 65t^{4} + 330t^{6} + O(t^{7})$

| QUIVER | HILBERT SERIES |
|---|---|

**0977 2102.06190-C2-l1-2**

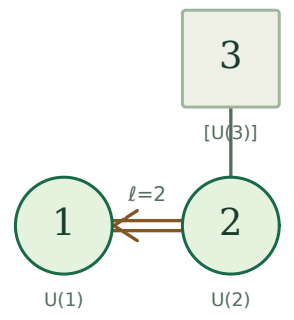


$1 + 4t^{2} + 6t^{3} + O(t^{4})$

**0978 2102.06190-C2-l1-3**

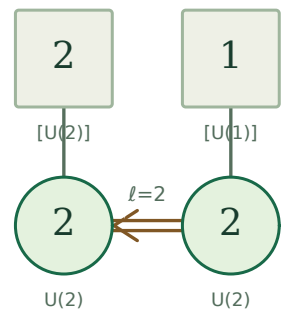


$1 + 4t^{2} + 6t^{3} + 14t^{4} + 30t^{5} + O(t^{6})$

**0979 2102.06190-C2-l1-4**

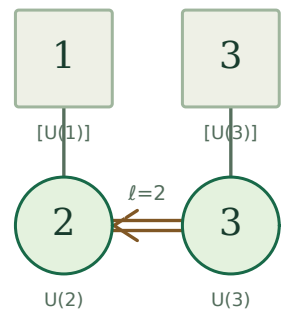


$1 + 4t^{2} + 6t^{3} + 14t^{4} + 30t^{5} + O(t^{6})$

**0980 2102.06190-C2-l1-5**

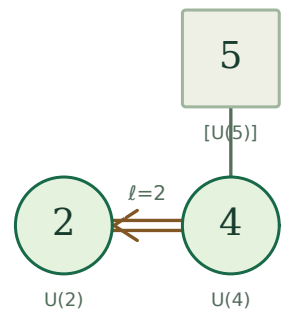


$1 + 4t^{2} + 6t^{3} + 14t^{4} + 30t^{5} + O(t^{6})$

**0981 2102.06190-C2-l1-6**

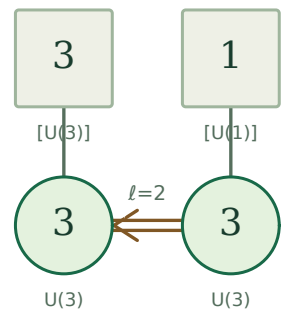


$1 + 4t^{2} + 6t^{3} + 14t^{4} + 30t^{5} + 61t^{6} + 114t^{7} + O(t^{8})$

**0982 2102.06190-C2-equal-rank__x3** / x = 3

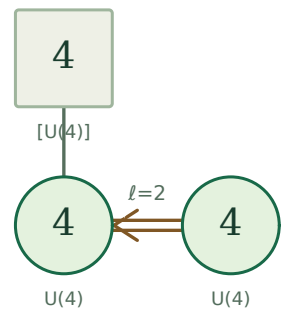


$1 + 10t^{2} + 65t^{4} + 330t^{6} + 1430t^{8} + O(t^{10})$

**0983 2102.06190-C2-equal-rank__x4** / x = 4

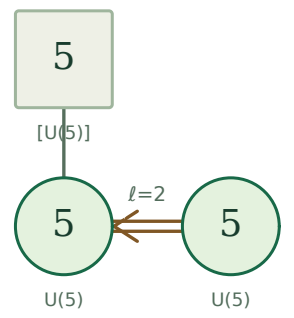


$1 + 10t^{2} + 65t^{4} + 330t^{6} + 1430t^{8} + 5512t^{10} + O(t^{12})$

**0984 2102.06190-C2-equal-rank__x5** / x = 5

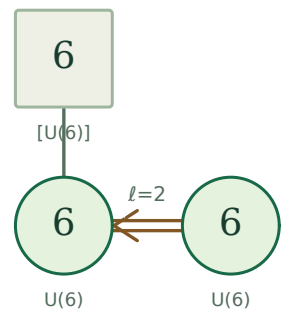


$1 + 10t^2 + 65t^4 + 330t^6 + 1430t^8 + 5512t^{10} + 19415t^{12} + O(t^{14})$

## 2103.05458 · Orthosymplectic Implosions (13)

**0985 2103.05458-D3**

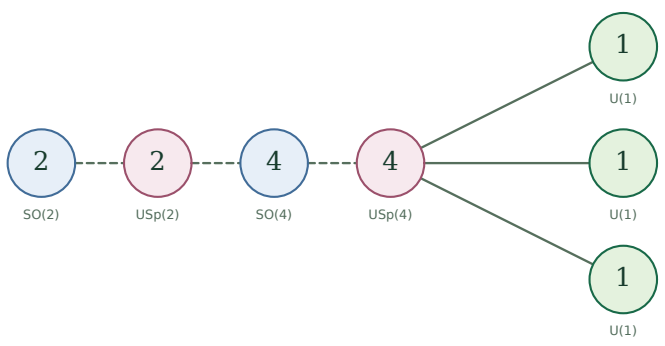


÷ $\mathbb{Z}_2$ diagonal

$1 + 18t^2 + 32t^3 + 206t^4 + 544t^5 + 1993t^6 + 5344t^7 + 15531t^8 + 39040t^9 + 98672t^{10} + O(t^{11})$

**Paper typo:** All explicitly printed terms through $t^{10}$ agree. A nonzero $t^{11}$ term is omitted, so the printed $O(t^{12})$ must become $O(t^{11})$, or the missing term must be restored.

**0986 2103.05458-D4**

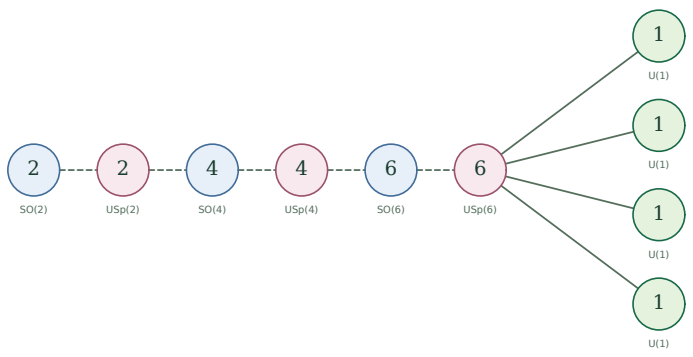

÷ $\mathbb{Z}_2$ diagonal

$1 + 32t^2 + 527t^4 + 6144t^6 + 57782t^8 + 466656t^{10} + 3339801t^{12} + 21550576t^{14} + 126756294t^{16} + 685145416t^{18} + 3426352669t^{20} + O(t^{22})$

**0987 2103.05458-D5**

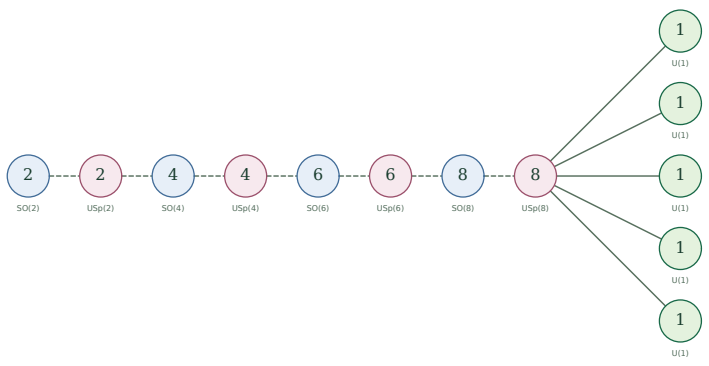

÷ $\mathbb{Z}_2$ diagonal

$1 + 50t^2 + 1274t^4 + 22050t^6 + 291649t^8 + 3145771t^{10} + 28843013t^{12} + 231437604t^{14} + 1660782225t^{16} + 10838595181t^{18} + 65196668471t^{20} + O(t^{22})$

**0988 2103.05458-D6**

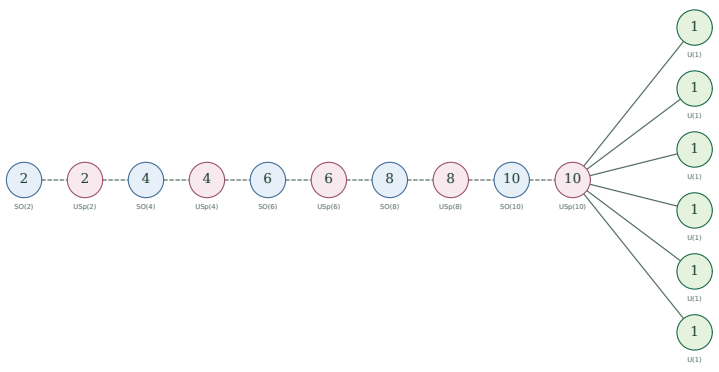

÷ $\mathbb{Z}_2$ diagonal

$1 + 72t^2 + 2627t^4 + 64752t^6 + 1212821t^8 + 18410088t^{10} + 235885925t^{12} + 2623730304t^{14} + 2048t^{15} + 25859417578t^{16} + 145408t^{17} + 229405181384t^{18} + 5232640t^{19} + 1854541984877t^{20} + 127229952t^{21} + O(t^{22})$

**0989 2103.05458-D7**

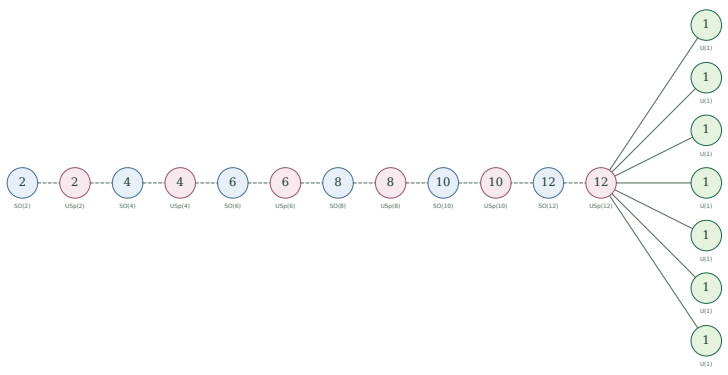

÷ $\mathbb{Z}_2$ diagonal

$1 + 98t^2 + 4850t^4 + 161602t^6 + 4078073t^8 + 83129872t^{10} + 1425752755t^{12} + O(t^{14})$

| QUIVER | HILBERT SERIES |
| --- | --- |

**0990 2103.05458-B2**

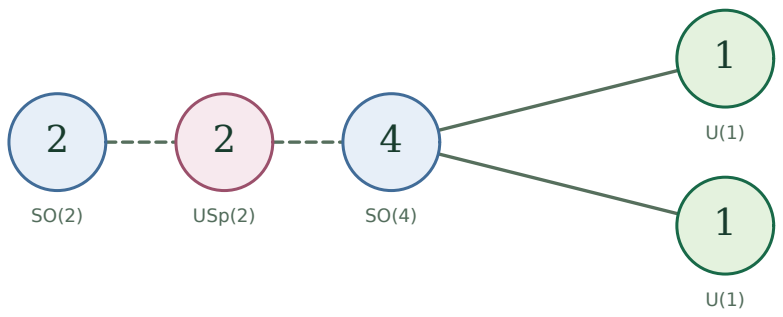


÷ $\mathbb{Z}_2$ diagonal

$1 + 12t^{2} + 16t^{3} + 97t^{4} + 176t^{5} + 612t^{6} + 1168t^{7} + 3054t^{8} + 5728t^{9} + 12640t^{10} + 22768t^{11} + 44842t^{12} + 77312t^{13} + 140220t^{14} + 231936t^{15} + 394955t^{16} + 629232t^{17} + 1019236t^{18} + 1570512t^{19} + 2442031t^{20} + O(t^{21})$

**Paper typo:** All explicitly printed terms through $t^{20}$ agree. A nonzero $t^{21}$ term is omitted, so the printed $O(t^{22})$ must become $O(t^{21})$, or the missing term must be restored.

**0991 2103.05458-B3**

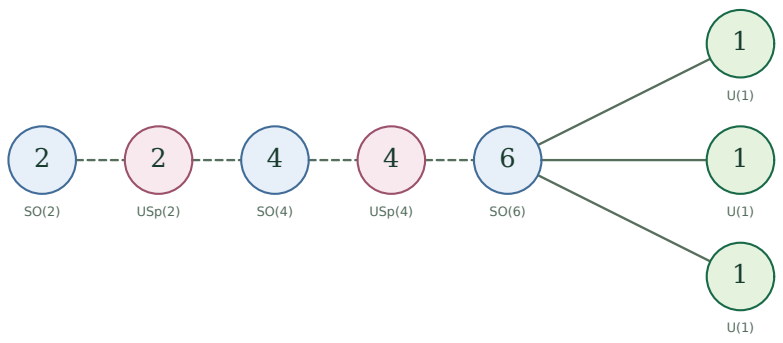


÷ $\mathbb{Z}_2$ diagonal

$1 + 24t^{2} + 299t^{4} + 2682t^{6} + 19687t^{8} + 125058t^{10} + 705840t^{12} + 3592368t^{14} + 16656892t^{16} + 70957310t^{18} + 279781688t^{20} + O(t^{22})$

**0992 2103.05458-B4**

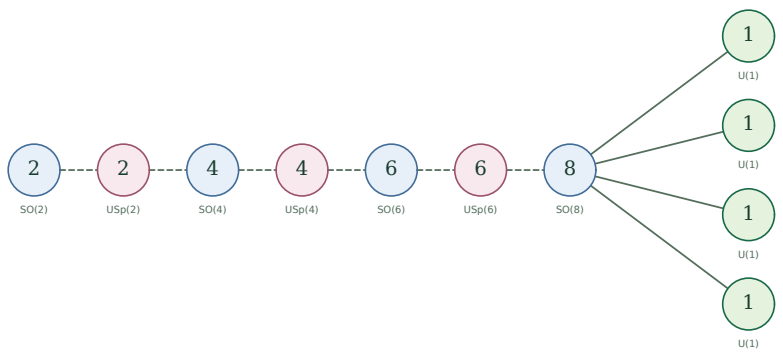


÷ $\mathbb{Z}_2$ diagonal

$1 + 40t^{2} + 819t^{4} + 11440t^{6} + 122661t^{8} + 1077552t^{10} + 8086902t^{12} + 53392192t^{14} + 316944489t^{16} + 1720172104t^{18} + 8648640839t^{20} + O(t^{22})$

**0993 2103.05458-B5**

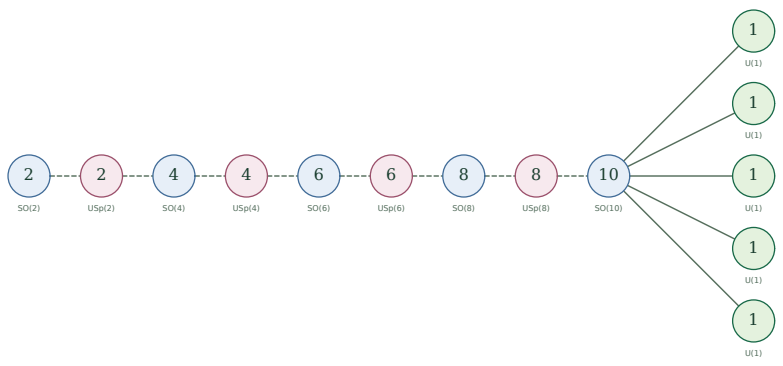


÷ $\mathbb{Z}_2$ diagonal

$1 + 60t^{2} + 1829t^{4} + 37760t^{6} + 593834t^{8} + 7586742t^{10} + 82007875t^{12} + 771321698t^{14} + 1024t^{15} + 6443119425t^{16} + 60416t^{17} + 48554821508t^{18} + 1811456t^{19} + 334196827069t^{20} + 36792320t^{21} + O(t^{22})$

**0994 2103.05458-B6**

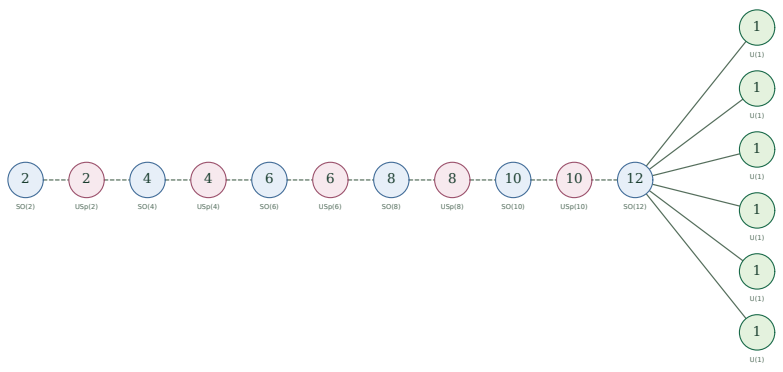


÷ $\mathbb{Z}_2$ diagonal

$1 + 84t^{2} + 3569t^{4} + 102256t^{6} + 2222324t^{8} + 39073328t^{10} + 578877666t^{12} + 7432109337t^{14} + 84404062467t^{16} + 861258507989t^{18} + 7994156173400t^{20} + 4096t^{21} + O(t^{22})$

**Paper typo:** The first explicit discrepancy is a difference of 13 at $t^{12}$. Two implementations agree after checking the source and global form; half-integral sectors cannot contribute at this order, and the original cause is unknown.

**0995 2103.05458-BSp4**

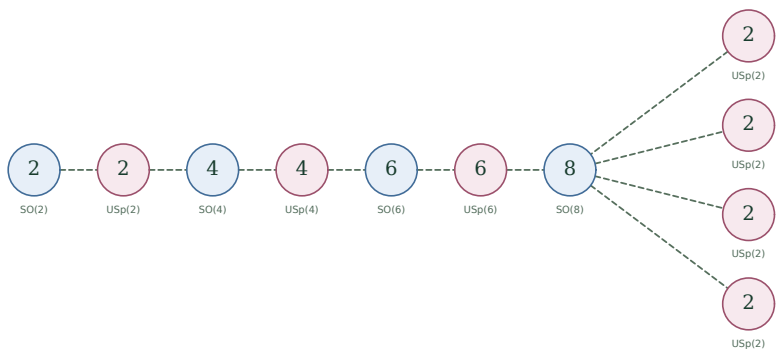


÷ $\mathbb{Z}_2$ diagonal

$1 + 52t^{2} + 1455t^{4} + 28834t^{6} + 449122t^{8} + 5793780t^{10} + 63853945t^{12} + 613989328t^{14} + 5232181818t^{16} + 40010832518t^{18} + 277431116267t^{20} + O(t^{22})$

| QUIVER | HILBERT SERIES |
| --- | --- |

**0996 2103.05458-BSp5**

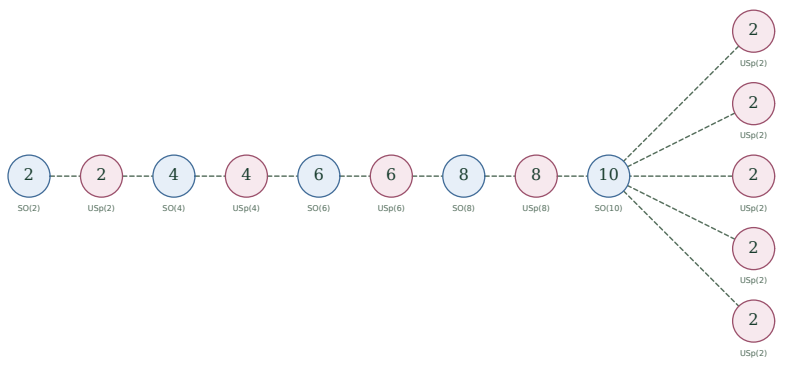

÷ $\mathbb{Z}_2$ diagonal

$1 + 55t^{2} + 1544t^{4} + 32t^{5} + 29535t^{6} + 1888t^{7} + 433464t^{8} + 56608t^{9} + 5209798t^{10} + 1151360t^{11} + 53454368t^{12} + 17885120t^{13} + 482022542t^{14} + 226457760t^{15} + 3904141695t^{16} + 2435618944t^{17} + 28892424245t^{18} + 22895407232t^{19} + 198088208252t^{20} + 192087708320t^{21} + O(t^{22})$

**0997 2103.05458-BSp6**

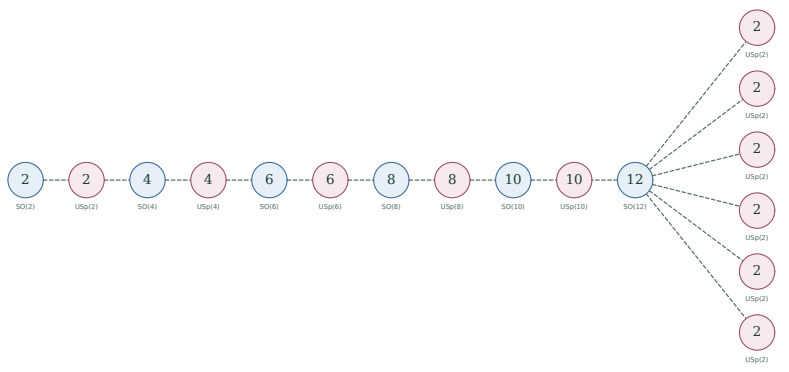

÷ $\mathbb{Z}_2$ diagonal

$1 + 78t^{2} + 3086t^{4} + 82550t^{6} + 1679237t^{8} + 64t^{9} + 27703312t^{10} + 5312t^{11} + 386049641t^{12} + 223040t^{13} + 4673378944t^{14} + 6315904t^{15} + 50165955517t^{16} + 135683648t^{17} + 485045344353t^{18} + 2358618624t^{19} + 4276770597522t^{20} + 34556791744t^{21} + O(t^{22})$

## 2105.08064 · Conformal Manifolds and 3d Mirrors of Argyres-Douglas theories (18)

**0998 2105.08064-pair-SO__m1** / m = 1

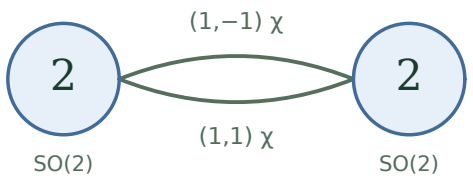


÷ $\mathbb{Z}_2$ diagonal

$1 + 4t + 10t^{2} + 20t^{3} + 35t^{4} + 56t^{5} + 84t^{6} + 120t^{7} + 165t^{8} + 220t^{9} + 286t^{10} + 364t^{11} + 455t^{12} + 560t^{13} + 680t^{14} + 816t^{15} + 969t^{16} + 1140t^{17} + 1330t^{18} + 1540t^{19} + 1771t^{20} + 2024t^{21} + 2300t^{22} + 2600t^{23} + 2925t^{24} + 3276t^{25} + 3654t^{26} + 4060t^{27} + 4495t^{28} + 4960t^{29} + 5456t^{30} + O(t^{31})$

**0999 2105.08064-pair-SO__m2** / m = 2

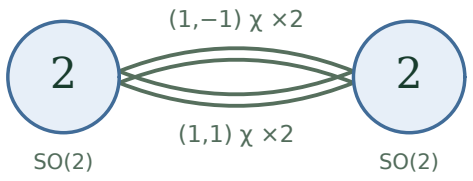


÷ $\mathbb{Z}_2$ diagonal

$1 + 6t^{2} + 19t^{4} + 44t^{6} + 85t^{8} + 146t^{10} + 231t^{12} + 344t^{14} + 489t^{16} + 670t^{18} + 891t^{20} + 1156t^{22} + 1469t^{24} + 1834t^{26} + 2255t^{28} + 2736t^{30} + O(t^{31})$

**1000 2105.08064-pair-SO__m3** / m = 3

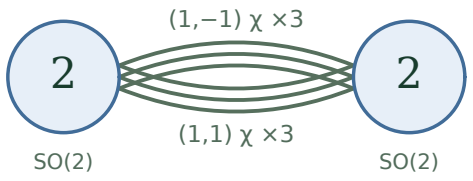


÷ $\mathbb{Z}_2$ diagonal

$1 + 2t^{2} + 4t^{3} + 3t^{4} + 8t^{5} + 12t^{6} + 12t^{7} + 21t^{8} + 28t^{9} + 30t^{10} + 44t^{11} + 55t^{12} + 60t^{13} + 80t^{14} + 96t^{15} + 105t^{16} + 132t^{17} + 154t^{18} + 168t^{19} + 203t^{20} + 232t^{21} + 252t^{22} + 296t^{23} + 333t^{24} + 360t^{25} + 414t^{26} + 460t^{27} + 495t^{28} + 560t^{29} + 616t^{30} + O(t^{31})$

**1001 2105.08064-pair-U__m2** / m = 2

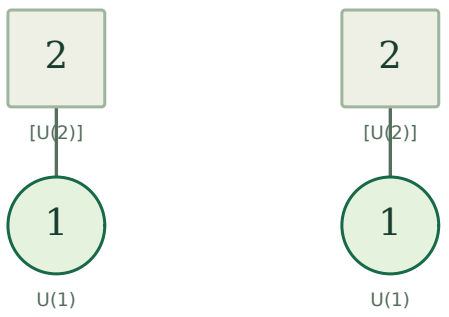


$1 + 6t^{2} + 19t^{4} + 44t^{6} + 85t^{8} + 146t^{10} + 231t^{12} + 344t^{14} + 489t^{16} + 670t^{18} + 891t^{20} + 1156t^{22} + 1469t^{24} + 1834t^{26} + 2255t^{28} + 2736t^{30} + O(t^{31})$

| QUIVER | HILBERT SERIES |
|---|---|

**1002 2105.08064-pair-U__m3** / m = 3

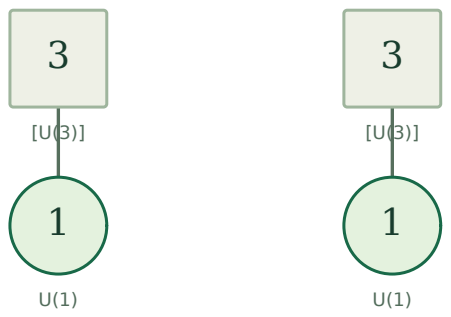


$1 + 2t^{2} + 4t^{3} + 3t^{4} + 8t^{5} + 12t^{6} + 12t^{7} + 21t^{8} + 28t^{9} + 30t^{10} + 44t^{11} + 55t^{12} + 60t^{13} + 80t^{14} + 96t^{15} + 105t^{16} + 132t^{17} + 154t^{18} + 168t^{19} + 203t^{20} + 232t^{21} + 252t^{22} + 296t^{23} + 333t^{24} + 360t^{25} + 414t^{26} + 460t^{27} + 495t^{28} + 560t^{29} + 616t^{30} + O(t^{31})$

**1003 2105.08064-pair-U__m4** / m = 4

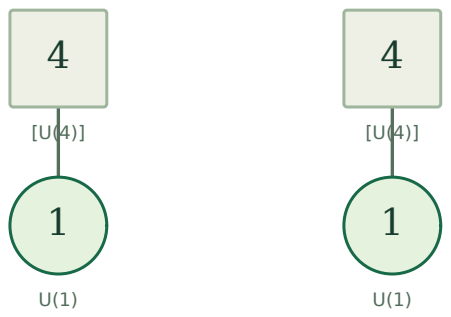


$1 + 2t^{2} + 7t^{4} + 12t^{6} + 25t^{8} + 38t^{10} + 63t^{12} + 88t^{14} + 129t^{16} + 170t^{18} + 231t^{20} + 292t^{22} + 377t^{24} + 462t^{26} + 575t^{28} + 688t^{30} + O(t^{31})$

**1004 2105.08064-triangle-SO__m1** / m = 1

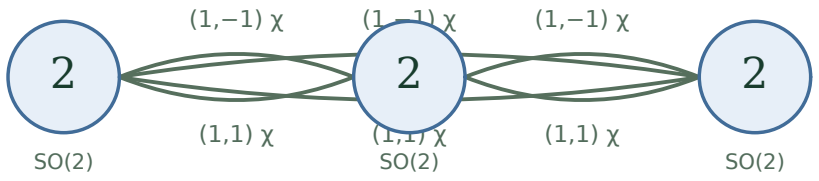


÷ $\mathbb{Z}_2$ diagonal

**Computed by Quiver Lab**

$1 + 3t^{2} + 8t^{3} + 12t^{4} + 24t^{5} + 48t^{6} + 72t^{7} + 117t^{8} + 184t^{9} + 267t^{10} + 384t^{11} + 548t^{12} + 744t^{13} + 1008t^{14} + 1344t^{15} + 1749t^{16} + 2256t^{17} + 2887t^{18} + 3624t^{19} + 4524t^{20} + 5600t^{21} + 6852t^{22} + 8328t^{23} + 10065t^{24} + 12048t^{25} + 14355t^{26} + 17008t^{27} + 20016t^{28} + 23448t^{29} + 27356t^{30} + O(t^{31})$

Source originally Claimed · no numerical literature comparison is available.

**1005 2105.08064-triangle-SO__m2** / m = 2

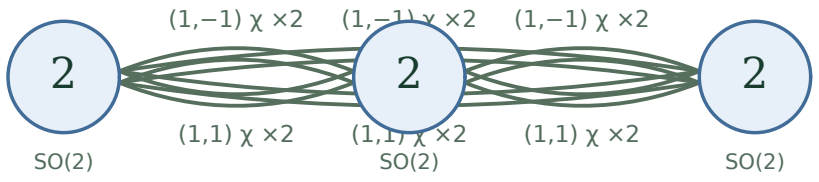


÷ $\mathbb{Z}_2$ diagonal

**Computed by Quiver Lab**

$1 + 3t^{2} + 6t^{4} + 18t^{6} + 45t^{8} + 87t^{10} + 164t^{12} + 300t^{14} + 501t^{16} + 799t^{18} + 1242t^{20} + 1854t^{22} + 2685t^{24} + 3807t^{26} + 5268t^{28} + 7148t^{30} + O(t^{31})$

Source originally Claimed · no numerical literature comparison is available.

**1006 2105.08064-triangle-SO__m3** / m = 3

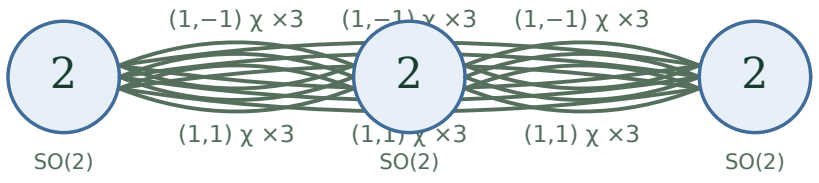


÷ $\mathbb{Z}_2$ diagonal

**Computed by Quiver Lab**

$1 + 3t^{2} + 6t^{4} + 10t^{6} + 15t^{8} + 8t^{9} + 21t^{10} + 24t^{11} + 34t^{12} + 48t^{13} + 54t^{14} + 80t^{15} + 81t^{16} + 120t^{17} + 135t^{18} + 168t^{19} + 216t^{20} + 248t^{21} + 324t^{22} + 360t^{23} + 465t^{24} + 504t^{25} + 639t^{26} + 712t^{27} + 846t^{28} + 984t^{29} + 1134t^{30} + O(t^{31})$

Source originally Claimed · no numerical literature comparison is available.

**1007 2105.08064-twoflavour-SO__N2** / N = 2

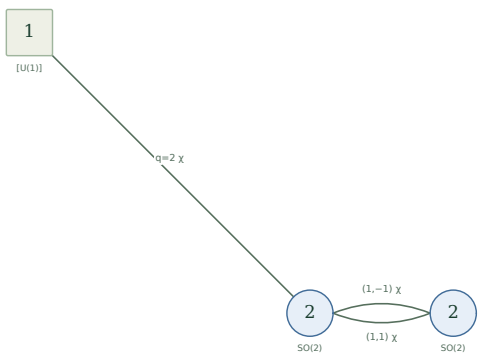

÷ $\mathbb{Z}_2$ diagonal

**Computed by Quiver Lab**

$1 + 8t^{2} + 27t^{4} + 64t^{6} + 125t^{8} + 216t^{10} + 343t^{12} + 512t^{14} + 729t^{16} + 1000t^{18} + 1331t^{20} + 1728t^{22} + 2197t^{24} + 2744t^{26} + 3375t^{28} + 4096t^{30} + O(t^{31})$

Source originally Claimed · no numerical literature comparison is available.

| QUIVER | HILBERT SERIES |
| --- | --- |

**1008 2105.08064-twoflavour-SO__N3 / N = 3**

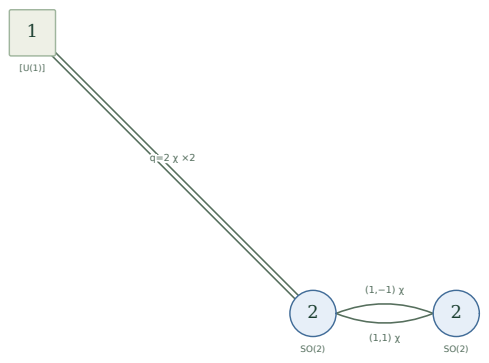

÷ $\mathbb{Z}_2$ diagonal

**Computed by Quiver Lab**

$$1 + 4t^2 + 4t^3 + 9t^4 + 12t^5 + 22t^6 + 24t^7 + 41t^8 + 48t^9 + 66t^{10} + 80t^{11} + 107t^{12} + 120t^{13} + 158t^{14} + 180t^{15} + 219t^{16} + 252t^{17} + 304t^{18} + 336t^{19} + 403t^{20} + 448t^{21} + 516t^{22} + 576t^{23} + 661t^{24} + 720t^{25} + 824t^{26} + 900t^{27} + 1005t^{28} + 1100t^{29} + 1226t^{30} + O(t^{31})$$

Source originally Claimed · no numerical literature comparison is available.

**1009 2105.08064-twoflavour-SO__N4 / N = 4**

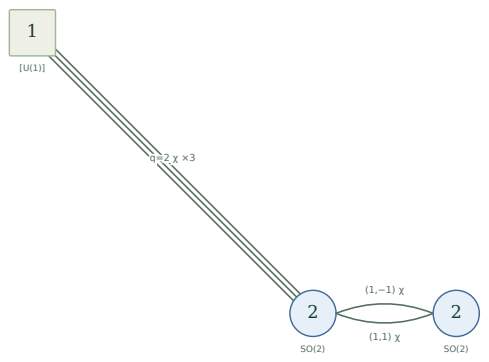

÷ $\mathbb{Z}_2$ diagonal

**Computed by Quiver Lab**

$$1 + 4t^2 + 13t^4 + 28t^6 + 55t^8 + 92t^{10} + 147t^{12} + 216t^{14} + 309t^{16} + 420t^{18} + 561t^{20} + 724t^{22} + 923t^{24} + 1148t^{26} + 1415t^{28} + 1712t^{30} + O(t^{31})$$

Source originally Claimed · no numerical literature comparison is available.

**1010 2105.08064-DSO6-k1-SO**

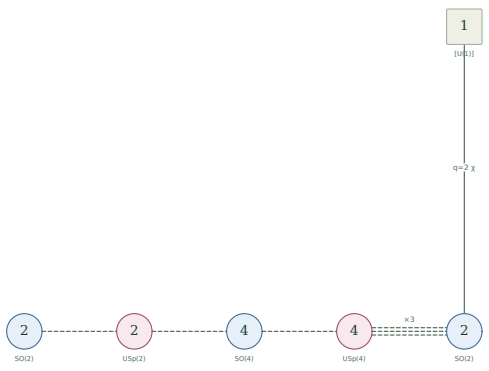

÷ $\mathbb{Z}_2$ diagonal

$$1 + 16t^2 + 143t^4 + O(t^6)$$

**1011 2105.08064-DSO6-k1-U**

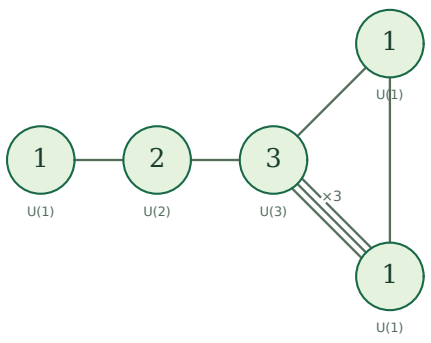


÷ U(1) diagonal

$$1 + 16t^2 + 143t^4 + O(t^6)$$

**1012 2105.08064-DSO6-k2-SO**

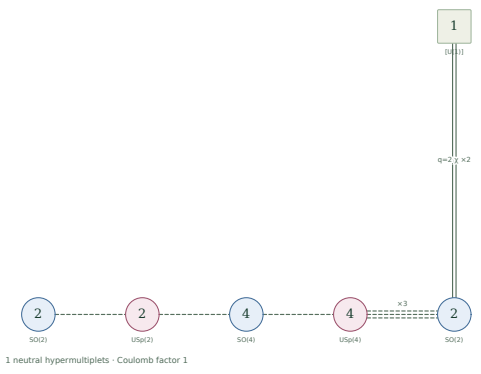

Overall diagonal $\mathbb{Z}_2$ quotient · integral and half-integral magnetic sectors · 1 neutral hypermultiplets · Coulomb factor 1

$$1 + 16t^2 + 135t^4 + 8t^5 + O(t^6)$$

**1013 2105.08064-DSO6-k2-U**

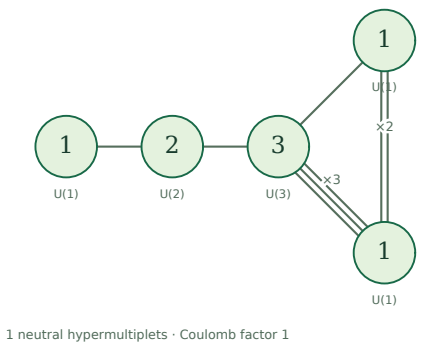


÷ U(1) diagonal · 1 neutral hypermultiplets · Coulomb factor 1

$$1 + 16t^2 + 135t^4 + 8t^5 + O(t^6)$$

| QUIVER | HILBERT SERIES |
| --- | --- |

**1014 2105.08064-DSO6-k3-SO**

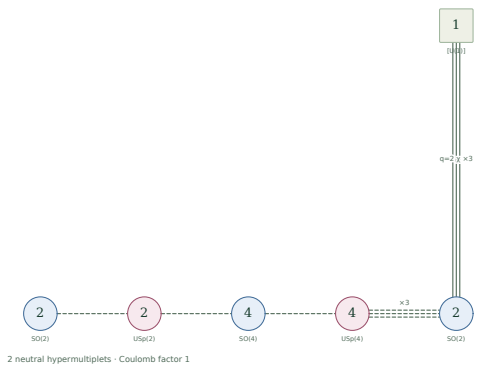

Overall diagonal $\mathbb{Z}_2$ quotient · integral and half-integral magnetic sectors · 2 neutral hypermultiplets · Coulomb factor 1

$1 + 16t^2 + 135t^4 + O(t^6)$

**1015 2105.08064-DSO6-k3-U**

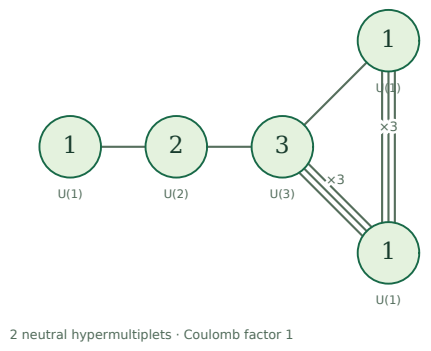


÷ U(1) diagonal · 2 neutral hypermultiplets · Coulomb factor 1

$1 + 16t^2 + 135t^4 + O(t^6)$

## 2107.00754 · Folding orthosymplectic quivers (88)

**1016 2107.00754-example2-OSp**

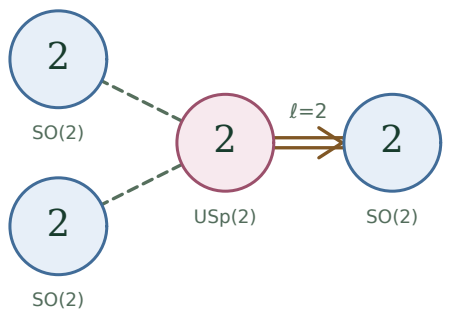


÷ $\mathbb{Z}_2$ diagonal

$1 + 21t^2 + 168t^4 + 825t^6 + 3003t^8 + 8918t^{10} + 22848t^{12} + 52326t^{14} + 109725t^{16} + 214291t^{18} + 394680t^{20} + 692055t^{22} + 1163799t^{24} + 1887900t^{26} + 2968064t^{28} + 4539612t^{30} + O(t^{31})$

**1017 2107.00754-example3-OSp**

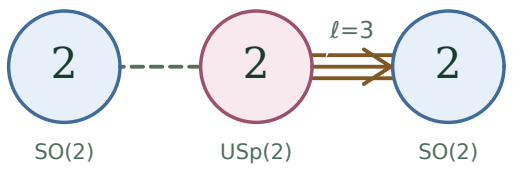


÷ $\mathbb{Z}_2$ diagonal

$1 + 14t^2 + 77t^4 + 273t^6 + 748t^8 + 1729t^{10} + 3542t^{12} + 6630t^{14} + 11571t^{16} + 19096t^{18} + 30107t^{20} + 45695t^{22} + 67158t^{24} + 96019t^{26} + 134044t^{28} + 183260t^{30} + O(t^{31})$

**1018 2107.00754-example4-U**

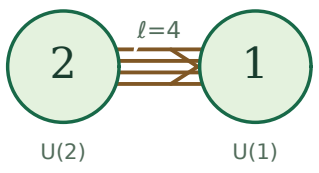


÷ U(1) diagonal

$1 + 8t^2 + 27t^4 + 64t^6 + 125t^8 + 216t^{10} + 343t^{12} + 512t^{14} + 729t^{16} + 1000t^{18} + 1331t^{20} + 1728t^{22} + 2197t^{24} + 2744t^{26} + 3375t^{28} + 4096t^{30} + O(t^{31})$

**1019 2107.00754-example4-OSp**

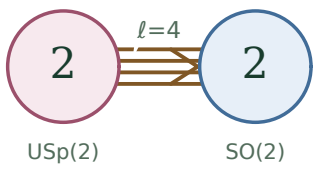


÷ $\mathbb{Z}_2$ diagonal

$1 + 8t^2 + 27t^4 + 64t^6 + 125t^8 + 216t^{10} + 343t^{12} + 512t^{14} + 729t^{16} + 1000t^{18} + 1331t^{20} + 1728t^{22} + 2197t^{24} + 2744t^{26} + 3375t^{28} + 4096t^{30} + O(t^{31})$

| QUIVER | HILBERT SERIES |
|---|---|

**1020 2107.00754-example6-U**

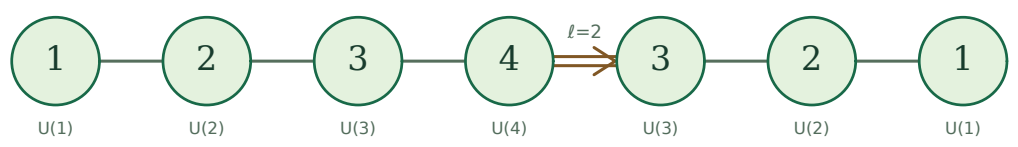


÷ U(1) diagonal

$1 + 30t^{2} + 80t^{3} + 584t^{4} + 2272t^{5} + 10279t^{6} + 38368t^{7} + 144475t^{8} + 491920t^{9} + 1631913t^{10} + 5083840t^{11} + 15298014t^{12} + 43954368t^{13} + 122087710t^{14} + 326672848t^{15} + 847435540t^{16} + 2130112112t^{17} + 5207287099t^{18} + 12386692096t^{19} + 28739898200t^{20} + 65098298896t^{21} + 144199194051t^{22} + 312651968240t^{23} + 664418400183t^{24} + 1385107838976t^{25} + 2835629028958t^{26} + 5705392404464t^{27} + 11292095461816t^{28} + 22000212594272t^{29} + 42224650188739t^{30} + O(t^{31})$

**1021 2107.00754-example6-OSp**

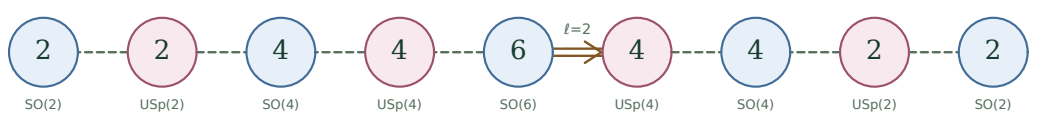


÷ $\mathbb{Z}_2$ diagonal

$1 + 30t^{2} + 80t^{3} + 584t^{4} + 2272t^{5} + 10279t^{6} + 38368t^{7} + 144475t^{8} + 491920t^{9} + 1631913t^{10} + 5083840t^{11} + 15298014t^{12} + 43954368t^{13} + 122087710t^{14} + 326672848t^{15} + 847435540t^{16} + 2130112112t^{17} + 5207287099t^{18} + 12386692096t^{19} + 28739898200t^{20} + 65098298896t^{21} + 144199194051t^{22} + 312651968240t^{23} + 664418400183t^{24} + 1385107838976t^{25} + 2835629028958t^{26} + 5705392404464t^{27} + 11292095461816t^{28} + 22000212594272t^{29} + 42224650188739t^{30} + O(t^{31})$

**1022 2107.00754-example7-U**

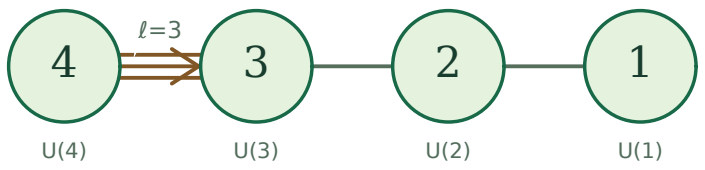


÷ U(1) diagonal

$1 + 15t^{2} + 40t^{3} + 170t^{4} + 560t^{5} + 1834t^{6} + 5168t^{7} + 14600t^{8} + 37480t^{9} + 93308t^{10} + 219760t^{11} + 500255t^{12} + 1090920t^{13} + 2309184t^{14} + 4723432t^{15} + 9405734t^{16} + 18215408t^{17} + 34451225t^{18} + 63639280t^{19} + 115149075t^{20} + 204164528t^{21} + 355443412t^{22} + 607988520t^{23} + 1023280599t^{24} + 1695677576t^{25} + 2769748298t^{26} + 4462138376t^{27} + 7096549344t^{28} + 11147892088t^{29} + 17310084300t^{30} + O(t^{31})$

**1023 2107.00754-example7-OSp**

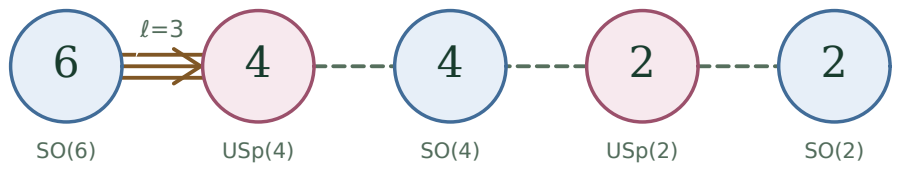


Overall diagonal $\mathbb{Z}_2$ quotient · integral and half-integral magnetic sectors

$1 + 15t^{2} + 40t^{3} + 170t^{4} + 560t^{5} + 1834t^{6} + 5168t^{7} + 14600t^{8} + 37480t^{9} + 93308t^{10} + 219760t^{11} + 500255t^{12} + 1090920t^{13} + 2309184t^{14} + 4723432t^{15} + 9405734t^{16} + 18215408t^{17} + 34451225t^{18} + 63639280t^{19} + 115149075t^{20} + 204164528t^{21} + 355443412t^{22} + 607988520t^{23} + 1023280599t^{24} + 1695677576t^{25} + 2769748298t^{26} + 4462138376t^{27} + 7096549344t^{28} + 11147892088t^{29} + 17310084300t^{30} + O(t^{31})$

**1024 2107.00754-foldedeven__n2** / n = 2

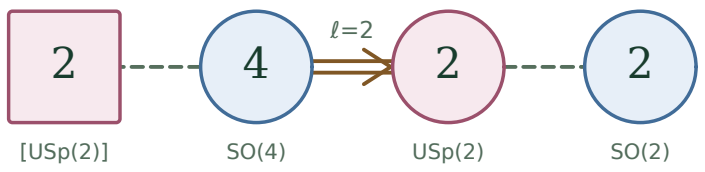


$1 + 15t^{2} + 104t^{4} + 475t^{6} + 1659t^{8} + 4802t^{10} + 12096t^{12} + 27378t^{14} + 56925t^{16} + 110473t^{18} + 202488t^{20} + 353717t^{22} + 593047t^{24} + 959700t^{26} + 1505792t^{28} + 2299284t^{30} + O(t^{31})$

**1025 2107.00754-foldedeven__n3** / n = 3

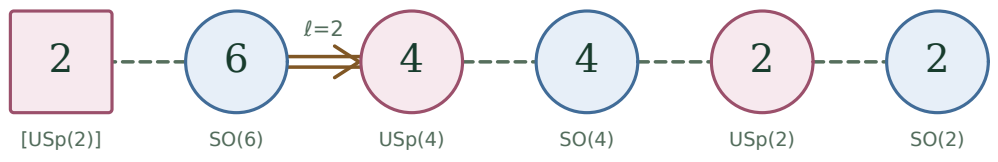


$1 + 35t^{2} + 594t^{4} + 6545t^{6} + 53129t^{8} + 341810t^{10} + 1828631t^{12} + 8415585t^{14} + 34162850t^{16} + 124699530t^{18} + 415477084t^{20} + 1278866250t^{22} + 3672384106t^{24} + 9917924380t^{26} + 25361162266t^{28} + 61753523170t^{30} + O(t^{31})$

| QUIVER | HILBERT SERIES |
|---|---|

**1026 2107.00754-foldedeven__n4** / n = 4

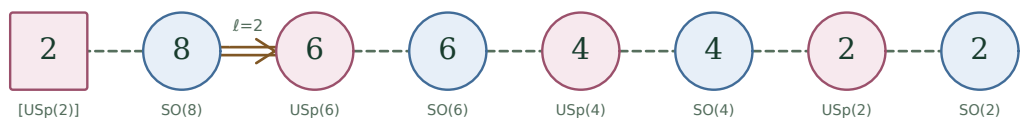


$1 + 63t^{2} + 1952t^{4} + 39711t^{6} + 597617t^{8} + 7109907t^{10} + 69808641t^{12} + 583160451t^{14} + 4240678926t^{16} + 27328486207t^{18} + 158333310185t^{20} + 834515097935t^{22} + 4040954681973t^{24} + 18127810474571t^{26} + 75878364607212t^{28} + 298179391866411t^{30} + O(t^{31})$

**1027 2107.00754-smallerorbitseven__k1_n3** / k = 1 · n = 3

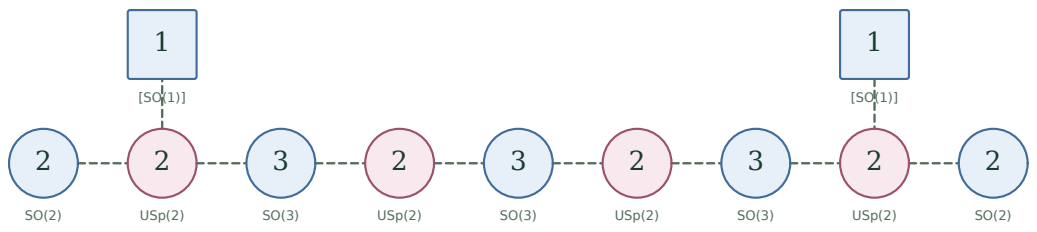


$1 + 66t^{2} + 1638t^{4} + 23100t^{6} + 222156t^{8} + 1613898t^{10} + 9447438t^{12} + 46562373t^{14} + 199377750t^{16} + 759230472t^{18} + 2617486872t^{20} + 8284996512t^{22} + 24347884704t^{24} + 67041815400t^{26} + 174263649912t^{28} + 430295153574t^{30} + O(t^{31})$

**1028 2107.00754-smallerorbitseven__k1_n4** / k = 1 · n = 4

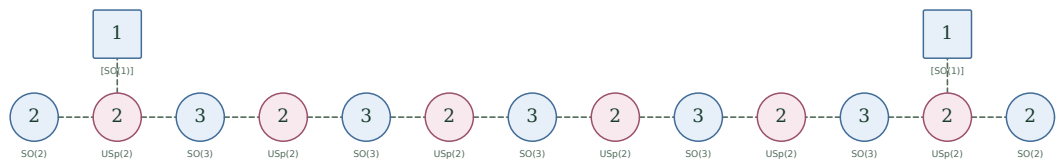

$1 + 120t^{2} + 5304t^{4} + 129675t^{6} + 2102100t^{8} + 25049024t^{10} + 234436800t^{12} + 1803991176t^{14} + 11799609900t^{16} + 67271339200t^{18} + 340870385856t^{20} + 1559094662580t^{22} + 6518164602264t^{24} + 25166658696000t^{26} + 90511667240000t^{28} + 305427205898100t^{30} + O(t^{31})$

**1029 2107.00754-smallerorbitseven__k1_n5** / k = 1 · n = 5

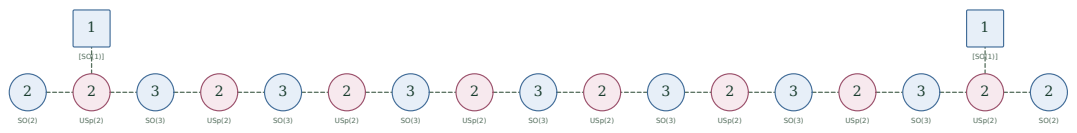

$1 + 190t^{2} + 13090t^{4} + 490314t^{6} + 11967150t^{8} + 211100526t^{10} + 2877823410t^{12} + 31767444060t^{14} + 293831778284t^{16} + 2337298236350t^{18} + 16318234210850t^{20} + 101638141829250t^{22} + 572319913551750t^{24} + 2945783209827150t^{26} + 13987926442487250t^{28} + 61758814626375525t^{30} + O(t^{31})$

**1030 2107.00754-smallerorbitseven__k2_n3** / k = 2 · n = 3

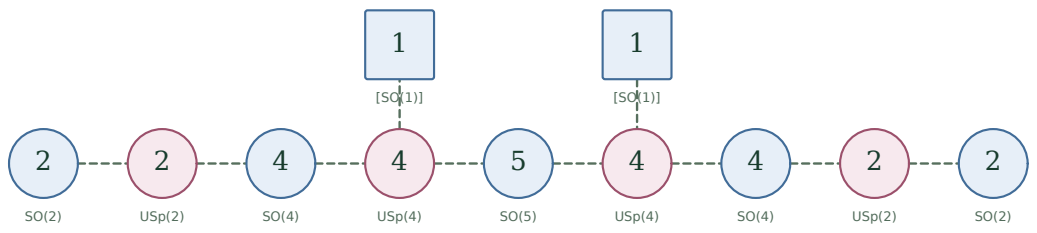


$1 + 66t^{2} + 2133t^{4} + 44121t^{6} + 656667t^{8} + 7544394t^{10} + 70199051t^{12} + 547856023t^{14} + 3682768464t^{16} + 21771081552t^{18} + 115079087904t^{20} + 551305855179t^{22} + 2420555949735t^{24} + 9831435299090t^{26} + 37232689377015t^{28} + 132361603081125t^{30} + O(t^{31})$

**1031 2107.00754-smallerorbitseven__k2_n4** / k = 2 · n = 4

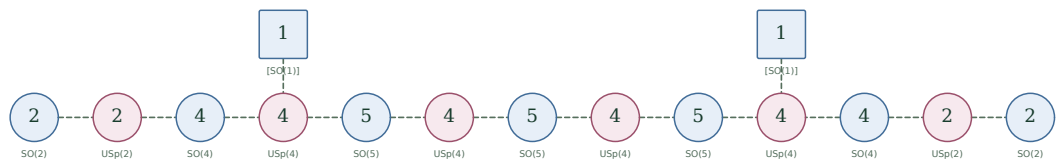

$1 + 120t^{2} + 7124t^{4} + 271047t^{6} + 7387620t^{8} + 154155456t^{10} + 2577219924t^{12} + 35694015201t^{14} + 420187051050t^{16} + 4290719033640t^{18} + 38638692550731t^{20} + 311052924254586t^{22} + 2264227733632540t^{24} + 15048073367304000t^{26} + 92069541164361500t^{28} + 522320251886076150t^{30} + O(t^{31})$

**1032 2107.00754-smallerorbitseven__k2_n5** / k = 2 · n = 5

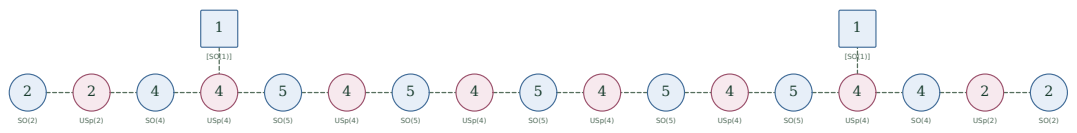

$1 + 190t^{2} + 17935t^{4} + 1082829t^{6} + 46614125t^{8} + 1526831982t^{10} + 39786415130t^{12} + 852350907254t^{14} + 15397486427444t^{16} + 239326687369500t^{18} + 3253736604652300t^{20} + 39223793818121250t^{22} + 424115914632272694t^{24} + 4153787717962721244t^{26} + 37161829656657424555t^{28} + 305937257452638544467t^{30} + O(t^{31})$

| QUIVER | HILBERT SERIES |
| --- | --- |

**1033 2107.00754-smallerorbitseven__k3_n4** / **k = 3 · n = 4**

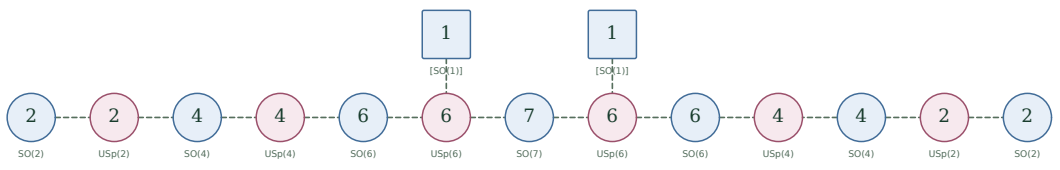

$1 + 120t^{2} + 7124t^{4} + 279055t^{6} + 8103660t^{8} + 185847246t^{10} + 3503515924t^{12} + 55833753259t^{14} + 768176387166t^{16} + 9276341165220t^{18} + 99652329885899t^{20} + 963086527970297t^{22} + 8453678576351336t^{24} + 67949594048770948t^{26} + 503712774883619875t^{28} + 3465359460566552222t^{30} + O(t^{31})$

**1034 2107.00754-smallerorbitseven__k3_n5** / **k = 3 · n = 5**

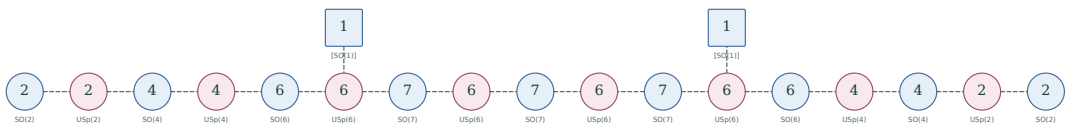

$1 + 190t^{2} + 17935t^{4} + 1121589t^{6} + 52156805t^{8} + 1918710117t^{10} + 58040887810t^{12} + 1482969123614t^{14} + 32650100915982t^{16} + 629222065863385t^{18} + 10750677808892872t^{20} + 164594045434941815t^{22} + 2278694158452689469t^{24} + 28751678727847257249t^{26} + 332910791495861291071t^{28} + 3558859763280765582422t^{30} + O(t^{31})$

**1035 2107.00754-smallerorbitseven__k3_n6** / **k = 3 · n = 6**

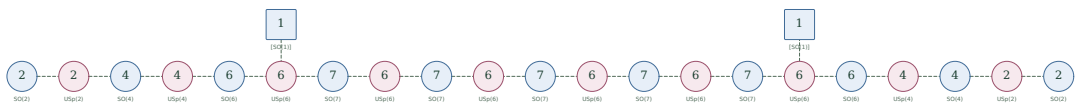

$1 + 276t^{2} + 37926t^{4} + 3459775t^{6} + 234997578t^{8} + 12631089699t^{10} + 557956043820t^{12} + 20790859185395t^{14} + 666335062867448t^{16} + 18650203615683738t^{18} + 461577497083919466t^{20} + 10206970474130480492t^{22} + 203469880246555396050t^{24} + 3684666882594299143708t^{26} + 61027042728032144359234t^{28} + 929958385873819402939647t^{30} + O(t^{31})$

**1036 2107.00754-smallfoldedeven__k1_n2** / **k = 1 · n = 2**

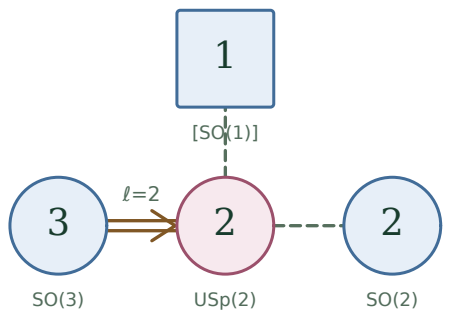


$1 + 15t^{2} + 84t^{4} + 300t^{6} + 825t^{8} + 1911t^{10} + 3920t^{12} + 7344t^{14} + 12825t^{16} + 21175t^{18} + 33396t^{20} + 50700t^{22} + 74529t^{24} + 106575t^{26} + 148800t^{28} + 203456t^{30} + O(t^{31})$

**1037 2107.00754-smallfoldedeven__k1_n3** / **k = 1 · n = 3**

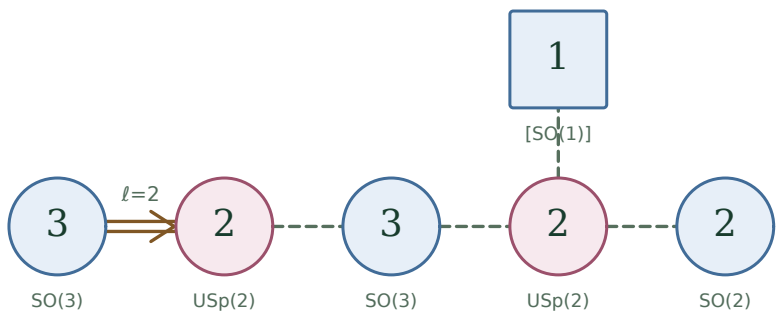


$1 + 35t^{2} + 405t^{4} + 2695t^{6} + 12740t^{8} + 47628t^{10} + 149940t^{12} + 413820t^{14} + 1029105t^{16} + 2351635t^{18} + 5010005t^{20} + 10061415t^{22} + 19211920t^{24} + 35119280t^{26} + 61799760t^{28} + 105163632t^{30} + O(t^{31})$

**1038 2107.00754-smallfoldedeven__k1_n4** / **k = 1 · n = 4**

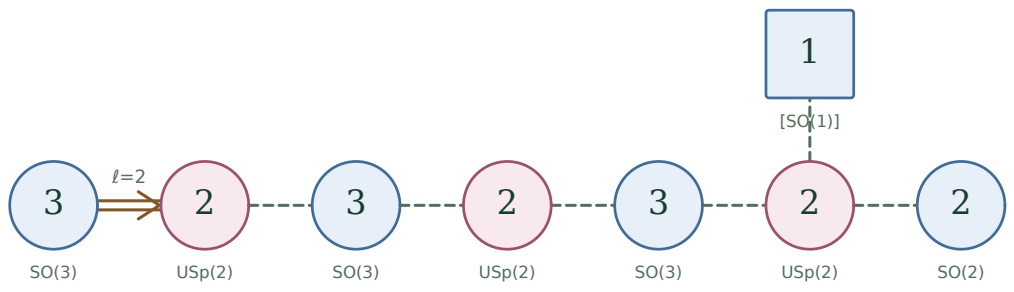


$1 + 63t^{2} + 1232t^{4} + 13104t^{6} + 94500t^{8} + 518364t^{10} + 2317392t^{12} + 8833968t^{14} + 29630601t^{16} + 89464375t^{18} + 247351104t^{20} + 634542272t^{22} + 1526183568t^{24} + 3470399856t^{26} + 7511688000t^{28} + 15564217536t^{30} + O(t^{31})$

| QUIVER | HILBERT SERIES |
| --- | --- |

**1039 2107.00754-smallfoldedeven__k2_n3** / k = 2 · n = 3

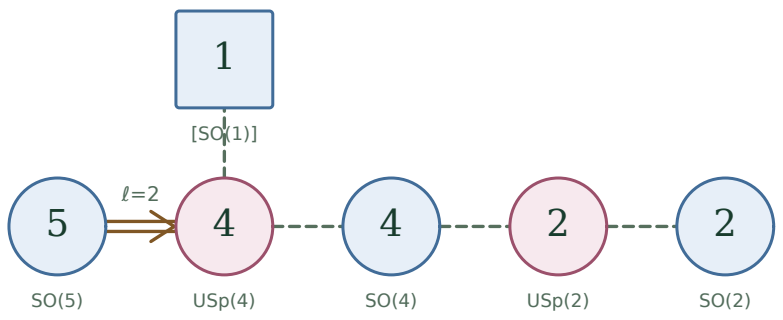


$1 + 35t^{2} + 594t^{4} + 6370t^{6} + 49160t^{8} + 295974t^{10} + 1467530t^{12} + 6225571t^{14} + 23234715t^{16} + 77907310t^{18} + 238527884t^{20} + 675409320t^{22} + 1786949696t^{24} + 4454518840t^{26} + 10534793580t^{28} + 23773252846t^{30} + O(t^{31})$

**1040 2107.00754-smallfoldedeven__k2_n4** / k = 2 · n = 4

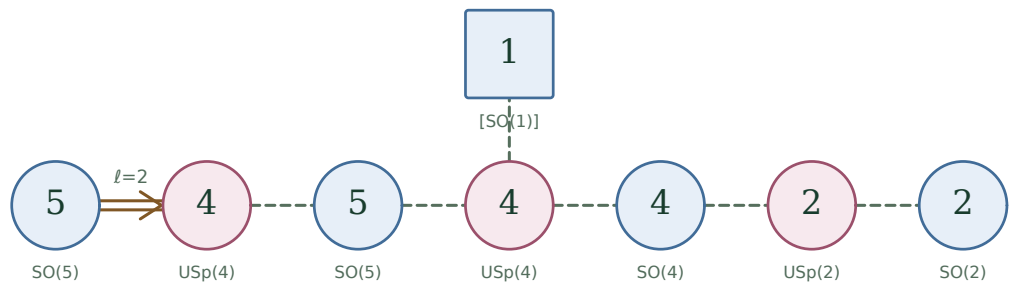


$1 + 63t^{2} + 1952t^{4} + 37359t^{6} + 499625t^{8} + 5061672t^{10} + 41075776t^{12} + 277953624t^{14} + 1616051178t^{16} + 8260334750t^{18} + 37793500416t^{20} + 157024907998t^{22} + 599433532362t^{24} + 2122924391224t^{26} + 7031665800000t^{28} + 21931885566920t^{30} + O(t^{31})$

**1041 2107.00754-smallfoldedeven__k2_n5** / k = 2 · n = 5

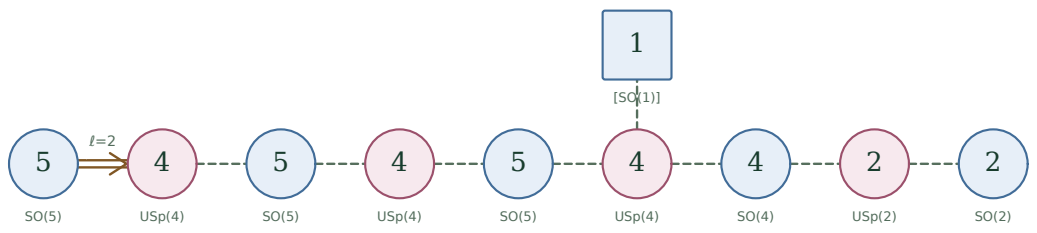

$1 + 99t^{2} + 4850t^{4} + 144474t^{6} + 2953026t^{8} + 44917158t^{10} + 538062042t^{12} + 5289218550t^{14} + 44005830792t^{16} + 317367523858t^{18} + 2021892183234t^{20} + 11553712008566t^{22} + 59963519465850t^{24} + 285610649271066t^{26} + 1259470023104634t^{28} + 5180400897760779t^{30} + O(t^{31})$

**1042 2107.00754-smallfoldedeven__k3_n4** / k = 3 · n = 4

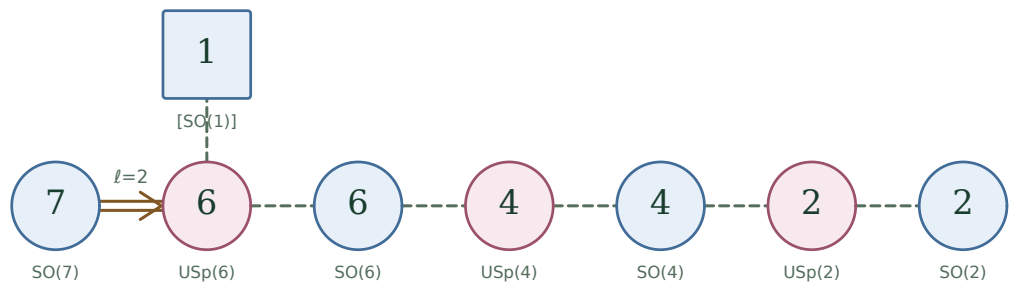


$1 + 63t^{2} + 1952t^{4} + 39711t^{6} + 595853t^{8} + 7032291t^{10} + 68095329t^{12} + 557807640t^{14} + 3957231964t^{16} + 24770625924t^{18} + 138900586216t^{20} + 706534481277t^{22} + 3294423774551t^{24} + 14206800103641t^{26} + 57091937681667t^{28} + 215202903453552t^{30} + O(t^{31})$

**1043 2107.00754-smallfoldedeven__k3_n5** / k = 3 · n = 5

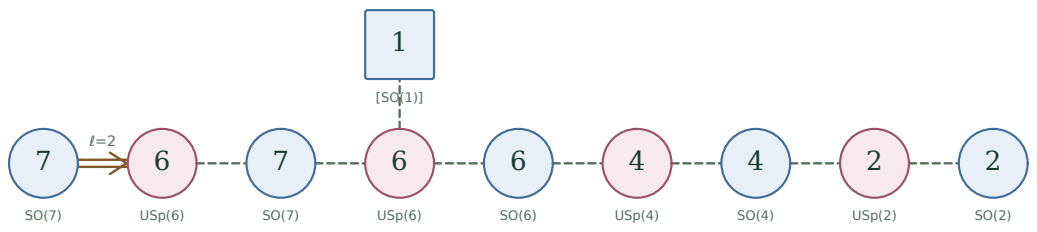

$1 + 99t^{2} + 4850t^{4} + 156849t^{6} + 3739526t^{8} + 69736161t^{10} + 1057652464t^{12} + 13428119100t^{14} + 145995610839t^{16} + 1384598204561t^{18} + 11630269875588t^{20} + 87637359444187t^{22} + 598860320540250t^{24} + 3745587572193937t^{26} + 21614188116483963t^{28} + 115875704871933278t^{30} + O(t^{31})$

**1044 2107.00754-smallfoldedeven__k3_n6** / k = 3 · n = 6

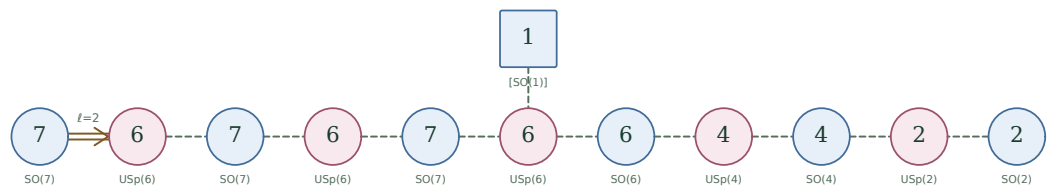

$1 + 143t^{2} + 10152t^{4} + 477191t^{6} + 16515212t^{8} + 445002272t^{10} + 9688251365t^{12} + 175269864770t^{14} + 2694520744124t^{16} + 35855909819444t^{18} + 419381877924790t^{20} + 4367667055351418t^{22} + 40951289981199532t^{24} + 348960576951028708t^{26} + 2724813543198949774t^{28} + 19636066474904260651t^{30} + O(t^{31})$

**1045 2107.00754-beforefoldodd__n3** / n = 3

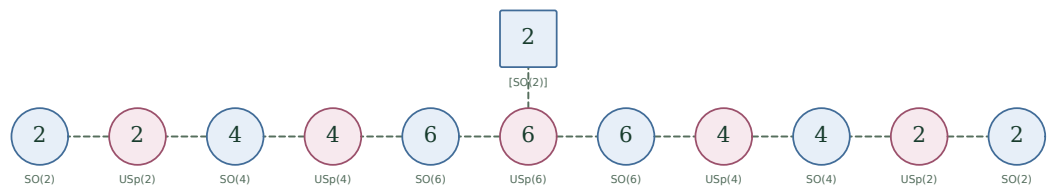

$1 + 91t^{2} + 4081t^{4} + 120315t^{6} + 2621801t^{8} + 45026982t^{10} + 634993345t^{12} + 7569785080t^{14} + 77969074288t^{16} + 705979238874t^{18} + 5699104231898t^{20} + 41500952990692t^{22} + 275331284195709t^{24} + 1678384572644549t^{26} + 9470240422216119t^{28} + 49779621546473959t^{30} + O(t^{31})$

**QUIVER** | **HILBERT SERIES**

**1046 2107.00754-beforefoldodd__n4** / n = 4

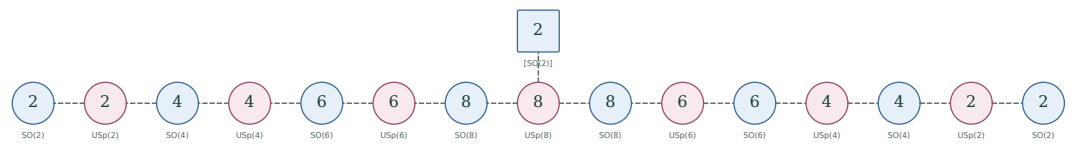

$1 + 153t^{2} + 11610t^{4} + 582692t^{6} + 21764556t^{8} + 645441057t^{10} + 15831830787t^{12} + 330402698756t^{14} + 5989511435922t^{16} + 95824060145797t^{18} + 1370165410208935t^{20} + 17691424781804432t^{22} + 208058194359516714t^{24} + 2245042097813936901t^{26} + 22368473245624143358t^{28} + 206934350075300887325t^{30} + O(t^{31})$

**1047 2107.00754-beforefoldodd__n5** / n = 5

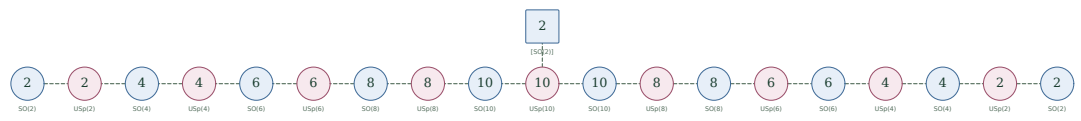

$1 + 231t^{2} + 26543t^{4} + 2022965t^{6} + 115058097t^{8} + 5209605478t^{10} + 195623194899t^{12} + 6266681727987t^{14} + 174843463698849t^{16} + 4316494900444914t^{18} + 95480171675065211t^{20} + 1911590673887590836t^{22} + 34931417541145743024t^{24} + 586734038991593032342t^{26} + 9113563596196986394497t^{28} + 131590000916203403008577t^{30} + O(t^{31})$

**1048 2107.00754-foldedodd__n1** / n = 1

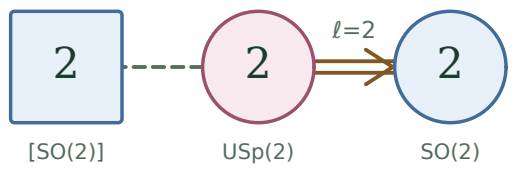


$1 + 8t^{2} + 27t^{4} + 64t^{6} + 125t^{8} + 216t^{10} + 343t^{12} + 512t^{14} + 729t^{16} + 1000t^{18} + 1331t^{20} + 1728t^{22} + 2197t^{24} + 2744t^{26} + 3375t^{28} + 4096t^{30} + O(t^{31})$

**1049 2107.00754-foldedodd__n2** / n = 2

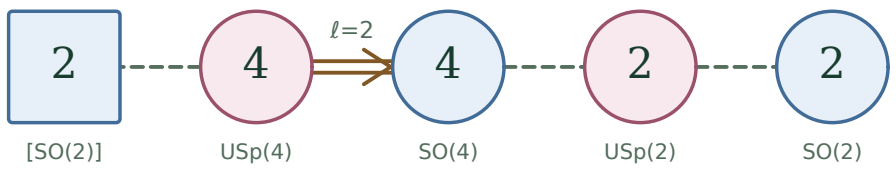


$1 + 24t^{2} + 275t^{4} + 2024t^{6} + 10976t^{8} + 47552t^{10} + 173448t^{12} + 552200t^{14} + 1574573t^{16} + 4098952t^{18} + 9884303t^{20} + 22330672t^{22} + 47691800t^{24} + 96988672t^{26} + 188934328t^{28} + 354281904t^{30} + O(t^{31})$

**1050 2107.00754-foldedodd__n3** / n = 3

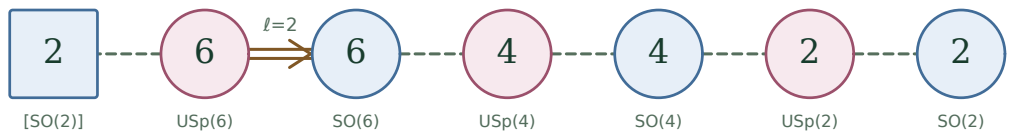


$1 + 48t^{2} + 1127t^{4} + 17296t^{6} + 195707t^{8} + 1747584t^{10} + 12877936t^{12} + 80859264t^{14} + 443181576t^{16} + 2160739328t^{18} + 9513900414t^{20} + 38300683648t^{22} + 142422703952t^{24} + 493390217344t^{26} + 1603925182320t^{28} + 4923173095040t^{30} + O(t^{31})$

**1051 2107.00754-smallerorbitsodd__k1_n3** / k = 1 · n = 3

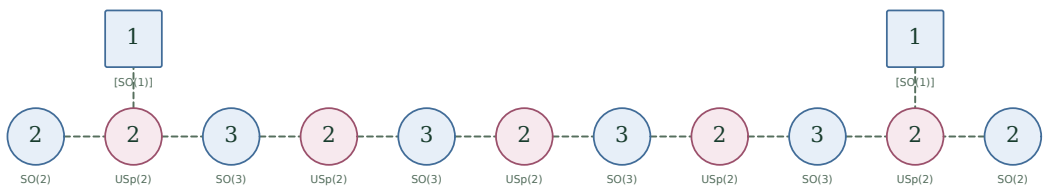

$1 + 91t^{2} + 3080t^{4} + 58344t^{6} + 741741t^{8} + 7014007t^{10} + 52676624t^{12} + 328671200t^{14} + 1760034276t^{16} + 8288708428t^{18} + 34981096384t^{20} + 134282273216t^{22} + 474507071220t^{24} + 1558691535324t^{26} + 4798442924704t^{28} + 13938796754960t^{30} + O(t^{31})$

**1052 2107.00754-smallerorbitsodd__k1_n4** / k = 1 · n = 4

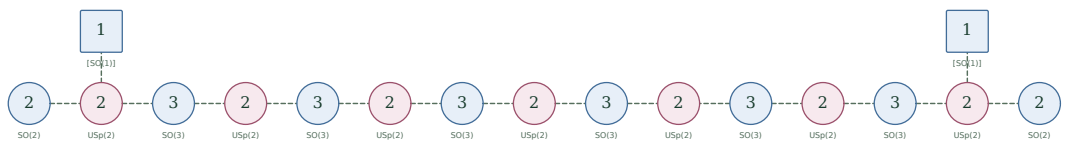

$1 + 153t^{2} + 8550t^{4} + 261800t^{6} + 5264424t^{8} + 77093640t^{10} + 878867496t^{12} + 8168960700t^{14} + 64037140200t^{16} + 434360020072t^{18} + 2600775492084t^{20} + 13967495576280t^{22} + 68159700635000t^{24} + 305476876935000t^{26} + 1268697624499800t^{28} + 4919891369426550t^{30} + O(t^{31})$

| QUIVER | HILBERT SERIES |
| --- | --- |
| **1053 2107.00754-smallerorbitsodd__k1_n5** / k = 1 · n = 5 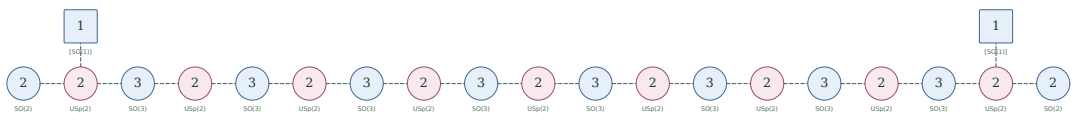 | $1 + 231t^{2} + 19228t^{4} + 864500t^{6} + 25162137t^{8} + 525967519t^{10} + 8445091424t^{12} + 109163508960t^{14} + 1175885335625t^{16} + 10836722099775t^{18} + 87226293854700t^{20} + 623468421821700t^{22} + 4011311605125825t^{24} + 23493653545044375t^{26} + 126449420246366400t^{28} + 630492483929616000t^{30} + O(t^{31})$ |
| **1054 2107.00754-smallerorbitsodd__k2_n3** / k = 2 · n = 3 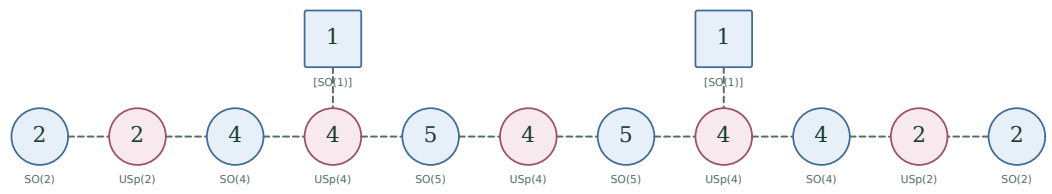 | $1 + 91t^{2} + 4081t^{4} + 117312t^{6} + 2421601t^{8} + 38416378t^{10} + 490596638t^{12} + 5217820608t^{14} + 47435963498t^{16} + 376261850510t^{18} + 2647425001221t^{20} + 16749811545376t^{22} + 96375453431621t^{24} + 509144439032075t^{26} + 2489826624206129t^{28} + 11349894794302400t^{30} + O(t^{31})$ |
| **1055 2107.00754-smallerorbitsodd__k2_n4** / k = 2 · n = 4 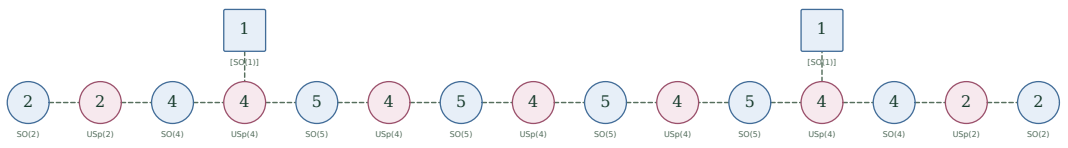 | $1 + 153t^{2} + 11610t^{4} + 564128t^{6} + 19588998t^{8} + 519047757t^{10} + 10976608959t^{12} + 191495537120t^{14} + 2827105695018t^{16} + 36042035541512t^{18} + 403375912084024t^{20} + 4017586314680640t^{22} + 36020667357733000t^{24} + 293564714813116560t^{26} + 2193116384146069494t^{28} + 15128041482713256192t^{30} + O(t^{31})$ |
| **1056 2107.00754-smallerorbitsodd__k2_n5** / k = 2 · n = 5 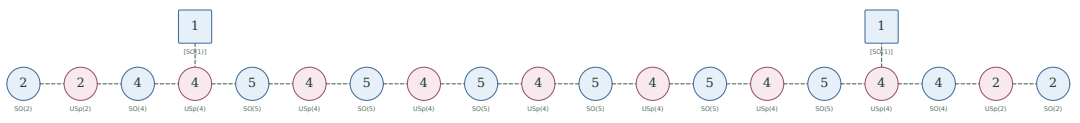 | $1 + 231t^{2} + 26543t^{4} + 1948352t^{6} + 101759427t^{8} + 4033774822t^{10} + 126862061469t^{12} + 3270727583760t^{14} + 70895962405125t^{16} + 1318278395924490t^{18} + 21376472503081059t^{20} + 306435045552718992t^{22} + 3928416167284375767t^{24} + 45482764442745570402t^{26} + 479639992007095262478t^{28} + 4641260181306109922176t^{30} + O(t^{31})$ |
| **1057 2107.00754-smallerorbitsodd__k3_n4** / k = 3 · n = 4 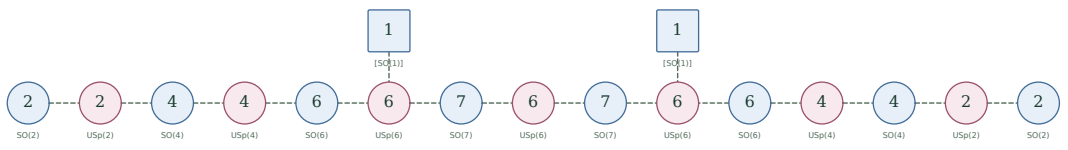 | $1 + 153t^{2} + 11610t^{4} + 582692t^{6} + 21720798t^{8} + 640150317t^{10} + 15513875877t^{12} + 317733399728t^{14} + 5612768150682t^{16} + 86901341331801t^{18} + 1194749860404637t^{20} + 14745449953907528t^{22} + 164889119612102892t^{24} + 1684026667783392636t^{26} + 15818160981462606517t^{28} + 137492023370270094723t^{30} + O(t^{31})$ |
| **1058 2107.00754-smallerorbitsodd__k3_n5** / k = 3 · n = 5 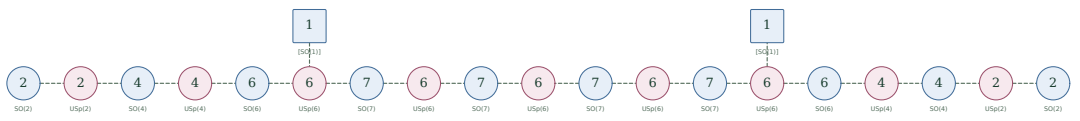 | $1 + 231t^{2} + 26543t^{4} + 2022965t^{6} + 114738327t^{8} + 5149467400t^{10} + 189996180891t^{12} + 5917360294834t^{14} + 158654871259203t^{16} + 3718989000072115t^{18} + 77179731842683119t^{20} + 1433033886939440071t^{22} + 24020410043018558565t^{24} + 366308460399118346725t^{26} + 5116910104483414146129t^{28} + 65867439451968885776045t^{30} + O(t^{31})$ |

| QUIVER | HILBERT SERIES |
|---|---|

**1059 2107.00754-smallerorbitsodd__k3_n6** / **k = 3 · n = 6**

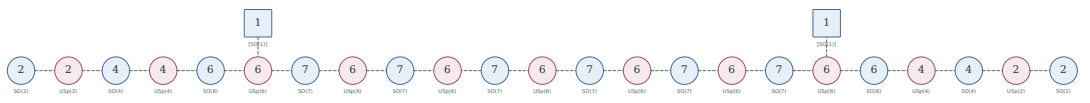

$1 + 325t^{2} + 52624t^{4} + 5660550t^{6} + 453508900t^{8} + 28750911605t^{10} + 1497363102924t^{12} + 65739923866395t^{14} + 2480320612887676t^{16} + 81643970692083475t^{18} + 2373733154937725569t^{20} + 61591099681369702820t^{22} + 1438852106872521022656t^{24} + 30496545079774362731830t^{26} + 590386903683675079154525t^{28} + 10501583450210522200856975t^{30} + O(t^{31})$

**1060 2107.00754-foldedoddd__k1_n2** / **k = 1 · n = 2**

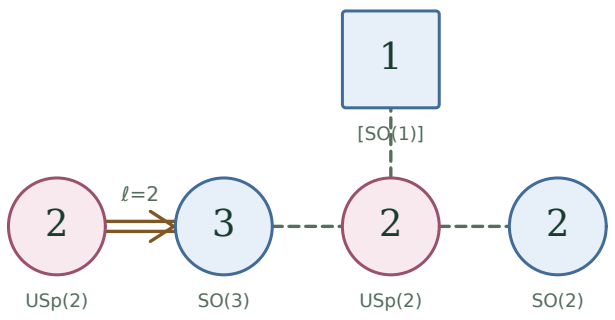


$1 + 24t^{2} + 200t^{4} + 1000t^{6} + 3675t^{8} + 10976t^{10} + 28224t^{12} + 64800t^{14} + 136125t^{16} + 266200t^{18} + 490776t^{20} + 861224t^{22} + 1449175t^{24} + 2352000t^{26} + 3699200t^{28} + 5659776t^{30} + O(t^{31})$

**1061 2107.00754-foldedoddd__k1_n3** / **k = 1 · n = 3**

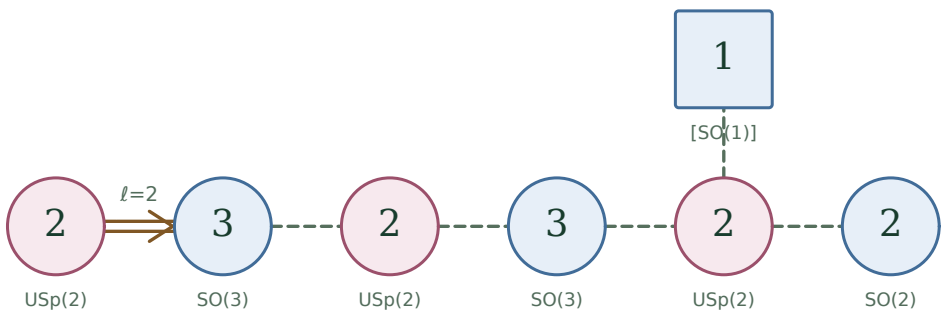


$1 + 48t^{2} + 735t^{4} + 6272t^{6} + 37044t^{8} + 169344t^{10} + 640332t^{12} + 2090880t^{14} + 6073353t^{16} + 16032016t^{18} + 39078039t^{20} + 89037312t^{22} + 191456720t^{24} + 391523328t^{26} + 766192176t^{28} + 1442244096t^{30} + O(t^{31})$

**1062 2107.00754-foldedoddd__k1_n4** / **k = 1 · n = 4**

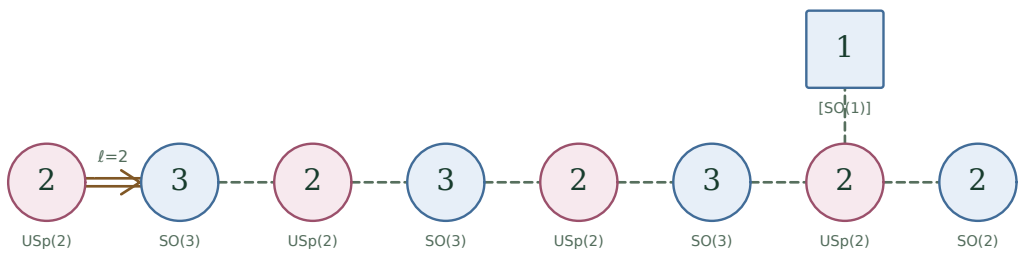

$1 + 80t^{2} + 1944t^{4} + 25200t^{6} + 217800t^{8} + 1411344t^{10} + 7361640t^{12} + 32391216t^{14} + 124227675t^{16} + 425339200t^{18} + 1323786464t^{20} + 3797876160t^{22} + 10155802176t^{24} + 25539739200t^{26} + 60844672800t^{28} + 138154965696t^{30} + O(t^{31})$

**1063 2107.00754-foldedoddd__k2_n3** / **k = 2 · n = 3**

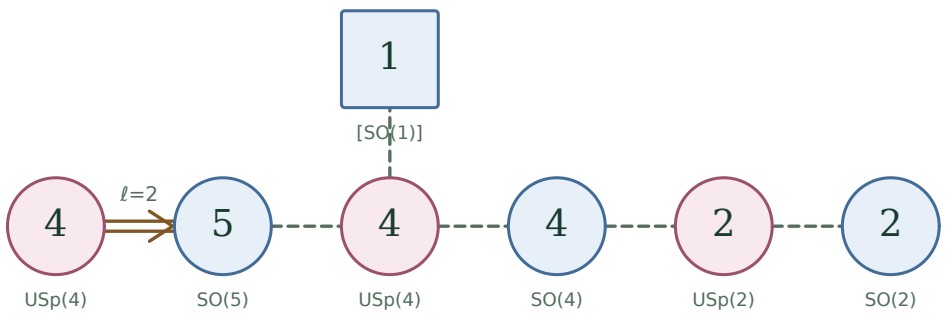


$1 + 48t^{2} + 1127t^{4} + 16512t^{6} + 171207t^{8} + 1361808t^{10} + 8780849t^{12} + 47740800t^{14} + 225332770t^{16} + 943986176t^{18} + 3571203102t^{20} + 12369015808t^{22} + 39659834718t^{24} + 118801081728t^{26} + 334991625762t^{28} + 894864072064t^{30} + O(t^{31})$

**1064 2107.00754-foldedoddd__k2_n4** / **k = 2 · n = 4**

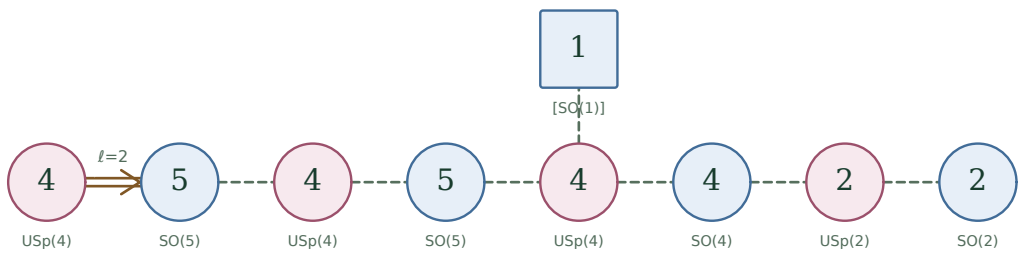

$1 + 80t^{2} + 3159t^{4} + 76400t^{6} + 1278244t^{8} + 16041888t^{10} + 159759756t^{12} + 1315045152t^{14} + 9224012700t^{16} + 56442438272t^{18} + 306931165684t^{20} + 1505538366048t^{22} + 6742847699836t^{24} + 27853273690400t^{26} + 107020216604916t^{28} + 385237309217904t^{30} + O(t^{31})$

**1065 2107.00754-foldedoddd__k2_n5** / **k = 2 · n = 5**

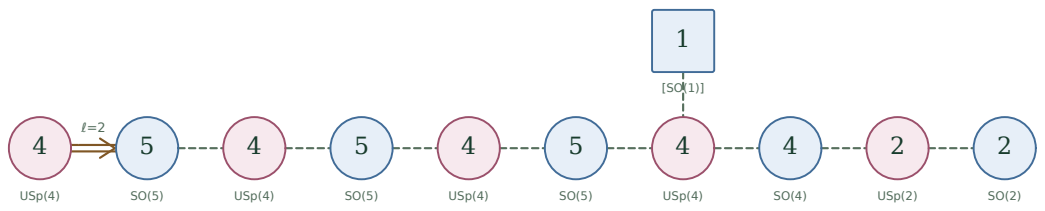

$1 + 120t^{2} + 7139t^{4} + 256640t^{6} + 6287645t^{8} + 113884232t^{10} + 1614301535t^{12} + 18665765888t^{14} + 181635105575t^{16} + 1523883200840t^{18} + 11236698445973t^{20} + 73961887592320t^{22} + 440158897812907t^{24} + 2393713676467960t^{26} + 12003366418926265t^{28} + 55928578777400320t^{30} + O(t^{31})$

| QUIVER | HILBERT SERIES |
|---|---|

**1066 2107.00754-foldedoddd__k3_n4** / k = 3 · n = 4

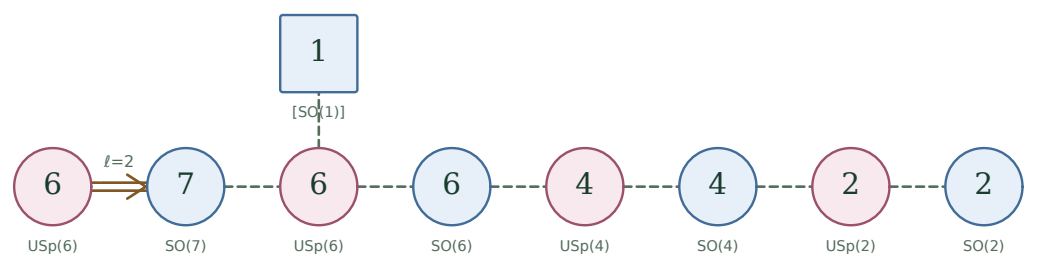


$1 + 80t^{2} + 3159t^{4} + 82160t^{6} + 1575919t^{8} + 23706512t^{10} + 291204865t^{12} + 3008679408t^{14} + 26754267421t^{16} + 208587397232t^{18} + 1447647171603t^{20} + 9057743823600t^{22} + 51641695853547t^{24} + 270741941997168t^{26} + 1315462567564365t^{28} + 5963559235740480t^{30} + O(t^{31})$

**1067 2107.00754-foldedoddd__k3_n5** / k = 3 · n = 5

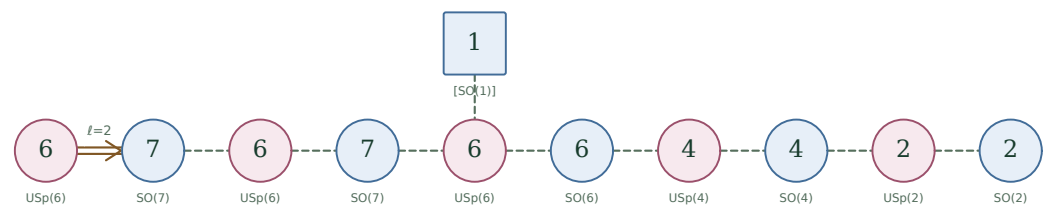


$1 + 120t^{2} + 7139t^{4} + 280840t^{6} + 8140034t^{8} + 184096296t^{10} + 3374368746t^{12} + 51569931864t^{14} + 672111235995t^{16} + 7608877202464t^{18} + 75976434599175t^{20} + 677807065747936t^{22} + 5462012972704560t^{24} + 40132105812921760t^{26} + 271046574623333385t^{28} + 1694622579041633376t^{30} + O(t^{31})$

**1068 2107.00754-foldedoddd__k3_n6** / k = 3 · n = 6

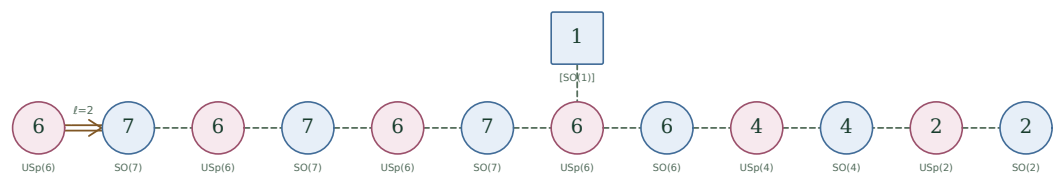


$1 + 168t^{2} + 14027t^{4} + 776216t^{6} + 31603508t^{8} + 999748736t^{10} + 25485004468t^{12} + 538243750616t^{14} + 9631130221350t^{16} + 148722491540128t^{18} + 2012617944625002t^{20} + 24181236268391176t^{22} + 260822158691744572t^{24} + 2549779460013802240t^{26} + 22779571898754213436t^{28} + 187331635878643376328t^{30} + O(t^{31})$

**1069 2107.00754-E8-folded**

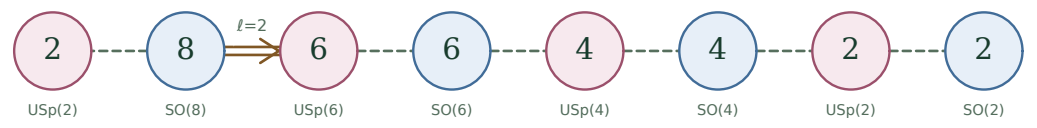


÷ ℤ₂ diagonal

**Computed by Quiver Lab**

$1 + 133t^{2} + 7371t^{4} + 238602t^{6} + 5248750t^{8} + 85709988t^{10} + 1101296924t^{12} + O(t^{13})$

Source originally Claimed · no numerical literature comparison is available.

**1070 2107.00754-E7-folded**

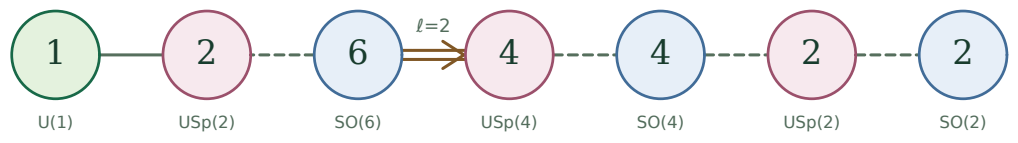


÷ ℤ₂ diagonal

**Computed by Quiver Lab**

$1 + 78t^{2} + 2430t^{4} + 43758t^{6} + 537966t^{8} + 4969107t^{10} + 36685506t^{12} + 225961450t^{14} + 1198006524t^{16} + 5597569328t^{18} + 23474156784t^{20} + 89644484592t^{22} + 315415779120t^{24} + 1032380107812t^{26} + 3168537039954t^{28} + 9180278955210t^{30} + O(t^{31})$

Source originally Claimed · no numerical literature comparison is available.

**1071 2107.00754-E6-folded**

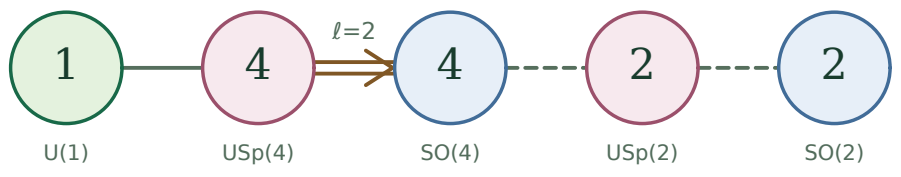


÷ ℤ₂ diagonal

**Computed by Quiver Lab**

$1 + 45t^{2} + 770t^{4} + 7644t^{6} + 52920t^{8} + 282744t^{10} + 1241460t^{12} + 4671810t^{14} + 15520791t^{16} + 46521475t^{18} + 127891764t^{20} + 326602640t^{22} + 782658240t^{24} + 1774339776t^{26} + 3830960880t^{28} + 7921074996t^{30} + O(t^{31})$

Source originally Claimed · no numerical literature comparison is available.

**1072 2107.00754-E5-folded**

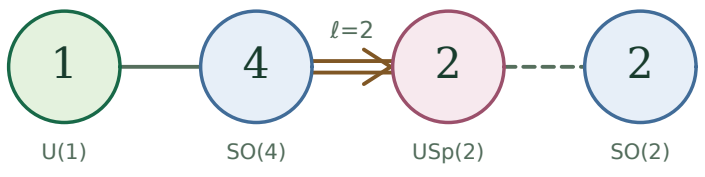


÷ ℤ₂ diagonal

**Computed by Quiver Lab**

$1 + 28t^{2} + 300t^{4} + 1925t^{6} + 8918t^{8} + 32928t^{10} + 102816t^{12} + 282150t^{14} + 698775t^{16} + 1591876t^{18} + 3383380t^{20} + 6782139t^{22} + 12931100t^{24} + 23609600t^{26} + 41505024t^{28} + 70570332t^{30} + O(t^{31})$

Source originally Claimed · no numerical literature comparison is available.

**1073 2107.00754-E4-folded**

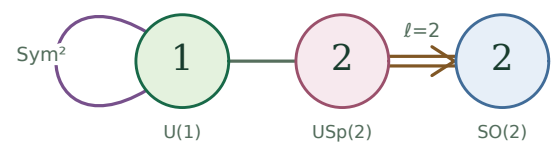


÷ ℤ₂ diagonal

**Computed by Quiver Lab**

$1 + 15t^{2} + 84t^{4} + 300t^{6} + 825t^{8} + 1911t^{10} + 3920t^{12} + 7344t^{14} + 12825t^{16} + 21175t^{18} + 33396t^{20} + 50700t^{22} + 74529t^{24} + 106575t^{26} + 148800t^{28} + 203456t^{30} + O(t^{31})$

Source originally Claimed · no numerical literature comparison is available.

**1074 2107.00754-E8-general-OSp__k2** / **k = 2**

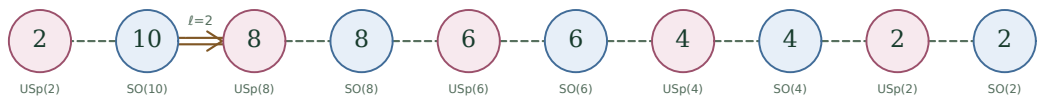


÷ ℤ₂ diagonal

$1 + 99t^{2} + 252t^{3} + 4851t^{4} + 21252t^{5} + 176352t^{6} + 891000t^{7} + 5209181t^{8} + 25516008t^{9} + 125039871t^{10} + 563803856t^{11} + 2460256303t^{12} + 10168819992t^{13} + 40491753456t^{14} + 154537296672t^{15} + 569259611721t^{16} + 2023009492528t^{17} + 6960511874371t^{18} + 23206866923844t^{19} + 75142821855432t^{20} + 236568232514372t^{21} + 725306404386888t^{22} + 2168090635081380t^{23} + 6326688036311568t^{24} + 18041368348542996t^{25} + 50328640600683742t^{26} + 137472185538130484t^{27} + 368010304559804250t^{28} + 966285257616252684t^{29} + 2490525359481666029t^{30} + O(t^{31})$

**1075 2107.00754-E8-general-OSp__k3** / **k = 3**

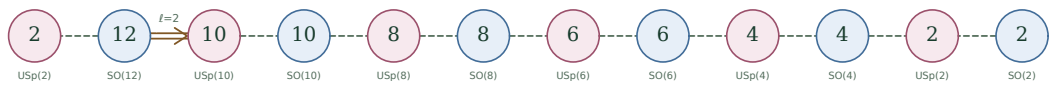


÷ ℤ₂ diagonal

$1 + 143t^{2} + 11077t^{4} + 592306t^{6} + 24068186t^{8} + 783494932t^{10} + 21145234824t^{12} + 484890980352t^{14} + 9626745774138t^{16} + 167962358012097t^{18} + 2607200215817288t^{20} + 36378523457743673t^{22} + 460308336968852387t^{24} + 5322259703587422849t^{26} + 56608628672484215983t^{28} + 557139971985813650908t^{30} + O(t^{31})$

**1076 2107.00754-E8-general-OSp__k4** / **k = 4**

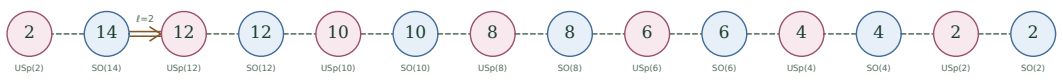

÷ ℤ₂ diagonal

$1 + 195t^{2} + 18915t^{4} + 3432t^{5} + 1217060t^{6} + 592020t^{7} + 58447349t^{8} + 50870820t^{9} + 2237654679t^{10} + 2903868968t^{11} + 71318298674t^{12} + 123910826520t^{13} + 1952082097966t^{14} + 4217754656856t^{15} + 46968166363169t^{16} + 119372012914832t^{17} + 1011019385236575t^{18} + 2891664296261988t^{19} + 19727291024880256t^{20} + 61253502585371252t^{21} + 352295216708070155t^{22} + 1153533499068344288t^{23} + 5798536401893335192t^{24} + 19567355183950198768t^{25} + 88421124672505260710t^{26} + 302154095202250249840t^{27} + 1254226577560277175754t^{28} + 4284305953978249615624t^{29} + 16605354359355036932085t^{30} + O(t^{31})$

**1077 2107.00754-E7-general-OSp__k2** / **k = 2**

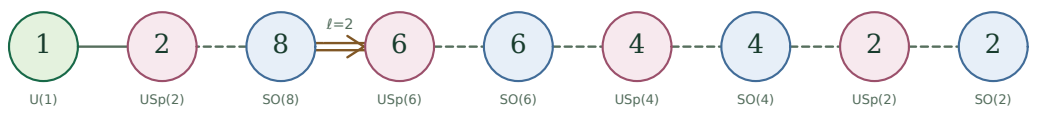


÷ ℤ₂ diagonal

$1 + 66t^{2} + 140t^{3} + 2147t^{4} + 7588t^{5} + 51247t^{6} + 204540t^{7} + 976067t^{8} + 3762804t^{9} + 14948834t^{10} + 53289144t^{11} + 187287550t^{12} + 615222064t^{13} + 1967063777t^{14} + 5991595360t^{15} + 17718900900t^{16} + 50436508216t^{17} + 139532644918t^{18} + 373925032912t^{19} + 975965396850t^{20} + 2478430652048t^{21} + 6144125348334t^{22} + 14868649472880t^{23} + 35202598762940t^{24} + 81572086565472t^{25} + 185293888997398t^{26} + 412834242155312t^{27} + 903249391673992t^{28} + 1941859162236744t^{29} + 4106010310832217t^{30} + O(t^{31})$

**1078 2107.00754-E7-general-OSp__k3** / **k = 3**

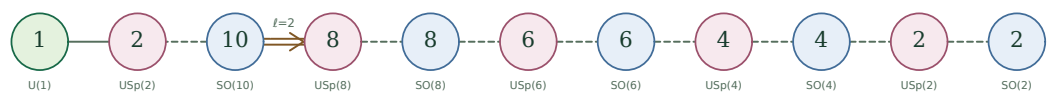


÷ ℤ₂ diagonal

$1 + 102t^{2} + 5657t^{4} + 215515t^{6} + 6196610t^{8} + 141799682t^{10} + 2675727869t^{12} + 42730311064t^{14} + 589248387684t^{16} + 7131043102939t^{18} + 76752089321454t^{20} + 742973421347096t^{22} + 6530505641975437t^{24} + 52551335797171266t^{26} + 389936421709734810t^{28} + 2684758370700917581t^{30} + O(t^{31})$

| QUIVER | HILBERT SERIES |
|---|---|
| **1079 2107.00754-E7-general-OSp__k4** / **k = 4** 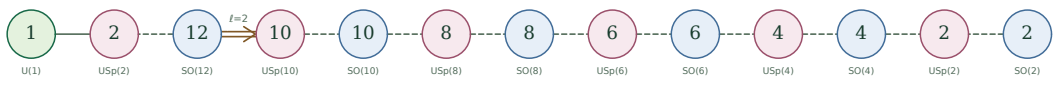 ÷ $\mathbb{Z}_2$ diagonal | $1 + 146t^{2} + 10587t^{4} + 1848t^{5} + 508515t^{6} + 233640t^{7} + 18205767t^{8} + 14704800t^{9} + 519042818t^{10} + 614545184t^{11} + 12306472042t^{12} + 19193904840t^{13} + 250297790033t^{14} + 478173189160t^{15} + 4468399237304t^{16} + 9906012823072t^{17} + 71234041302331t^{18} + 175684088025808t^{19} + 1027151738083790t^{20} + 2725434791958680t^{21} + 13525057039563030t^{22} + 37602085972629144t^{23} + 163803786901448415t^{24} + 467483930580770128t^{25} + 1834930782068644075t^{26} + 5293140165291555232t^{27} + 19099201579968686768t^{28} + 55060797941984181472t^{29} + 185450727164346850007t^{30} + O(t^{31})$ |
| **1080 2107.00754-E6-general-OSp__k2** / **k = 2** 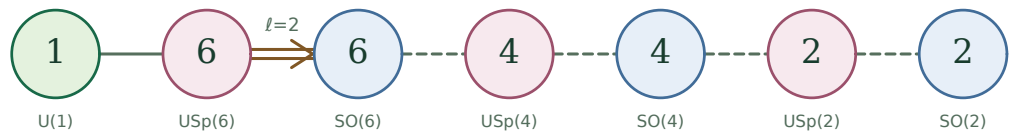  ÷ $\mathbb{Z}_2$ diagonal | $1 + 49t^{2} + 70t^{3} + 1176t^{4} + 2716t^{5} + 19452t^{6} + 52332t^{7} + 246519t^{8} + 678678t^{9} + 2499867t^{10} + 6690740t^{11} + 20927291t^{12} + 53463410t^{13} + 148550839t^{14} + 360491726t^{15} + 914505299t^{16} + 2109700334t^{17} + 4974669287t^{18} + 10942823976t^{19} + 24285278637t^{20} + 51129126834t^{21} + 107776208197t^{22} + 218003271808t^{23} + 439527670831t^{24} + 857239856116t^{25} + 1662177744535t^{26} + 3136124182588t^{27} + 5874142376745t^{28} + 10753184869364t^{29} + 19527645532265t^{30} + O(t^{31})$ |
| **1081 2107.00754-E6-general-OSp__k3** / **k = 3** 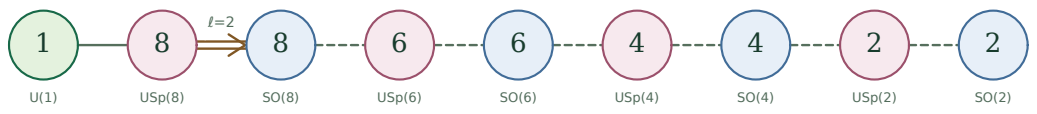  ÷ $\mathbb{Z}_2$ diagonal | $1 + 81t^{2} + 3492t^{4} + 102284t^{6} + 2242647t^{8} + 38894201t^{10} + 553885641t^{12} + 6659017595t^{14} + 69063153688t^{16} + 628761773260t^{18} + 5097353973445t^{20} + 37240758765813t^{22} + 247689707046066t^{24} + 1512795147936312t^{26} + 8548482176451218t^{28} + 44984990556833690t^{30} + O(t^{31})$ |
| **1082 2107.00754-E6-general-OSp__k4** / **k = 4** 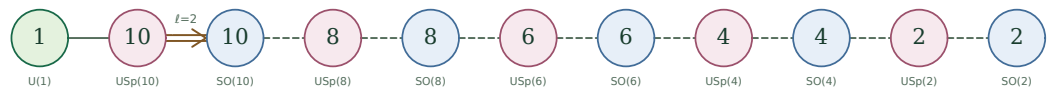 ÷ $\mathbb{Z}_2$ diagonal | $1 + 121t^{2} + 7260t^{4} + 924t^{5} + 288100t^{6} + 95304t^{7} + 8509809t^{8} + 4888884t^{9} + 199771241t^{10} + 166407670t^{11} + 3888123371t^{12} + 4230821672t^{13} + 64631072131t^{14} + 85763312998t^{15} + 937937812450t^{16} + 1445018020380t^{17} + 12084010027242t^{18} + 20831834348402t^{19} + 140029121682979t^{20} + 262522308559588t^{21} + 1474656129387523t^{22} + 2940058432098436t^{23} + 14231863432378297t^{24} + 29647687959519916t^{25} + 126744796293629569t^{26} + 272084050644477302t^{27} + 1047673139590027980t^{28} + 2292623575773712676t^{29} + 8078480316613211988t^{30} + O(t^{31})$ |
| **1083 2107.00754-E5-general-OSp__k2** / **k = 2** 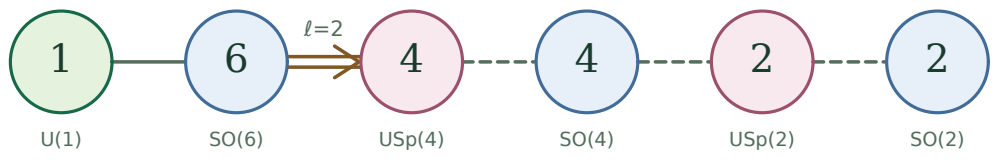  ÷ $\mathbb{Z}_2$ diagonal | $1 + 36t^{2} + 40t^{3} + 630t^{4} + 1120t^{5} + 7525t^{6} + 15624t^{7} + 68592t^{8} + 147800t^{9} + 502074t^{10} + 1071840t^{11} + 3061139t^{12} + 6355752t^{13} + 16006440t^{14} + 32099424t^{15} + 73454610t^{16} + 142067064t^{17} + 301360808t^{18} + 562670200t^{19} + 1122071704t^{20} + 2026486520t^{21} + 3838433310t^{22} + 6721370080t^{23} + 12186830890t^{24} + 20740686792t^{25} + 36216358230t^{26} + 60043907000t^{27} + 101458175136t^{28} + 164215723320t^{29} + 269563015292t^{30} + O(t^{31})$ |
| **1084 2107.00754-E5-general-OSp__k3** / **k = 3** 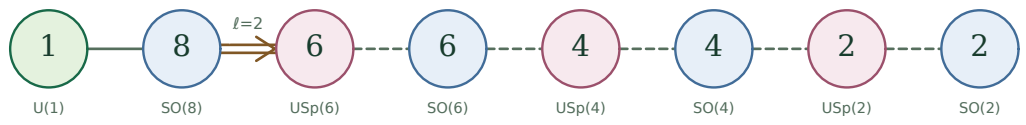  ÷ $\mathbb{Z}_2$ diagonal | $1 + 64t^{2} + 2156t^{4} + 49035t^{6} + 832376t^{8} + 11170539t^{10} + 123240632t^{12} + 1150637073t^{14} + 9298241055t^{16} + 66216202892t^{18} + 421718414235t^{20} + 2431558547462t^{22} + 12823770808393t^{24} + 62403806228172t^{26} + 282310404939196t^{28} + 1195031191158327t^{30} + O(t^{31})$ |

| QUIVER | HILBERT SERIES |
|---|---|

**1085 2107.00754-E5-general-OSp__k4** / k = 4

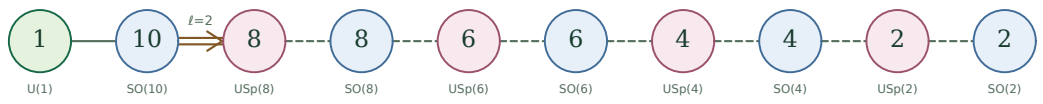


÷ ℤ₂ diagonal

$1 + 100t^{2} + 4950t^{4} + 504t^{5} + 161799t^{6} + 42504t^{7} + 3931025t^{8} + 1782000t^{9} + 75806477t^{10} + 49568400t^{11} + 1210470062t^{12} + 1030094296t^{13} + 16488114686t^{14} + 17076084984t^{15} + 195859119114t^{16} + 235463885184t^{17} + 2063657881647t^{18} + 2780836351328t^{19} + 19545881300449t^{20} + 28743047350224t^{21} + 168214318765762t^{22} + 264386685513152t^{23} + 1327044082602609t^{24} + 2193083840747160t^{25} + 9667515347165193t^{26} + 16583185605706312t^{27} + 65443087305513482t^{28} + 115336207672605232t^{29} + 413879942453904528t^{30} + O(t^{31})$

**1086 2107.00754-E4-2l-general-OSp__l0_k2** / l = 0 · k = 2

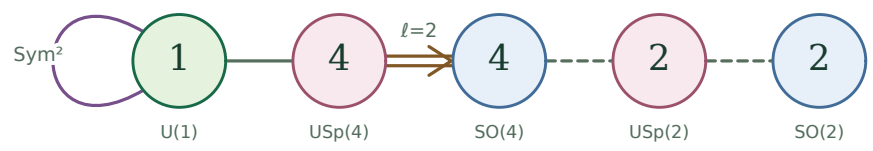


÷ ℤ₂ diagonal

$1 + 25t^{2} + 20t^{3} + 300t^{4} + 370t^{5} + 2424t^{6} + 3430t^{7} + 14840t^{8} + 21910t^{9} + 73172t^{10} + 109040t^{11} + 303920t^{12} + 450570t^{13} + 1098980t^{14} + 1610332t^{15} + 3544013t^{16} + 5120190t^{17} + 10382825t^{18} + 14779270t^{19} + 28037108t^{20} + 39323328t^{21} + 70582700t^{22} + 97595830t^{23} + 167187230t^{24} + 228069114t^{25} + 375416042t^{26} + 505645480t^{27} + 804119660t^{28} + 1070215640t^{29} + 1651503236t^{30} + O(t^{31})$

**1087 2107.00754-E4-2l-general-OSp__l0_k3** / l = 0 · k = 3

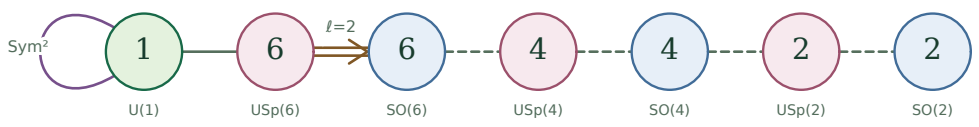


÷ ℤ₂ diagonal

$1 + 49t^{2} + 1246t^{4} + 21188t^{6} + 267491t^{8} + 2664549t^{10} + 21835695t^{12} + 151838053t^{14} + 917609211t^{16} + 4911883473t^{18} + 23649622875t^{20} + 103720463973t^{22} + 418719128391t^{24} + 1569734773005t^{26} + 5505817595111t^{28} + 18183934074729t^{30} + O(t^{31})$

**1088 2107.00754-E4-2l-general-OSp__l0_k4** / l = 0 · k = 4

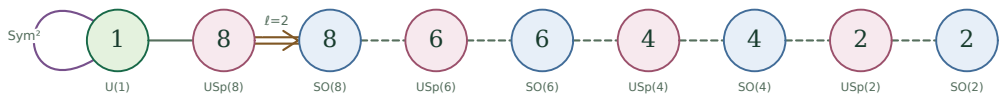


÷ ℤ₂ diagonal

$1 + 81t^{2} + 3240t^{4} + 252t^{5} + 85400t^{6} + 16884t^{7} + 1670139t^{8} + 561924t^{9} + 25884899t^{10} + 12404502t^{11} + 331737219t^{12} + 204626610t^{13} + 3622330151t^{14} + 2694655404t^{15} + 34458668506t^{16} + 29551440150t^{17} + 290537533930t^{18} + 277998097158t^{19} + 2201124204757t^{20} + 2293077759666t^{21} + 15151681903365t^{22} + 16867925449596t^{23} + 95644564199250t^{24} + 112154474845866t^{25} + 557986078855818t^{26} + 681459047092236t^{27} + 3028605160640594t^{28} + 3818315684800488t^{29} + 15382428517646558t^{30} + O(t^{31})$

**1089 2107.00754-E4-2l-general-OSp__l1_k2** / l = 1 · k = 2

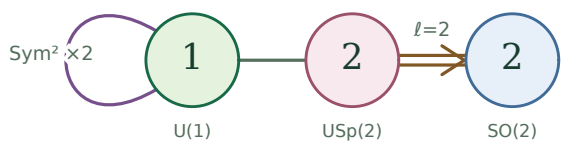


÷ ℤ₂ diagonal

$1 + 9t^{2} + 6t^{3} + 36t^{4} + 36t^{5} + 112t^{6} + 120t^{7} + 285t^{8} + 320t^{9} + 621t^{10} + 720t^{11} + 1234t^{12} + 1428t^{13} + 2262t^{14} + 2618t^{15} + 3873t^{16} + 4488t^{17} + 6321t^{18} + 7272t^{19} + 9882t^{20} + 11312t^{21} + 14874t^{22} + 16968t^{23} + 21747t^{24} + 24648t^{25} + 30963t^{26} + 34918t^{27} + 43038t^{28} + 48348t^{29} + 58674t^{30} + O(t^{31})$

**1090 2107.00754-E4-2l-general-OSp__l1_k3** / l = 1 · k = 3

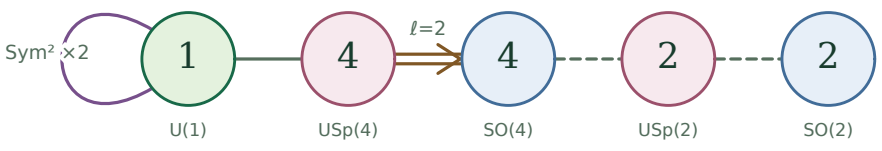


÷ ℤ₂ diagonal

$1 + 25t^{2} + 320t^{4} + 2694t^{6} + 16830t^{8} + 83952t^{10} + 351250t^{12} + 1276030t^{14} + 4128223t^{16} + 12122795t^{18} + 32792670t^{20} + 82665780t^{22} + 196017310t^{24} + 440530630t^{26} + 944254180t^{28} + 1940446198t^{30} + O(t^{31})$

| QUIVER | HILBERT SERIES |
|---|---|

**1091 2107.00754-E4-2l-general-OSp__l1_k4** / l = 1 · k = 4

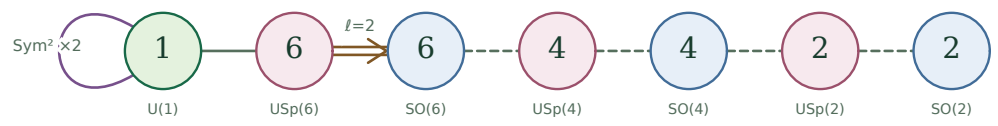


÷ ℤ₂ diagonal

$1 + 49t^{2} + 1176t^{4} + 70t^{5} + 18472t^{6} + 2716t^{7} + 214179t^{8} + 52332t^{9} + 1962743t^{10} + 670446t^{11} + 14872039t^{12} + 6449660t^{13} + 96237067t^{14} + 49868042t^{15} + 544918739t^{16} + 323784356t^{17} + 2751152271t^{18} + 1820368592t^{19} + 12571033213t^{20} + 9064892298t^{21} + 52617902281t^{22} + 40680903270t^{23} + 203749915821t^{24} + 166787382580t^{25} + 735934858505t^{26} + 631596701512t^{27} + 2496752855617t^{28} + 2229029098758t^{29} + 8003456811157t^{30} + O(t^{31})$

**1092 2107.00754-E4-2l-general-OSp__l2_k3** / l = 2 · k = 3

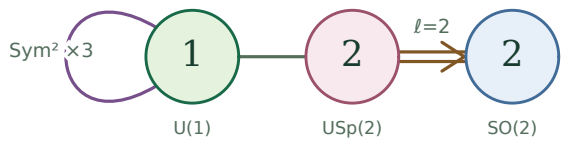


÷ ℤ₂ diagonal

$1 + 9t^{2} + 42t^{4} + 136t^{6} + 357t^{8} + 801t^{10} + 1614t^{12} + 2982t^{14} + 5163t^{16} + 8463t^{18} + 13284t^{20} + 20082t^{22} + 29435t^{24} + 41979t^{26} + 58500t^{28} + 79844t^{30} + O(t^{31})$

**1093 2107.00754-E4-2l-general-OSp__l2_k4** / l = 2 · k = 4

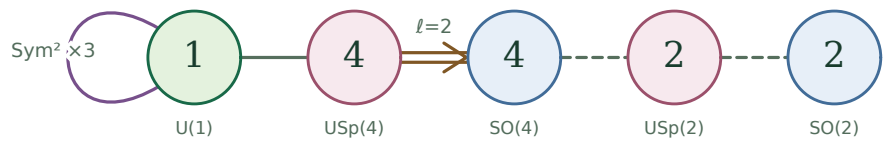


÷ ℤ₂ diagonal

$1 + 25t^{2} + 300t^{4} + 20t^{5} + 2324t^{6} + 370t^{7} + 13300t^{8} + 3430t^{9} + 60952t^{10} + 21560t^{11} + 235840t^{12} + 104280t^{13} + 798820t^{14} + 416480t^{15} + 2429713t^{16} + 1436190t^{17} + 6760325t^{18} + 4407380t^{19} + 17446408t^{20} + 12297230t^{21} + 42210700t^{22} + 31690890t^{23} + 96561600t^{24} + 76341472t^{25} + 210296232t^{26} + 173519320t^{27} + 438478780t^{28} + 374923600t^{29} + 879406564t^{30} + O(t^{31})$

**1094 2107.00754-E4-2l-general-OSp__l2_k5** / l = 2 · k = 5

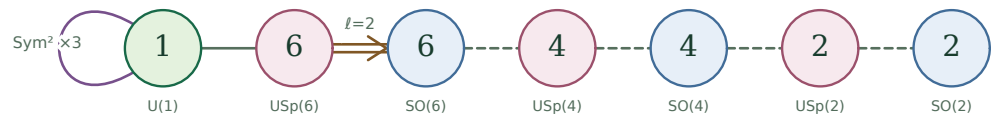


÷ ℤ₂ diagonal

$1 + 49t^{2} + 1176t^{4} + 18542t^{6} + 216895t^{8} + 2014095t^{10} + 15511125t^{12} + 102180963t^{14} + 589278453t^{16} + 3029209575t^{18} + 14082058757t^{20} + 59903447765t^{22} + 235469802075t^{24} + 862357469267t^{26} + 2963069215567t^{28} + 9609475324823t^{30} + O(t^{31})$

**1095 2107.00754-E3-2l-general-OSp__l0_k1** / l = 0 · k = 1

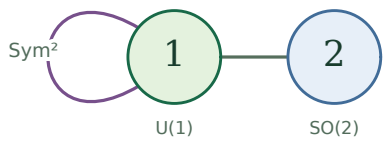


÷ ℤ₂ diagonal

$1 + 8t^{2} + 27t^{4} + 64t^{6} + 125t^{8} + 216t^{10} + 343t^{12} + 512t^{14} + 729t^{16} + 1000t^{18} + 1331t^{20} + 1728t^{22} + 2197t^{24} + 2744t^{26} + 3375t^{28} + 4096t^{30} + O(t^{31})$

**1096 2107.00754-E3-2l-general-OSp__l0_k2** / l = 0 · k = 2

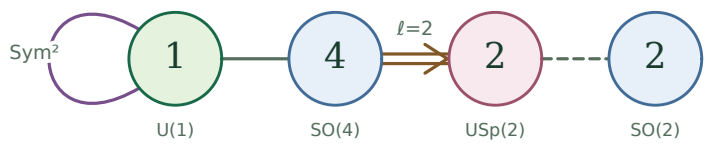


÷ ℤ₂ diagonal

$1 + 16t^{2} + 12t^{3} + 120t^{4} + 140t^{5} + 635t^{6} + 840t^{7} + 2644t^{8} + 3628t^{9} + 9114t^{10} + 12628t^{11} + 27202t^{12} + 37464t^{13} + 72402t^{14} + 98504t^{15} + 175347t^{16} + 235392t^{17} + 392890t^{18} + 520056t^{19} + 824682t^{20} + 1076616t^{21} + 1636911t^{22} + 2109624t^{23} + 3096672t^{24} + 3942864t^{25} + 5618522t^{26} + 7073484t^{27} + 9826146t^{28} + 12242956t^{29} + 16635534t^{30} + O(t^{31})$

**QUIVER** | **HILBERT SERIES**

**1097 2107.00754-E3-2l-general-OSp__l0_k3** / l = 0 · k = 3

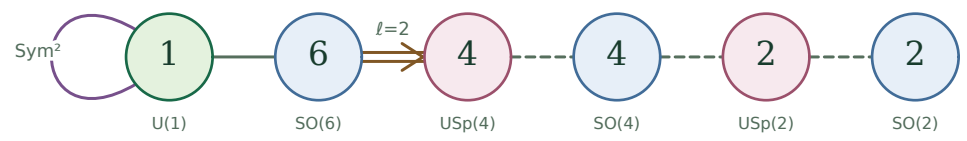


÷ $\mathbb{Z}_2$ diagonal

$1 + 36t^{2} + 670t^{4} + 8295t^{6} + 76278t^{8} + 556242t^{10} + 3363345t^{12} + 17417948t^{14} + 79176858t^{16} + 322010706t^{18} + 1189618112t^{20} + 4041495806t^{22} + 12754038944t^{24} + 37701311274t^{26} + 105127729978t^{28} + 278172637452t^{30} + O(t^{31})$

**1098 2107.00754-E3-2l-general-OSp__l1_k2** / l = 1 · k = 2

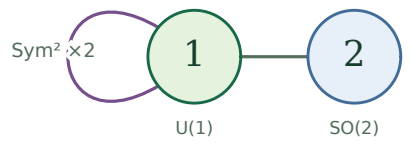


÷ $\mathbb{Z}_2$ diagonal

$1 + 4t^{2} + 4t^{3} + 9t^{4} + 12t^{5} + 22t^{6} + 24t^{7} + 41t^{8} + 48t^{9} + 66t^{10} + 80t^{11} + 107t^{12} + 120t^{13} + 158t^{14} + 180t^{15} + 219t^{16} + 252t^{17} + 304t^{18} + 336t^{19} + 403t^{20} + 448t^{21} + 516t^{22} + 576t^{23} + 661t^{24} + 720t^{25} + 824t^{26} + 900t^{27} + 1005t^{28} + 1100t^{29} + 1226t^{30} + O(t^{31})$

**1099 2107.00754-E3-2l-general-OSp__l1_k3** / l = 1 · k = 3

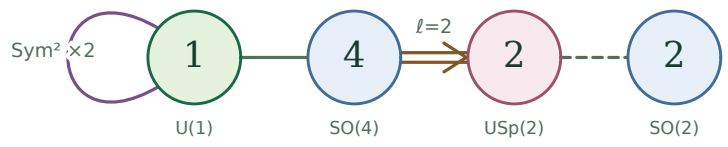


÷ $\mathbb{Z}_2$ diagonal

$1 + 16t^{2} + 132t^{4} + 735t^{6} + 3134t^{8} + 10974t^{10} + 33070t^{12} + 88502t^{14} + 215217t^{16} + 483580t^{18} + 1016976t^{20} + 2021631t^{22} + 3828884t^{24} + 6952892t^{26} + 12168444t^{28} + 20612724t^{30} + O(t^{31})$

**1100 2107.00754-E3-2l-general-OSp__l1_k4** / l = 1 · k = 4

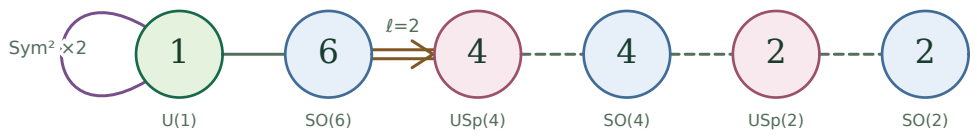


÷ $\mathbb{Z}_2$ diagonal

$1 + 36t^{2} + 630t^{4} + 40t^{5} + 7175t^{6} + 1120t^{7} + 60304t^{8} + 15624t^{9} + 402464t^{10} + 145840t^{11} + 2239033t^{12} + 1030680t^{13} + 10746290t^{14} + 5910256t^{15} + 45631342t^{16} + 28750960t^{17} + 174710900t^{18} + 122371672t^{19} + 612052486t^{20} + 466097280t^{21} + 1984652864t^{22} + 1615941584t^{23} + 6012364420t^{24} + 5167220608t^{25} + 17146741728t^{26} + 15399808424t^{27} + 46328786906t^{28} + 43139588904t^{29} + 119228571686t^{30} + O(t^{31})$

**1101 2107.00754-E3-2l-general-OSp__l2_k3** / l = 2 · k = 3

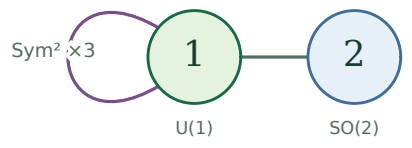


÷ $\mathbb{Z}_2$ diagonal

$1 + 4t^{2} + 13t^{4} + 28t^{6} + 55t^{8} + 92t^{10} + 147t^{12} + 216t^{14} + 309t^{16} + 420t^{18} + 561t^{20} + 724t^{22} + 923t^{24} + 1148t^{26} + 1415t^{28} + 1712t^{30} + O(t^{31})$

**1102 2107.00754-E3-2l-general-OSp__l2_k4** / l = 2 · k = 4

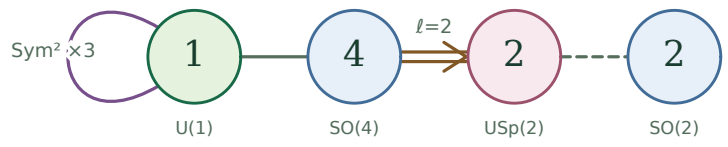


÷ $\mathbb{Z}_2$ diagonal

$1 + 16t^{2} + 120t^{4} + 12t^{5} + 595t^{6} + 140t^{7} + 2254t^{8} + 840t^{9} + 7096t^{10} + 3528t^{11} + 19542t^{12} + 11760t^{13} + 48588t^{14} + 33364t^{15} + 111295t^{16} + 84028t^{17} + 238120t^{18} + 192960t^{19} + 480886t^{20} + 411348t^{21} + 924273t^{22} + 824460t^{23} + 1701868t^{24} + 1568728t^{25} + 3017950t^{26} + 2855088t^{27} + 5176326t^{28} + 5000100t^{29} + 8618346t^{30} + O(t^{31})$

| QUIVER | HILBERT SERIES |
|---|---|

**1103 2107.00754-E3-2l-general-OSp__l2_k5** / l = 2 · k = 5

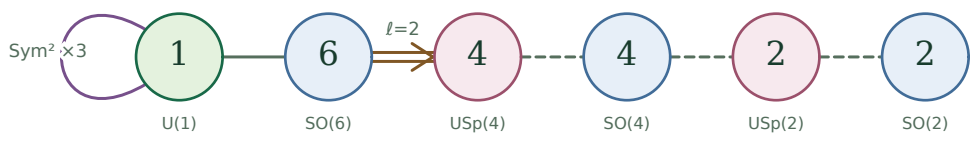


÷ ℤ₂ diagonal

$1 + 36t^{2} + 630t^{4} + 7215t^{6} + 61424t^{8} + 417738t^{10} + 2376935t^{12} + 11685298t^{14} + 50817436t^{16} + 199042632t^{18} + 712153360t^{20} + 2354093840t^{22} + 7256551922t^{24} + 21020521798t^{26} + 57594502526t^{28} + 150085167088t^{30} + O(t^{31})$

## 2107.09077 · Magnetic quivers from brane webs with O7+-planes (13)

**1104 2107.09077-NSL-N7-v2**

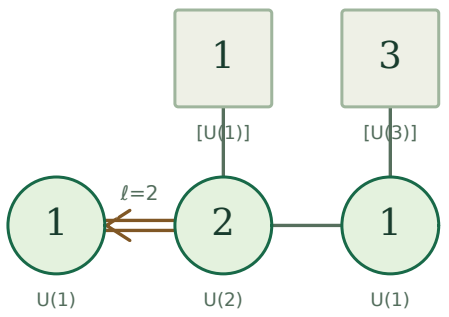


$1 + 11t^{2} + 60t^{4} + 10t^{5} + 225t^{6} + 80t^{7} + 665t^{8} + 350t^{9} + 1694t^{10} + O(t^{11})$

**1105 2107.09077-NSL-N8-v2**

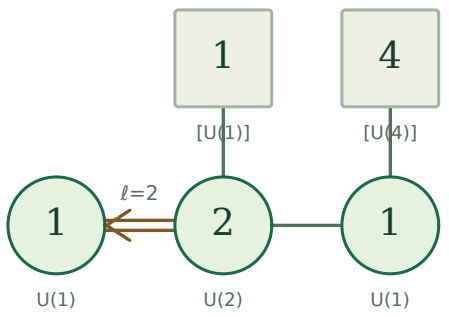


$1 + 11t^{2} + 60t^{4} + 10t^{5} + 225t^{6} + 80t^{7} + 665t^{8} + 350t^{9} + 1694t^{10} + O(t^{11})$

**Paper typo:** The N=8 series repeats the preceding N=7 row. All allowed monopole and Casimir degrees are even, so the printed 10t⁵ must vanish; a copy error is inferred, not author-confirmed.

**1106 2107.09077-NSL-N9-v2**

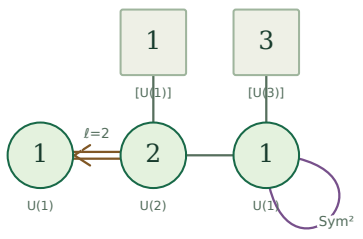


$1 + 11t^{2} + 60t^{4} + 225t^{6} + 10t^{7} + 665t^{8} + 80t^{9} + 1666t^{10} + O(t^{11})$

**1107 2107.09077-NSL-N10-v2**

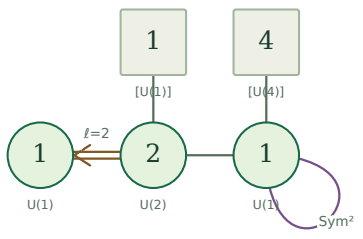


$1 + 11t^{2} + 60t^{4} + 225t^{6} + 10t^{7} + 665t^{8} + 80t^{9} + 1666t^{10} + O(t^{11})$

**Paper typo:** The N=10 series repeats the preceding N=9 row. All allowed monopole and Casimir degrees are even, so the printed 10t⁷ must vanish; a copy error is inferred, not author-confirmed.

**1108 2107.09077-NSL-N9-v4**

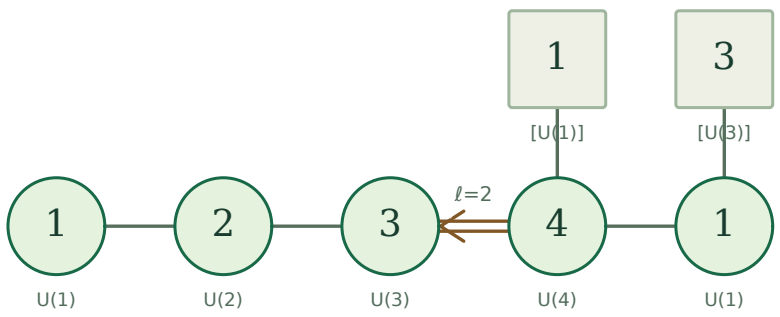


$1 + 37t^{2} + 675t^{4} + 8130t^{6} + 84t^{7} + 73047t^{8} + O(t^{9})$

**1109 2107.09077-NSL-N10-v4**

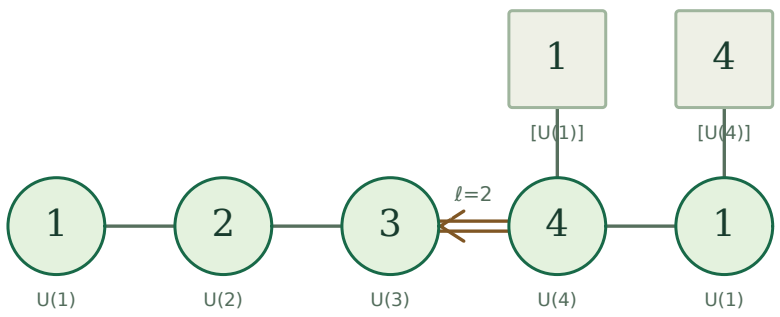


$1 + 37t^{2} + 675t^{4} + 8130t^{6} + 73131t^{8} + 526815t^{10} + O(t^{11})$

| QUIVER | HILBERT SERIES |
|---|---|

**1110 2107.09077-NSL-N11-v4**

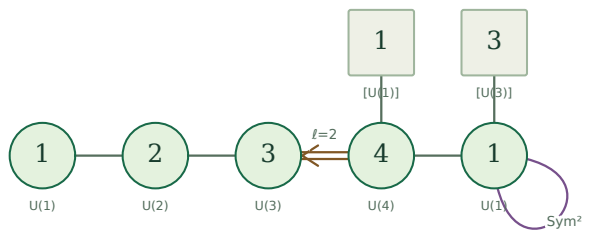


$1 + 37t^2 + 675t^4 + 8130t^6 + 73047t^8 + O(t^9)$

**1111 2107.09077-NSL-N12-v4**

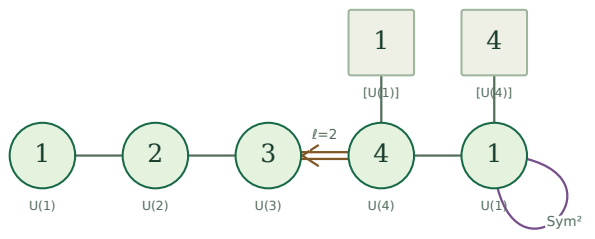


$1 + 37t^2 + 675t^4 + 8130t^6 + 73047t^8 + 524505t^{10} + O(t^{11})$

**1112 2107.09077-NSL-edge-N4**

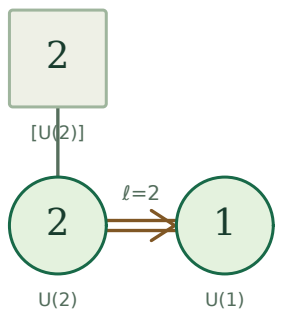


Source fixed magnetic charge · node n0

$1 + 6t^2 + 19t^4 + 44t^6 + 85t^8 + 146t^{10} + 231t^{12} + 344t^{14} + 489t^{16} + 670t^{18} + 891t^{20} + 1156t^{22} + 1469t^{24} + 1834t^{26} + 2255t^{28} + 2736t^{30} + O(t^{31})$

**1113 2107.09077-NSL-edge-N5**

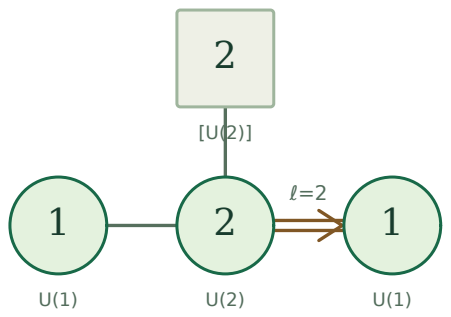


Source fixed magnetic charge · node n1

$1 + 6t^2 + 6t^3 + 20t^4 + 28t^5 + 65t^6 + 80t^7 + 166t^8 + 208t^9 + 354t^{10} + 456t^{11} + 711t^{12} + 876t^{13} + 1308t^{14} + 1594t^{15} + 2226t^{16} + 2712t^{17} + 3647t^{18} + 4344t^{19} + 5712t^{20} + 6736t^{21} + 8576t^{22} + 10072t^{23} + 12561t^{24} + 14552t^{25} + 17902t^{26} + 20582t^{27} + 24852t^{28} + 28452t^{29} + 33913t^{30} + O(t^{31})$

**1114 2107.09077-NSL-edge-N6**

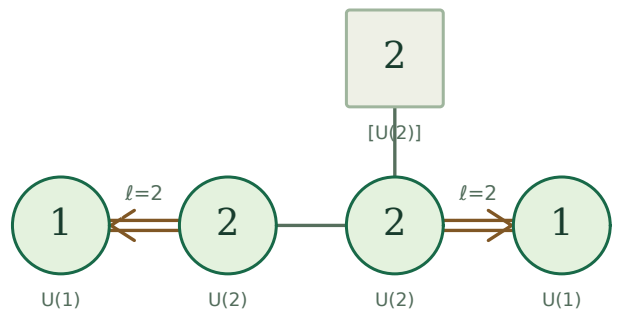


Source fixed magnetic charge · node n2

$1 + 13t^2 + 100t^4 + 527t^6 + 2174t^8 + 7425t^{10} + 21997t^{12} + 58102t^{14} + 139937t^{16} + 312042t^{18} + 652377t^{20} + 1290655t^{22} + 2435092t^{24} + 4407814t^{26} + 7694006t^{28} + 13004276t^{30} + O(t^{31})$

**1115 2107.09077-NSL-top-N4**

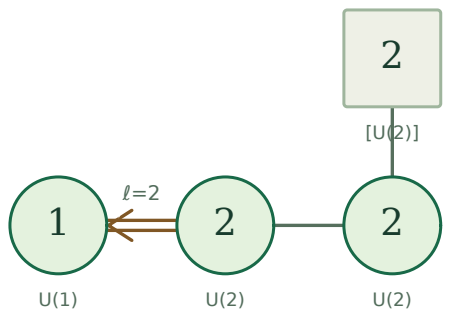


Source fixed magnetic charge · node n2

**Computed by Quiver Lab**

$1 + 15t^2 + 104t^4 + 475t^6 + 1659t^8 + 4802t^{10} + 12096t^{12} + 27378t^{14} + 56925t^{16} + 110473t^{18} + 202488t^{20} + 353717t^{22} + 593047t^{24} + 959700t^{26} + 1505792t^{28} + 2299284t^{30} + O(t^{31})$

Source originally Claimed · no numerical literature comparison is available.

**QUIVER** | **HILBERT SERIES**

**1116 2107.09077-NSL-top-N6**

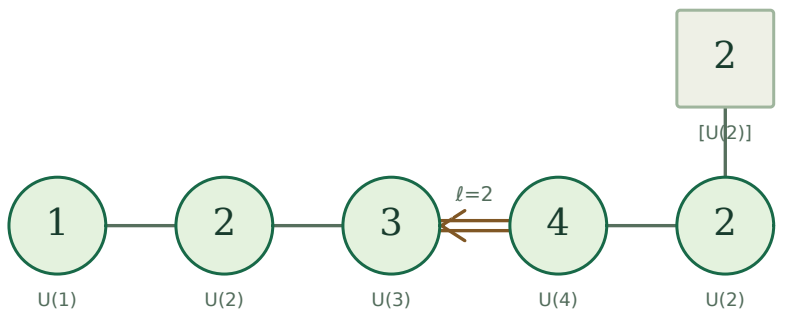


Source fixed magnetic charge · node n4

$1 + 36t^{2} + 681t^{4} + 8688t^{6} + 83376t^{8} + 640695t^{10} + 4110730t^{12} + O(t^{13})$

## 2109.07237 · Coulomb Branch Global Symmetry and Quiver Addition (35)

**1117 2109.07237-adjoint-1__n2 / n = 2**

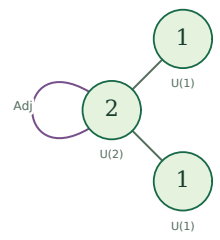

÷ U(1) diagonal

$1 + 10t^{2} + 49t^{4} + 165t^{6} + 440t^{8} + 1001t^{10} + 2030t^{12} + 3774t^{14} + 6555t^{16} + 10780t^{18} + 16951t^{20} + 25675t^{22} + 37674t^{24} + 53795t^{26} + 75020t^{28} + 102476t^{30} + O(t^{31})$

**1118 2109.07237-adjoint-1__n4 / n = 4**

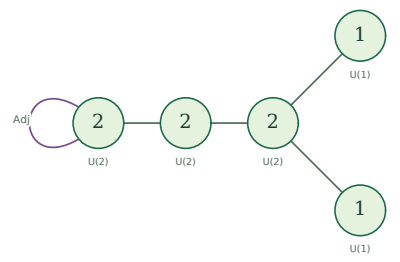

÷ U(1) diagonal

$1 + 36t^{2} + 539t^{4} + 4914t^{6} + 32130t^{8} + 164934t^{10} + 703494t^{12} + 2590731t^{14} + 8465886t^{16} + 25050025t^{18} + 68162094t^{20} + 172632824t^{22} + 410895576t^{24} + 926309736t^{26} + 1990597320t^{28} + 4099503726t^{30} + O(t^{31})$

**1119 2109.07237-adjoint-1__n5 / n = 5**

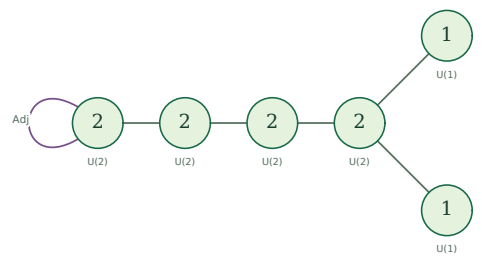

÷ U(1) diagonal

$1 + 55t^{2} + 1209t^{4} + 15675t^{6} + 141933t^{8} + 986271t^{10} + 5582577t^{12} + 26808639t^{14} + 112469500t^{16} + 421331196t^{18} + 1433385668t^{20} + 4487706444t^{22} + 13069085172t^{24} + 35711816700t^{26} + 92227253316t^{28} + 226471133460t^{30} + O(t^{31})$

**1120 2109.07237-adjoint-3__n2 / n = 2**

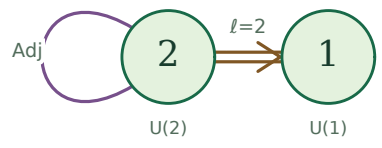


÷ U(1) diagonal

$1 + 6t^{2} + 19t^{4} + 44t^{6} + 85t^{8} + 146t^{10} + 231t^{12} + 344t^{14} + 489t^{16} + 670t^{18} + 891t^{20} + 1156t^{22} + 1469t^{24} + 1834t^{26} + 2255t^{28} + 2736t^{30} + O(t^{31})$

**1121 2109.07237-adjoint-3__n3 / n = 3**

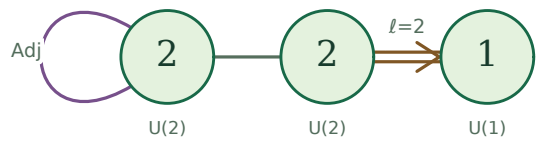


÷ U(1) diagonal

$1 + 15t^{2} + 104t^{4} + 475t^{6} + 1659t^{8} + 4802t^{10} + 12096t^{12} + 27378t^{14} + 56925t^{16} + 110473t^{18} + 202488t^{20} + 353717t^{22} + 593047t^{24} + 959700t^{26} + 1505792t^{28} + 2299284t^{30} + O(t^{31})$

| QUIVER | HILBERT SERIES |
|---|---|

**1122 2109.07237-adjoint-3__n4 / n = 4**

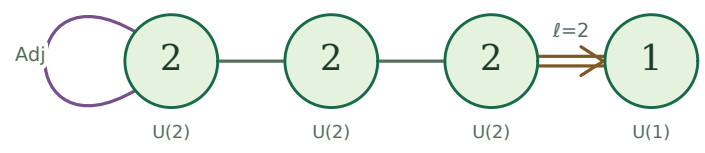


÷ U(1) diagonal

$1 + 28t^{2} + 335t^{4} + 2492t^{6} + 13524t^{8} + 58464t^{10} + 212652t^{12} + 675180t^{14} + 1920633t^{16} + 4989556t^{18} + 12010999t^{20} + 27095432t^{22} + 57795920t^{24} + 117412608t^{26} + 228513456t^{28} + 428166216t^{30} + O(t^{31})$

**1123 2109.07237-adjoint-4__n3 / n = 3**

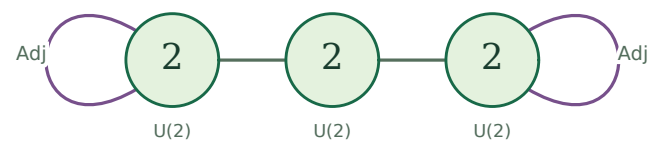


÷ U(1) diagonal

$1 + 15t^{2} + 125t^{4} + 685t^{6} + 2898t^{8} + 10052t^{10} + 30108t^{12} + 80130t^{14} + 194095t^{16} + 434643t^{18} + 911729t^{20} + 1808443t^{22} + 3419220t^{24} + 6199760t^{26} + 10837296t^{28} + 18338580t^{30} + O(t^{31})$

**1124 2109.07237-adjoint-4__n5 / n = 5**

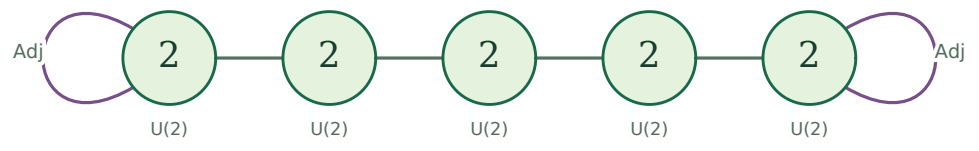


÷ U(1) diagonal

$1 + 45t^{2} + 879t^{4} + 10515t^{6} + 89925t^{8} + 599049t^{10} + 3283995t^{12} + 15383511t^{14} + 63283220t^{16} + 233359412t^{18} + 783757052t^{20} + 2427929868t^{22} + 7008323508t^{24} + 19008310980t^{26} + 48779719692t^{28} + 119133894924t^{30} + O(t^{31})$

**1125 2109.07237-adjoint-4__n6 / n = 6**

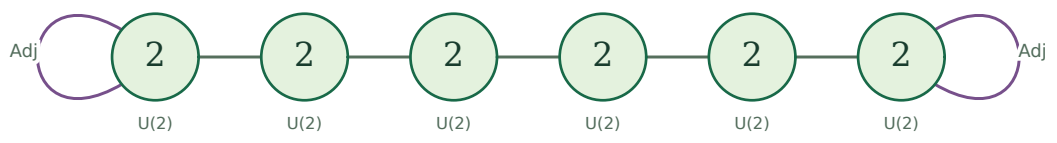


÷ U(1) diagonal

$1 + 66t^{2} + 1793t^{4} + 28963t^{6} + 326898t^{8} + 2821819t^{10} + 19734572t^{12} + 116345944t^{14} + 595157420t^{16} + 2699810256t^{18} + 11046275708t^{20} + 41317874468t^{22} + 142840248656t^{24} + 460539931060t^{26} + 1395257784712t^{28} + 3997323783244t^{30} + O(t^{31})$

**1126 2109.07237-adjoint-6__n2 / n = 2**

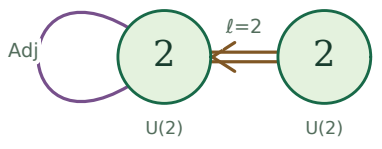


÷ U(1) diagonal

$1 + 9t^{2} + 48t^{4} + 160t^{6} + 435t^{8} + 987t^{10} + 2016t^{12} + 3744t^{14} + 6525t^{16} + 10725t^{18} + 16896t^{20} + 25584t^{22} + 37583t^{24} + 53655t^{26} + 74880t^{28} + 102272t^{30} + O(t^{31})$

**1127 2109.07237-adjoint-6__n3 / n = 3**

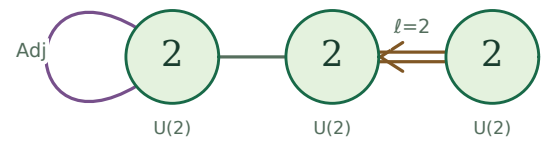


÷ U(1) diagonal

$1 + 25t^{2} + 255t^{4} + 1575t^{6} + 7140t^{8} + 25956t^{10} + 80220t^{12} + 218460t^{14} + 538065t^{16} + 1220505t^{18} + 2585583t^{20} + 5169255t^{22} + 9835280t^{24} + 17926160t^{26} + 31469040t^{28} + 53442288t^{30} + O(t^{31})$

| QUIVER | HILBERT SERIES |
| --- | --- |

**1128 2109.07237-adjoint-6__n4 / n = 4**

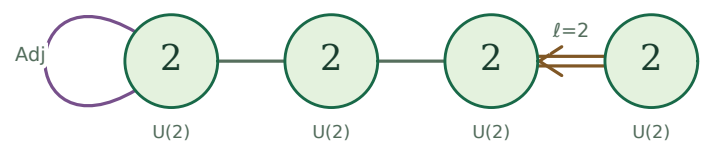


÷ U(1) diagonal

$1 + 49t^{2} + 840t^{4} + 8232t^{6} + 56280t^{8} + 297528t^{10} + 1295448t^{12} + 4842552t^{14} + 16002987t^{16} + 47762715t^{18} + 130850720t^{20} + 333211424t^{22} + 796618368t^{24} + 1802433024t^{26} + 3885147360t^{28} + 8021738592t^{30} + O(t^{31})$

**1129 2109.07237-T4-1**

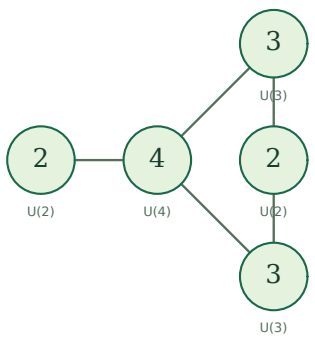


÷ U(1) diagonal

**Computed by Quiver Lab**

$1 + 36t^{2} + 750t^{4} + 11058t^{6} + 125895t^{8} + 1161611t^{10} + 8984030t^{12} + 59744441t^{14} + 348605188t^{16} + 1814397442t^{18} + 8538940498t^{20} + 36752437896t^{22} + 146063865357t^{24} + 540398120762t^{26} + 1874267169779t^{28} + 6130713555747t^{30} + O(t^{31})$

Source originally Claimed · no numerical literature comparison is available.

**1130 2109.07237-T4-5**

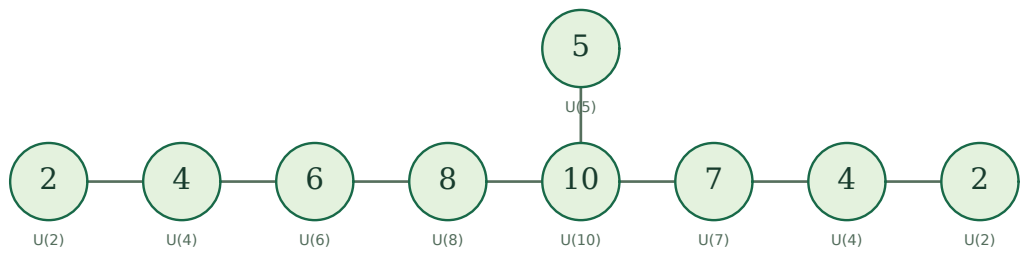


÷ U(1) diagonal

**Computed by Quiver Lab**

$1 + 108t^{2} + 256t^{3} + 6386t^{4} + 27648t^{5} + 297060t^{6} + 1603584t^{7} + 11911885t^{8} + 67143936t^{9} + 414249264t^{10} + 2264500224t^{11} + 12493799521t^{12} + O(t^{13})$

Source originally Claimed · no numerical literature comparison is available.

**1131 2109.07237-T5-1**

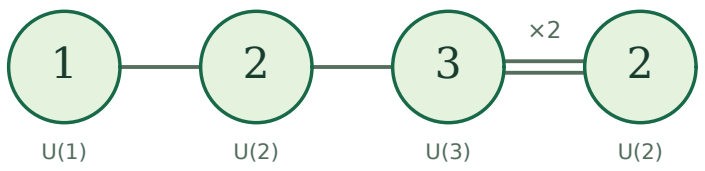


÷ U(1) diagonal

**Computed by Quiver Lab**

$1 + 21t^{2} + 230t^{4} + 1722t^{6} + 9870t^{8} + 46206t^{10} + 184338t^{12} + 645774t^{14} + 2031015t^{16} + 5832475t^{18} + 15496740t^{20} + 38499356t^{22} + 90199212t^{24} + 200695404t^{26} + 426573180t^{28} + 870367428t^{30} + O(t^{31})$

Source originally Claimed · no numerical literature comparison is available.

**1132 2109.07237-T5-4**

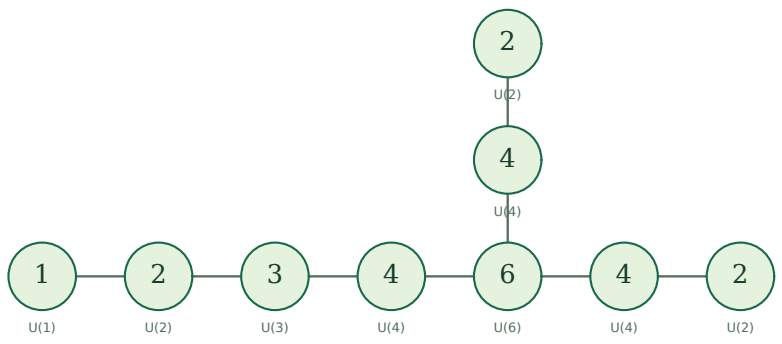


÷ U(1) diagonal

**Computed by Quiver Lab**

$1 + 56t^{2} + 160t^{3} + 1840t^{4} + 8800t^{5} + 54116t^{6} + 272000t^{7} + 1366365t^{8} + 6354720t^{9} + 28520413t^{10} + 121655296t^{11} + 499542996t^{12} + 1969751360t^{13} + 7495354624t^{14} + 27545098560t^{15} + 98014600407t^{16} + 338149311008t^{17} + 1133105273870t^{18} + 3692705161920t^{19} + 11720207852894t^{20} + O(t^{21})$

Source originally Claimed · no numerical literature comparison is available.

**1133 2109.07237-T5-5**

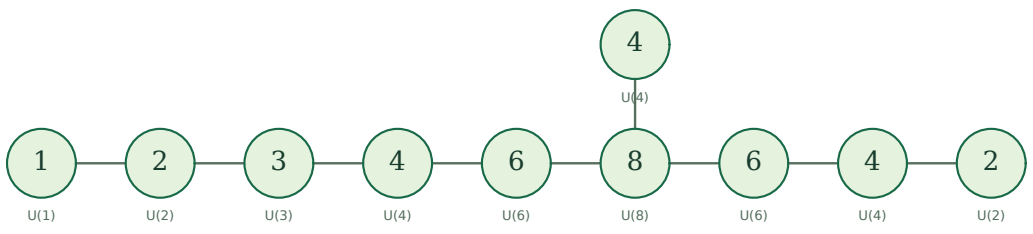


÷ U(1) diagonal

**Computed by Quiver Lab**

$1 + 87t^{2} + 256t^{3} + 4289t^{4} + 22016t^{5} + 179228t^{6} + 1040384t^{7} + 6519257t^{8} + 36442624t^{9} + 200610013t^{10} + 1042029312t^{11} + 5240252277t^{12} + O(t^{13})$

Source originally Claimed · no numerical literature comparison is available.

| QUIVER | HILBERT SERIES |
|---|---|

**1134 2109.07237-T6-3**

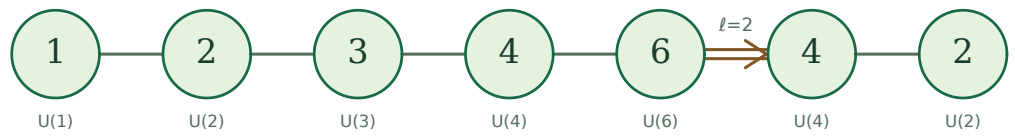


÷ U(1) diagonal

**Computed by Quiver Lab**

$1 + 42t^{2} + 112t^{3} + 1049t^{4} + O(t^{5})$

Source originally Claimed · no numerical literature comparison is available.

**1135 2109.07237-T6-4**

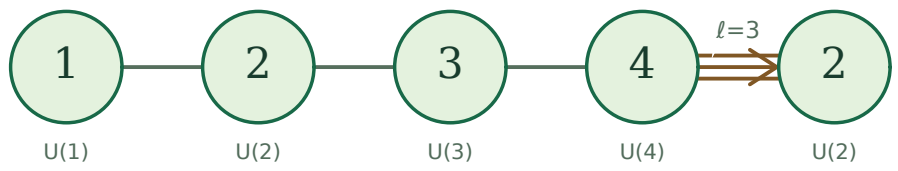


÷ U(1) diagonal

**Computed by Quiver Lab**

$1 + 24t^{2} + 32t^{3} + 320t^{4} + 736t^{5} + 3402t^{6} + 9088t^{7} + 30474t^{8} + 80944t^{9} + 229402t^{10} + 577968t^{11} + 1467816t^{12} + O(t^{13})$

Source originally Claimed · no numerical literature comparison is available.

**1136 2109.07237-T7-1**

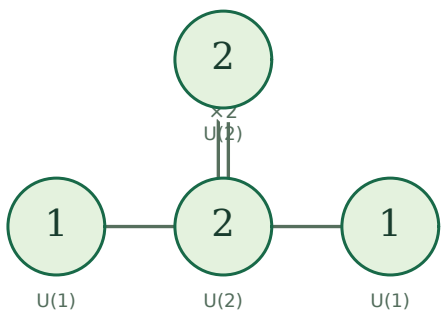


÷ U(1) diagonal

**Computed by Quiver Lab**

$1 + 13t^{2} + 100t^{4} + 527t^{6} + 2174t^{8} + 7425t^{10} + 21997t^{12} + 58102t^{14} + 139937t^{16} + 312042t^{18} + 652377t^{20} + 1290655t^{22} + 2435092t^{24} + 4407814t^{26} + 7694006t^{28} + 13004276t^{30} + O(t^{31})$

Source originally Claimed · no numerical literature comparison is available.

**1137 2109.07237-T7-4**

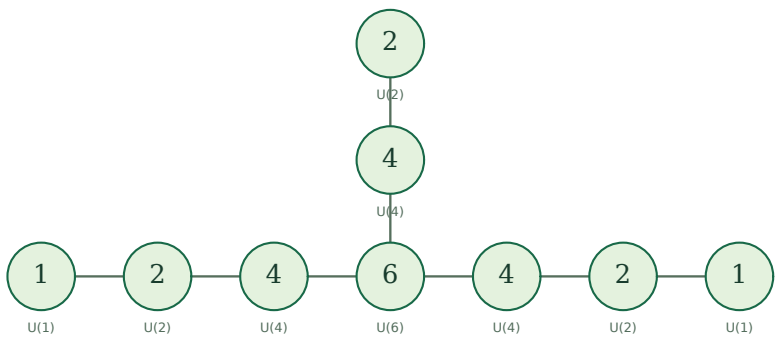


÷ U(1) diagonal

**Computed by Quiver Lab**

$1 + 56t^{2} + 128t^{3} + 1840t^{4} + 7040t^{5} + 49820t^{6} + 216064t^{7} + 1137320t^{8} + 4856960t^{9} + 21742985t^{10} + 87564160t^{11} + 351282417t^{12} + 1321740928t^{13} + 4869802204t^{14} + 17146526208t^{15} + 58829384985t^{16} + 194748713728t^{17} + 627960774339t^{18} + 1964708252800t^{19} + 5994524355559t^{20} + O(t^{21})$

Source originally Claimed · no numerical literature comparison is available.

**1138 2109.07237-T7-5**

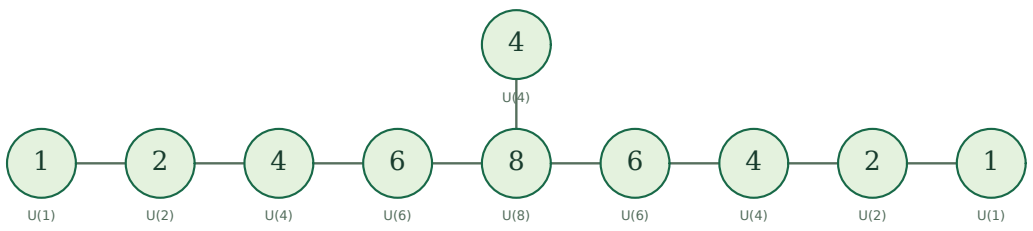


÷ U(1) diagonal

**Computed by Quiver Lab**

$1 + 89t^{2} + 216t^{3} + 4448t^{4} + 19008t^{5} + 177571t^{6} + 908944t^{7} + 6045435t^{8} + 31449920t^{9} + 175347495t^{10} + 872083792t^{11} + 4346535883t^{12} + O(t^{13})$

Source originally Claimed · no numerical literature comparison is available.

**1139 2109.07237-T12-3**

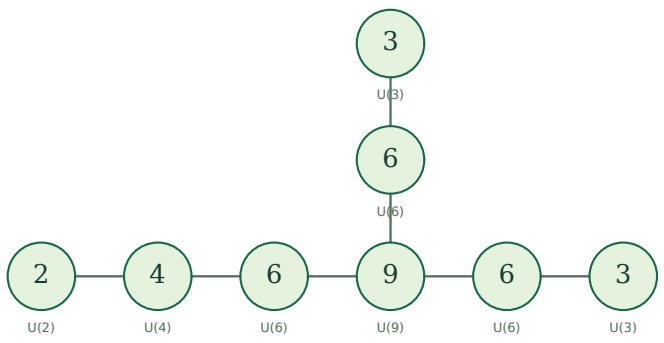


÷ U(1) diagonal

**Computed by Quiver Lab**

$1 + 49t^{2} + 140t^{3} + 1497t^{4} + 7280t^{5} + 44002t^{6} + 226500t^{7} + 1163330t^{8} + 5632556t^{9} + 26501895t^{10} + 119910188t^{11} + 526306116t^{12} + 2238143356t^{13} + 9250281803t^{14} + 37185315132t^{15} + 145627527485t^{16} + 556162933872t^{17} + 2073694301595t^{18} + 7555641796500t^{19} + 26926460568071t^{20} + O(t^{21})$

Source originally Claimed · no numerical literature comparison is available.

| QUIVER | HILBERT SERIES |
|---|---|

**1140 2109.07237-T12-4**

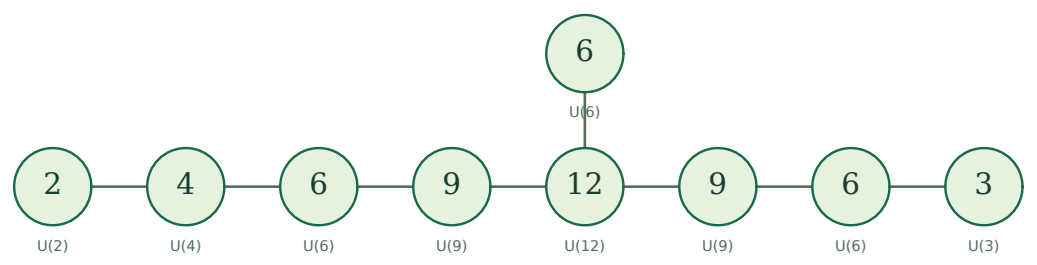


÷ U(1) diagonal

**Computed by Quiver Lab**

$1 + 80t^2 + 224t^3 + 3729t^4 + 18720t^5 + 152859t^6 + 890688t^7 + 5685332t^8 + 32604544t^9 + 185992108t^{10} + 1009579040t^{11} + 5348937897t^{12} + O(t^{13})$

Source originally Claimed · no numerical literature comparison is available.

**1141 2109.07237-T13-3**

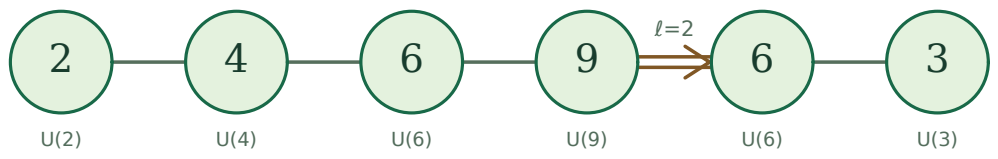


÷ U(1) diagonal

**Computed by Quiver Lab**

$1 + 35t^2 + 98t^3 + 804t^4 + O(t^5)$

Source originally Claimed · no numerical literature comparison is available.

**1142 2109.07237-T13-4**

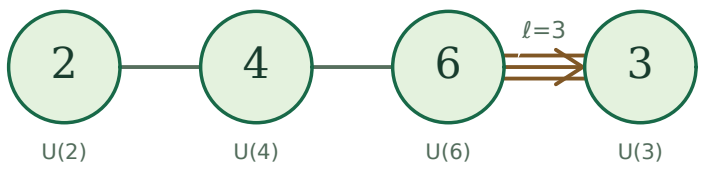


÷ U(1) diagonal

**Computed by Quiver Lab**

$1 + 17t^2 + 28t^3 + 201t^4 + 534t^5 + 2170t^6 + 6378t^7 + 20757t^8 + O(t^9)$

Source originally Claimed · no numerical literature comparison is available.

**1143 2109.07237-T14-3**

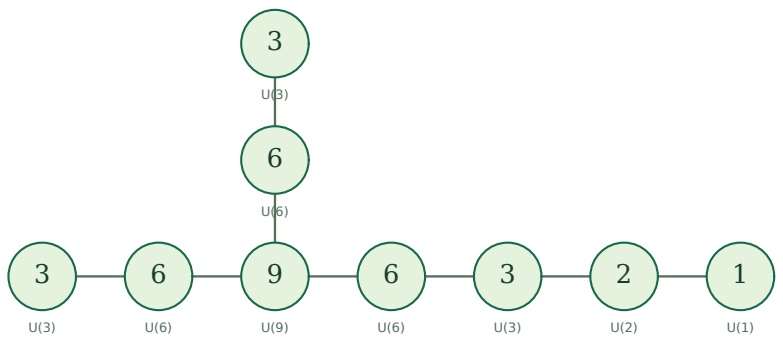


÷ U(1) diagonal

**Computed by Quiver Lab**

$1 + 92t^2 + 4823t^4 + 183631t^6 + 5565572t^8 + 140730709t^{10} + 3054674545t^{12} + 58032681082t^{14} + 978918127161t^{16} + O(t^{17})$

Source originally Claimed · no numerical literature comparison is available.

**1144 2109.07237-T14-4**

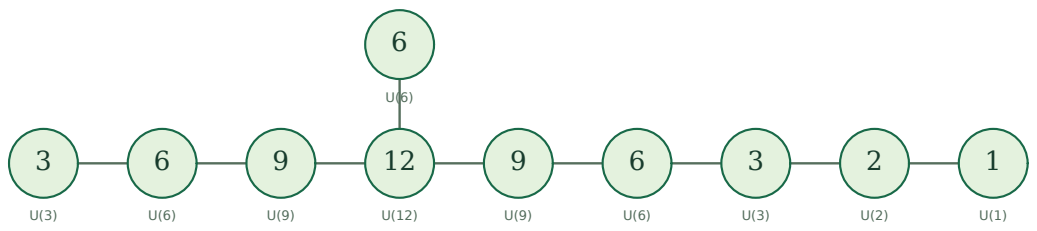


÷ U(1) diagonal

**Computed by Quiver Lab**

$1 + 147t^2 + 11808t^4 + 676046t^6 + 30547811t^8 + 1148212331t^{10} + 37055553367t^{12} + O(t^{13})$

Source originally Claimed · no numerical literature comparison is available.

**1145 2109.07237-T15-3**

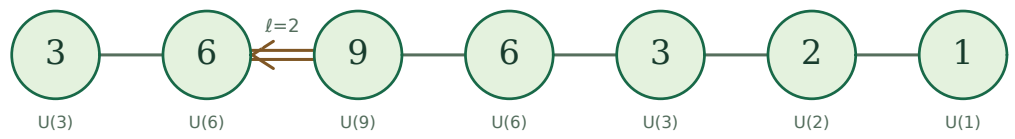


÷ U(1) diagonal

**Computed by Quiver Lab**

$1 + 66t^2 + 2574t^4 + O(t^5)$

Source originally Claimed · no numerical literature comparison is available.

| QUIVER | HILBERT SERIES |
|---|---|

**1146 2109.07237-T15-4**

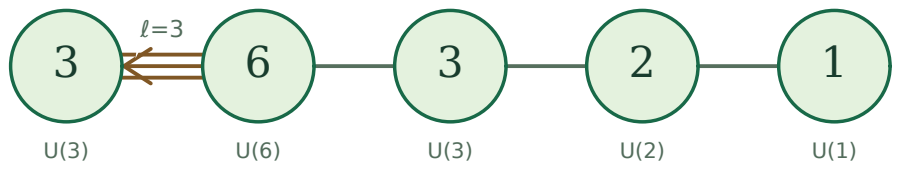


÷ U(1) diagonal

**Computed by Quiver Lab**

$1 + 28t^2 + 503t^4 + O(t^5)$

Source originally Claimed · no numerical literature comparison is available.

**1147 2109.07237-T20-3**

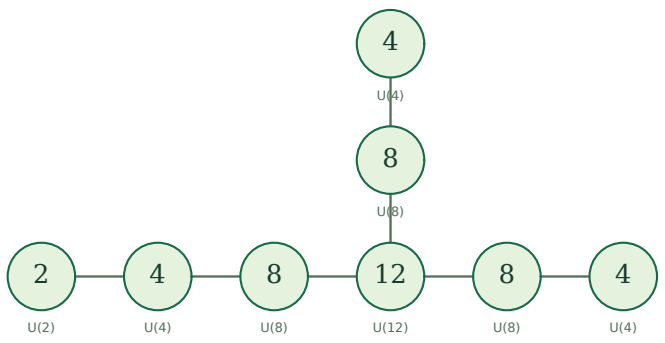


÷ U(1) diagonal

**Computed by Quiver Lab**

$1 + 86t^2 + 4221t^4 + 152285t^6 + 4451204t^8 + 110791183t^{10} + 2418579839t^{12} + 47204337508t^{14} + 834797952529t^{16} + O(t^{17})$

Source originally Claimed · no numerical literature comparison is available.

**1148 2109.07237-T21-4**

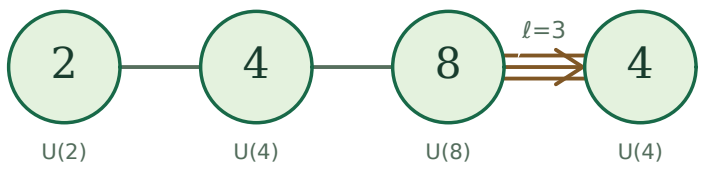


÷ U(1) diagonal

**Computed by Quiver Lab**

$1 + 22t^2 + 349t^4 + O(t^5)$

Source originally Claimed · no numerical literature comparison is available.

**1149 2109.07237-double-e6**

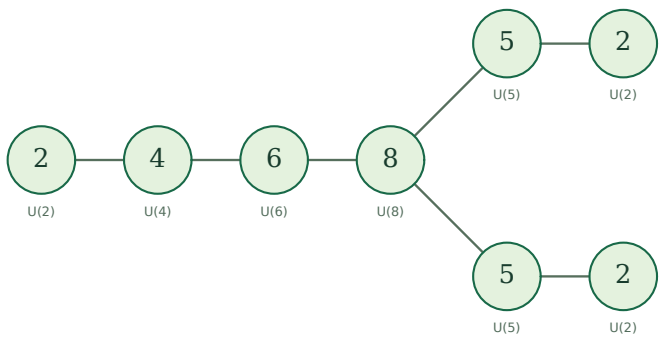


÷ U(1) diagonal

$1 + 79t^2 + O(t^4)$

**Paper typo:** Independent enumeration finds 128 degree-three monopoles, giving $1+79t^2+128t^3+O(t^4)$. If only the reported degree-two computation is retained, the remainder must be $O(t^3)$; the degree-two symmetry result is unaffected.

**1150 2109.07237-A1-G2-example**

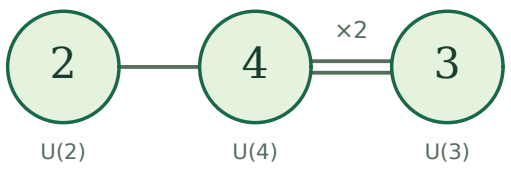


÷ U(1) diagonal

**Computed by Quiver Lab**

$1 + 14t^2 + 132t^4 + 917t^6 + 5180t^8 + 24688t^{10} + 102579t^{12} + 379402t^{14} + 1271213t^{16} + 3909758t^{18} + 11160849t^{20} + 29835161t^{22} + 75251028t^{24} + 180207701t^{26} + 411946685t^{28} + 903024590t^{30} + O(t^{31})$

Source originally Claimed · no numerical literature comparison is available.

**1151 2109.07237-d4-on-B5**

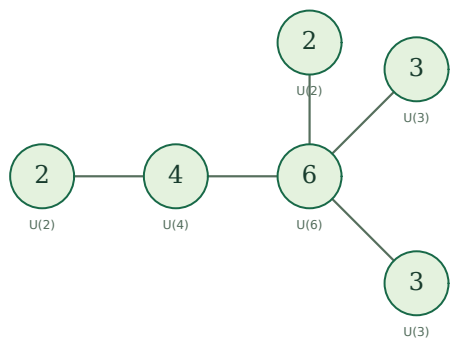


÷ U(1) diagonal

**Computed by Quiver Lab**

$1 + 55t^2 + 1705t^4 + 37378t^6 + 633776t^8 + 8725210t^{10} + 100713964t^{12} + 998047633t^{14} + 8650007961t^{16} + 66563264691t^{18} + 460525181622t^{20} + 2895181938333t^{22} + 16689260274006t^{24} + 88906292201572t^{26} + 440671382390051t^{28} + 2044416094231852t^{30} + O(t^{31})$

Source originally Claimed · no numerical literature comparison is available.

## 2109.07493 · Higgs branches of Argyres-Douglas theories as quiver varieties (2)

### 1152 2109.07493-D4SU6-T

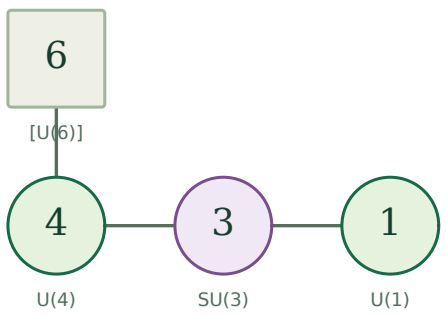


**Computed by Quiver Lab**

$1 + 3t^{2} + 8t^{3} + 15t^{4} + 34t^{5} + 76t^{6} + 142t^{7} + 282t^{8} + 528t^{9} + 951t^{10} + 1686t^{11} + 2930t^{12} + 4932t^{13} + 8206t^{14} + 13380t^{15} + 21424t^{16} + 33812t^{17} + 52615t^{18} + 80652t^{19} + 122189t^{20} + O(t^{21})$

Source originally Claimed · no numerical literature comparison is available.

### 1153 2109.07493-D9SU3-T

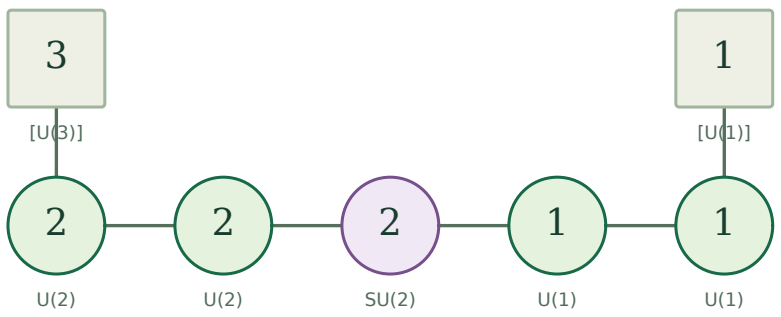


**Computed by Quiver Lab**

$1 + 12t^{2} + 28t^{3} + 101t^{4} + 288t^{5} + 785t^{6} + 1904t^{7} + 4563t^{8} + 10032t^{9} + 21409t^{10} + 43596t^{11} + 85795t^{12} + 162912t^{13} + 301173t^{14} + 540440t^{15} + 947318t^{16} + 1622016t^{17} + 2720149t^{18} + 4472212t^{19} + 7224090t^{20} + 11470616t^{21} + 17934140t^{22} + 27629100t^{23} + 41987002t^{24} + 62985232t^{25} + 93354126t^{26} + 136783800t^{27} + 198277970t^{28} + 284496240t^{29} + 404289002t^{30} + O(t^{31})$

Source originally Claimed · no numerical literature comparison is available.

## 2110.06940 · Conformal manifolds and 3d mirrors of (Dn,Dm) theories (5)

### 1154 2110.06940-520-n3

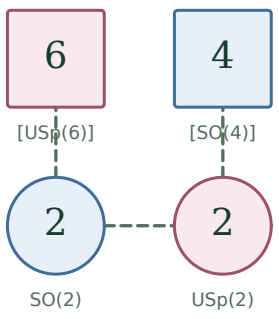


$1 + 2t^{2} + 5t^{4} + 8t^{6} + 17t^{8} + 26t^{10} + 41t^{12} + O(t^{13})$

### 1155 2110.06940-521-n3

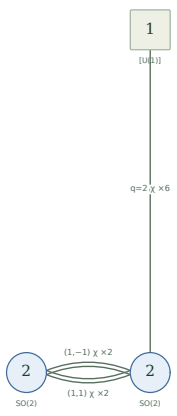

÷ $\mathbb{Z}_2$ diagonal

$1 + 2t^{2} + 5t^{4} + 8t^{6} + 17t^{8} + 26t^{10} + 41t^{12} + O(t^{13})$

### 1156 2110.06940-522-n3

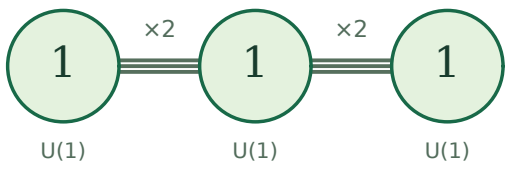


÷ U(1) diagonal

$1 + 2t^{2} + 5t^{4} + 8t^{6} + 17t^{8} + 26t^{10} + 41t^{12} + O(t^{13})$

### 1157 2110.06940-545-m2

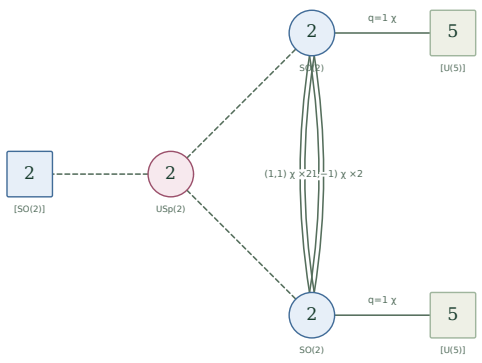

$1 + 3t^{2} + 8t^{4} + 16t^{6} + 29t^{8} + 47t^{10} + O(t^{11})$

| QUIVER | HILBERT SERIES |
|---|---|

**1158 2110.06940-546-m2**

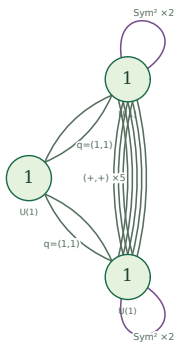

Central Z2 quotient · phases n0:1, n1:1, n2:1

$1 + 3t^{2} + 8t^{4} + 16t^{6} + 29t^{8} + 47t^{10} + O(t^{11})$

## 2110.11365 · Magnetic quivers for rank 2 theories (5)

**1159 2110.11365-kn-3**

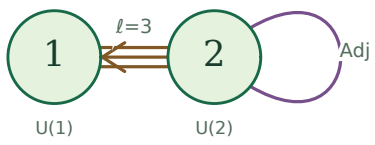


$1 + 3t^{2} + 4t^{3} + 6t^{4} + 10t^{5} + 17t^{6} + 18t^{7} + 31t^{8} + 38t^{9} + 48t^{10} + 62t^{11} + 81t^{12} + 90t^{13} + 119t^{14} + 138t^{15} + 162t^{16} + 192t^{17} + 229t^{18} + 252t^{19} + 303t^{20} + 340t^{21} + 384t^{22} + 436t^{23} + 497t^{24} + 540t^{25} + 619t^{26} + 680t^{27} + 750t^{28} + 830t^{29} + 921t^{30} + O(t^{31})$

**1160 2110.11365-kn-4**

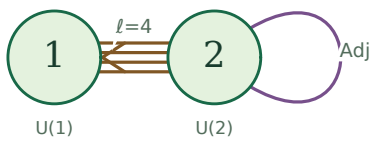


Ungauge long U(2) · node n1 · fix one flux

$1 + 3t^{2} + 11t^{4} + 21t^{6} + 44t^{8} + 70t^{10} + 116t^{12} + 166t^{14} + 243t^{16} + 325t^{18} + 441t^{20} + 563t^{22} + 726t^{24} + 896t^{26} + 1114t^{28} + 1340t^{30} + O(t^{31})$

**Paper typo:** After correcting the import to ungauge the long U(2) node, the printed numerator still repeats $t^8$. Its first $t^8$ should be $t^6$, yielding numerator $1+t^2+4t^4+t^6+t^8$ and coefficient 22 rather than 21 at $t^6$.

**1161 2110.11365-kn-6**

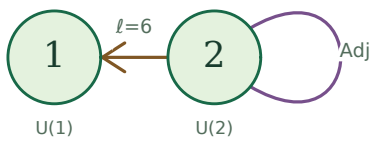


$1 + 3t^{2} + 6t^{4} + 17t^{6} + 31t^{8} + 48t^{10} + 81t^{12} + 119t^{14} + 162t^{16} + 229t^{18} + 303t^{20} + 384t^{22} + 497t^{24} + 619t^{26} + 750t^{28} + 921t^{30} + O(t^{31})$

**1162 2110.11365-G312**

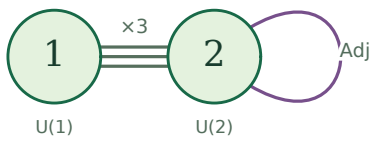


÷ U(1) diagonal

$1 + t^{2} + 2t^{3} + 2t^{4} + 4t^{5} + 7t^{6} + 6t^{7} + 11t^{8} + 14t^{9} + 16t^{10} + 22t^{11} + 29t^{12} + 30t^{13} + 41t^{14} + 48t^{15} + 54t^{16} + 66t^{17} + 79t^{18} + 84t^{19} + 103t^{20} + 116t^{21} + 128t^{22} + 148t^{23} + 169t^{24} + 180t^{25} + 209t^{26} + 230t^{27} + 250t^{28} + 280t^{29} + 311t^{30} + O(t^{31})$

**1163 2110.11365-G412**

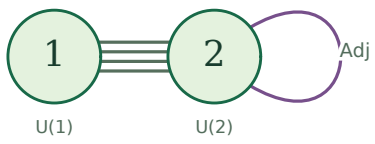


÷ U(1) diagonal

$1 + t^{2} + 4t^{4} + 6t^{6} + 14t^{8} + 19t^{10} + 33t^{12} + 44t^{14} + 67t^{16} + 85t^{18} + 118t^{20} + 146t^{22} + 192t^{24} + 231t^{26} + 291t^{28} + 344t^{30} + O(t^{31})$

## 2111.00026 · Balanced B and D-type orthosymplectic quivers - magnetic quivers for product theories (99)

**1164 2111.00026-T8-1__n3_k1_l0** / n = 3 · k = 1 · l = 0

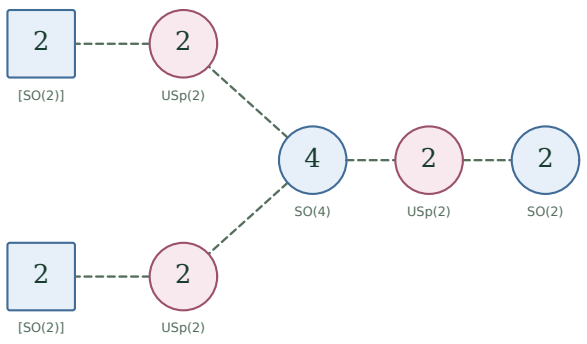

$1 + 30t^{2} + 393t^{4} + 3120t^{6} + 17706t^{8} + 78972t^{10} + 293770t^{12} + 948336t^{14} + 2731755t^{16} + 7166042t^{18} + 17382963t^{20} + O(t^{21})$

**1165 2111.00026-T8-2__n4_k1_l0** / n = 4 · k = 1 · l = 0

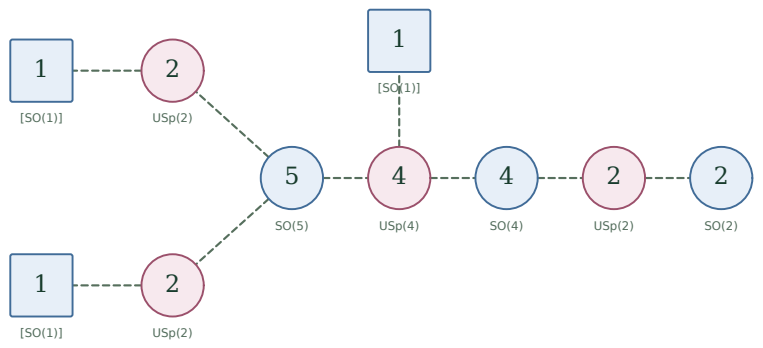

$1 + 56t^{2} + 1384t^{4} + 20650t^{6} + 215636t^{8} + 1720264t^{10} + 11106025t^{12} + 60413096t^{14} + O(t^{15})$

**1166 2111.00026-T8-3__n5_k1_l0** / n = 5 · k = 1 · l = 0

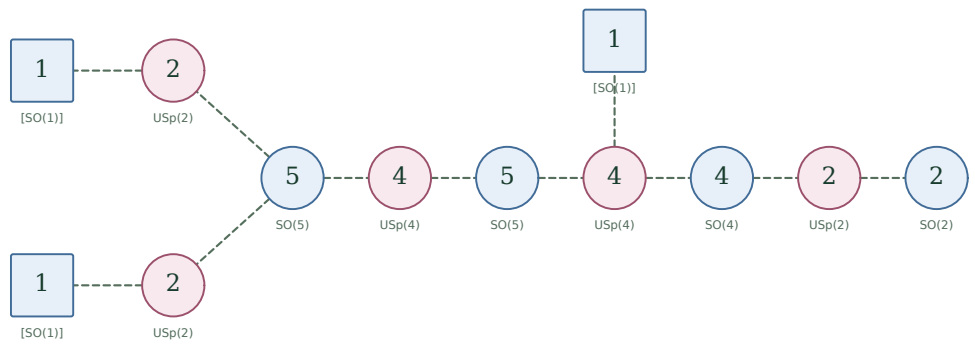

$1 + 90t^{2} + 3565t^{4} + 84588t^{6} + 1386700t^{8} + 17100048t^{10} + 167857416t^{12} + 1365541740t^{14} + O(t^{15})$

**1167 2111.00026-T8-4__n5_k2_l0** / n = 5 · k = 2 · l = 0

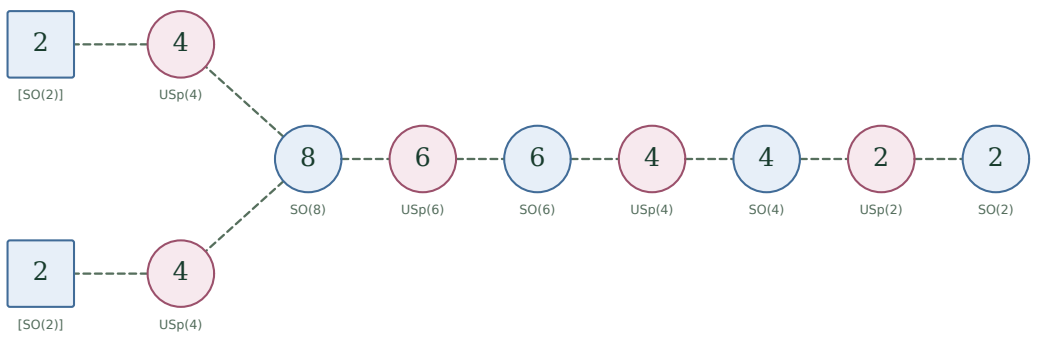

$1 + 90t^{2} + 3985t^{4} + 115368t^{6} + 2454040t^{8} + 40922106t^{10} + 557904916t^{12} + 6407626218t^{14} + 63421757476t^{16} + 550795731040t^{18} + 4259031312565t^{20} + O(t^{21})$

**1168 2111.00026-T8-5__n6_k1_l0** / n = 6 · k = 1 · l = 0

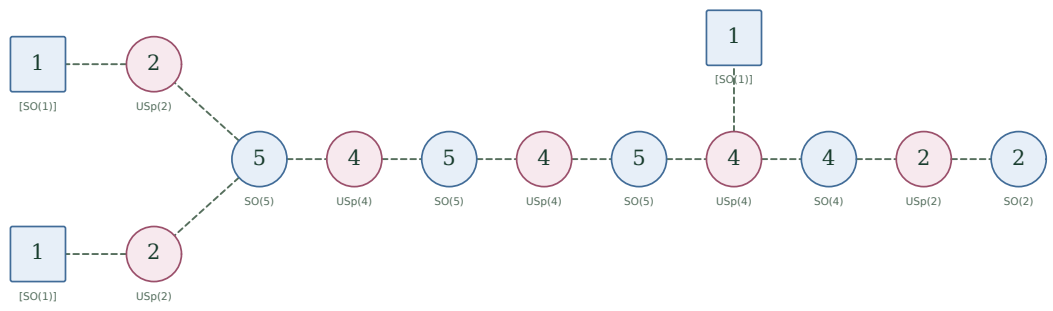

$1 + 132t^{2} + 8622t^{4} + 369798t^{6} + 11686995t^{8} + 289989018t^{10} + 5884262173t^{12} + 100491980996t^{14} + O(t^{15})$

**Paper typo:** The stated SQCD mirror and direct quiver sum give 7632. The printed factor includes an extra 495-dimensional representation, inconsistent with the stated minimal-orbit specialization.

**1169 2111.00026-T8-6__n6_k1_l1** / n = 6 · k = 1 · l = 1

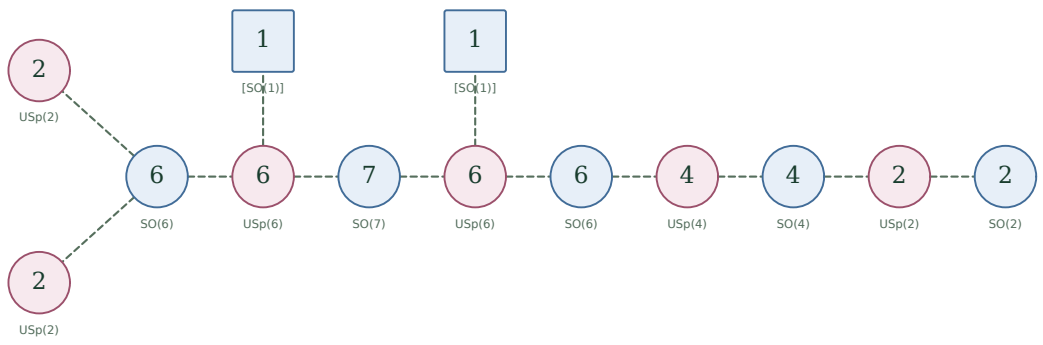

$1 + 132t^{2} + 8127t^{4} + 316107t^{6} + 8809263t^{8} + 188703108t^{10} + 3252768155t^{12} + 46622001052t^{14} + 569616000252t^{16} + O(t^{17})$

**1170 2111.00026-T8-7__n6_k1_l2** / n = 6 · k = 1 · l = 2

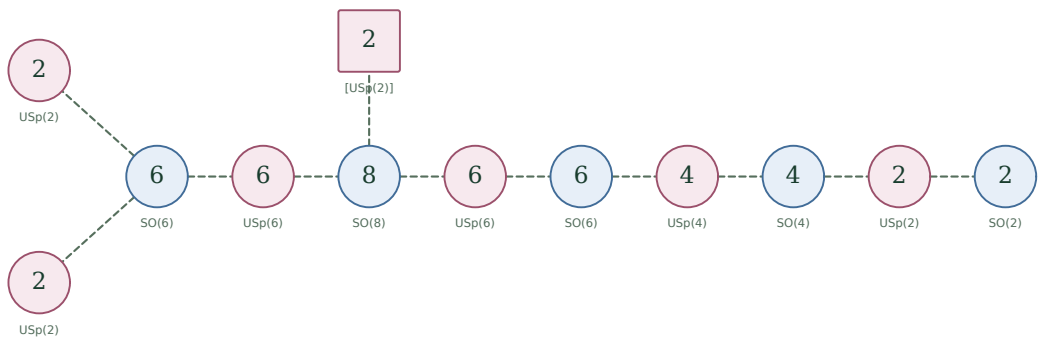

$1 + 132t^{2} + 8127t^{4} + 316569t^{6} + 8861205t^{8} + 191368881t^{10} + O(t^{11})$

**QUIVER** | **HILBERT SERIES**

**1171 2111.00026-T8-8__n7_k3_l0 / n = 7 · k = 3 · l = 0**

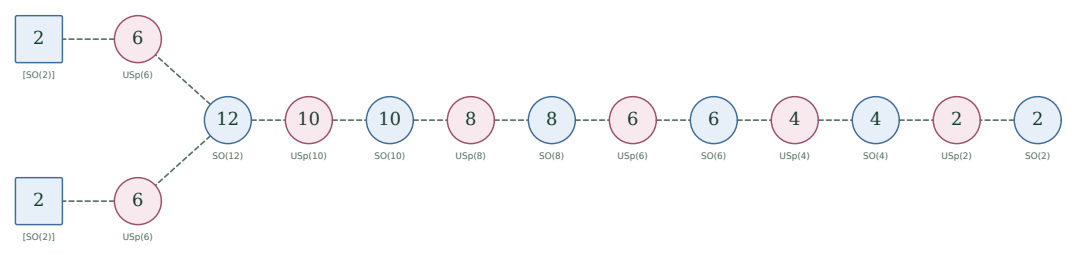

$1 + 182t^{2} + 16443t^{4} + 983372t^{6} + 43795493t^{8} + 1549232776t^{10} + 45339736401t^{12} + 1129102560664t^{14} + 24425137877287t^{16} + 466288937297638t^{18} + 7954868879934934t^{20} + O(t^{21})$

**1172 2111.00026-T8-9__n8_k2_l2 / n = 8 · k = 2 · l = 2**

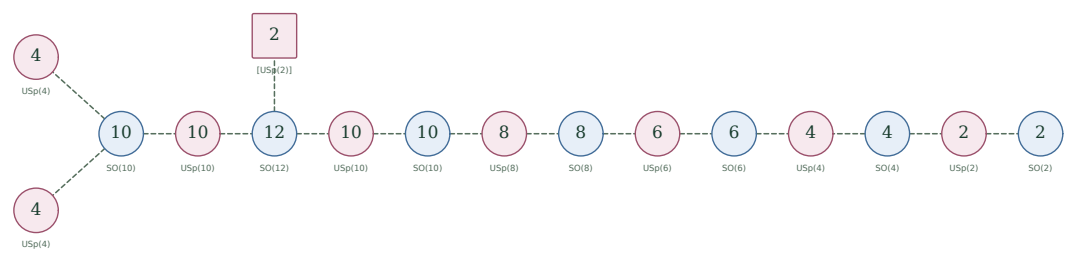

$1 + 240t^{2} + 28648t^{4} + 2259862t^{6} + 132261331t^{8} + 6119250745t^{10} + O(t^{11})$

**1173 2111.00026-T8-10__n9_k1_l2 / n = 9 · k = 1 · l = 2**

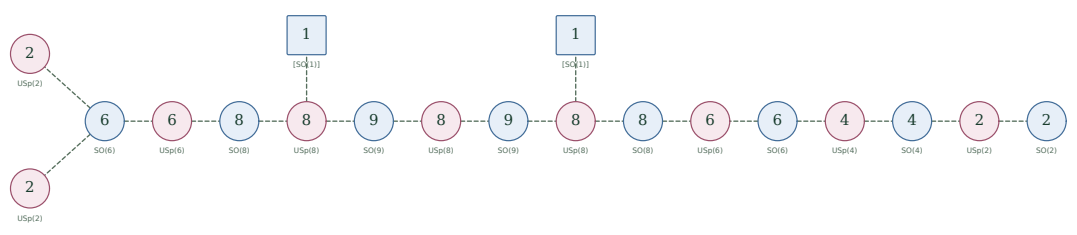

$1 + 306t^{2} + 43569t^{4} + 3928972t^{6} + 255457998t^{8} + 12867497523t^{10} + 525512619934t^{12} + 17956377133055t^{14} + O(t^{15})$

**1174 2111.00026-T9-1__k2_l0 / k = 2 · l = 0**

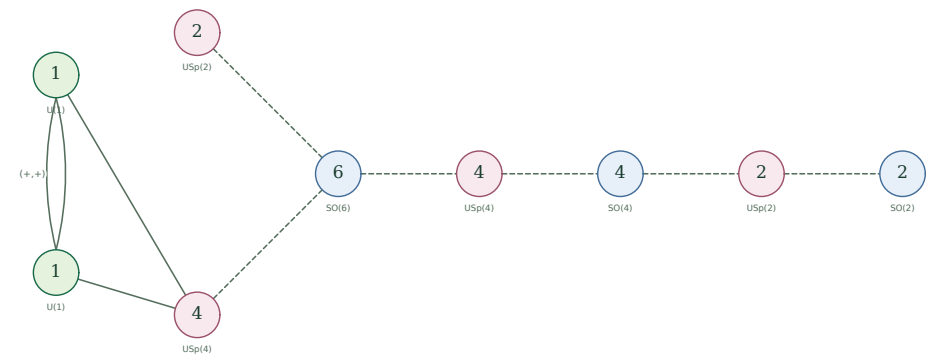

÷ $\mathbb{Z}_2$ diagonal

$1 + 58t^{2} + 32t^{3} + 1569t^{4} + 1600t^{5} + 27220t^{6} + 37856t^{7} + 348124t^{8} + 577920t^{9} + 3540936t^{10} + 6502720t^{11} + 29981572t^{12} + 58099072t^{13} + 217747736t^{14} + 432151456t^{15} + 1385881186t^{16} + 2764473568t^{17} + 7858110900t^{18} + 15572468640t^{19} + 40224531398t^{20} + O(t^{21})$

**1175 2111.00026-T9-2__k3_l1 / k = 3 · l = 1**

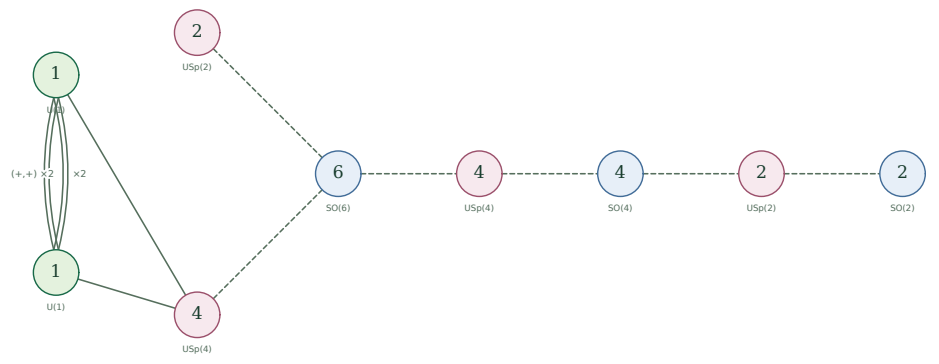

÷ $\mathbb{Z}_2$ diagonal

$1 + 58t^{2} + 1601t^{4} + 28424t^{6} + 369156t^{8} + 3771348t^{10} + 31759488t^{12} + 227801304t^{14} + 1425775758t^{16} + 7933063516t^{18} + 39822908626t^{20} + O(t^{21})$

**1176 2111.00026-T9-3__k3_l0 / k = 3 · l = 0**

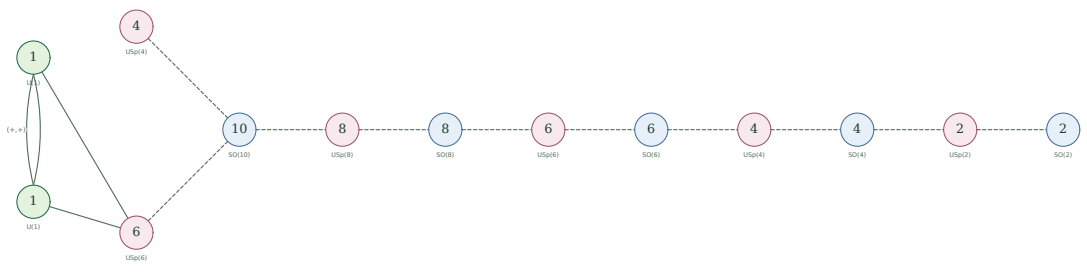

÷ $\mathbb{Z}_2$ diagonal

$1 + 134t^{2} + 9017t^{4} + 403982t^{6} + 13508026t^{8} + 358747158t^{10} + O(t^{11})$

| QUIVER | HILBERT SERIES |
|---|---|

**1177 2111.00026-T9-4__k4_l1** / **k = 4 · l = 1**

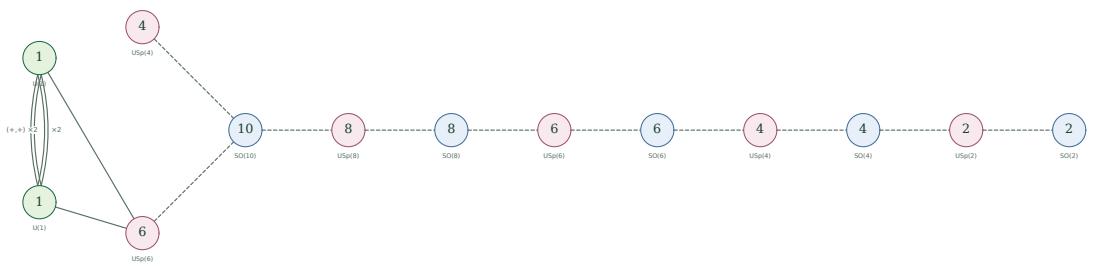

÷ $\mathbb{Z}_2$ diagonal

$1 + 134t^{2} + 8889t^{4} + 128t^{5} + 388366t^{6} + 15616t^{7} + 12558722t^{8} + 943360t^{9} + 320512966t^{10} + O(t^{11})$

**1178 2111.00026-T9-5__k1_l0** / **k = 1 · l = 0**

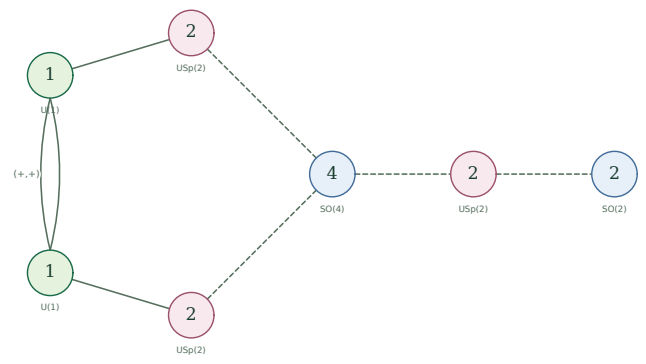


÷ $\mathbb{Z}_2$ diagonal

$1 + 48t^{2} + 976t^{4} + 11600t^{6} + 95350t^{8} + 598352t^{10} + 3053296t^{12} + 13224752t^{14} + 50129875t^{16} + 170108000t^{18} + 525728128t^{20} + O(t^{21})$

**1179 2111.00026-T9-6__k2_l1** / **k = 2 · l = 1**

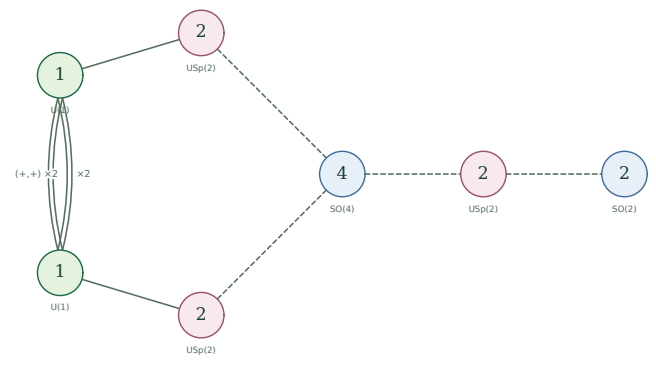


÷ $\mathbb{Z}_2$ diagonal

$1 + 32t^{2} + 16t^{3} + 456t^{4} + 416t^{5} + 4104t^{6} + 4960t^{7} + 27490t^{8} + 38400t^{9} + 148792t^{10} + 223840t^{11} + 681924t^{12} + 1062432t^{13} + 2729368t^{14} + 4305376t^{15} + 9754099t^{16} + 15374976t^{17} + 31652168t^{18} + 49482864t^{19} + 94506536t^{20} + O(t^{21})$

**1180 2111.00026-T9-7__k2_l0** / **k = 2 · l = 0**

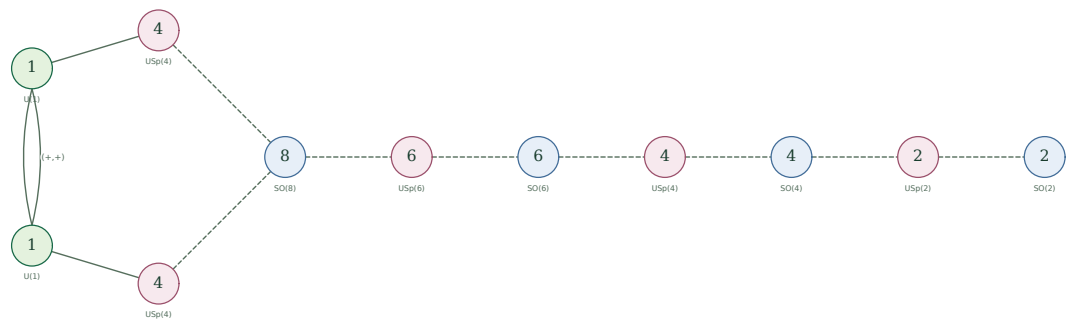

÷ $\mathbb{Z}_2$ diagonal

$1 + 92t^{2} + 64t^{3} + 4168t^{4} + 5248t^{5} + 125140t^{6} + 211968t^{7} + 2809296t^{8} + 5635008t^{9} + 50260590t^{10} + O(t^{11})$

**1181 2111.00026-T9-8__k3_l1** / **k = 3 · l = 1**

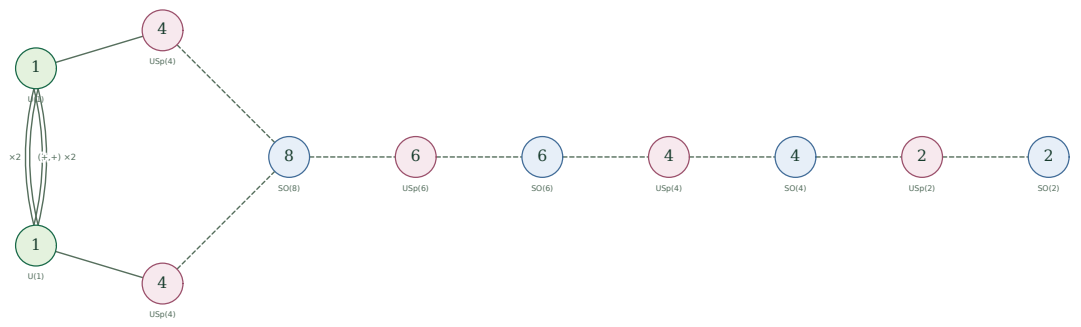

÷ $\mathbb{Z}_2$ diagonal

$1 + 92t^{2} + 4232t^{4} + 128860t^{6} + 2910592t^{8} + 51921078t^{10} + O(t^{11})$

**1182 2111.00026-T9-9__k1** / **k = 1**

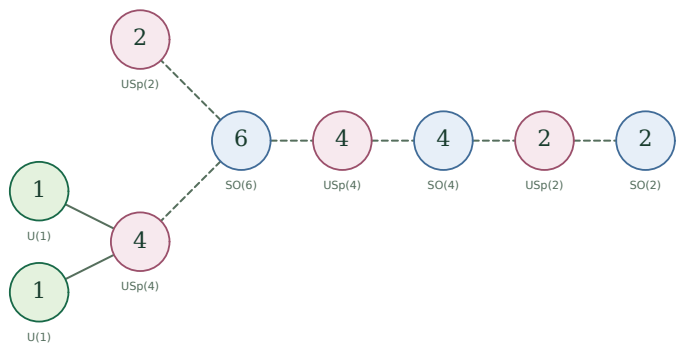


÷ $\mathbb{Z}_2$ diagonal

$1 + 90t^{2} + 3565t^{4} + 84588t^{6} + 1386700t^{8} + 17100048t^{10} + 167857416t^{12} + 1365541740t^{14} + 9486469554t^{16} + 57589566980t^{18} + 311107661634t^{20} + O(t^{21})$

| QUIVER | HILBERT SERIES |
|---|---|
| **1183 2111.00026-T9-10__k2** / k = 2 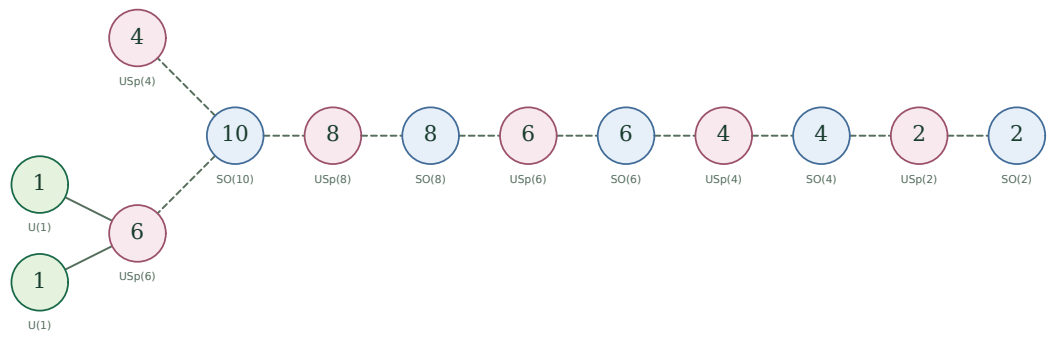  ÷ $\mathbb{Z}_2$ diagonal | $1 + 134t^{2} + 128t^{3} + 8889t^{4} + 15616t^{5} + 394310t^{6} + 943360t^{7} + 13220746t^{8} + 37713280t^{9} + 357061474t^{10} + 1123982464t^{11} + 8061468875t^{12} + 26656882432t^{13} + 155821066386t^{14} + 524237477376t^{15} + 2622606331104t^{16} + 8795126835840t^{17} + 38944260460754t^{18} + 128525072279296t^{19} + 515800796930805t^{20} + O(t^{21})$ |
| **1184 2111.00026-T9-11__k1** / k = 1 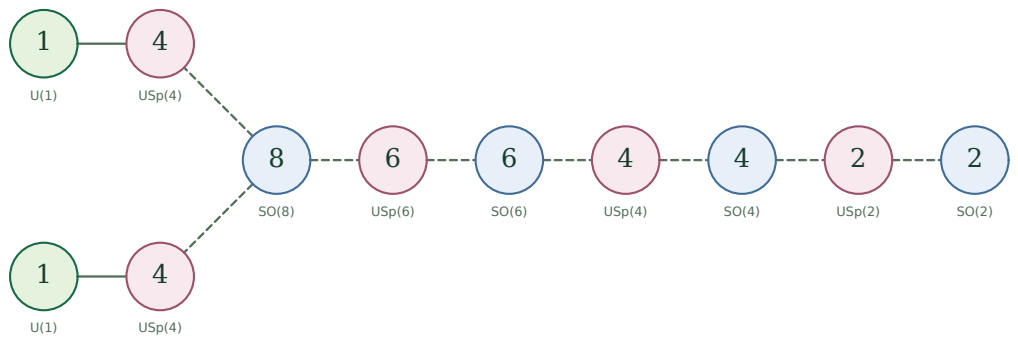  ÷ $\mathbb{Z}_2$ diagonal | $1 + 156t^{2} + 10944t^{4} + 466596t^{6} + 13807080t^{8} + 306524790t^{10} + 5377829028t^{12} + 77405354312t^{14} + 940221343776t^{16} + 9853246779220t^{18} + 90680857312617t^{20} + O(t^{21})$ |
| **1185 2111.00026-T9-12__k2** / k = 2 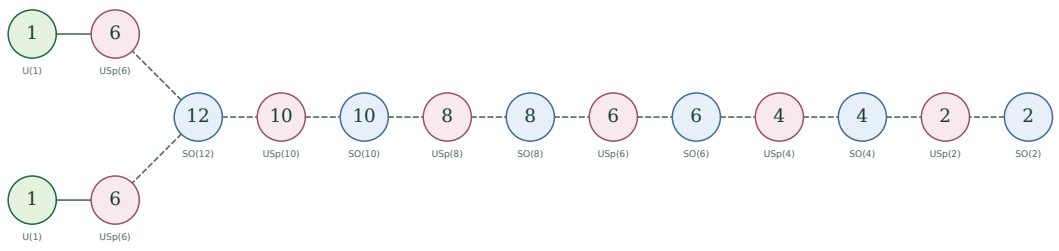  ÷ $\mathbb{Z}_2$ diagonal | $1 + 184t^{2} + 256t^{3} + 16810t^{4} + 43520t^{5} + 1040056t^{6} + 3674112t^{7} + 49469875t^{8} + 206409216t^{9} + 1925805206t^{10} + 8703438848t^{11} + 63418779630t^{12} + 294056655104t^{13} + 1801033108182t^{14} + 8290154332416t^{15} + 44696692490453t^{16} + 200425546653440t^{17} + 979719926063866t^{18} + 4237436679956480t^{19} + 19144937799760589t^{20} + O(t^{21})$ |
| **1186 2111.00026-T9-13__k1** / k = 1 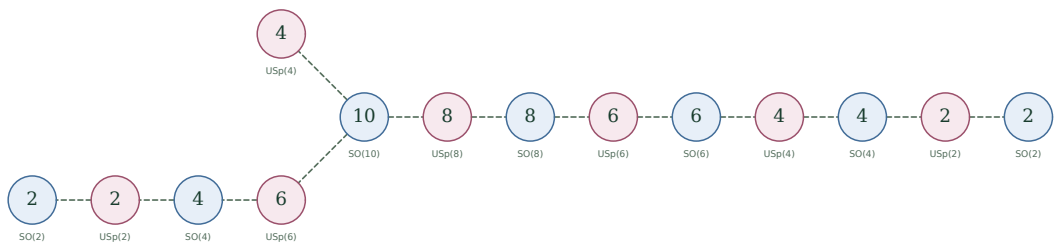  ÷ $\mathbb{Z}_2$ diagonal | $1 + 266t^{2} + 32431t^{4} + 2437890t^{6} + 128297273t^{8} + 5085058160t^{10} + 159309437560t^{12} + 4084414731904t^{14} + 87979394611180t^{16} + 1625455257184024t^{18} + 26192135424825720t^{20} + O(t^{21})$ |
| **1187 2111.00026-T9-14__k1** / k = 1 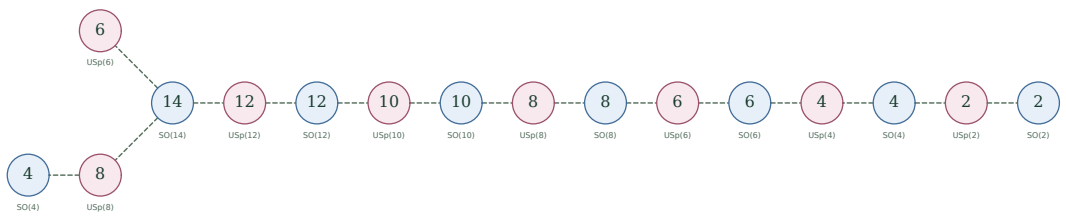  ÷ $\mathbb{Z}_2$ diagonal | $1 + 496t^{2} + 115504t^{4} + 16918250t^{6} + 1761796000t^{8} + 139749232560t^{10} + O(t^{11})$ |
| **1188 2111.00026-T10-1__n5_k1_l1** / n = 5 · k = 1 · l = 1 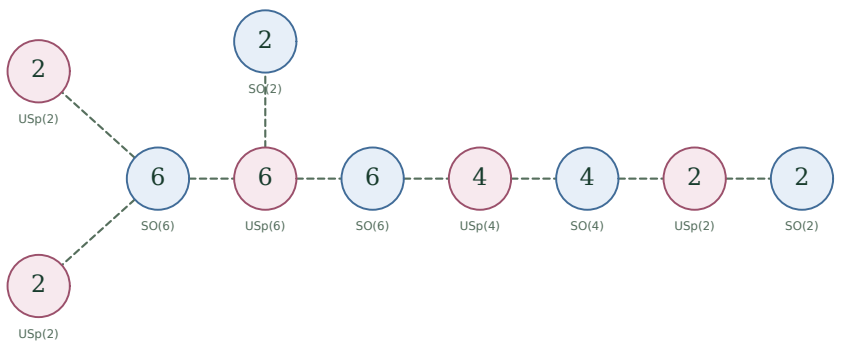  ÷ $\mathbb{Z}_2$ diagonal | $1 + 123t^{2} + 6710t^{4} + 220812t^{6} + 5027328t^{8} + 85856661t^{10} + 1160906385t^{12} + 12919480684t^{14} + 121836107745t^{16} + O(t^{17})$ |

| QUIVER | HILBERT SERIES |
|---|---|

**1189 2111.00026-T10-2__n7_k1_l2** / **n = 7 · k = 1 · l = 2**

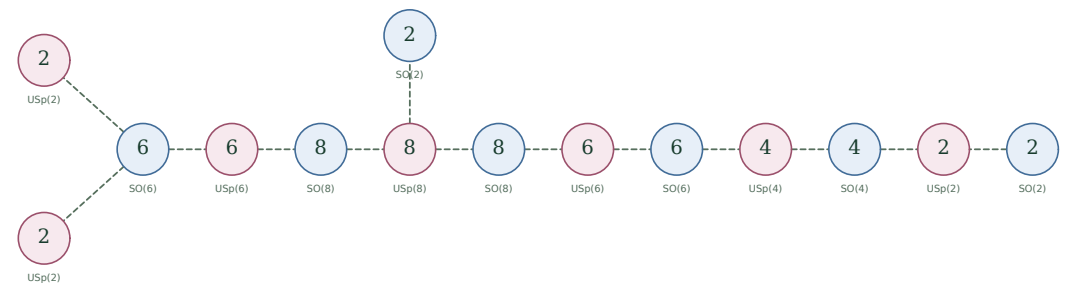


÷ ℤ₂ diagonal

$1 + 183t^2 + 128t^3 + 15625t^4 + 21632t^5 + 849367t^6 + 1687168t^7 + 33584330t^8 + 83001984t^9 + 1039787073t^{10} + 2940462720t^{11} + 26377189292t^{12} + 80780840064t^{13} + 564910951617t^{14} + 1804412218752t^{15} + 10430788358641t^{16} + 33865771104384t^{17} + 168680900002512t^{18} + 547066306438912t^{19} + 2419061167200157t^{20} + O(t^{21})$

**1190 2111.00026-T10-3__n7_k2_l1** / **n = 7 · k = 2 · l = 1**

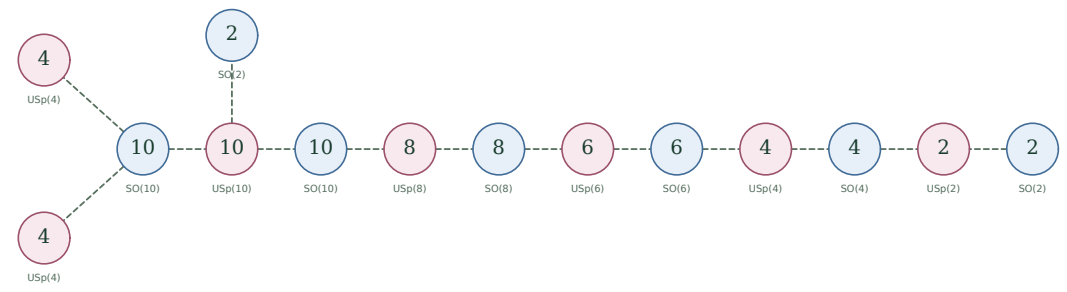


÷ ℤ₂ diagonal

$1 + 183t^2 + 128t^3 + 16626t^4 + 21632t^5 + 1000427t^6 + 1815296t^7 + 44866419t^8 + 100543872t^9 + 1599857948t^{10} + O(t^{11})$

**1191 2111.00026-T11-1__n6_k2_l1** / **n = 6 · k = 2 · l = 1**

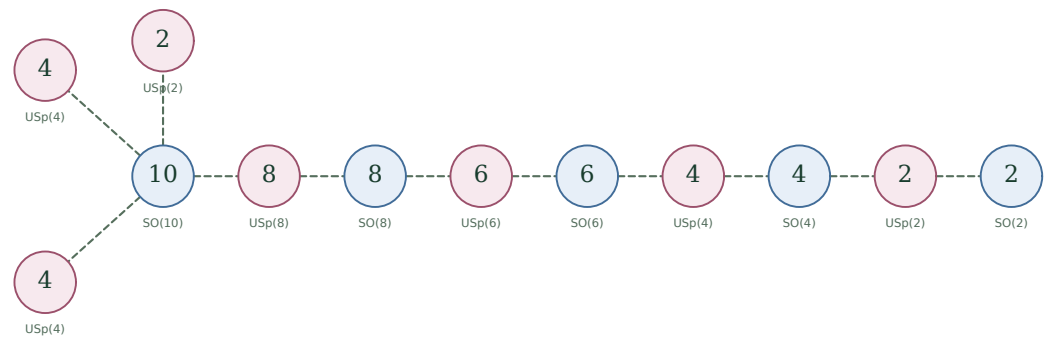


÷ ℤ₂ diagonal

$1 + 32t + 594t^2 + 8096t^3 + 89341t^4 + 840192t^5 + 6950889t^6 + 51677472t^7 + 350595927t^8 + 2195533120t^9 + 12804871178t^{10} + O(t^{11})$

**1192 2111.00026-T11-2__n8_k3_l1** / **n = 8 · k = 3 · l = 1**

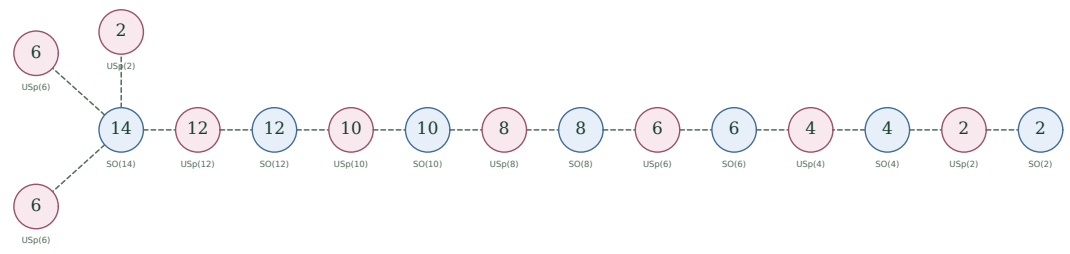


÷ ℤ₂ diagonal

$1 + 368t^2 + 63884t^4 + 7048932t^6 + 560375300t^8 + O(t^9)$

**1193 2111.00026-T11-3__n6_k1_l2** / **n = 6 · k = 1 · l = 2**

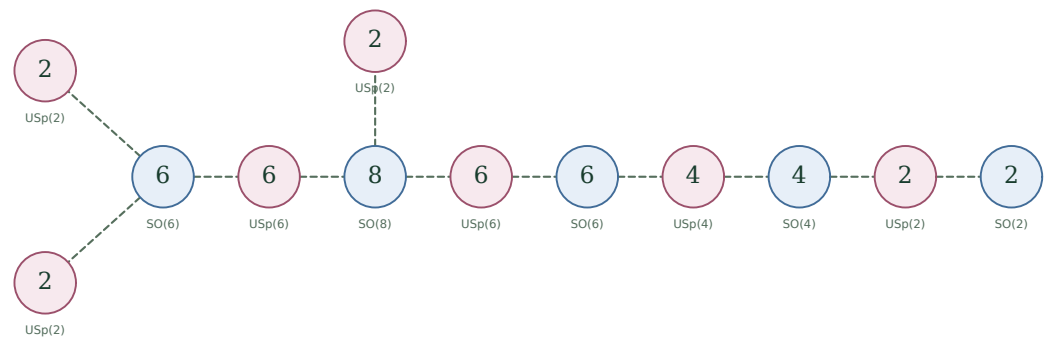


÷ ℤ₂ diagonal

$1 + 32t + 594t^2 + 8096t^3 + 88846t^4 + 824352t^5 + 6668508t^6 + 48042720t^7 + 313144128t^8 + 1869228064t^9 + 10318091234t^{10} + O(t^{11})$

**1194 2111.00026-T11-4__n8_k2_l2** / **n = 8 · k = 2 · l = 2**

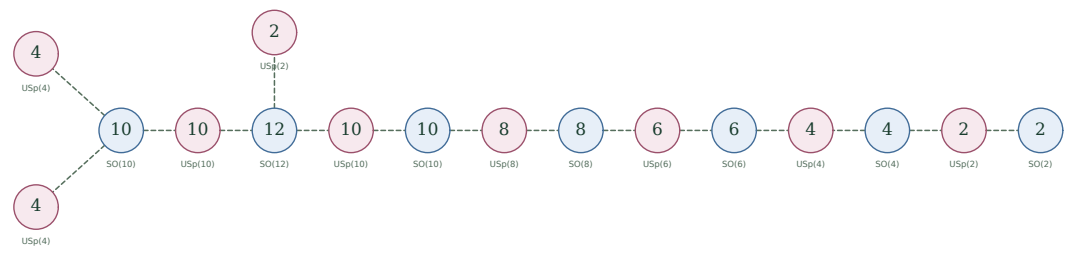


÷ ℤ₂ diagonal

$1 + 368t^2 + 63884t^4 + 7040924t^6 + 557673276t^8 + 34004993996t^{10} + O(t^{11})$

**1195 2111.00026-T12-1__k1 / k = 1**

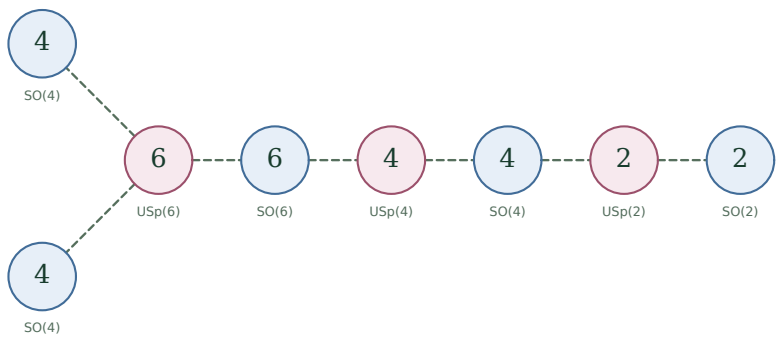


÷ $\mathbb{Z}_2$ diagonal

$1 + 104t^{2} + 4810t^{4} + 134264t^{6} + 2597465t^{8} + 37800384t^{10} + 437076562t^{12} + 4175738528t^{14} + 33943481936t^{16} + 240150358656t^{18} + 1505698109306t^{20} + O(t^{21})$

**1196 2111.00026-T12-2__k2 / k = 2**

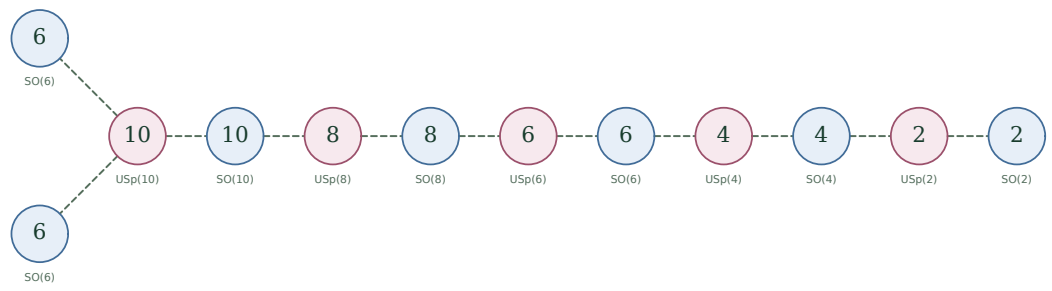

÷ $\mathbb{Z}_2$ diagonal

$1 + 156t^{2} + 128t^{3} + 12064t^{4} + 18304t^{5} + 620840t^{6} + 1297792t^{7} + 23982530t^{8} + 60845696t^{9} + 741714140t^{10} + 2122578432t^{11} + 19100059850t^{12} + 58780377472t^{13} + 420308771752t^{14} + 1346411945088t^{15} + 8051191269563t^{16} + O(t^{17})$

**1197 2111.00026-T13-1__k1 / k = 1**

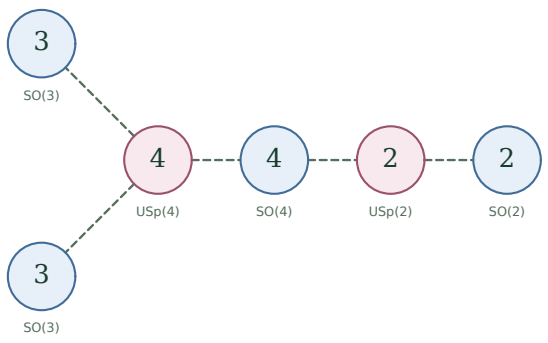


$1 + 42t^{2} + 777t^{4} + 8706t^{6} + 68880t^{8} + 421162t^{10} + 2109885t^{12} + 9015666t^{14} + 33826779t^{16} + 113879276t^{18} + 349750894t^{20} + O(t^{21})$

**1198 2111.00026-T13-2__k2 / k = 2**

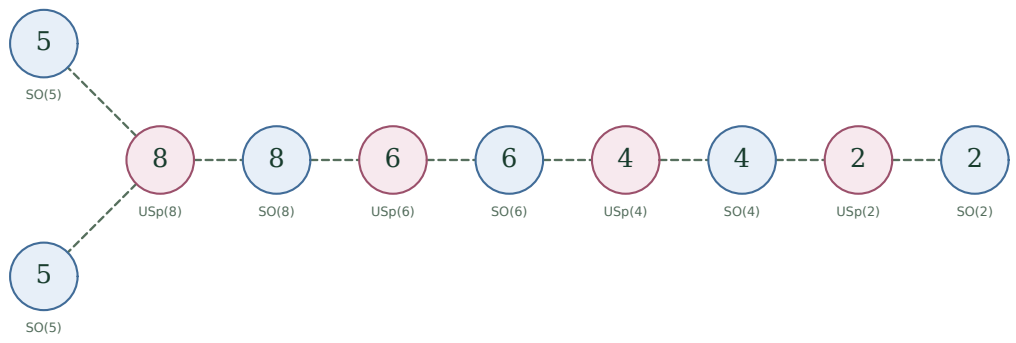


$1 + 110t^{2} + 5973t^{4} + 212606t^{6} + 5570246t^{8} + 114525004t^{10} + 1925660177t^{12} + 27266117052t^{14} + 332379773462t^{16} + 3549706986122t^{18} + 33687210855500t^{20} + O(t^{21})$

**1199 2111.00026-T13-3__k3 / k = 3**

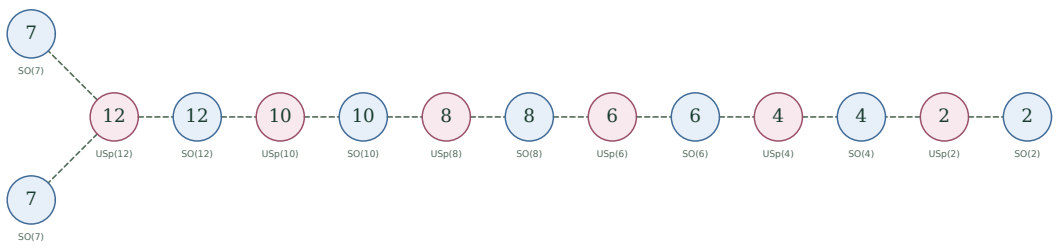

$1 + 210t^{2} + 21915t^{4} + 1515458t^{6} + O(t^{7})$

**1200 2111.00026-T14-1__k1 / k = 1**

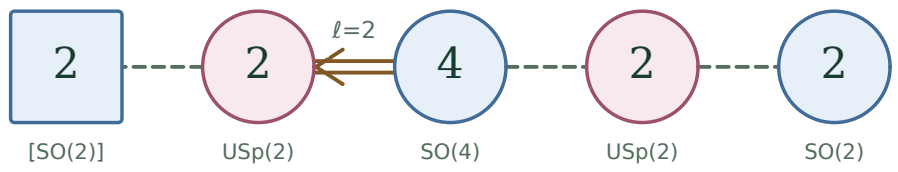


$1 + 25t^{2} + 269t^{4} + 1749t^{6} + 8190t^{8} + 30478t^{10} + 95710t^{12} + 263758t^{14} + 655303t^{16} + 1496495t^{18} + 3186755t^{20} + O(t^{21})$

**1201 2111.00026-T14-2__k2 / k = 2**

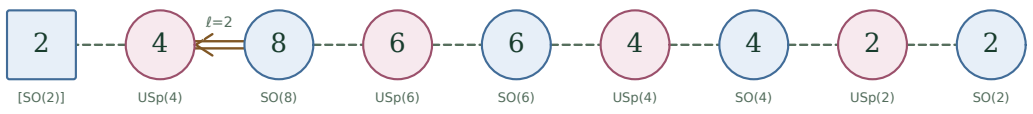

$1 + 81t^{2} + 3221t^{4} + 83585t^{6} + 1591224t^{8} + 23721865t^{10} + 288966230t^{12} + 2965143530t^{14} + 26230232434t^{16} + 203743162413t^{18} + 1410507872500t^{20} + O(t^{21})$

| QUIVER | HILBERT SERIES |
|---|---|

**1202 2111.00026-T14-3__k3** / k = 3

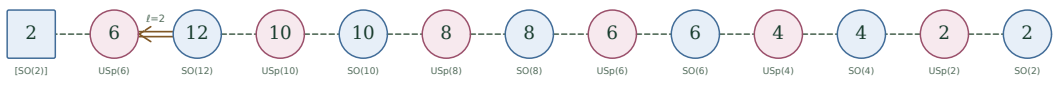


$1 + 169t^{2} + 14169t^{4} + 785875t^{6} + 32440512t^{8} + 1063059375t^{10} + O(t^{11})$

**1203 2111.00026-T15-1__n4_k1_l0** / n = 4 · k = 1 · l = 0

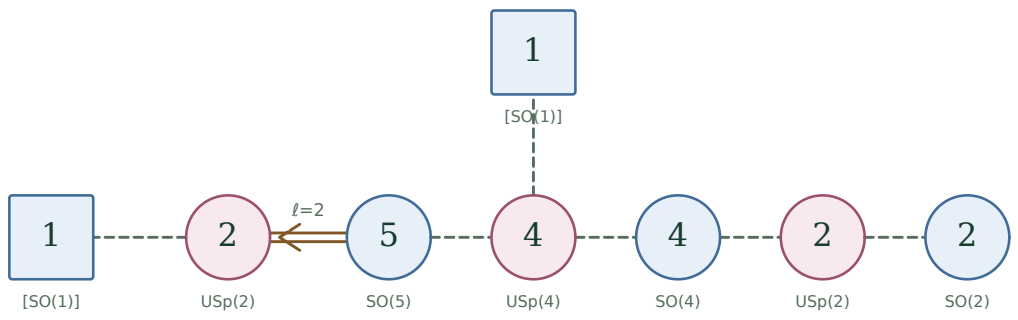


$1 + 49t^{2} + 1056t^{4} + 13754t^{6} + 125846t^{8} + 884108t^{10} + 5054105t^{12} + 24478785t^{14} + O(t^{15})$

**1204 2111.00026-T15-2__n5_k1_l0** / n = 5 · k = 1 · l = 0

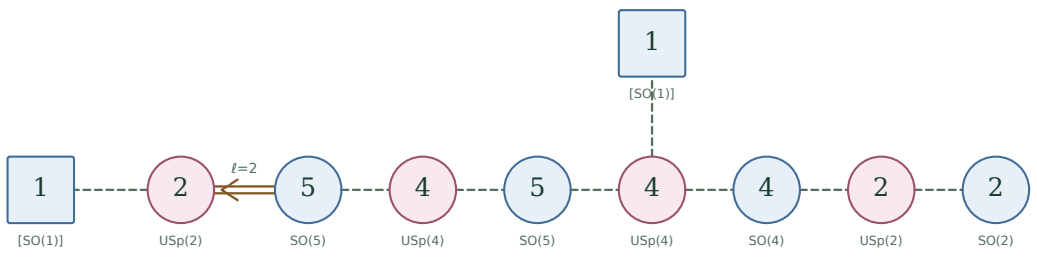


$1 + 81t^{2} + 2885t^{4} + 61643t^{6} + 912366t^{8} + 10189480t^{10} + 90890304t^{12} + 674183898t^{14} + O(t^{15})$

**1205 2111.00026-T15-3__n6_k2_l0** / n = 6 · k = 2 · l = 0

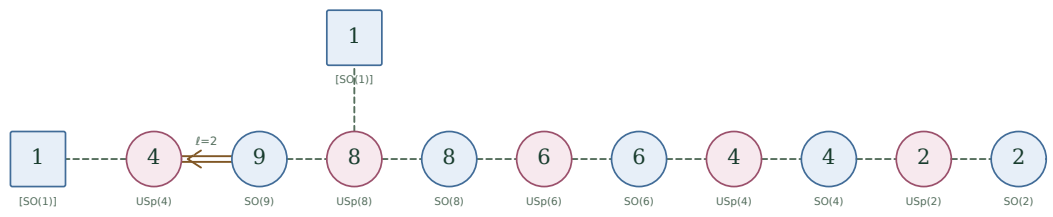


$1 + 121t^{2} + 7237t^{4} + 283953t^{6} + 8203712t^{8} + 186007152t^{10} + 3448403056t^{12} + 53812917025t^{14} + O(t^{15})$

**1206 2111.00026-T15-4__n6_k1_l1** / n = 6 · k = 1 · l = 1

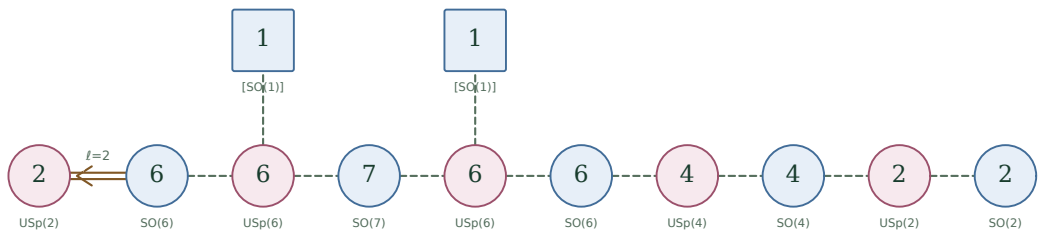


$1 + 121t^{2} + 6907t^{4} + 250590t^{6} + 6536574t^{8} + 131359647t^{10} + 2127959063t^{12} + 28706249847t^{14} + 330553477948t^{16} + O(t^{17})$

**1207 2111.00026-T15-5__n9_k1_l2** / n = 9 · k = 1 · l = 2

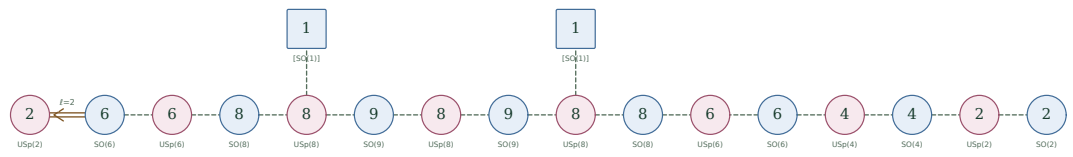


$1 + 289t^{2} + 39201t^{4} + 3385651t^{6} + 211578124t^{8} + O(t^{9})$

| QUIVER | HILBERT SERIES |
|---|---|

**1208 2111.00026-T16-1__k1 / k = 1**

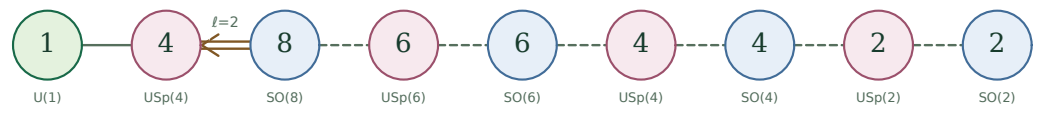


÷ $\mathbb{Z}_2$ diagonal

$1 + 130t^2 + 7539t^4 + 264628t^6 + 6438276t^8 + 117582633t^{10} + 1700156133t^{12} + 20221864235t^{14} + 203634547831t^{16} + 1775507329154t^{18} + 13646329994778t^{20} + O(t^{21})$

**1209 2111.00026-T16-2__k2 / k = 2**

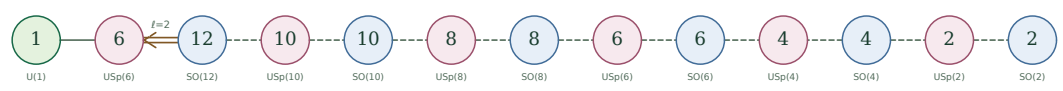


÷ $\mathbb{Z}_2$ diagonal

$1 + 170t^2 + 192t^3 + 14339t^4 + 30016t^5 + 811838t^6 + 2328704t^7 + 34914970t^8 + 119841920t^9 + 1216054981t^{10} + 4608606912t^{11} + 35579828088t^{12} + 141340385344t^{13} + 894676190640t^{14} + 3601291105024t^{15} + 19640751761433t^{16} + 78397263587712t^{17} + 380968232656361t^{18} + 1488093619535168t^{19} + 6594271153605775t^{20} + O(t^{21})$

**1210 2111.00026-T17-1__k1_l1 / k = 1 · l = 1**

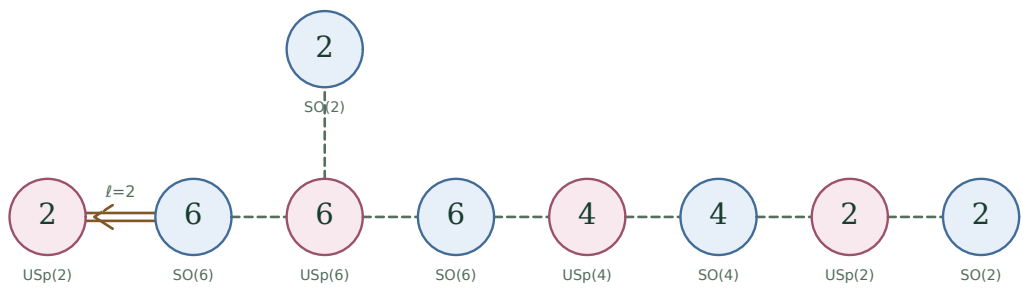


÷ $\mathbb{Z}_2$ diagonal

$1 + 114t^2 + 5733t^4 + 173852t^6 + 3651348t^8 + 57617325t^{10} + 721201692t^{12} + 7444836763t^{14} + 65256772383t^{16} + 496825085316t^{18} + 3345139658705t^{20} + O(t^{21})$

**1211 2111.00026-T17-2__k2_l1 / k = 2 · l = 1**

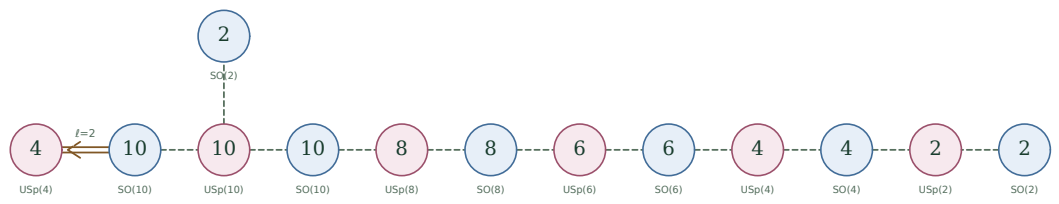


÷ $\mathbb{Z}_2$ diagonal

$1 + 170t^2 + 128t^3 + 13624t^4 + 19968t^5 + 700114t^6 + 1454336t^7 + 26309217t^8 + 67287040t^9 + 777105979t^{10} + O(t^{11})$

**Paper typo:** Both folded and product quivers give 14339 at t⁴. The printed factor uses 2275 for a different rank-two SO(13) orbit, rather than 2990 for the rank-four orbit identified by the diagram and text.

**1212 2111.00026-T18-1__k2_l1 / k = 2 · l = 1**

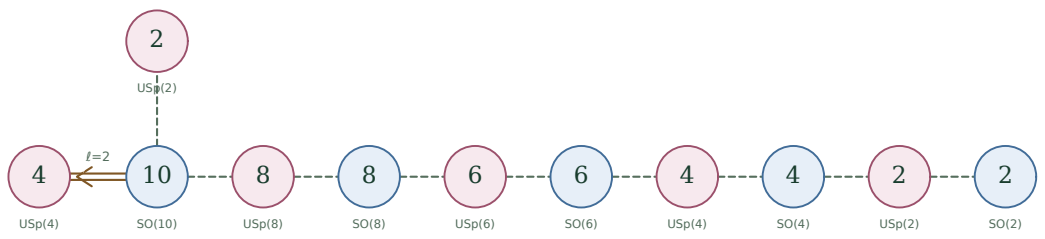


÷ $\mathbb{Z}_2$ diagonal

$1 + 32t + 583t^2 + 7744t^3 + 82874t^4 + 753280t^5 + 6008089t^6 + 42982688t^7 + 280199502t^8 + O(t^9)$

**1213 2111.00026-T18-2__k3_l1 / k = 3 · l = 1**

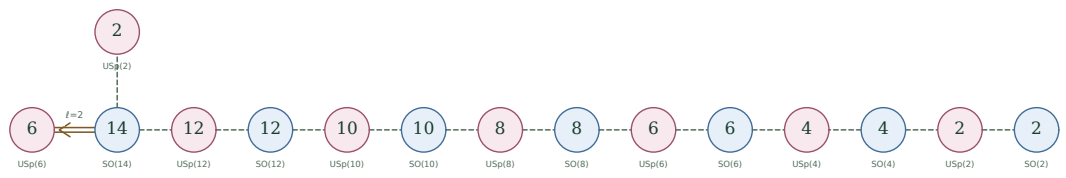


÷ $\mathbb{Z}_2$ diagonal

$1 + 353t^2 + 58485t^4 + 6134489t^6 + 462119507t^8 + O(t^9)$

| QUIVER | HILBERT SERIES |
| --- | --- |

**1214 2111.00026-T19-1__k1** / **k = 1**

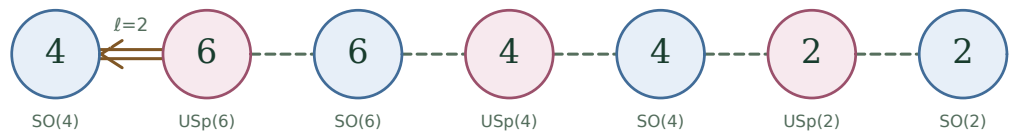


÷ $\mathbb{Z}_2$ diagonal

$1 + 88t^2 + 3420t^4 + 80028t^6 + 1298687t^8 + 15893460t^{10} + 155107049t^{12} + 1256071828t^{14} + 8694268479t^{16} + 52624765960t^{18} + 283597751451t^{20} + O(t^{21})$

**1215 2111.00026-T19-2__k2** / **k = 2**

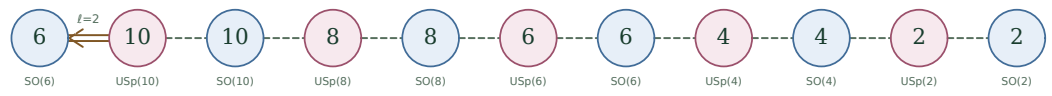


÷ $\mathbb{Z}_2$ diagonal

$1 + 144t^2 + 96t^3 + 10271t^4 + 12608t^5 + 485497t^6 + 820352t^7 + 17130124t^8 + 35239616t^9 + 481174913t^{10} + O(t^{11})$

**1216 2111.00026-T20-1__k1** / **k = 1**

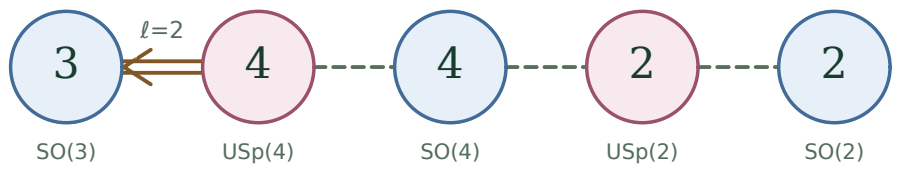


$1 + 36t^2 + 567t^4 + 5409t^6 + 36615t^8 + 192899t^{10} + 839021t^{12} + 3136395t^{14} + 10368906t^{16} + 30964325t^{18} + 84878874t^{20} + O(t^{21})$

**1217 2111.00026-T20-2__k2** / **k = 2**

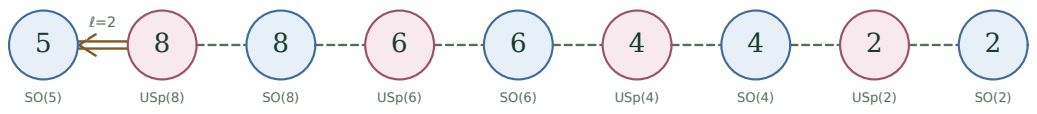


$1 + 100t^2 + 4929t^4 + 159047t^6 + 3773635t^8 + 70214649t^{10} + 1068107756t^{12} + 13683068037t^{14} + 150962851740t^{16} + 1460049948432t^{18} + 12558512227814t^{20} + O(t^{21})$

**1218 2111.00026-T20-3__k3** / **k = 3**

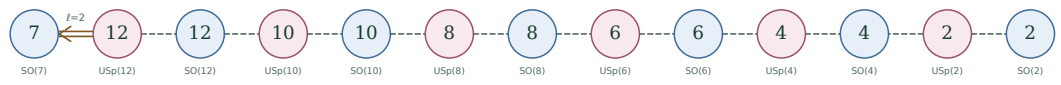


$1 + 196t^2 + 19081t^4 + 1230319t^6 + 59106675t^8 + 2256496369t^{10} + 71299059243t^{12} + 1917659317006t^{14} + 44815111648324t^{16} + 924438389586125t^{18} + 17043068970722031t^{20} + O(t^{21})$

**1219 2111.00026-T21-1__p2_q1_k1_n3** / **p = 2 · q = 1 · k = 1 · n = 3**

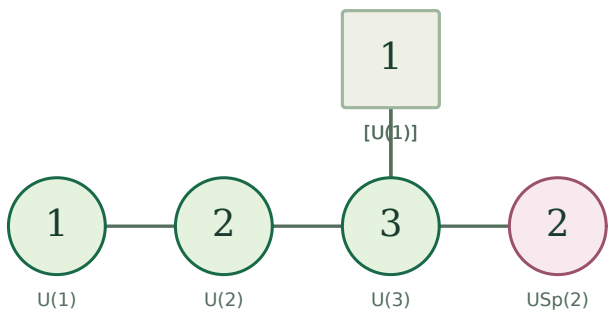


$1 + 30t^2 + 413t^4 + 3595t^6 + 22845t^8 + 115073t^{10} + 484367t^{12} + 1766649t^{14} + 5731324t^{16} + 16864225t^{18} + 45687006t^{20} + O(t^{21})$

**1220 2111.00026-T21-2__p2_q1_k2_n3** / **p = 2 · q = 1 · k = 2 · n = 3**

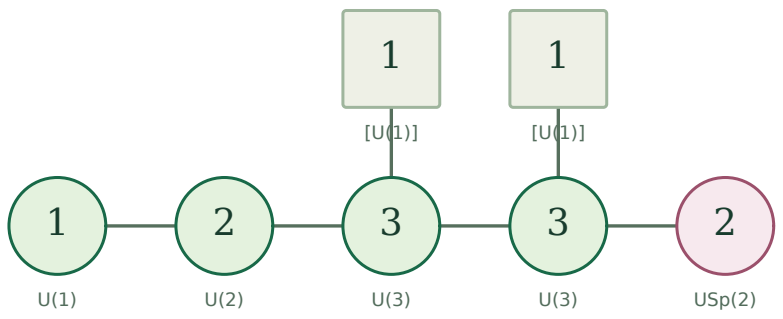


$1 + 70t^2 + 40t^3 + 2415t^4 + 2520t^5 + 55560t^6 + 78624t^7 + 965011t^8 + 1631240t^9 + 13512030t^{10} + O(t^{11})$

**Paper typo:** The printed first-line coefficient is 70 at $t^2$, while its own product and SQCD factors give 48, confirmed by direct monopoles. The two Table 21 first-line expansions are swapped between their parameter rows.

**QUIVER** | **HILBERT SERIES**

**1221 2111.00026-T21-3__p3_q2_k1_n5** / p = 3 · q = 2 · k = 1 · n = 5

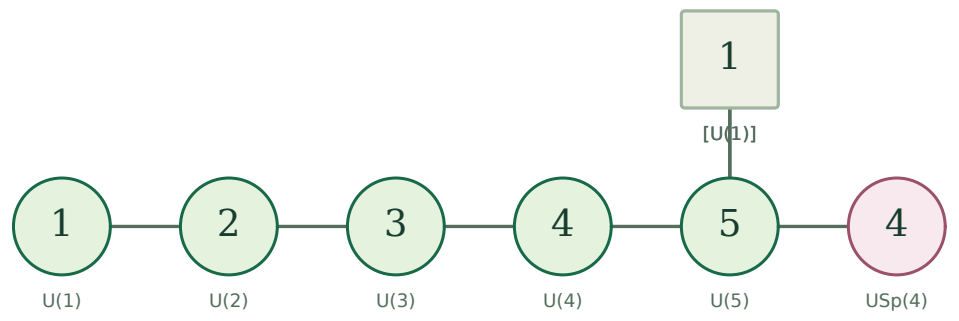


$1 + 48t^{2} + 1051t^{4} + 14424t^{6} + 142227t^{8} + 1089952t^{10} + 6836169t^{12} + 36400128t^{14} + 169074122t^{16} + 699481456t^{18} + 2619981678t^{20} + O(t^{21})$

**Paper typo:** The printed first-line coefficient is 48 at t², while its own product and SQCD factors give 70, confirmed by direct monopoles. The two Table 21 first-line expansions are swapped between their parameter rows.

**1222 2111.00026-T22-1__k2** / k = 2

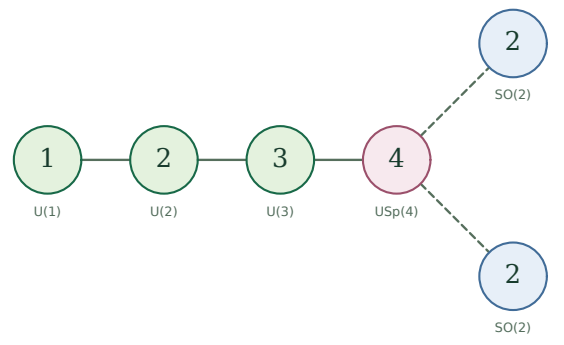


÷ $\mathbb{Z}_2$ diagonal

$1 + 56t^{2} + 1384t^{4} + 20650t^{6} + 215636t^{8} + 1720264t^{10} + 11106025t^{12} + 60413096t^{14} + 285191074t^{16} + 1194750560t^{18} + 4519533676t^{20} + O(t^{21})$

**1223 2111.00026-T22-2__k3** / k = 3

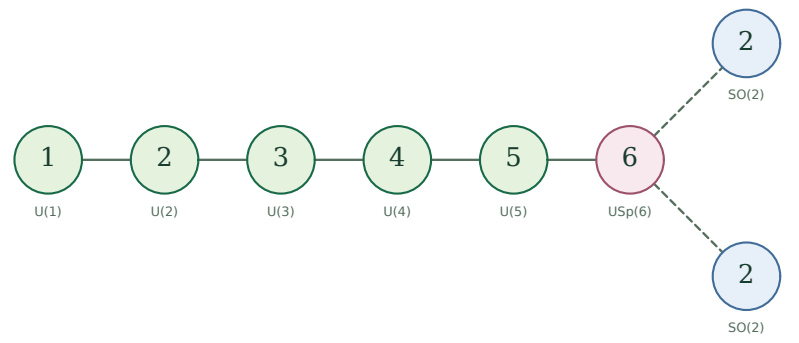


÷ $\mathbb{Z}_2$ diagonal

$1 + 72t^{2} + 80t^{3} + 2556t^{4} + 5120t^{5} + 62010t^{6} + 162288t^{7} + 1165484t^{8} + 3433728t^{9} + 17928592t^{10} + O(t^{11})$

**1224 2111.00026-T23-1__k2** / k = 2

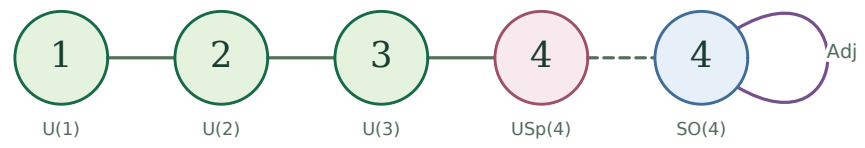


÷ $\mathbb{Z}_2$ diagonal

$1 + 42t^{2} + 831t^{4} + 10500t^{6} + 96727t^{8} + 700938t^{10} + 4198425t^{12} + 21520704t^{14} + 96852826t^{16} + 390254420t^{18} + 1429637622t^{20} + O(t^{21})$

**1225 2111.00026-T23-2__k3** / k = 3

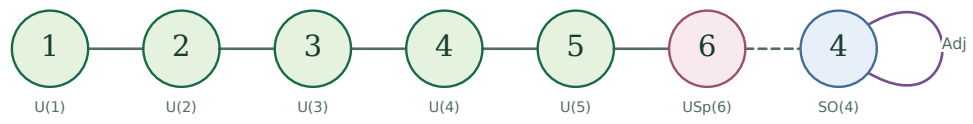


÷ $\mathbb{Z}_2$ diagonal

$1 + 70t^{2} + 40t^{3} + 2415t^{4} + 2520t^{5} + 55560t^{6} + 78624t^{7} + 965011t^{8} + 1631240t^{9} + 13512030t^{10} + O(t^{11})$

**1226 2111.00026-product-T9-1__k2_l0** / k = 2 · l = 0

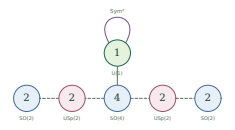

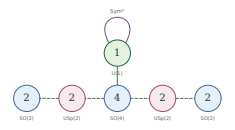

÷ Z2 diagonal on {c0_n0, c0_n1, c0_n2, c0_n3, c0_n4, c0_n5} · ÷ Z2 diagonal on {c1_n0, c1_n1, c1_n2, c1_n3, c1_n4, c1_n5}

$1 + 58t^{2} + 32t^{3} + 1569t^{4} + 1600t^{5} + 27220t^{6} + 37856t^{7} + 348124t^{8} + 577920t^{9} + 3540936t^{10} + 6502720t^{11} + 29981572t^{12} + 58099072t^{13} + 217747736t^{14} + 432151456t^{15} + 1385881186t^{16} + 2764473568t^{17} + 7858110900t^{18} + 15572468640t^{19} + 40224531398t^{20} + O(t^{21})$

| QUIVER | HILBERT SERIES |
|---|---|
| **1227 2111.00026-product-T9-2__k3_l1** / k = 3 · l = 1 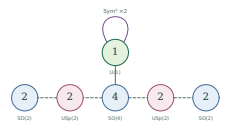 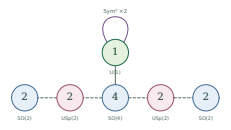 ÷ Z2 diagonal on {c0_n0, c0_n1, c0_n2, c0_n3, c0_n4, c0_n5} · ÷ Z2 diagonal on {c1_n0, c1_n1, c1_n2, c1_n3, c1_n4, c1_n5} | $1 + 58t^{2} + 1601t^{4} + 28424t^{6} + 369156t^{8} + 3771348t^{10} + 31759488t^{12} + 227801304t^{14} + 1425775758t^{16} + 7933063516t^{18} + 39822908626t^{20} + O(t^{21})$ |
| **1228 2111.00026-product-T9-3__k3_l0** / k = 3 · l = 0 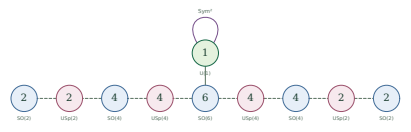 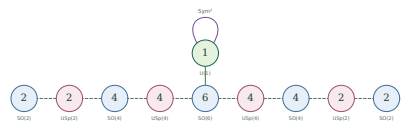 ÷ Z2 diagonal on {c0_n0, c0_n1, c0_n2, c0_n3, c0_n4, c0_n5, c0_n6, c0_n7, c0_n8, c0_n9} · ÷ Z2 diagonal on {c1_n0, c1_n1, c1_n2, c1_n3, c1_n4, c1_n5, c1_n6, c1_n7, c1_n8, c1_n9} | $1 + 134t^{2} + 9017t^{4} + 403982t^{6} + 13508026t^{8} + 358747158t^{10} + O(t^{11})$ |
| **1229 2111.00026-product-T9-4__k4_l1** / k = 4 · l = 1 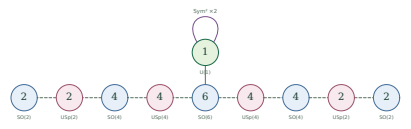 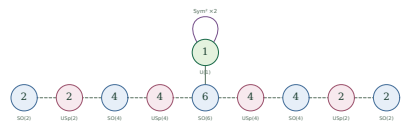 ÷ Z2 diagonal on {c0_n0, c0_n1, c0_n2, c0_n3, c0_n4, c0_n5, c0_n6, c0_n7, c0_n8, c0_n9} · ÷ Z2 diagonal on {c1_n0, c1_n1, c1_n2, c1_n3, c1_n4, c1_n5, c1_n6, c1_n7, c1_n8, c1_n9} | $1 + 134t^{2} + 8889t^{4} + 128t^{5} + 388366t^{6} + 15616t^{7} + 12558722t^{8} + 943360t^{9} + 320512966t^{10} + O(t^{11})$ |
| **1230 2111.00026-product-T9-5__k1_l0** / k = 1 · l = 0 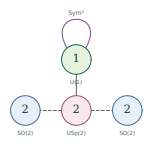 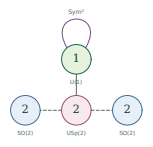 ÷ Z2 diagonal on {c0_n0, c0_n1, c0_n2, c0_n3} · ÷ Z2 diagonal on {c1_n0, c1_n1, c1_n2, c1_n3} | $1 + 48t^{2} + 976t^{4} + 11600t^{6} + 95350t^{8} + 598352t^{10} + 3053296t^{12} + 13224752t^{14} + 50129875t^{16} + 170108000t^{18} + 525728128t^{20} + O(t^{21})$ |
| **1231 2111.00026-product-T9-6__k2_l1** / k = 2 · l = 1 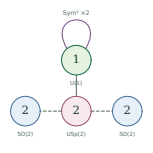 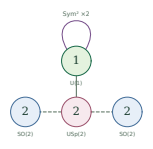 ÷ Z2 diagonal on {c0_n0, c0_n1, c0_n2, c0_n3} · ÷ Z2 diagonal on {c1_n0, c1_n1, c1_n2, c1_n3} | $1 + 32t^{2} + 16t^{3} + 456t^{4} + 416t^{5} + 4104t^{6} + 4960t^{7} + 27490t^{8} + 38400t^{9} + 148792t^{10} + 223840t^{11} + 681924t^{12} + 1062432t^{13} + 2729368t^{14} + 4305376t^{15} + 9754099t^{16} + 15374976t^{17} + 31652168t^{18} + 49482864t^{19} + 94506536t^{20} + O(t^{21})$ |

| QUIVER | HILBERT SERIES |
|---|---|
| **1232 2111.00026-product-T9-7__k2_l0** / **k = 2 · l = 0** 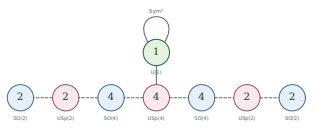 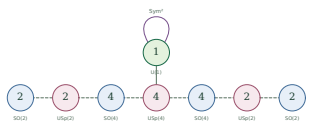 ÷ Z2 diagonal on {c0_n0, c0_n1, c0_n2, c0_n3, c0_n4, c0_n5, c0_n6, c0_n7} · ÷ Z2 diagonal on {c1_n0, c1_n1, c1_n2, c1_n3, c1_n4, c1_n5, c1_n6, c1_n7} | $1 + 92t^{2} + 64t^{3} + 4168t^{4} + 5248t^{5} + 125140t^{6} + 211968t^{7} + 2809296t^{8} + 5635008t^{9} + 50260590t^{10} + O(t^{11})$ |
| **1233 2111.00026-product-T9-8__k3_l1** / **k = 3 · l = 1** 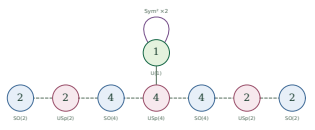 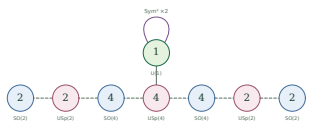 ÷ Z2 diagonal on {c0_n0, c0_n1, c0_n2, c0_n3, c0_n4, c0_n5, c0_n6, c0_n7} · ÷ Z2 diagonal on {c1_n0, c1_n1, c1_n2, c1_n3, c1_n4, c1_n5, c1_n6, c1_n7} | $1 + 92t^{2} + 4232t^{4} + 128860t^{6} + 2910592t^{8} + 51921078t^{10} + O(t^{11})$ |
| **1234 2111.00026-product-T9-9__k1** / **k = 1** 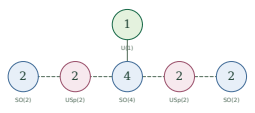 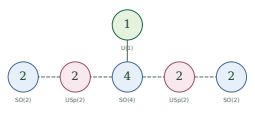 ÷ Z2 diagonal on {c0_n0, c0_n1, c0_n2, c0_n3, c0_n4, c0_n5} · ÷ Z2 diagonal on {c1_n0, c1_n1, c1_n2, c1_n3, c1_n4, c1_n5} | $1 + 90t^{2} + 3565t^{4} + 84588t^{6} + 1386700t^{8} + 17100048t^{10} + 167857416t^{12} + 1365541740t^{14} + 9486469554t^{16} + 57589566980t^{18} + 311107661634t^{20} + O(t^{21})$ |
| **1235 2111.00026-product-T9-10__k2** / **k = 2** 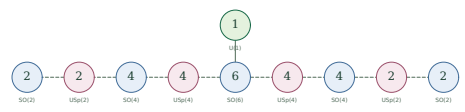 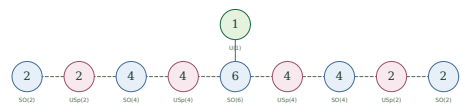 ÷ Z2 diagonal on {c0_n0, c0_n1, c0_n2, c0_n3, c0_n4, c0_n5, c0_n6, c0_n7, c0_n8, c0_n9} · ÷ Z2 diagonal on {c1_n0, c1_n1, c1_n2, c1_n3, c1_n4, c1_n5, c1_n6, c1_n7, c1_n8, c1_n9} | $1 + 134t^{2} + 128t^{3} + 8889t^{4} + 15616t^{5} + 394310t^{6} + 943360t^{7} + 13220746t^{8} + 37713280t^{9} + 357061474t^{10} + 1123982464t^{11} + 8061468875t^{12} + 26656882432t^{13} + 155821066386t^{14} + 524237477376t^{15} + 2622606331104t^{16} + 8795126835840t^{17} + 38944260460754t^{18} + 128525072279296t^{19} + 515800796930805t^{20} + O(t^{21})$ |
| **1236 2111.00026-product-T9-11__k1** / **k = 1** 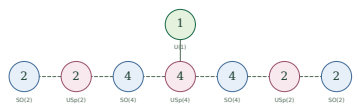 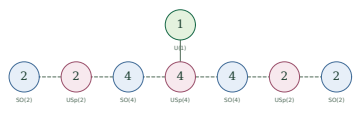 ÷ Z2 diagonal on {c0_n0, c0_n1, c0_n2, c0_n3, c0_n4, c0_n5, c0_n6, c0_n7} · ÷ Z2 diagonal on {c1_n0, c1_n1, c1_n2, c1_n3, c1_n4, c1_n5, c1_n6, c1_n7} | $1 + 156t^{2} + 10944t^{4} + 466596t^{6} + 13807080t^{8} + 306524790t^{10} + 5377829028t^{12} + 77405354312t^{14} + 940221343776t^{16} + 9853246779220t^{18} + 90680857312617t^{20} + O(t^{21})$ |

| QUIVER | HILBERT SERIES |
|---|---|

**1237 2111.00026-product-T9-12__k2** / **k = 2**

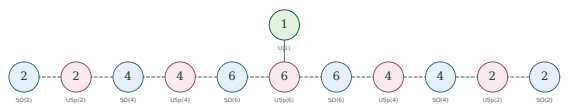

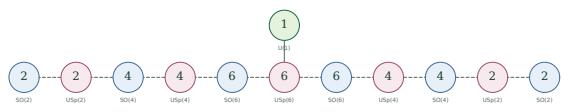

÷ Z2 diagonal on {c0_n0, c0_n1, c0_n2, c0_n3, c0_n4, c0_n5, c0_n6, c0_n7, c0_n8, c0_n9, c0_n10, c0_n11} · ÷ Z2 diagonal on {c1_n0, c1_n1, c1_n2, c1_n3, c1_n4, c1_n5, c1_n6, c1_n7, c1_n8, c1_n9, c1_n10, c1_n11}

$1 + 184t^{2} + 256t^{3} + 16810t^{4} + 43520t^{5} + 1040056t^{6} + 3674112t^{7} + 49469875t^{8} + 206409216t^{9} + 1925805206t^{10} + 8703438848t^{11} + 63418779630t^{12} + 294056655104t^{13} + 1801033108182t^{14} + 8290154332416t^{15} + 44696692490453t^{16} + 200425546653440t^{17} + 979719926063866t^{18} + 4237436679956480t^{19} + 19144937799760589t^{20} + O(t^{21})$

**1238 2111.00026-product-T9-13__k1** / **k = 1**

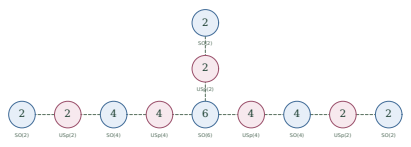

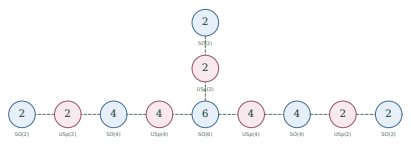

÷ Z2 diagonal on {c0_n0, c0_n1, c0_n2, c0_n3, c0_n4, c0_n5, c0_n6, c0_n7, c0_n8, c0_n9, c0_n10} · ÷ Z2 diagonal on {c1_n0, c1_n1, c1_n2, c1_n3, c1_n4, c1_n5, c1_n6, c1_n7, c1_n8, c1_n9, c1_n10}

$1 + 266t^{2} + 32431t^{4} + 2437890t^{6} + 128297273t^{8} + 5085058160t^{10} + 159309437560t^{12} + 4084414731904t^{14} + 87979394611180t^{16} + 1625455257184024t^{18} + 26192135424825720t^{20} + O(t^{21})$

**1239 2111.00026-product-T9-14__k1** / **k = 1**

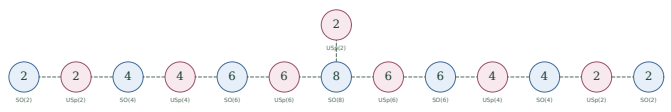

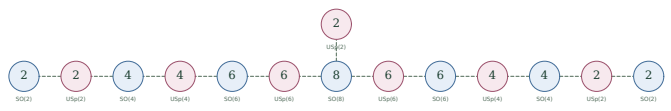

÷ Z2 diagonal on {c0_n0, c0_n1, c0_n2, c0_n3, c0_n4, c0_n5, c0_n6, c0_n7, c0_n8, c0_n9, c0_n10, c0_n11, c0_n12, c0_n13} · ÷ Z2 diagonal on {c1_n0, c1_n1, c1_n2, c1_n3, c1_n4, c1_n5, c1_n6, c1_n7, c1_n8, c1_n9, c1_n10, c1_n11, c1_n12, c1_n13}

$1 + 496t^{2} + 115504t^{4} + 16918250t^{6} + 1761796000t^{8} + 139749232560t^{10} + O(t^{11})$

**1240 2111.00026-product-T10-1__n5_k1_l1** / **n = 5 · k = 1 · l = 1**

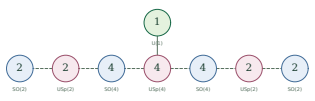



÷ Z2 diagonal on {c0_n0, c0_n1, c0_n2, c0_n3, c0_n4, c0_n5, c0_n6, c0_n7}

$1 + 123t^{2} + 6710t^{4} + 220812t^{6} + 5027328t^{8} + 85856661t^{10} + 1160906385t^{12} + 12919480684t^{14} + 121836107745t^{16} + O(t^{17})$

**1241 2111.00026-product-T10-2__n7_k1_l2** / **n = 7 · k = 1 · l = 2**

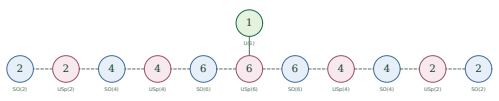

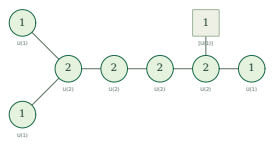

÷ Z2 diagonal on {c0_n0, c0_n1, c0_n2, c0_n3, c0_n4, c0_n5, c0_n6, c0_n7, c0_n8, c0_n9, c0_n10, c0_n11}

$1 + 183t^{2} + 128t^{3} + 15625t^{4} + 21632t^{5} + 849367t^{6} + 1687168t^{7} + 33584330t^{8} + 83001984t^{9} + 1039787073t^{10} + 2940462720t^{11} + 26377189292t^{12} + 80780840064t^{13} + 564910951617t^{14} + 1804412218752t^{15} + 10430788358641t^{16} + 33865771104384t^{17} + 168680900002512t^{18} + 547066306438912t^{19} + 2419061167200157t^{20} + O(t^{21})$

| QUIVER | HILBERT SERIES |
|---|---|

**1242 2111.00026-product-T10-3__n7_k2_l1** / n = 7 · k = 2 · l = 1

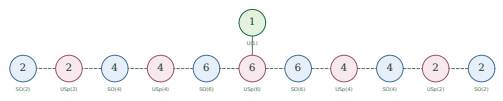

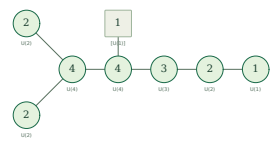

÷ Z2 diagonal on {c0_n0, c0_n1, c0_n2, c0_n3, c0_n4, c0_n5, c0_n6, c0_n7, c0_n8, c0_n9, c0_n10, c0_n11}

$1 + 183t^{2} + 128t^{3} + 16626t^{4} + 21632t^{5} + 1000427t^{6} + 1815296t^{7} + 44866419t^{8} + 100543872t^{9} + 1599857948t^{10} + O(t^{11})$

**1243 2111.00026-product-T11-1__n6_k2_l1** / n = 6 · k = 2 · l = 1

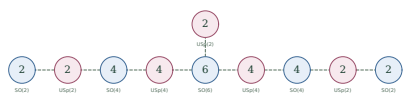

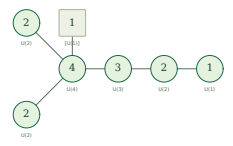

÷ Z2 diagonal on {c0_n0, c0_n1, c0_n2, c0_n3, c0_n4, c0_n5, c0_n6, c0_n7, c0_n8, c0_n9}

$1 + 32t + 594t^{2} + 8096t^{3} + 89341t^{4} + 840192t^{5} + 6950889t^{6} + 51677472t^{7} + 350595927t^{8} + 2195533120t^{9} + 12804871178t^{10} + O(t^{11})$

**1244 2111.00026-product-T11-2__n8_k3_l1** / n = 8 · k = 3 · l = 1

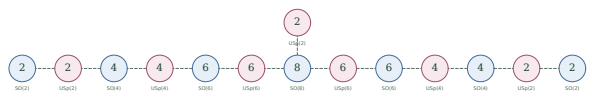

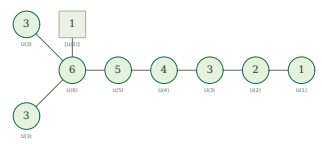

÷ Z2 diagonal on {c0_n0, c0_n1, c0_n2, c0_n3, c0_n4, c0_n5, c0_n6, c0_n7, c0_n8, c0_n9, c0_n10, c0_n11, c0_n12, c0_n13}

$1 + 368t^{2} + 63884t^{4} + 7048932t^{6} + 560375300t^{8} + O(t^{9})$

**1245 2111.00026-product-T11-3__n6_k1_l2** / n = 6 · k = 1 · l = 2

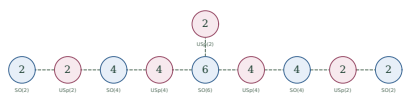

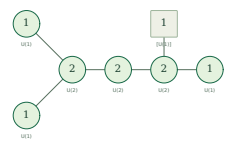

÷ Z2 diagonal on {c0_n0, c0_n1, c0_n2, c0_n3, c0_n4, c0_n5, c0_n6, c0_n7, c0_n8, c0_n9}

$1 + 32t + 594t^{2} + 8096t^{3} + 88846t^{4} + 824352t^{5} + 6668508t^{6} + 48042720t^{7} + 313144128t^{8} + 1869228064t^{9} + 10318091234t^{10} + O(t^{11})$

**1246 2111.00026-product-T11-4__n8_k2_l2** / n = 8 · k = 2 · l = 2

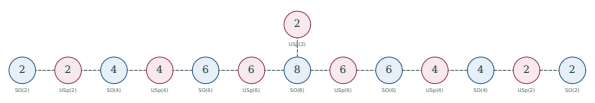

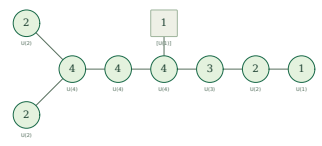

÷ Z2 diagonal on {c0_n0, c0_n1, c0_n2, c0_n3, c0_n4, c0_n5, c0_n6, c0_n7, c0_n8, c0_n9, c0_n10, c0_n11, c0_n12, c0_n13}

$1 + 368t^{2} + 63884t^{4} + 7040924t^{6} + 557673276t^{8} + 34004993996t^{10} + O(t^{11})$

**1247 2111.00026-product-T12-1__k1** / k = 1

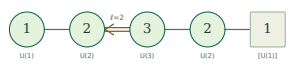

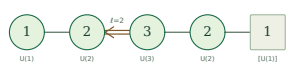

$1 + 104t^{2} + 4810t^{4} + 134264t^{6} + 2597465t^{8} + 37800384t^{10} + 437076562t^{12} + 4175738528t^{14} + 33943481936t^{16} + 240150358656t^{18} + 1505698109306t^{20} + O(t^{21})$

| QUIVER | HILBERT SERIES |
|---|---|

**1248 2111.00026-product-T12-2__k2** / k = 2

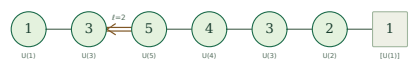

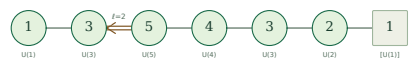

$1 + 156t^{2} + 128t^{3} + 12064t^{4} + 18304t^{5} + 620840t^{6} + 1297792t^{7} + 23982530t^{8} + 60845696t^{9} + 741714140t^{10} + 2122578432t^{11} + 19100059850t^{12} + 58780377472t^{13} + 420308771752t^{14} + 1346411945088t^{15} + 8051191269563t^{16} + O(t^{17})$

---

**1249 2111.00026-product-T16-1__k1** / k = 1

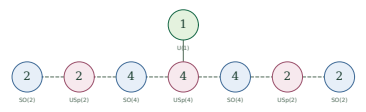

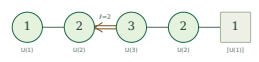

÷ Z2 diagonal on {c0_n0, c0_n1, c0_n2, c0_n3, c0_n4, c0_n5, c0_n6, c0_n7}

$1 + 130t^{2} + 7539t^{4} + 264628t^{6} + 6438276t^{8} + 117582633t^{10} + 1700156133t^{12} + 20221864235t^{14} + 203634547831t^{16} + 1775507329154t^{18} + 13646329994778t^{20} + O(t^{21})$

---

**1250 2111.00026-product-T16-2__k2** / k = 2

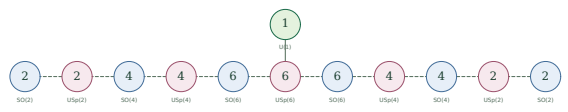

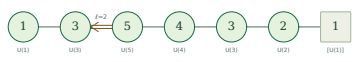

÷ Z2 diagonal on {c0_n0, c0_n1, c0_n2, c0_n3, c0_n4, c0_n5, c0_n6, c0_n7, c0_n8, c0_n9, c0_n10, c0_n11}

$1 + 170t^{2} + 192t^{3} + 14339t^{4} + 30016t^{5} + 811838t^{6} + 2328704t^{7} + 34914970t^{8} + 119841920t^{9} + 1216054981t^{10} + 4608606912t^{11} + 35579828088t^{12} + 141340385344t^{13} + 894676190640t^{14} + 3601291105024t^{15} + 19640751761433t^{16} + 78397263587712t^{17} + 380968232656361t^{18} + 1488093619535168t^{19} + 6594271153605775t^{20} + O(t^{21})$

---

**1251 2111.00026-product-T17-1__k1_l1** / k = 1 · l = 1

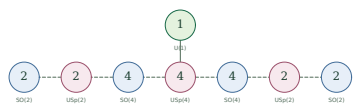

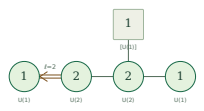

÷ Z2 diagonal on {c0_n0, c0_n1, c0_n2, c0_n3, c0_n4, c0_n5, c0_n6, c0_n7}

$1 + 114t^{2} + 5733t^{4} + 173852t^{6} + 3651348t^{8} + 57617325t^{10} + 721201692t^{12} + 7444836763t^{14} + 65256772383t^{16} + 496825085316t^{18} + 3345139658705t^{20} + O(t^{21})$

---

**1252 2111.00026-product-T17-2__k2_l1** / k = 2 · l = 1

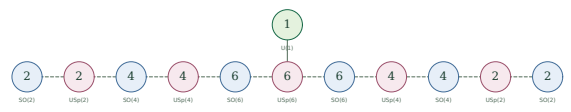

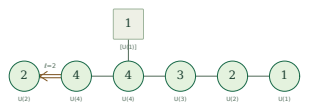

÷ Z2 diagonal on {c0_n0, c0_n1, c0_n2, c0_n3, c0_n4, c0_n5, c0_n6, c0_n7, c0_n8, c0_n9, c0_n10, c0_n11}

$1 + 170t^{2} + 128t^{3} + 13624t^{4} + 19968t^{5} + 700114t^{6} + 1454336t^{7} + 26309217t^{8} + 67287040t^{9} + 777105979t^{10} + O(t^{11})$

**Paper typo:** The actual graph components give 4173+92×78+2990=14339 at $t^4$. The printed 2275 belongs to a different rank-two orbit, whereas the diagram and text identify the rank-four orbit.

---

**1253 2111.00026-product-T18-1__k2_l1** / k = 2 · l = 1

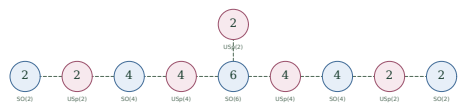

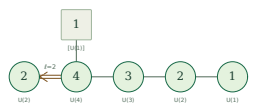

÷ Z2 diagonal on {c0_n0, c0_n1, c0_n2, c0_n3, c0_n4, c0_n5, c0_n6, c0_n7, c0_n8, c0_n9}

$1 + 32t + 583t^{2} + 7744t^{3} + 82874t^{4} + 753280t^{5} + 6008089t^{6} + 42982688t^{7} + 280199502t^{8} + O(t^{9})$

| QUIVER | HILBERT SERIES |
| --- | --- |

**1254 2111.00026-product-T18-2__k3_l1** / **k = 3 · l = 1**

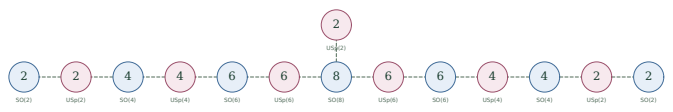

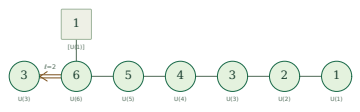

÷ Z2 diagonal on {c0_n0, c0_n1, c0_n2, c0_n3, c0_n4, c0_n5, c0_n6, c0_n7, c0_n8, c0_n9, c0_n10, c0_n11, c0_n12, c0_n13}

$1 + 353t^{2} + 58485t^{4} + 6134489t^{6} + 462119507t^{8} + O(t^{9})$

**1255 2111.00026-product-T19-1__k1** / **k = 1**

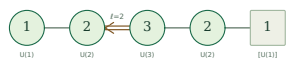

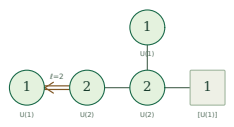

$1 + 88t^{2} + 3420t^{4} + 80028t^{6} + 1298687t^{8} + 15893460t^{10} + 155107049t^{12} + 1256071828t^{14} + 8694268479t^{16} + 52624765960t^{18} + 283597751451t^{20} + O(t^{21})$

**1256 2111.00026-product-T19-2__k2** / **k = 2**

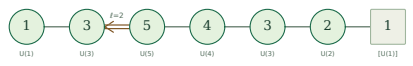

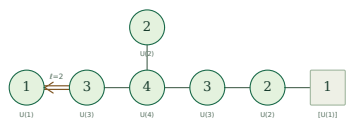

$1 + 144t^{2} + 96t^{3} + 10271t^{4} + 12608t^{5} + 485497t^{6} + 820352t^{7} + 17130124t^{8} + 35239616t^{9} + 481174913t^{10} + O(t^{11})$

**1257 2111.00026-product-T22-1__k2** / **k = 2**





÷ U(1) diagonal on {c0_n0, c0_n1, c0_n2, c0_n3, c0_n4} · ÷ U(1) diagonal on {c1_n0, c1_n1, c1_n2, c1_n3, c1_n4}

$1 + 56t^{2} + 1384t^{4} + 20650t^{6} + 215636t^{8} + 1720264t^{10} + 11106025t^{12} + 60413096t^{14} + 285191074t^{16} + 1194750560t^{18} + 4519533676t^{20} + O(t^{21})$

**1258 2111.00026-product-T22-2__k3** / **k = 3**

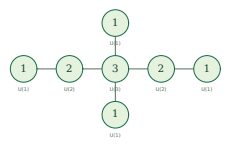

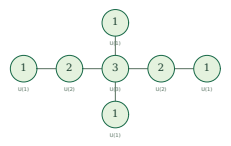

÷ U(1) diagonal on {c0_n0, c0_n1, c0_n2, c0_n3, c0_n4, c0_n5, c0_n6} · ÷ U(1) diagonal on {c1_n0, c1_n1, c1_n2, c1_n3, c1_n4, c1_n5, c1_n6}

$1 + 72t^{2} + 80t^{3} + 2556t^{4} + 5120t^{5} + 62010t^{6} + 162288t^{7} + 1165484t^{8} + 3433728t^{9} + 17928592t^{10} + O(t^{11})$

**1259 2111.00026-product-T23-1__k2** / **k = 2**





÷ U(1) diagonal on {c0_n0, c0_n1, c0_n2, c0_n3} · ÷ U(1) diagonal on {c1_n0, c1_n1, c1_n2, c1_n3}

$1 + 42t^{2} + 831t^{4} + 10500t^{6} + 96727t^{8} + 700938t^{10} + 4198425t^{12} + 21520704t^{14} + 96852826t^{16} + 390254420t^{18} + 1429637622t^{20} + O(t^{21})$

| QUIVER | HILBERT SERIES |
|---|---|

**1260 2111.00026-product-T23-2__k3 / k = 3**

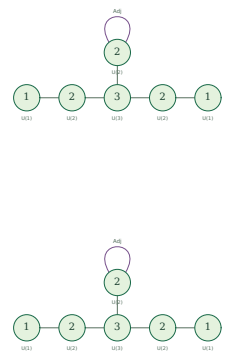

÷ U(1) diagonal on {c0_n0, c0_n1, c0_n2, c0_n3, c0_n4, c0_n5} ·
÷ U(1) diagonal on {c1_n0, c1_n1, c1_n2, c1_n3, c1_n4, c1_n5}

$1 + 70t^{2} + 40t^{3} + 2415t^{4} + 2520t^{5} + 55560t^{6} + 78624t^{7} + 965011t^{8} + 1631240t^{9} + 13512030t^{10} + O(t^{11})$

**1261 2111.00026-free-E8**

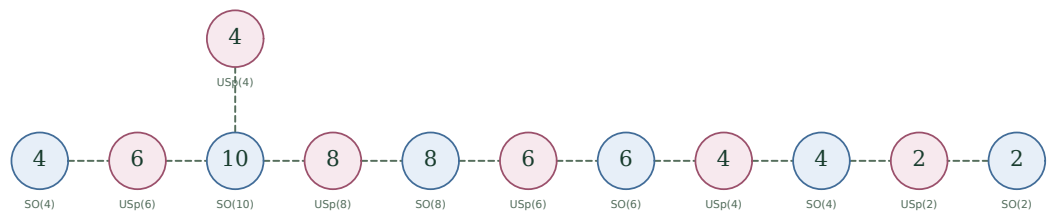


Overall diagonal $\mathbb{Z}_2$ quotient · integral and half-integral magnetic sectors

$1 + 64t + 2080t^{2} + 45760t^{3} + 766480t^{4} + 10424128t^{5} + 119877472t^{6} + 1198774720t^{7} + 10639125640t^{8} + 85113005120t^{9} + 621324937376t^{10} + 4179822305984t^{11} + 26123889412400t^{12} + 152724276564800t^{13} + 839983521106400t^{14} + 4367914309753280t^{15} + 21566576904406820t^{16} + 101489773667796800t^{17} + 456703981505085600t^{18} + 1971038235969316800t^{19} + 8179808679272664720t^{20} + 32719234717090658880t^{21} + 126415225043304818400t^{22} + 472683015379313668800t^{23} + 1713475930750012049400t^{24} + 6031435276240042413888t^{25} + 20646066907129375955232t^{26} + 68820223023764586517440t^{27} + 223665724827234906181680t^{28} + 709560230486400392024640t^{29} + 2199636714507841215276384t^{30} + O(t^{31})$

**Paper typo:** The caption PE[10t] contradicts the $H^{32}$ identification in the table and text. Independent monopoles find 64 degree-one operators, supporting the latter at the checked order; this does not prove a free branch to all orders.

**1262 2111.00026-AppendixB**

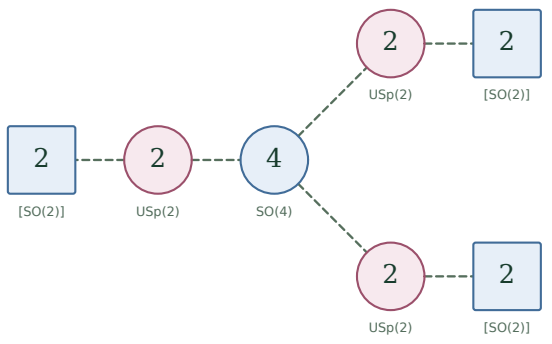


$1 + 13t^{2} + 119t^{4} + 691t^{6} + 3064t^{8} + 10968t^{10} + 33632t^{12} + 91040t^{14} + 223405t^{16} + 505257t^{18} + 1068195t^{20} + 2132015t^{22} + 4051464t^{24} + 7376616t^{26} + 12939016t^{28} + 21958312t^{30} + O(t^{31})$

## 2111.02831 · Magnetic quivers and line defects - On a duality between 3d N=4 unitary and orthosymplectic quivers (4)

**1263 2111.02831-mirror-U-k2-Nf6**

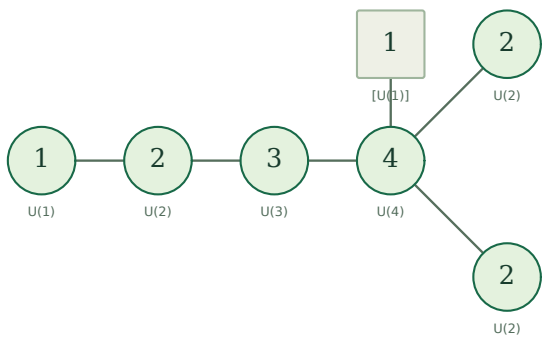


**Computed by Quiver Lab**

$1 + 66t^{2} + 2133t^{4} + 44121t^{6} + 656667t^{8} + 7544394t^{10} + 70199051t^{12} + O(t^{13})$

Source originally Claimed · no numerical literature comparison is available.

**1264 2111.02831-E5-B**

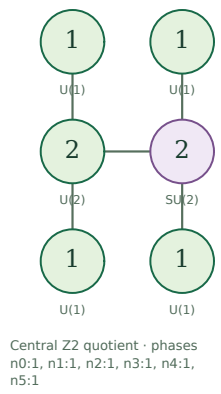


Central Z2 quotient · phases n0:1, n1:1, n2:1, n3:1, n4:1, n5:1

$1 + 45t^{2} + 770t^{4} + 7644t^{6} + 52920t^{8} + 282744t^{10} + O(t^{11})$

| QUIVER | HILBERT SERIES |
| --- | --- |

**1265 2111.02831-E7-B**

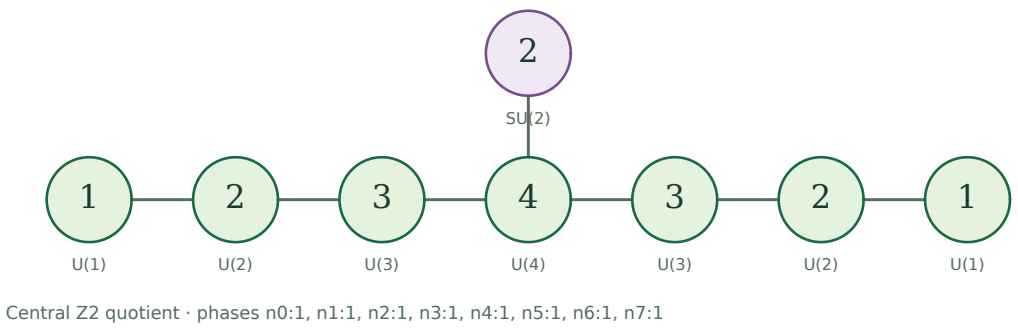


Central Z2 quotient · phases n0:1, n1:1, n2:1, n3:1, n4:1, n5:1, n6:1, n7:1

$1 + 133t^{2} + 7371t^{4} + O(t^{5})$

**1266 2111.02831-E7-C**

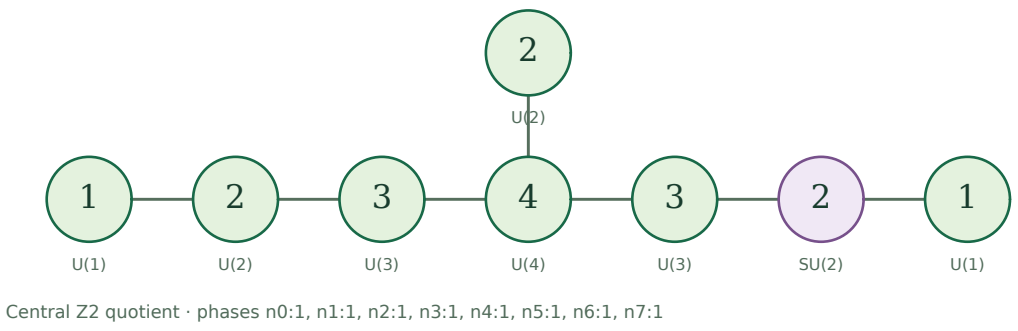


Central Z2 quotient · phases n0:1, n1:1, n2:1, n3:1, n4:1, n5:1, n6:1, n7:1

$1 + 133t^{2} + 7371t^{4} + O(t^{5})$

## 2111.04745 · Higgs branches of U/SU quivers via brane locking (13)

**1267 2111.04745-T3-electric-2**

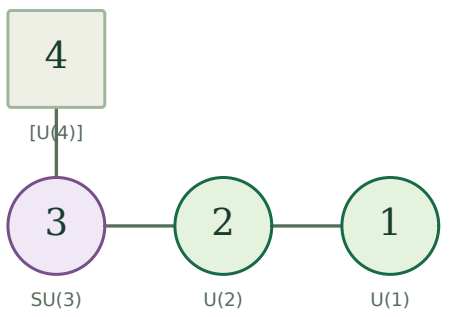


$1 + 8t^{2} + 44t^{4} + 183t^{6} + 636t^{8} + 1916t^{10} + 5164t^{12} + 12692t^{14} + 28897t^{16} + 61644t^{18} + 124368t^{20} + 239047t^{22} + 440416t^{24} + 781640t^{26} + 1341960t^{28} + 2236608t^{30} + O(t^{31})$

**1268 2111.04745-T3-magnetic-2**

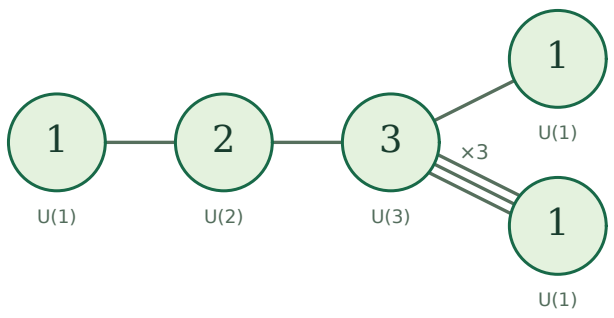


÷ U(1) diagonal

$1 + 16t^{2} + 8t^{3} + 135t^{4} + 120t^{5} + 819t^{6} + 952t^{7} + 4011t^{8} + 5352t^{9} + 16737t^{10} + 23960t^{11} + 61375t^{12} + 90800t^{13} + 201993t^{14} + 302312t^{15} + 606128t^{16} + 906600t^{17} + 1679157t^{18} + 2492848t^{19} + 4338408t^{20} + 6369008t^{21} + 10542396t^{22} + 15276048t^{23} + 24265138t^{24} + 34678504t^{25} + 53217012t^{26} + 75005776t^{27} + 111775532t^{28} + 155411752t^{29} + 225818646t^{30} + O(t^{31})$

**1269 2111.04745-T3-electric-4**

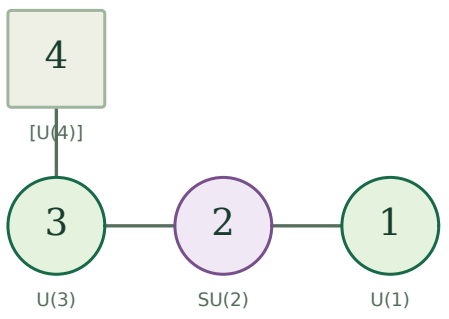


$1 + 6t^{2} + 36t^{4} + 145t^{6} + 518t^{8} + 1554t^{10} + 4234t^{12} + 10406t^{14} + 23811t^{16} + 50824t^{18} + 102806t^{20} + 197715t^{22} + 364812t^{24} + 647756t^{26} + 1113132t^{28} + 1855884t^{30} + O(t^{31})$

**1270 2111.04745-T3-electric-5**

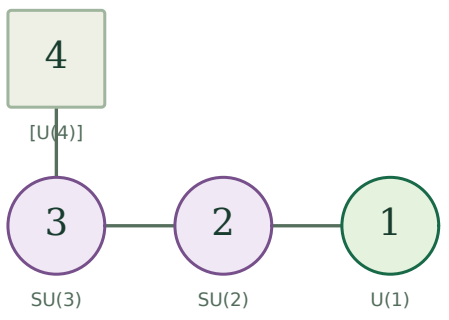


$1 + 3t^{2} + 14t^{4} + 41t^{6} + 117t^{8} + 282t^{10} + 642t^{12} + 1324t^{14} + 2601t^{16} + 4791t^{18} + 8480t^{20} + 14347t^{22} + 23515t^{24} + 37262t^{26} + 57566t^{28} + 86640t^{30} + O(t^{31})$

| QUIVER | HILBERT SERIES |
|---|---|

**1271 2111.04745-T3-magnetic-5**

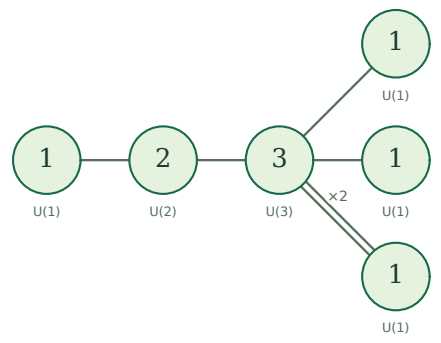

÷ U(1) diagonal

$1 + 17t^{2} + 16t^{3} + 164t^{4} + 256t^{5} + 1213t^{6} + 2240t^{7} + 7386t^{8} + 14192t^{9} + 37963t^{10} + 72352t^{11} + 168452t^{12} + 312848t^{13} + 659123t^{14} + 1185552t^{15} + 2315053t^{16} + 4029024t^{17} + 7407354t^{18} + 12489552t^{19} + 21857188t^{20} + 35779632t^{21} + 60084030t^{22} + 95713952t^{23} + 155174094t^{24} + 241111120t^{25} + 379180468t^{26} + 575928064t^{27} + 881925572t^{28} + 1312015360t^{29} + 1962370964t^{30} + O(t^{31})$

**1272 2111.04745-T3-electric-6**

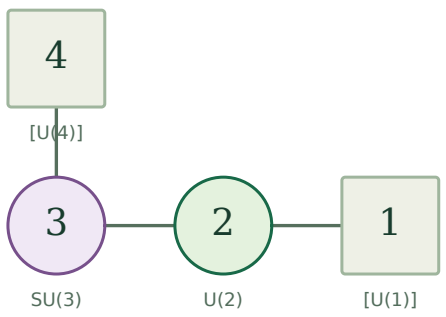


$1 + 3t^{2} + 14t^{4} + 41t^{6} + 117t^{8} + 282t^{10} + 642t^{12} + 1324t^{14} + 2601t^{16} + 4791t^{18} + 8480t^{20} + 14347t^{22} + 23515t^{24} + 37262t^{26} + 57566t^{28} + 86640t^{30} + O(t^{31})$

**1273 2111.04745-T3-electric-7**

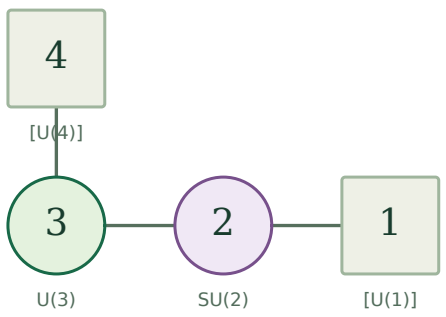


$1 + 3t^{2} + 14t^{4} + 41t^{6} + 117t^{8} + 282t^{10} + 642t^{12} + 1324t^{14} + 2601t^{16} + 4791t^{18} + 8480t^{20} + 14347t^{22} + 23515t^{24} + 37262t^{26} + 57566t^{28} + 86640t^{30} + O(t^{31})$

**1274 2111.04745-T3-electric-8**

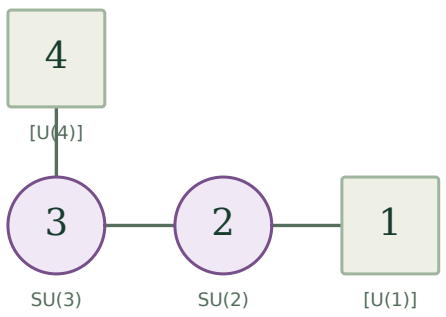


$1 + 5t^{4} + 7t^{6} + 20t^{8} + 35t^{10} + 72t^{12} + 116t^{14} + 205t^{16} + 312t^{18} + 493t^{20} + 715t^{22} + 1052t^{24} + 1459t^{26} + 2040t^{28} + 2732t^{30} + O(t^{31})$

**1275 2111.04745-T3-magnetic-8**

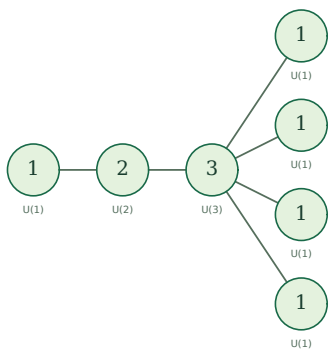

÷ U(1) diagonal

$1 + 18t^{2} + 32t^{3} + 206t^{4} + 544t^{5} + 1993t^{6} + 5344t^{7} + 15531t^{8} + 39040t^{9} + 98672t^{10} + 230016t^{11} + 526784t^{12} + 1144288t^{13} + 2428175t^{14} + 4957120t^{15} + 9884608t^{16} + 19122112t^{17} + 36191166t^{18} + 66817184t^{19} + 120941522t^{20} + 214368128t^{21} + 373286116t^{22} + 638386304t^{23} + 1074571672t^{24} + 1780471968t^{25} + 2908454698t^{26} + 4685262752t^{27} + 7451734878t^{28} + 11705290208t^{29} + 18176095340t^{30} + O(t^{31})$

**1276 2111.04745-427-electric**

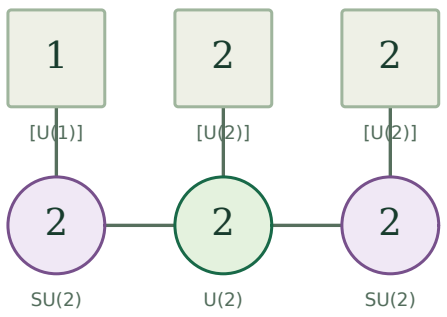


$1 + 2t^{2} + 12t^{4} + 33t^{6} + 101t^{8} + 242t^{10} + 572t^{12} + 1185t^{14} + 2375t^{16} + 4401t^{18} + 7884t^{20} + 13401t^{22} + 22139t^{24} + 35207t^{26} + 54680t^{28} + 82526t^{30} + O(t^{31})$

**1277 2111.04745-427-magnetic**

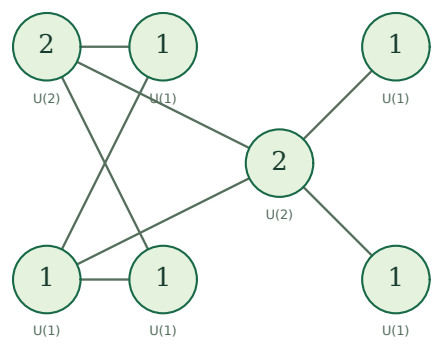


÷ U(1) diagonal

$1 + 6t + 33t^2 + 152t^3 + 624t^4 + 2336t^5 + 8105t^6 + 26254t^7 + 80086t^8 + 231334t^9 + 635973t^{10} + 1671080t^{11} + 4212606t^{12} + 10221596t^{13} + 23943059t^{14} + 54283374t^{15} + 119398753t^{16} + 255325672t^{17} + 531836847t^{18} + 1080935242t^{19} + 2147027711t^{20} + 4173592864t^{21} + 7950244432t^{22} + 14858121288t^{23} + 27272961278t^{24} + 49217437380t^{25} + 87402285977t^{26} + 152866165274t^{27} + 263526600911t^{28} + 448101623152t^{29} + 752071539989t^{30} + O(t^{31})$

**Paper typo:** A proposed repair moves $(1-t)^3$ from denominator to numerator, removing a surplus factor of three free hypermultiplets. The corrected series matches QuiverLab through $t^{30}$ and an independent sum through $t^6$; the quiver's complex dimension is 16, while the printed pole order is 22.

**1278 2111.04745-440**

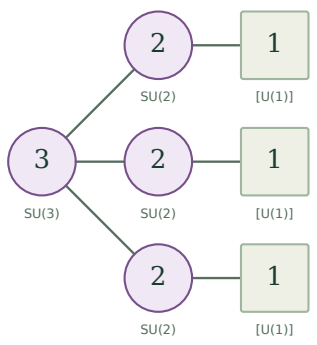


$1 + 15t^4 + 34t^6 + 165t^8 + 453t^{10} + 1345t^{12} + 3289t^{14} + 7824t^{16} + 16835t^{18} + 34810t^{20} + 67545t^{22} + 126374t^{24} + 226078t^{26} + 392190t^{28} + 657782t^{30} + O(t^{31})$

**1279 2111.04745-441**

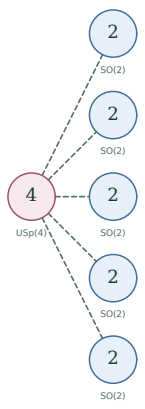


÷ $\mathbb{Z}_2$ diagonal

$1 + 6t^2 + 75t^4 + 453t^6 + 2581t^8 + 11457t^{10} + 45217t^{12} + 155463t^{14} + 485500t^{16} + 1382449t^{18} + 3657034t^{20} + 9044358t^{22} + 21130198t^{24} + 46888222t^{26} + 99471862t^{28} + 202604194t^{30} + O(t^{31})$

## 2112.10825 · Partial Implosions and Quivers (2)

**1280 2112.10825-SL5-partial**

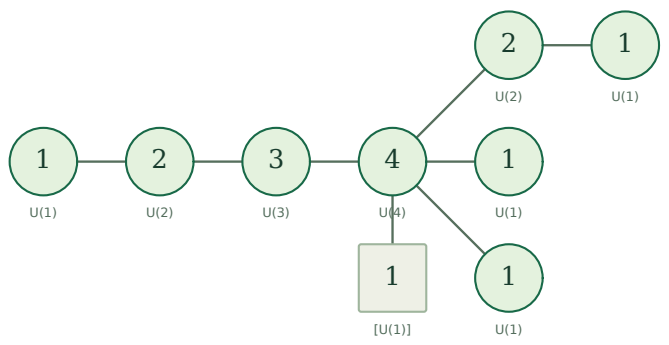


$1 + 30t^2 + 494t^4 + O(t^5)$

**Paper typo:** The printed expansion omits 20 unrefined degree-three monopoles when jumping from $t^2$ to $t^4$. All explicitly printed coefficients through $t^4$ agree; this is an omitted term, not an explicitly printed zero.

**1281 2112.10825-SO8-partial**

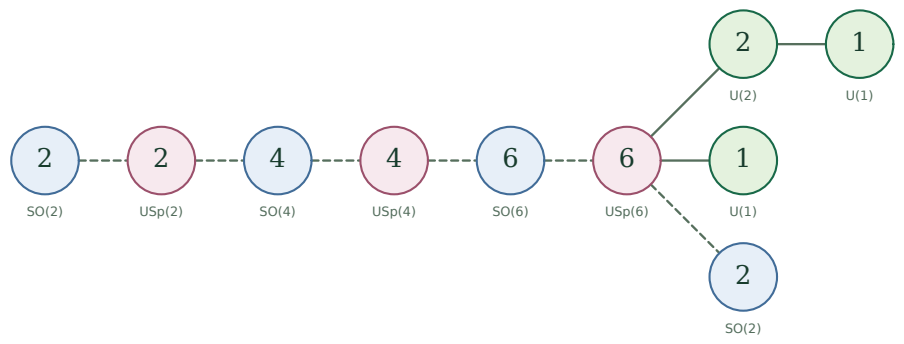


÷ $\mathbb{Z}_2$ diagonal

$1 + 34t^2 + 594t^4 + O(t^5)$

## 2203.01951 · Exploring the orthosymplectic zoo (32)

**1282 2203.01951-T3-05**

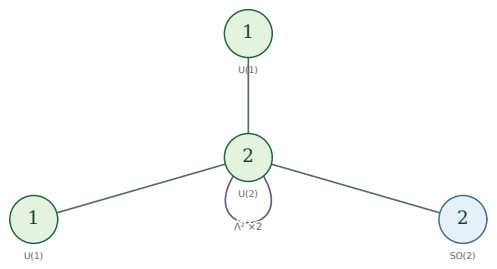


÷ $\mathbb{Z}_2$ diagonal

$1 + 16t^2 + 132t^4 + 735t^6 + 3134t^8 + 10974t^{10} + O(t^{11})$

| QUIVER | HILBERT SERIES |
| --- | --- |

**1283 2203.01951-T3-07**

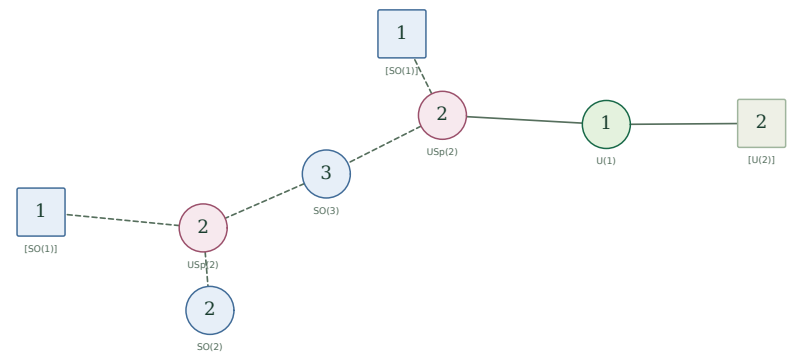


$1 + 16t^{2} + 132t^{4} + 735t^{6} + 3134t^{8} + 10974t^{10} + O(t^{11})$

**1284 2203.01951-T3-09**

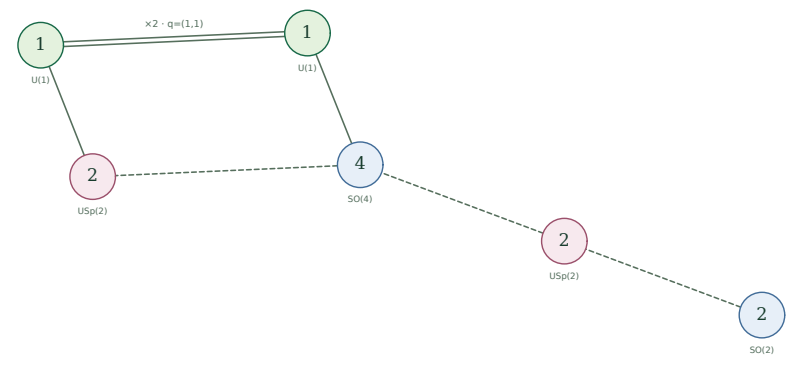


÷ $\mathbb{Z}_2$ diagonal

$1 + 25t^{2} + 320t^{4} + 2694t^{6} + 16830t^{8} + 83952t^{10} + 351250t^{12} + 1276030t^{14} + 4128223t^{16} + 12122795t^{18} + 32792670t^{20} + O(t^{21})$

**1285 2203.01951-T3-10**

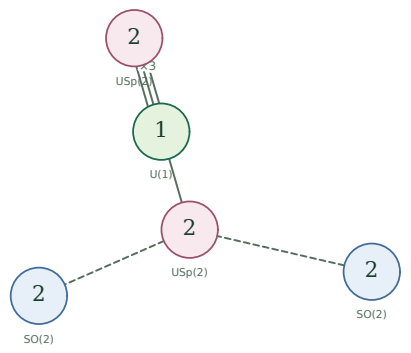


÷ $\mathbb{Z}_2$ diagonal

$1 + 25t^{2} + 235t^{4} + 1375t^{6} + 5980t^{8} + 21126t^{10} + O(t^{11})$

**1286 2203.01951-T3-11**

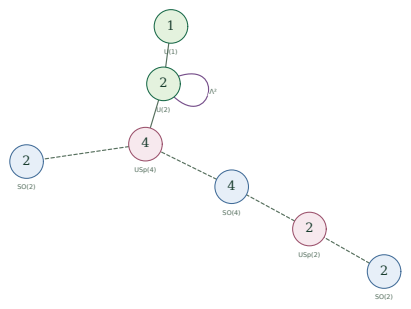

÷ $\mathbb{Z}_2$ diagonal

$1 + 36t^{2} + 670t^{4} + 8295t^{6} + 76278t^{8} + 556242t^{10} + 3363345t^{12} + 17417948t^{14} + 79176858t^{16} + 322010706t^{18} + 1189618112t^{20} + O(t^{21})$

**1287 2203.01951-T3-12**

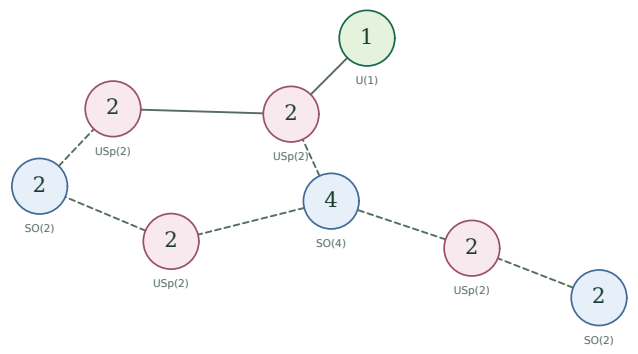


÷ $\mathbb{Z}_2$ diagonal

$1 + 36t^{2} + 660t^{4} + 7798t^{6} + 66870t^{8} + 448662t^{10} + 2478994t^{12} + 11698815t^{14} + 48442662t^{16} + 179656880t^{18} + 606366432t^{20} + O(t^{21})$

**1288 2203.01951-T3-13**

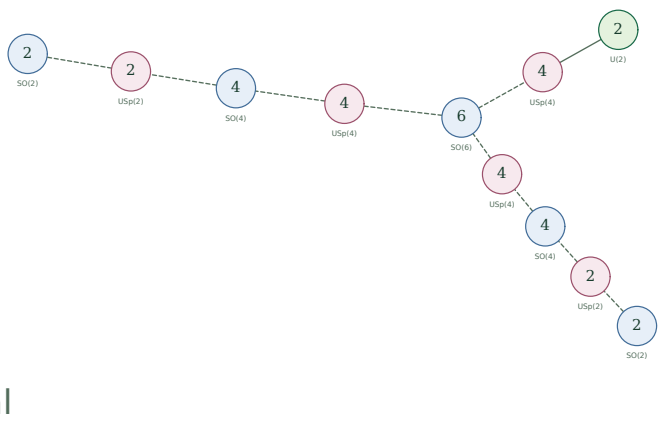

÷ $\mathbb{Z}_2$ diagonal

$1 + 69t^{2} + 2630t^{4} + 68469t^{6} + 1334326t^{8} + 20554428t^{10} + 259940921t^{12} + 2776383765t^{14} + 25605813759t^{16} + 207577975304t^{18} + 1500882475500t^{20} + O(t^{21})$

| QUIVER | HILBERT SERIES |
|---|---|
| **1289 2203.01951-T3-19** 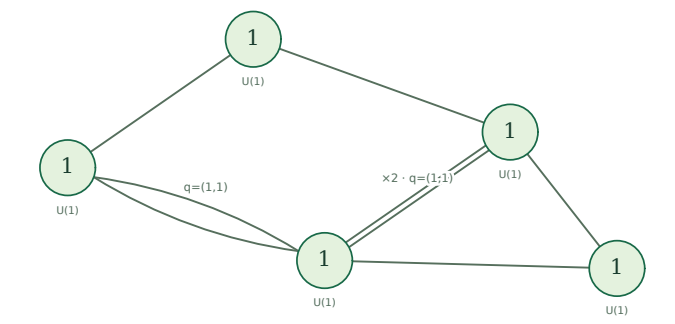 Central Z2 quotient · phases A:1, B:1, C:1, D:1, E:1 | $1 + 13t^2 + 14t^3 + 93t^4 + 134t^5 + 499t^6 + 738t^7 + 2092t^8 + 3068t^9 + 7208t^{10} + O(t^{11})$ |
| **1290 2203.01951-T3-22** 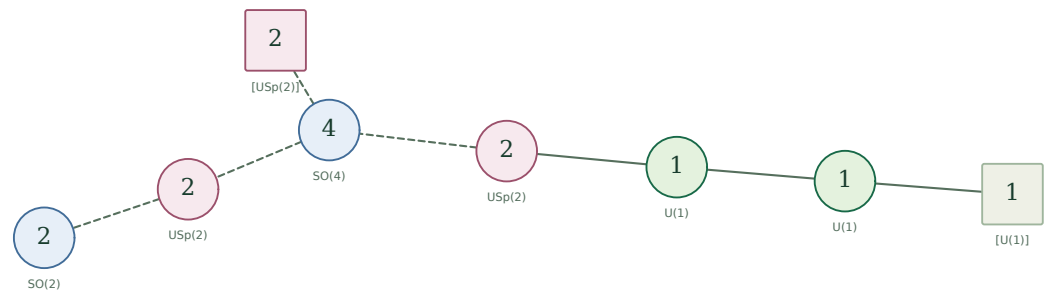 | $1 + 19t^2 + 24t^3 + 188t^4 + 368t^5 + 1396t^6 + 2968t^7 + 8302t^8 + 17168t^9 + 40474t^{10} + 79648t^{11} + 167230t^{12} + 312656t^{13} + 603338t^{14} + 1074896t^{15} + 1943919t^{16} + 3316912t^{17} + 5693149t^{18} + 9351600t^{19} + 15372660t^{20} + O(t^{21})$ |
| **1291 2203.01951-T3-24** 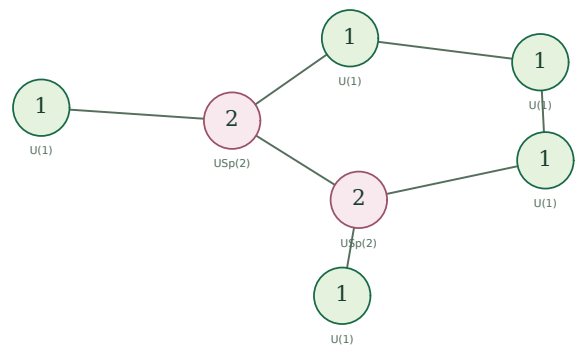 ÷ $\mathbb{Z}_2$ diagonal | $1 + 19t^2 + 24t^3 + 188t^4 + 368t^5 + 1396t^6 + 2968t^7 + 8302t^8 + 17168t^9 + 40474t^{10} + O(t^{11})$ |
| **1292 2203.01951-T3-32** 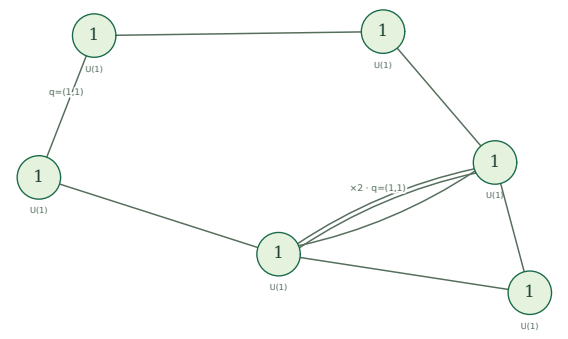 Central Z2 quotient · phases A:1, B:1, C:1, D:1, E:1, F:1 | $1 + 24t^2 + 8t^3 + 255t^4 + 144t^5 + 1716t^6 + 1256t^7 + 8594t^8 + 7312t^9 + 34872t^{10} + 32640t^{11} + 120628t^{12} + 120336t^{13} + 367968t^{14} + 383440t^{15} + 1013621t^{16} + 1088720t^{17} + 2565512t^{18} + 2814744t^{19} + 6045369t^{20} + O(t^{21})$ |
| **1293 2203.01951-T3-35** 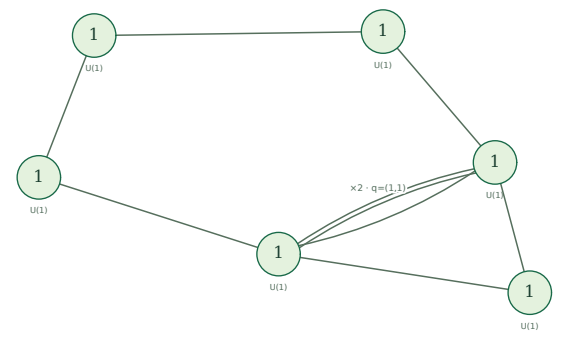 Central Z2 quotient · phases A:1, B:1, C:1, D:1, E:1, F:1 | $1 + 22t^2 + 16t^3 + 215t^4 + 240t^5 + 1408t^6 + 1824t^7 + 7074t^8 + 9680t^9 + 29072t^{10} + 40560t^{11} + 102124t^{12} + 142832t^{13} + 316400t^{14} + 439792t^{15} + 884077t^{16} + 1216592t^{17} + 2265934t^{18} + 3082080t^{19} + 5398337t^{20} + O(t^{21})$ |
| **1294 2203.01951-T3-42** 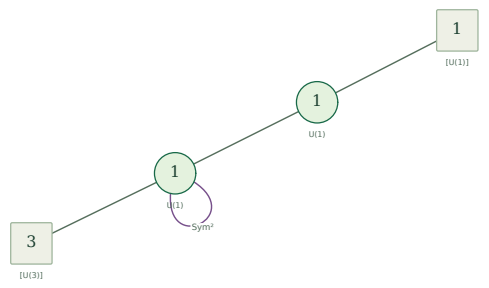 | $1 + 4t^2 + 9t^4 + 20t^6 + 37t^8 + 60t^{10} + 95t^{12} + 140t^{14} + 195t^{16} + 268t^{18} + 355t^{20} + 456t^{22} + 581t^{24} + 724t^{26} + 885t^{28} + 1076t^{30} + O(t^{31})$ |

| QUIVER | HILBERT SERIES |
|---|---|
| **1295 2203.01951-T3-45** | $1 + 11t^{2} + 60t^{4} + 235t^{6} + 745t^{8} + 2016t^{10} + O(t^{11})$ |

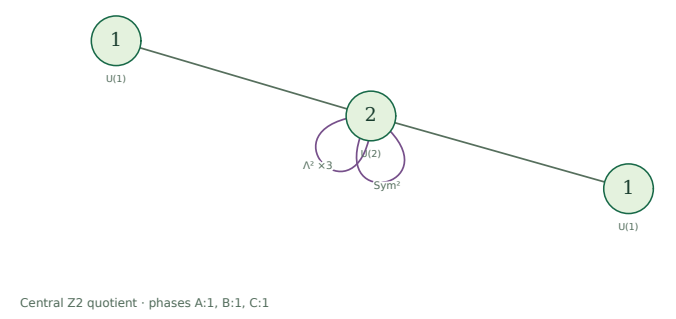


Central Z2 quotient · phases A:1, B:1, C:1

| | |
|---|---|
| **1296 2203.01951-T3-46** | $1 + 14t^{2} + 98t^{4} + 487t^{6} + 1926t^{8} + 6396t^{10} + 18552t^{12} + 48296t^{14} + 115005t^{16} + 254274t^{18} + 528046t^{20} + O(t^{21})$ |

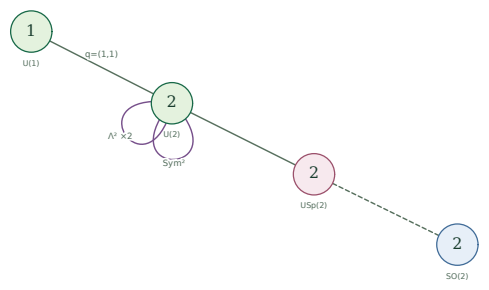

÷ ℤ₂ diagonal

| | |
|---|---|
| **1297 2203.01951-T3-49** | $1 + 21t^{2} + 247t^{4} + 2042t^{6} + 13118t^{8} + 69176t^{10} + 311061t^{12} + 1225801t^{14} + 4322339t^{16} + 13861430t^{18} + 40957078t^{20} + O(t^{21})$ |

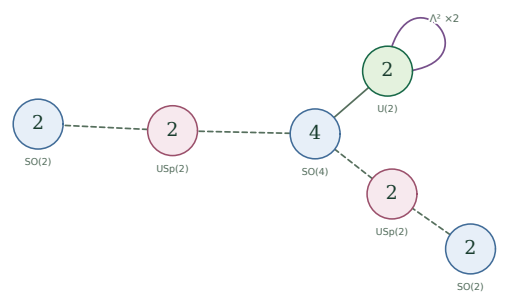

÷ ℤ₂ diagonal

| | |
|---|---|
| **1298 2203.01951-T3-50** | $1 + 10t^{2} + 54t^{4} + 210t^{6} + 659t^{8} + 1772t^{10} + 4236t^{12} + 9228t^{14} + 18645t^{16} + 35398t^{18} + 63778t^{20} + O(t^{21})$ |

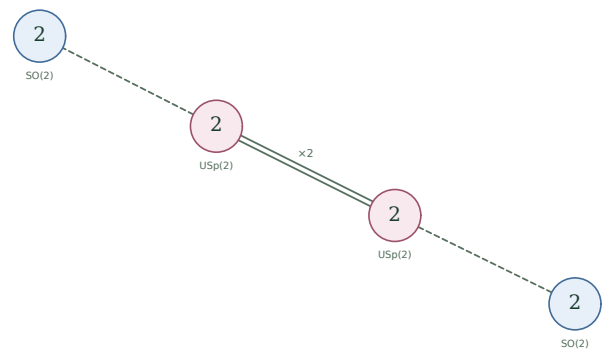

÷ ℤ₂ diagonal

| | |
|---|---|
| **1299 2203.01951-T3-52** | $1 + 10t^{2} + 54t^{4} + 210t^{6} + 659t^{8} + 1772t^{10} + 4236t^{12} + 9228t^{14} + 18645t^{16} + 35398t^{18} + 63778t^{20} + O(t^{21})$ |

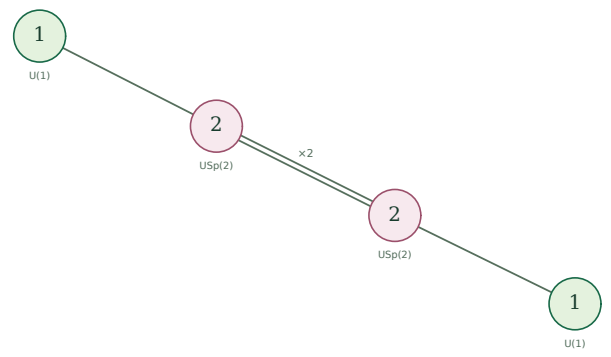

÷ ℤ₂ diagonal

| | |
|---|---|
| **1300 2203.01951-T3-56** | $1 + 10t^{2} + 55t^{4} + 215t^{6} + 679t^{8} + 1831t^{10} + 4390t^{12} + 9579t^{14} + 19385t^{16} + 36839t^{18} + 66437t^{20} + O(t^{21})$ |

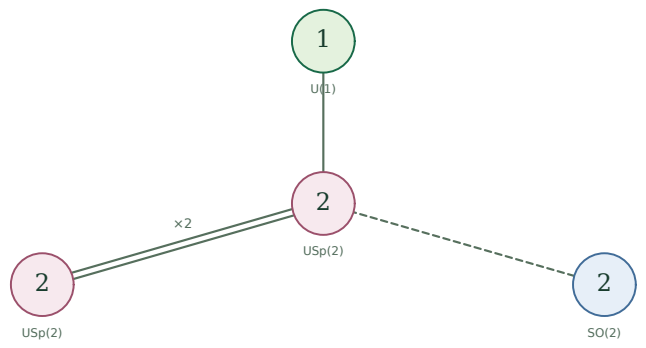


÷ ℤ₂ diagonal

| QUIVER | HILBERT SERIES |
|---|---|
| **1301 2203.01951-U-21** | $1 + 20t^2 + 28t^3 + 220t^4 + 456t^5 + 1905t^6 + 4132t^7 + 13026t^8 + 27600t^9 + 72438t^{10} + O(t^{11})$ |

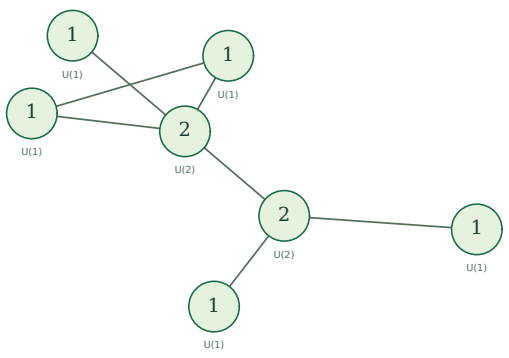


÷ U(1) diagonal

| | |
|---|---|
| **1302 2203.01951-U-22** | $1 + 19t^2 + 24t^3 + 188t^4 + 368t^5 + 1396t^6 + 2968t^7 + 8302t^8 + 17168t^9 + 40474t^{10} + 79648t^{11} + 167230t^{12} + 312656t^{13} + 603338t^{14} + 1074896t^{15} + 1943919t^{16} + 3316912t^{17} + 5693149t^{18} + 9351600t^{19} + 15372660t^{20} + O(t^{21})$ |

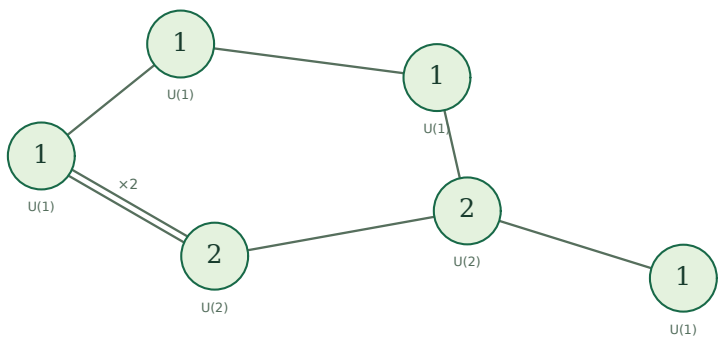


÷ U(1) diagonal

| | |
|---|---|
| **1303 2203.01951-U-25** | $1 + 29t^2 + 52t^3 + 484t^4 + 1398t^5 + 6568t^6 + 20378t^7 + 72624t^8 + 214202t^9 + 651868t^{10} + O(t^{11})$ |

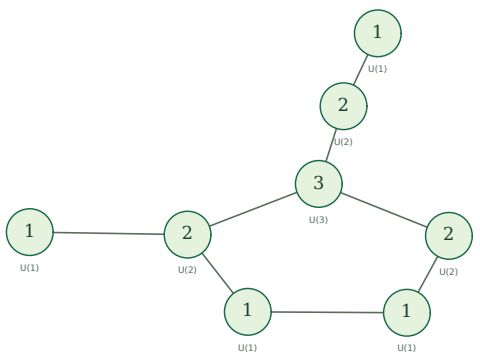

÷ U(1) diagonal

| | |
|---|---|
| **1304 2203.01951-U-26** | $1 + 40t^2 + 84t^3 + 929t^4 + 3192t^5 + 17752t^6 + 66556t^7 + 281028t^8 + 1007048t^9 + 3650679t^{10} + O(t^{11})$ |

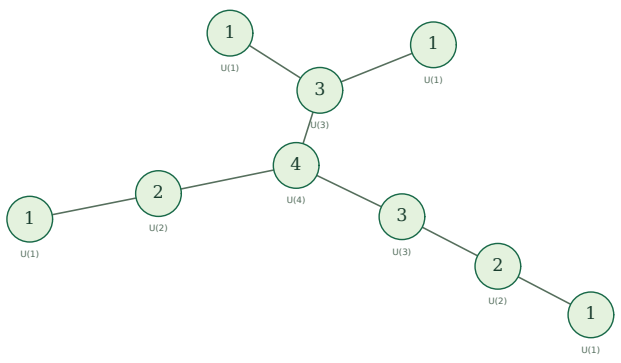

÷ U(1) diagonal

| | |
|---|---|
| **1305 2203.01951-U-27** | $1 + 78t^2 + 224t^3 + 3500t^4 + 16960t^5 + 131941t^6 + O(t^7)$ |

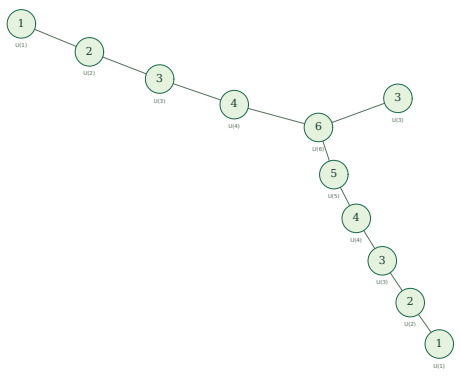

÷ U(1) diagonal

| | |
|---|---|
| **1306 2203.01951-U-28** | $1 + 18t^2 + 205t^4 + 1591t^6 + 9499t^8 + 45959t^{10} + 188535t^{12} + 675381t^{14} + 2163400t^{16} + 6305809t^{18} + 16961842t^{20} + O(t^{21})$ |

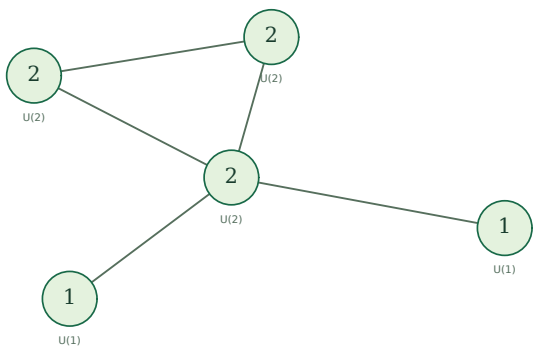


÷ U(1) diagonal

| QUIVER | HILBERT SERIES |
|---|---|

**1307 2203.01951-U-29**

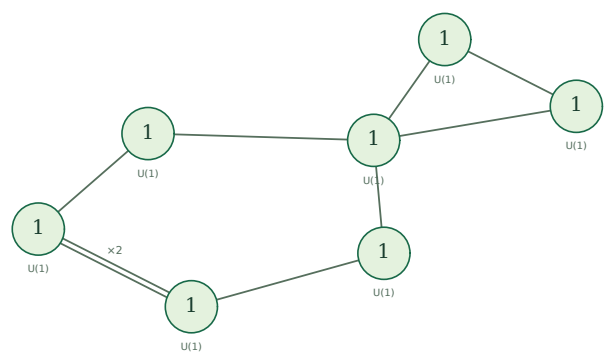

÷ U(1) diagonal

$1 + 24t^{2} + 8t^{3} + 255t^{4} + 144t^{5} + 1716t^{6} + 1256t^{7} + 8594t^{8} + 7312t^{9} + 34872t^{10} + 32640t^{11} + 120628t^{12} + 120336t^{13} + 367968t^{14} + 383440t^{15} + 1013621t^{16} + 1088720t^{17} + 2565512t^{18} + 2814744t^{19} + 6045369t^{20} + O(t^{21})$

**1308 2203.01951-U-30**

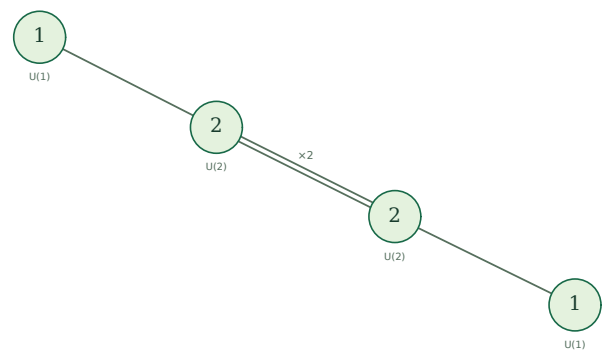

÷ U(1) diagonal

$1 + 11t^{2} + 8t^{3} + 65t^{4} + 80t^{5} + 295t^{6} + 432t^{7} + 1122t^{8} + 1720t^{9} + 3666t^{10} + 5640t^{11} + 10564t^{12} + 16024t^{13} + 27460t^{14} + 40752t^{15} + 65445t^{16} + 94888t^{17} + 144935t^{18} + 205440t^{19} + 301571t^{20} + O(t^{21})$

**1309 2203.01951-U-34**

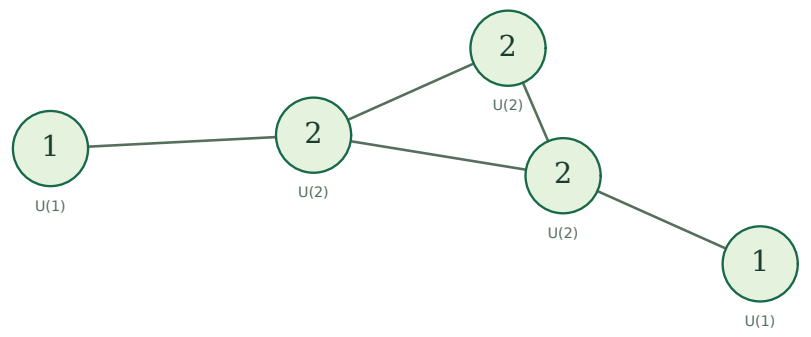

÷ U(1) diagonal

$1 + 14t^{2} + 16t^{3} + 119t^{4} + 208t^{5} + 806t^{6} + 1568t^{7} + 4480t^{8} + 8736t^{9} + 20886t^{10} + 39552t^{11} + 83778t^{12} + O(t^{13})$

**1310 2203.01951-U-37**

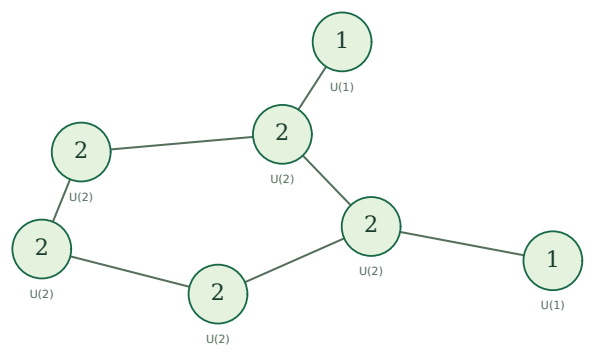

÷ U(1) diagonal

$1 + 26t^{2} + 32t^{3} + 425t^{4} + 800t^{5} + 5189t^{6} + 11328t^{7} + 50074t^{8} + 113120t^{9} + 394773t^{10} + O(t^{11})$

**1311 2203.01951-U-38**

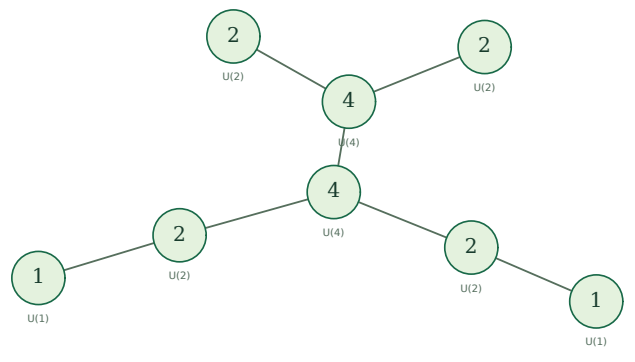

÷ U(1) diagonal

$1 + 39t^{2} + 64t^{3} + 921t^{4} + 2432t^{5} + 16966t^{6} + 52096t^{7} + 254322t^{8} + 801152t^{9} + 3163614t^{10} + O(t^{11})$

**1312 2203.01951-U-40**

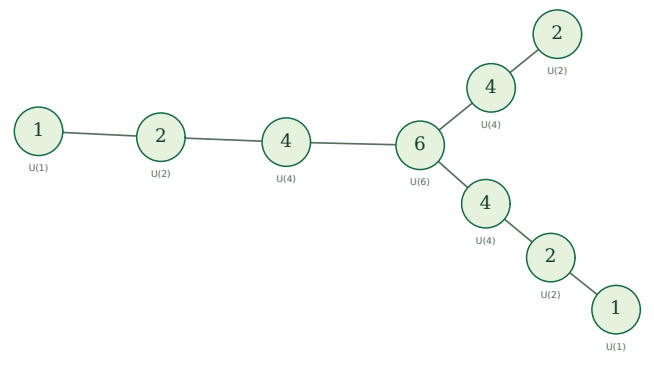

÷ U(1) diagonal

$1 + 56t^{2} + 128t^{3} + 1840t^{4} + 7040t^{5} + 49820t^{6} + 216064t^{7} + 1137320t^{8} + 4856960t^{9} + 21742985t^{10} + O(t^{11})$

| QUIVER | HILBERT SERIES |
| --- | --- |

**1313 2203.01951-U-41**

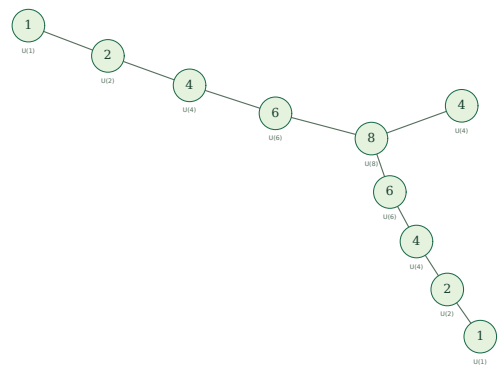


÷ U(1) diagonal

$1 + 89t^2 + 216t^3 + 4448t^4 + 19008t^5 + 177571t^6 + O(t^7)$

## 2203.16550 · Dynamical consequences of 1-form symmetries and the exceptional Argyres-Douglas theories (8)

**1314 2203.16550-star-product**

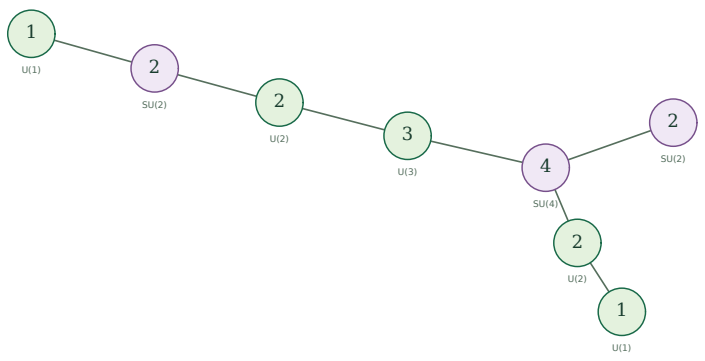


$1 + 58t^2 + 256t^3 + O(t^4)$

**1315 2203.16550-star-Z2**

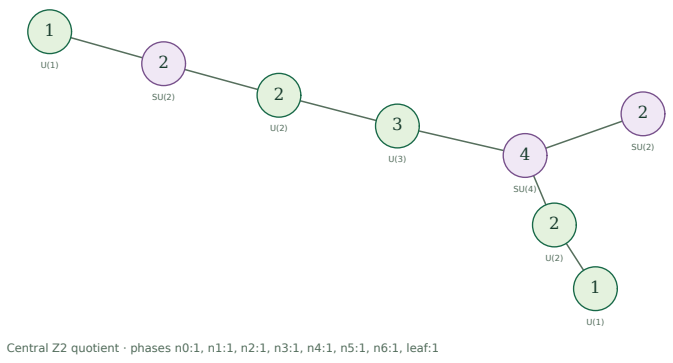


Central Z2 quotient · phases n0:1, n1:1, n2:1, n3:1, n4:1, n5:1, n6:1, leaf:1

$1 + 8t + 82t^2 + 616t^3 + O(t^4)$

**1316 2203.16550-E6-SO2-n2** / n = 2

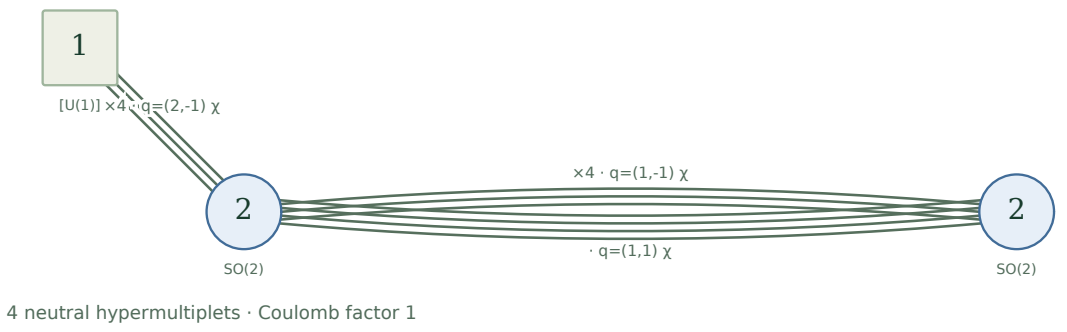


Overall diagonal $\mathbb{Z}_2$ quotient · integral and half-integral magnetic sectors · 4 neutral hypermultiplets · Coulomb factor 1

**Computed by Quiver Lab**

$1 + 2t^2 + 3t^4 + 4t^6 + 11t^8 + 18t^{10} + 25t^{12} + 32t^{14} + 51t^{16} + 70t^{18} + 89t^{20} + 108t^{22} + 145t^{24} + 182t^{26} + 219t^{28} + 256t^{30} + O(t^{31})$

Source originally Claimed · no numerical literature comparison is available.

**1317 2203.16550-E6-SO2-n6** / n = 6

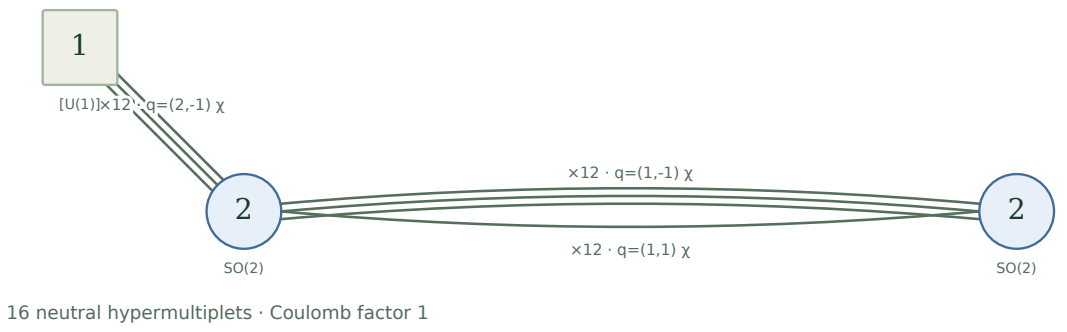


Overall diagonal $\mathbb{Z}_2$ quotient · integral and half-integral magnetic sectors · 16 neutral hypermultiplets · Coulomb factor 1

**Computed by Quiver Lab**

$1 + 2t^2 + 3t^4 + 4t^6 + 5t^8 + 6t^{10} + 7t^{12} + 8t^{14} + 9t^{16} + 10t^{18} + 11t^{20} + 12t^{22} + 19t^{24} + 26t^{26} + 33t^{28} + 40t^{30} + O(t^{31})$

Source originally Claimed · no numerical literature comparison is available.

**1318 2203.16550-E6-SO2-n10** / n = 10

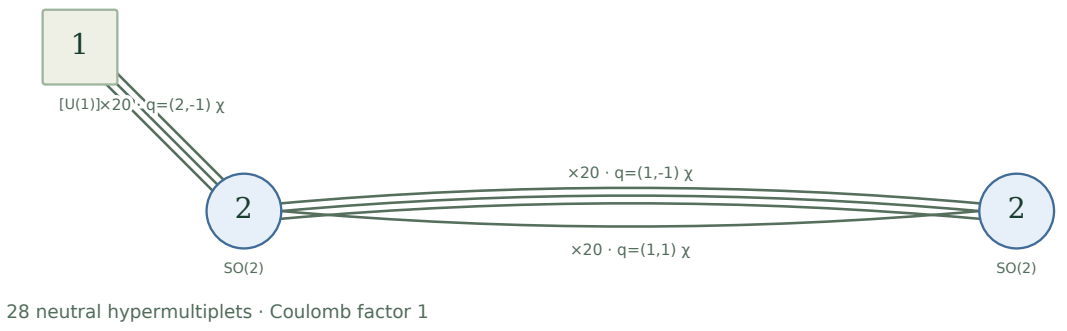


Overall diagonal $\mathbb{Z}_2$ quotient · integral and half-integral magnetic sectors · 28 neutral hypermultiplets · Coulomb factor 1

**Computed by Quiver Lab**

$1 + 2t^2 + 3t^4 + 4t^6 + 5t^8 + 6t^{10} + 7t^{12} + 8t^{14} + 9t^{16} + 10t^{18} + 11t^{20} + 12t^{22} + 13t^{24} + 14t^{26} + 15t^{28} + 16t^{30} + O(t^{31})$

Source originally Claimed · no numerical literature comparison is available.

| QUIVER | HILBERT SERIES |
| --- | --- |

**1319 2203.16550-E7-SQED-n3** / n = 3

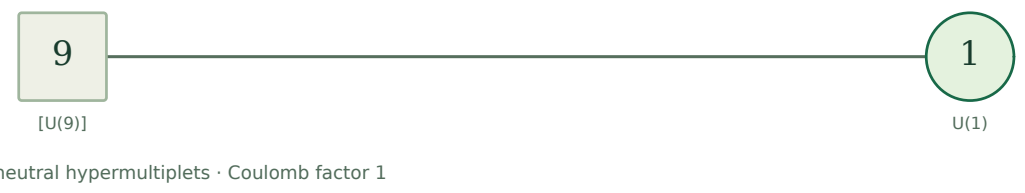


9 neutral hypermultiplets · Coulomb factor 1

**Computed by Quiver Lab**

$1 + t^2 + t^4 + t^6 + t^8 + 2t^9 + t^{10} + 2t^{11} + t^{12} + 2t^{13} + t^{14} + 2t^{15} + t^{16} + 2t^{17} + 3t^{18} + 2t^{19} + 3t^{20} + 2t^{21} + 3t^{22} + 2t^{23} + 3t^{24} + 2t^{25} + 3t^{26} + 4t^{27} + 3t^{28} + 4t^{29} + 3t^{30} + O(t^{31})$

Source originally Claimed · no numerical literature comparison is available.

**1320 2203.16550-E7-SQED-n6** / n = 6

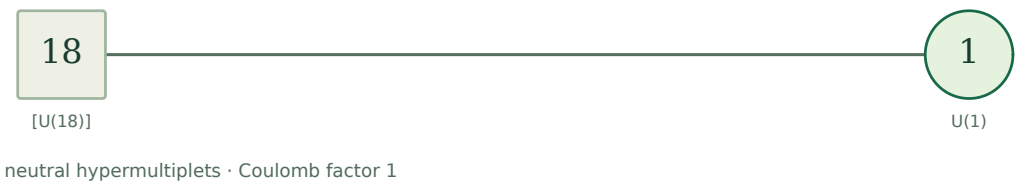


21 neutral hypermultiplets · Coulomb factor 1

**Computed by Quiver Lab**

$1 + t^2 + t^4 + t^6 + t^8 + t^{10} + t^{12} + t^{14} + t^{16} + 3t^{18} + 3t^{20} + 3t^{22} + 3t^{24} + 3t^{26} + 3t^{28} + 3t^{30} + O(t^{31})$

Source originally Claimed · no numerical literature comparison is available.

**1321 2203.16550-E7-SQED-n12** / n = 12

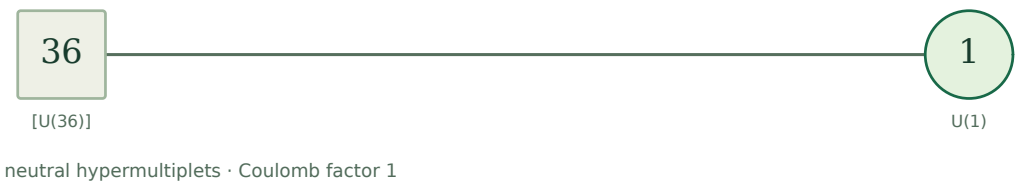


45 neutral hypermultiplets · Coulomb factor 1

**Computed by Quiver Lab**

$1 + t^2 + t^4 + t^6 + t^8 + t^{10} + t^{12} + t^{14} + t^{16} + t^{18} + t^{20} + t^{22} + t^{24} + t^{26} + t^{28} + t^{30} + O(t^{31})$

Source originally Claimed · no numerical literature comparison is available.

## 2206.05362 · Dualities and flavored indices of M2-brane SCFTs (5)

**1322 2206.05362-O2plus-sym-F1**

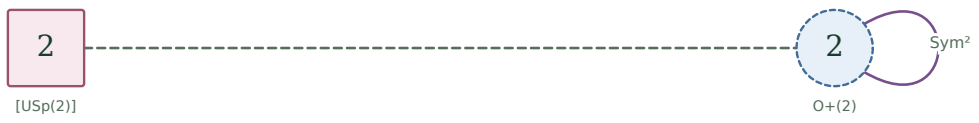


g: θ = +

$1 + t^4 + t^6 + 2t^8 + t^{10} + 3t^{12} + 2t^{14} + 3t^{16} + 3t^{18} + 4t^{20} + 3t^{22} + 5t^{24} + 4t^{26} + 5t^{28} + 5t^{30} + O(t^{31})$

**1323 2206.05362-O2plus-sym-F2**

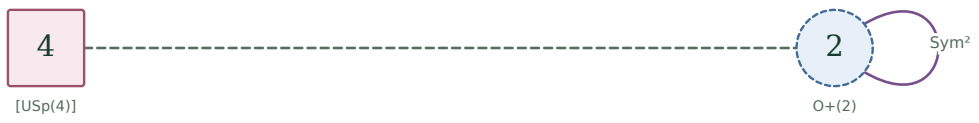


g: θ = +

$1 + t^4 + 2t^8 + t^{10} + 2t^{12} + t^{14} + 3t^{16} + 2t^{18} + 3t^{20} + 2t^{22} + 4t^{24} + 3t^{26} + 4t^{28} + 3t^{30} + O(t^{31})$

**1324 2206.05362-O3plus-sym-F1**

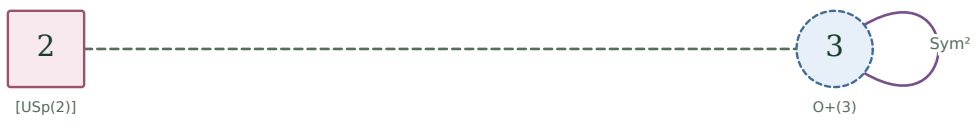


g: θ = +

$1 + t^4 + t^6 + 2t^8 + t^{10} + 3t^{12} + 2t^{14} + 3t^{16} + 3t^{18} + 4t^{20} + 3t^{22} + 5t^{24} + 4t^{26} + 5t^{28} + 5t^{30} + O(t^{31})$

| QUIVER | HILBERT SERIES |
| --- | --- |

**1325 2206.05362-O3plus-sym-F2**

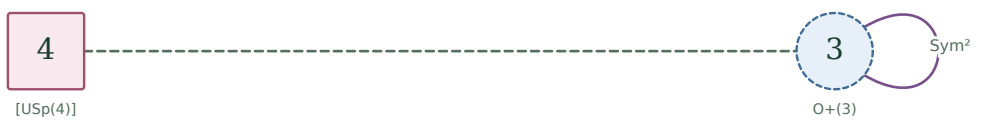


g: θ = +

$1 + t^{4} + 2t^{8} + t^{10} + 2t^{12} + t^{14} + 3t^{16} + 2t^{18} + 3t^{20} + 2t^{22} + 4t^{24} + 3t^{26} + 4t^{28} + 3t^{30} + O(t^{31})$

**1326 2206.05362-O4plus-sym-F1**

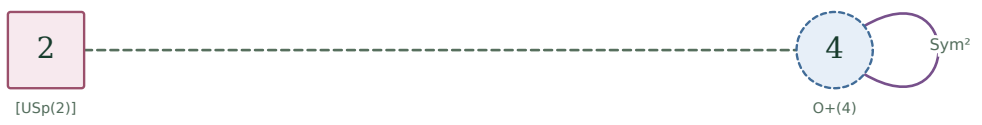


g: θ = +

$1 + t^{4} + t^{6} + 3t^{8} + 2t^{10} + 6t^{12} + 5t^{14} + 10t^{16} + 10t^{18} + 16t^{20} + 16t^{22} + 26t^{24} + 26t^{26} + 36t^{28} + 39t^{30} + O(t^{31})$

## 2207.05764 · Higgs, Coulomb, and Hall-Littlewood (7)

**1327 2207.05764-A2-2full-2null**

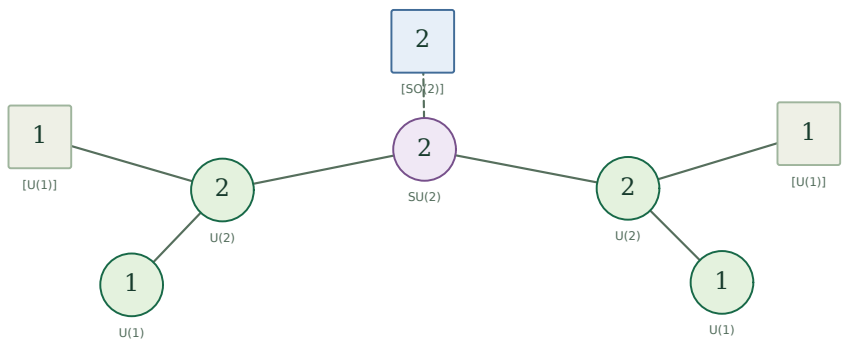


$1 + 16t^{2} + 135t^{4} + 862t^{6} + O(t^{8})$

**1328 2207.05764-A2-full-minimal-2null**

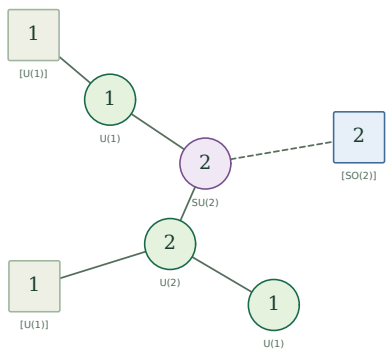

**Computed by Quiver Lab**

$1 + 9t^{2} + 2t^{3} + 53t^{4} + 34t^{5} + 229t^{6} + 216t^{7} + 826t^{8} + 942t^{9} + 2616t^{10} + 3214t^{11} + 7406t^{12} + 9362t^{13} + 19014t^{14} + 24260t^{15} + 44861t^{16} + 57332t^{17} + 98553t^{18} + 125570t^{19} + 203799t^{20} + 258090t^{21} + 399945t^{22} + 502622t^{23} + 749870t^{24} + 934796t^{25} + 1350620t^{26} + 1670204t^{27} + 2347978t^{28} + 2880986t^{29} + 3955506t^{30} + O(t^{31})$

Source originally Claimed · no numerical literature comparison is available.

**1329 2207.05764-A2-minimal-minimal-2null**

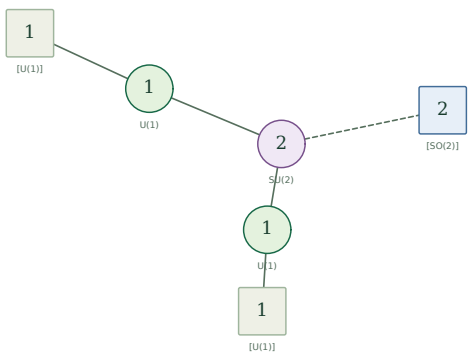

**Computed by Quiver Lab**

$1 + 3t^{2} + 8t^{3} + 12t^{4} + 24t^{5} + 48t^{6} + 72t^{7} + 117t^{8} + 184t^{9} + 267t^{10} + 384t^{11} + 548t^{12} + 744t^{13} + 1008t^{14} + 1344t^{15} + 1749t^{16} + 2256t^{17} + 2887t^{18} + 3624t^{19} + 4524t^{20} + 5600t^{21} + 6852t^{22} + 8328t^{23} + 10065t^{24} + 12048t^{25} + 14355t^{26} + 17008t^{27} + 20016t^{28} + 23448t^{29} + 27356t^{30} + O(t^{31})$

Source originally Claimed · no numerical literature comparison is available.

**1330 2207.05764-SU2-3kminus2F-singlets-k3** / **k = 3**

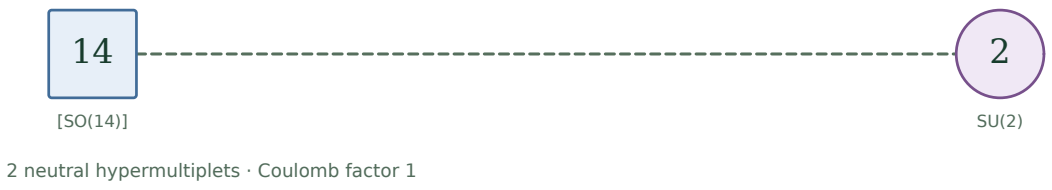


2 neutral hypermultiplets · Coulomb factor 1

$1 + t^{4} + t^{8} + t^{10} + 2t^{12} + t^{14} + 2t^{16} + t^{18} + 3t^{20} + 2t^{22} + 3t^{24} + 2t^{26} + 3t^{28} + 3t^{30} + O(t^{31})$

| QUIVER | HILBERT SERIES |
|---|---|

**1331 2207.05764-SU2-3kminus2F-singlets-k4** / **k = 4**

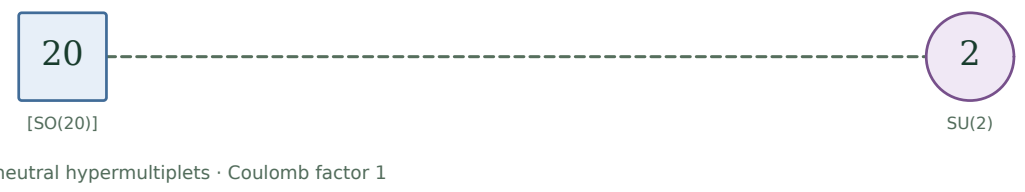


3 neutral hypermultiplets · Coulomb factor 1

$1 + t^4 + t^8 + t^{12} + 2t^{16} + t^{18} + 2t^{20} + t^{22} + 2t^{24} + t^{26} + 2t^{28} + t^{30} + O(t^{31})$

**1332 2207.05764-SU2-3kminus2F-singlets-k5** / **k = 5**

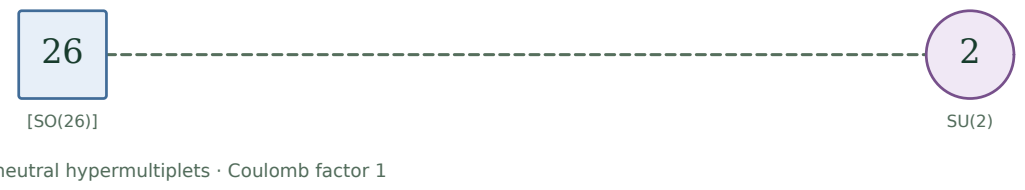


4 neutral hypermultiplets · Coulomb factor 1

$1 + t^4 + t^8 + t^{12} + t^{16} + t^{20} + t^{22} + 2t^{24} + t^{26} + 2t^{28} + t^{30} + O(t^{31})$

**1333 2207.05764-D4-23-four**

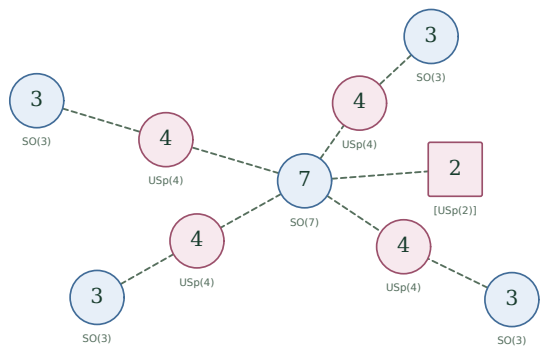


$1 + 12t^2 + 99t^4 + 616t^6 + 3312t^8 + O(t^{10})$

## 2208.07270 · Magnetic quivers and negatively charged branes (54)

**1334 2208.07270-T3-2-quot**

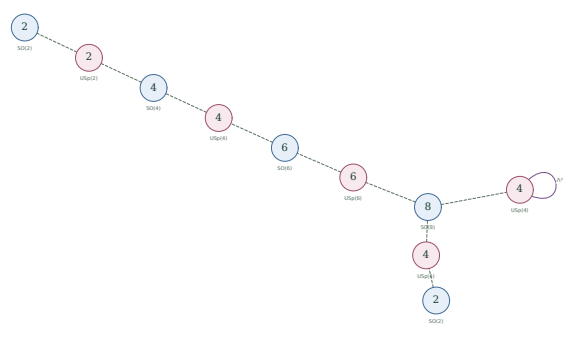

÷ $\mathbb{Z}_2$ diagonal

$1 + 78t^2 + 3080t^4 + 81432t^6 + 1613534t^8 + 25483029t^{10} + 333669414t^{12} + 3724134634t^{14} + 36177376300t^{16} + 310936089880t^{18} + 2396082860652t^{20} + 16738666045297t^{22} + 106997492233476t^{24} + 630843369418369t^{26} + 3454286851716519t^{28} + 17672434006327084t^{30} + O(t^{31})$

**1335 2208.07270-T3-2-integral**

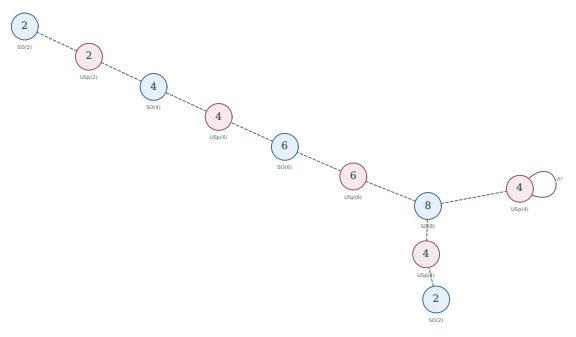

$1 + 46t^2 + 1608t^4 + 41208t^6 + O(t^7)$

**1336 2208.07270-T3-1pow2-quot**

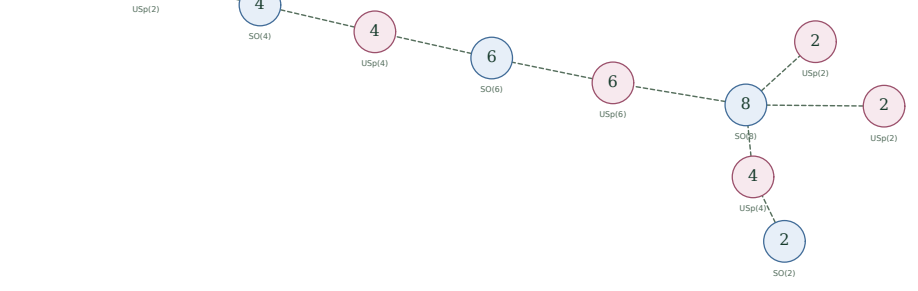

÷ $\mathbb{Z}_2$ diagonal

$1 + 78t^2 + 3158t^4 + 86787t^6 + 1797641t^8 + 29702895t^{10} + 406169555t^{12} + 4720049996t^{14} + 47575754291t^{16} + 422801389413t^{18} + 3357794934897t^{20} + 24101809777891t^{22} + 157869469048851t^{24} + 951454962040858t^{26} + 5314125074537538t^{28} + 27678826736346693t^{30} + O(t^{31})$

| QUIVER | HILBERT SERIES |
|---|---|

**1337 2208.07270-T3-1pow2-integral**

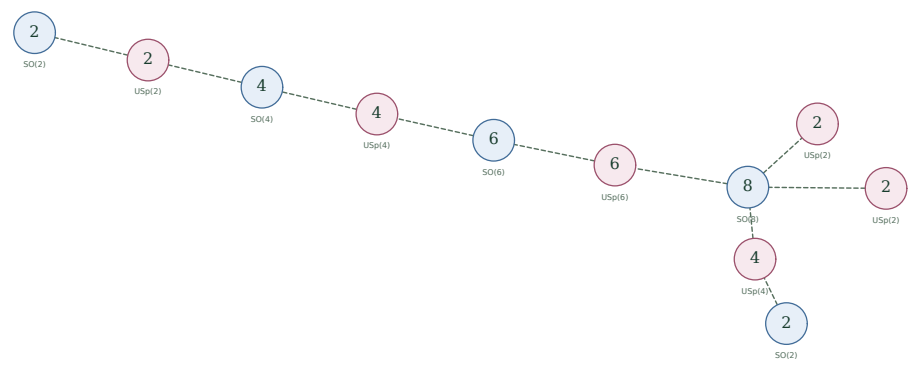


$1 + 46t^2 + 1654t^4 + 43971t^6 + O(t^7)$

**1338 2208.07270-T4-2-quot**

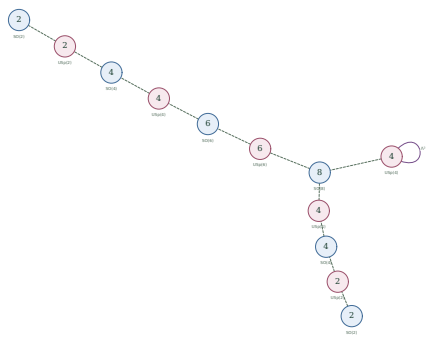

÷ $\mathbb{Z}_2$ diagonal

$1 + 46t^2 + 64t^3 + 1135t^4 + 2944t^5 + 21631t^6 + 71744t^7 + O(t^8)$

**1339 2208.07270-T4-2-integral**

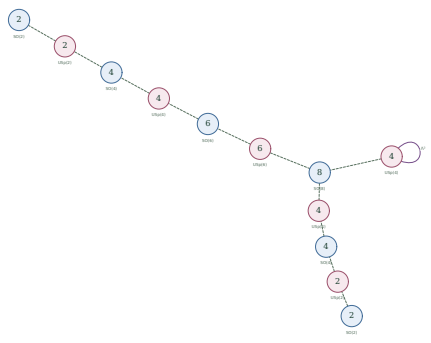

$1 + 46t^2 + 1135t^4 + 21631t^6 + O(t^7)$

**1340 2208.07270-T4-1pow2-quot**

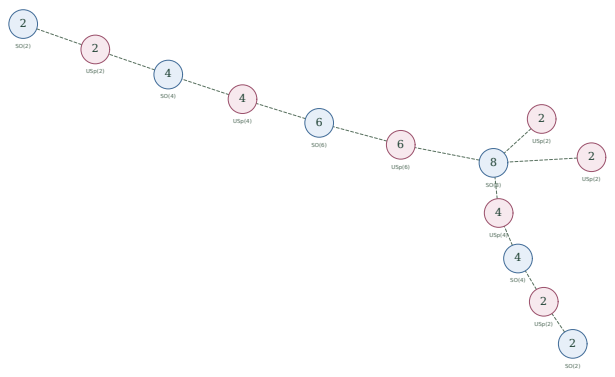

÷ $\mathbb{Z}_2$ diagonal

$1 + 46t^2 + 64t^3 + 1145t^4 + 3008t^5 + 22271t^6 + 75200t^7 + O(t^8)$

**1341 2208.07270-T4-1pow2-integral**

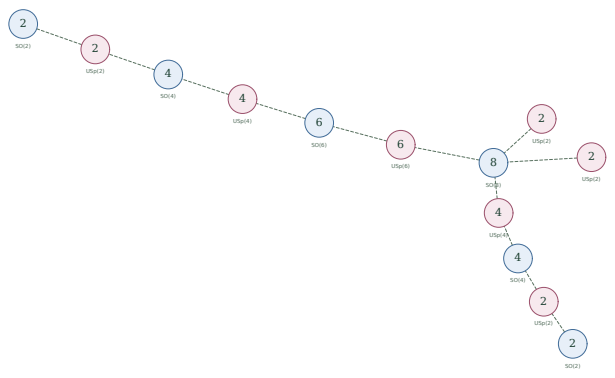

$1 + 46t^2 + 1145t^4 + 22271t^6 + O(t^7)$

**1342 2208.07270-T6-3-quot**

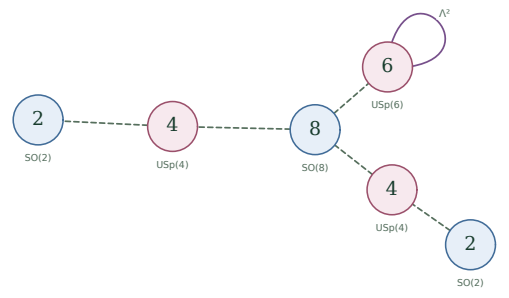


÷ $\mathbb{Z}_2$ diagonal

$1 + 16t^2 + 200t^4 + 1864t^6 + 14629t^8 + 98284t^{10} + 582712t^{12} + O(t^{13})$

**1343 2208.07270-T6-3-integral**

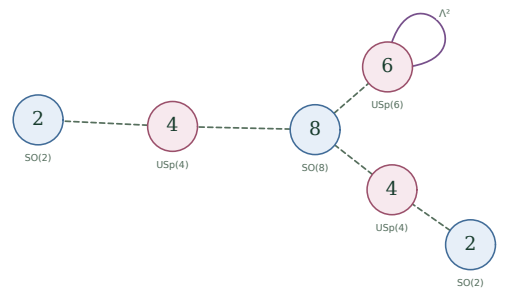


$1 + 8t^2 + 104t^4 + 928t^6 + 7349t^8 + 49108t^{10} + 291552t^{12} + O(t^{13})$

| QUIVER | HILBERT SERIES |
|---|---|

**1344 2208.07270-T6-2-1-quot**

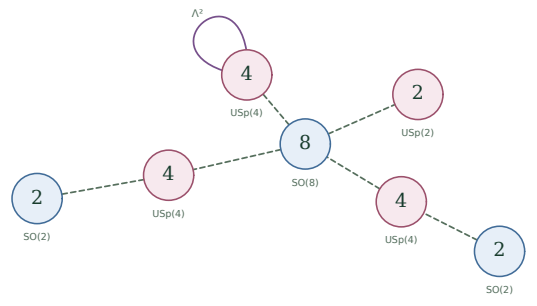

÷ $\mathbb{Z}_2$ diagonal

$1 + 16t^{2} + 216t^{4} + 2182t^{6} + 18667t^{8} + 136080t^{10} + 869924t^{12} + O(t^{13})$

**1345 2208.07270-T6-2-1-integral**

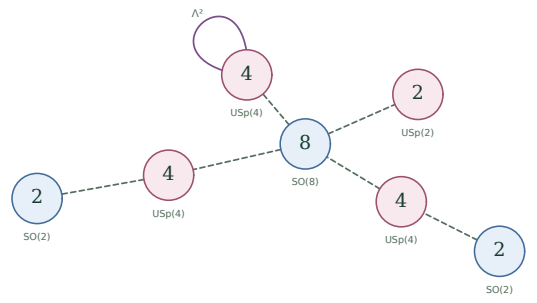

$1 + 8t^{2} + 112t^{4} + 1086t^{6} + 9371t^{8} + 67992t^{10} + 435188t^{12} + O(t^{13})$

**1346 2208.07270-T6-1pow3-quot**

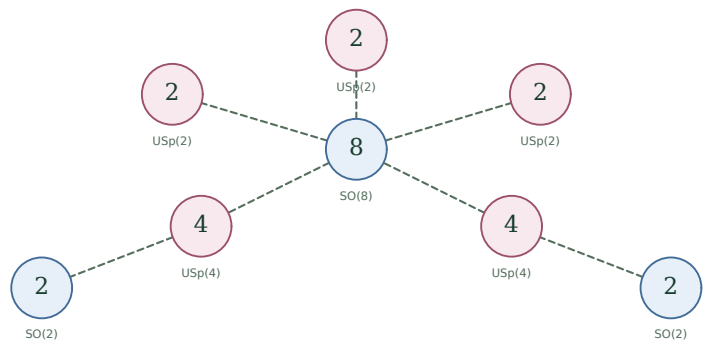

÷ $\mathbb{Z}_2$ diagonal

$1 + 16t^{2} + 232t^{4} + 2501t^{6} + 22825t^{8} + 176140t^{10} + 1183373t^{12} + 7014254t^{14} + 37246002t^{16} + 179227588t^{18} + 789798018t^{20} + 3215266755t^{22} + 12185764959t^{24} + 43281556009t^{26} + 144908209896t^{28} + 459653559364t^{30} + O(t^{31})$

**1347 2208.07270-T6-1pow3-integral**

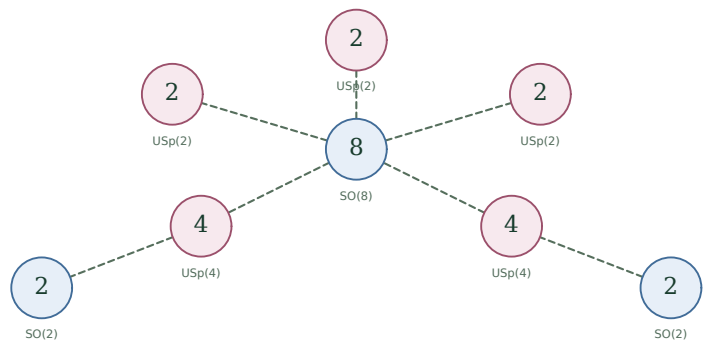

$1 + 8t^{2} + 120t^{4} + 1245t^{6} + 11449t^{8} + 88012t^{10} + 591909t^{12} + O(t^{13})$

**1348 2208.07270-T7-3-quot**

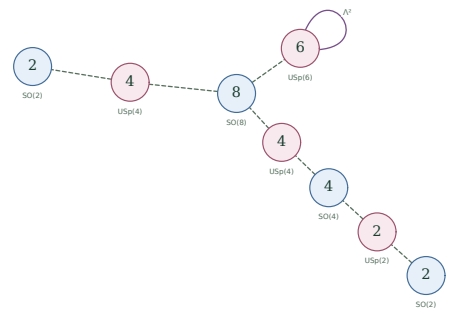

÷ $\mathbb{Z}_2$ diagonal

$1 + 14t^{2} + 16t^{3} + 135t^{4} + 272t^{5} + 1147t^{6} + O(t^{7})$

**1349 2208.07270-T7-3-integral**

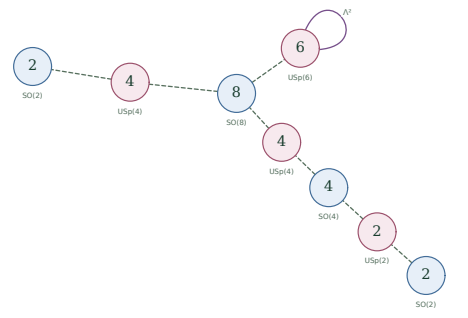

$1 + 14t^{2} + 135t^{4} + 1147t^{6} + O(t^{7})$

**1350 2208.07270-T7-2-1-quot**

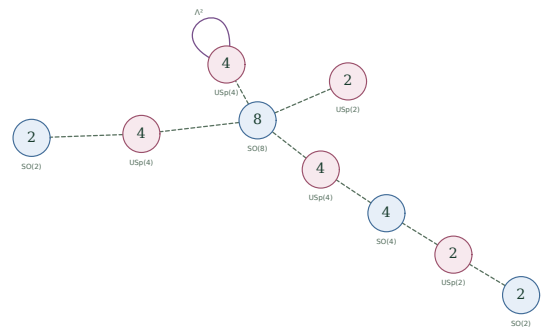

÷ $\mathbb{Z}_2$ diagonal

$1 + 14t^{2} + 16t^{3} + 139t^{4} + 288t^{5} + 1231t^{6} + O(t^{7})$

| QUIVER | HILBERT SERIES |
| --- | --- |

**1351 2208.07270-T7-2-1-integral**

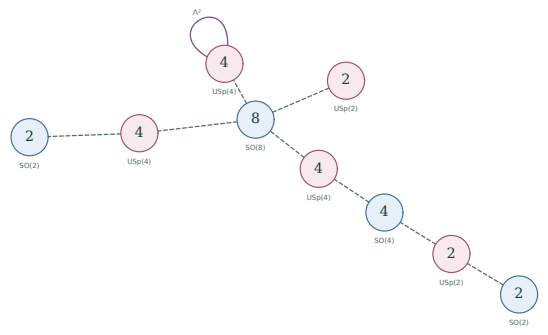

$1 + 14t^2 + 139t^4 + 1231t^6 + O(t^7)$

**1352 2208.07270-T7-1pow3-quot**

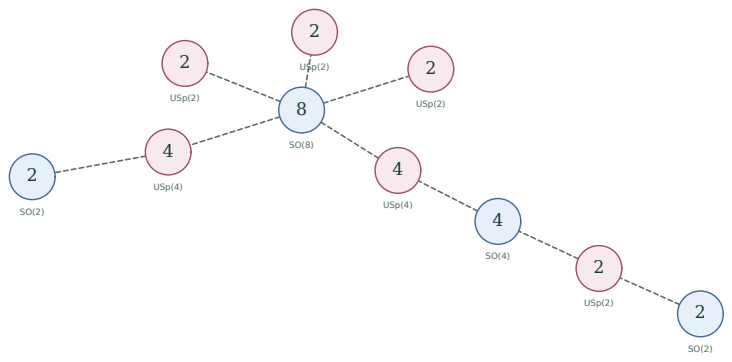

÷ $\mathbb{Z}_2$ diagonal

$1 + 14t^2 + 16t^3 + 143t^4 + 304t^5 + 1315t^6 + O(t^7)$

**1353 2208.07270-T7-1pow3-integral**

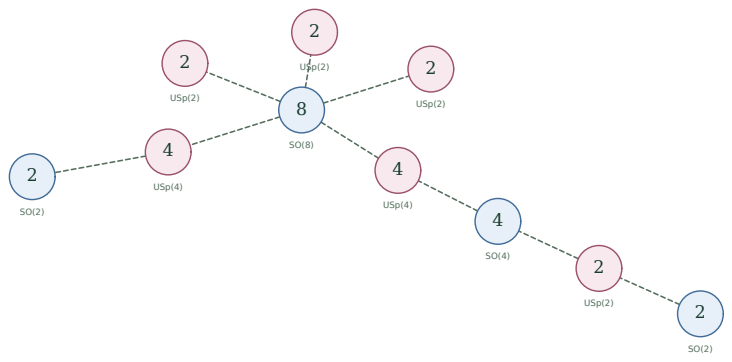

$1 + 14t^2 + 143t^4 + 1315t^6 + O(t^7)$

**1354 2208.07270-T8-3-quot**

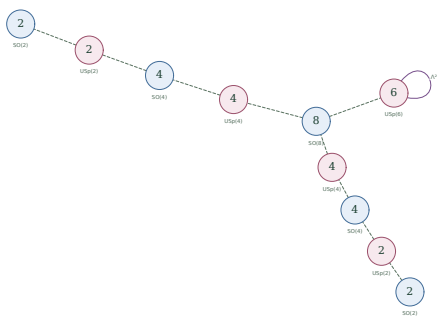

÷ $\mathbb{Z}_2$ diagonal

$1 + 21t^2 + 288t^4 + 3071t^6 + O(t^7)$

**1355 2208.07270-T8-3-integral**

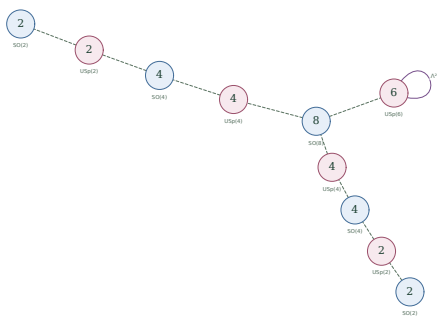

$1 + 21t^2 + 256t^4 + 2335t^6 + O(t^7)$

**1356 2208.07270-T8-2-1-quot**

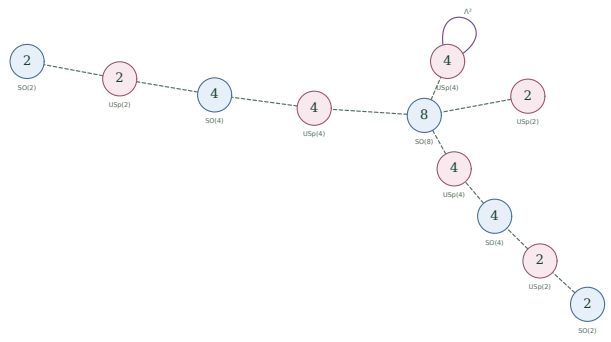

÷ $\mathbb{Z}_2$ diagonal

$1 + 21t^2 + 291t^4 + 3179t^6 + O(t^7)$

**1357 2208.07270-T8-2-1-integral**

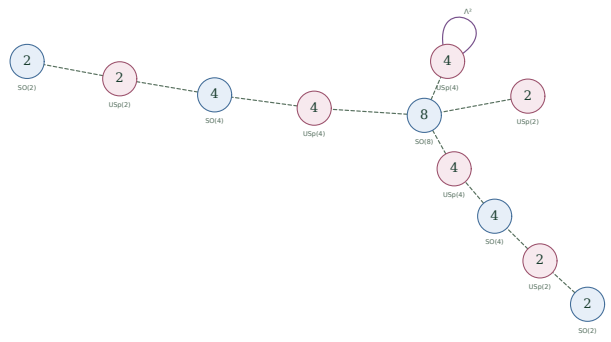

$1 + 21t^2 + 259t^4 + 2411t^6 + O(t^7)$

| QUIVER | HILBERT SERIES |
|---|---|

**1358 2208.07270-T8-1pow3-quot**

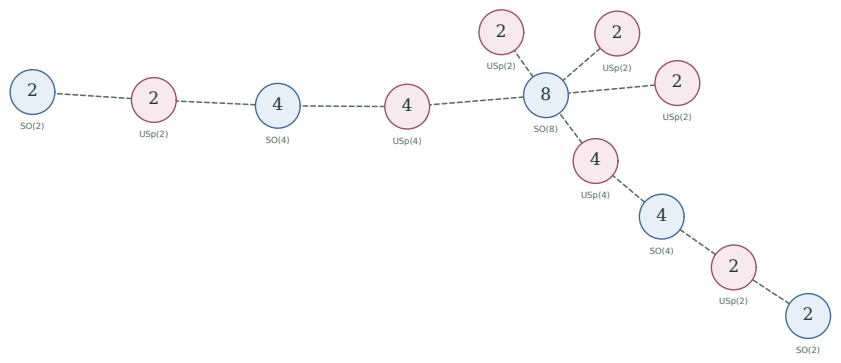

÷ ℤ₂ diagonal

$1 + 21t^2 + 294t^4 + 3287t^6 + O(t^7)$

**1359 2208.07270-T8-1pow3-integral**

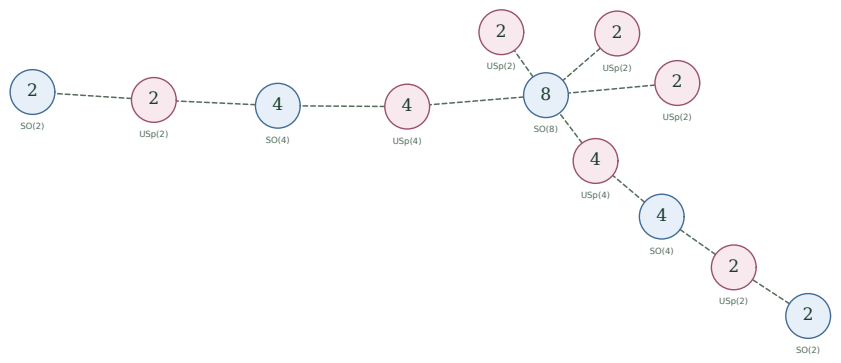

$1 + 21t^2 + 262t^4 + 2487t^6 + O(t^7)$

**1360 2208.07270-T9-2-quot**

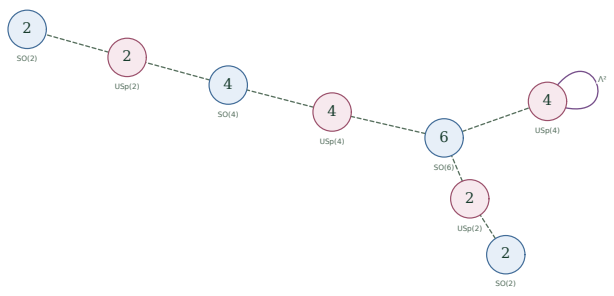

÷ ℤ₂ diagonal

$1 + 16t + 157t^2 + 1152t^3 + 6927t^4 + 35760t^5 + 163335t^6 + 673728t^7 + 2547854t^8 + O(t^9)$

**1361 2208.07270-T9-2-integral**

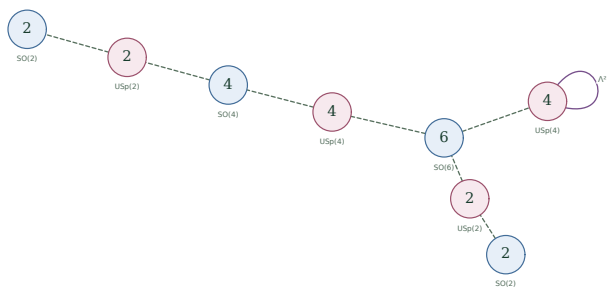

$1 + 157t^2 + 6927t^4 + 163335t^6 + 2547854t^8 + O(t^9)$

**1362 2208.07270-T9-1pow2-quot**

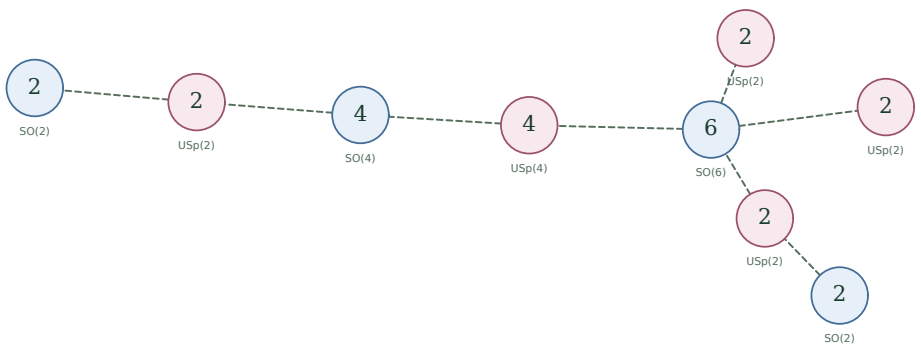


÷ ℤ₂ diagonal

$1 + 16t + 164t^2 + 1264t^3 + 7984t^4 + 43152t^5 + 205517t^6 + 880256t^7 + 3443224t^8 + O(t^9)$

**1363 2208.07270-T9-1pow2-integral**

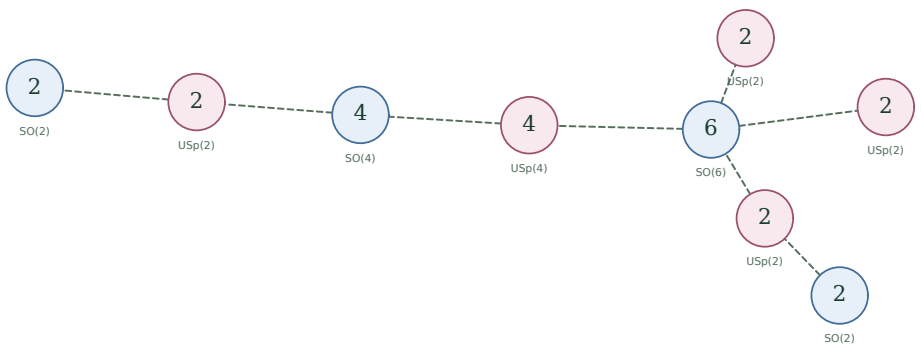


$1 + 164t^2 + 7984t^4 + 205517t^6 + 3443224t^8 + O(t^9)$

**1364 2208.07270-T10-3-quot**

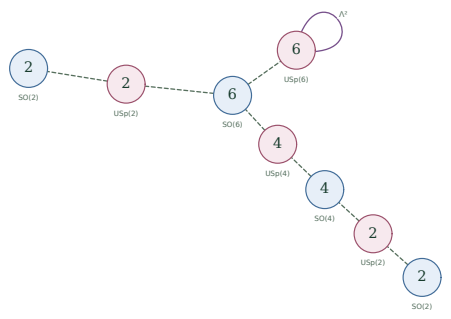

÷ ℤ₂ diagonal

$1 + 35t^2 + 660t^4 + 8743t^6 + 90244t^8 + O(t^9)$

**QUIVER** | **HILBERT SERIES**

**1365 2208.07270-T10-3-integral**

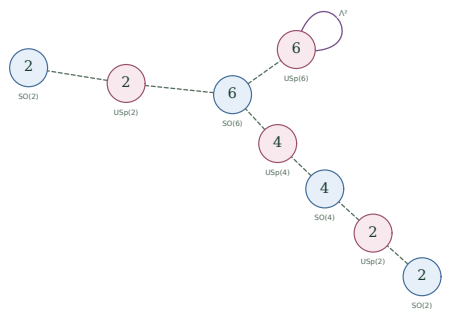

$1 + 19t^2 + 340t^4 + 4391t^6 + 45220t^8 + O(t^9)$

**1366 2208.07270-T10-1pow3-quot**

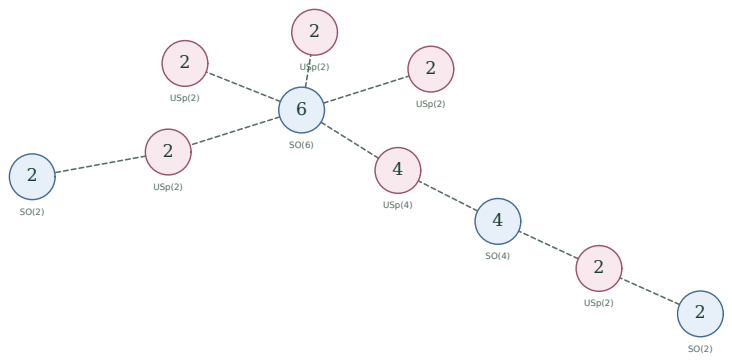

÷ $\mathbb{Z}_2$ diagonal

$1 + 37t^2 + 792t^4 + 12180t^6 + 145838t^8 + O(t^9)$

**1367 2208.07270-T10-1pow3-integral**

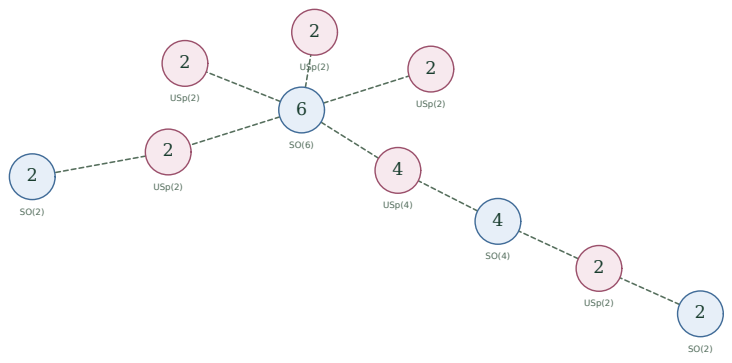

$1 + 21t^2 + 408t^4 + 6116t^6 + 73022t^8 + O(t^9)$

**1368 2208.07270-T11-3-quot**

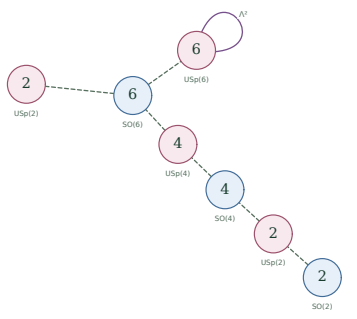

Overall diagonal $\mathbb{Z}_2$ quotient · integral and half-integral magnetic sectors

$1 + 8t + 52t^2 + 264t^3 + 1182t^4 + 4720t^5 + 17321t^6 + 58984t^7 + 188673t^8 + O(t^9)$

**1369 2208.07270-T11-3-integral**

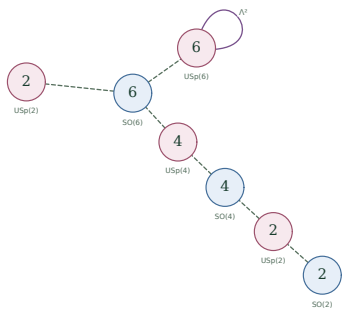

$1 + 52t^2 + 1182t^4 + 17321t^6 + 188673t^8 + O(t^9)$

**1370 2208.07270-T11-1pow3-quot**

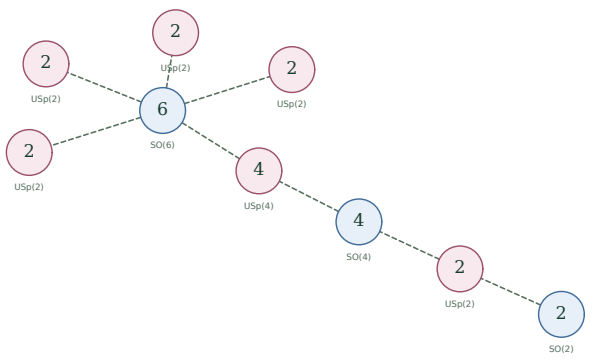

Overall diagonal $\mathbb{Z}_2$ quotient · integral and half-integral magnetic sectors

$1 + 8t + 54t^2 + 296t^3 + 1440t^4 + 6296t^5 + 25257t^6 + 93840t^7 + 325958t^8 + O(t^9)$

| QUIVER | HILBERT SERIES |
|---|---|

**1371 2208.07270-T11-1pow3-integral**

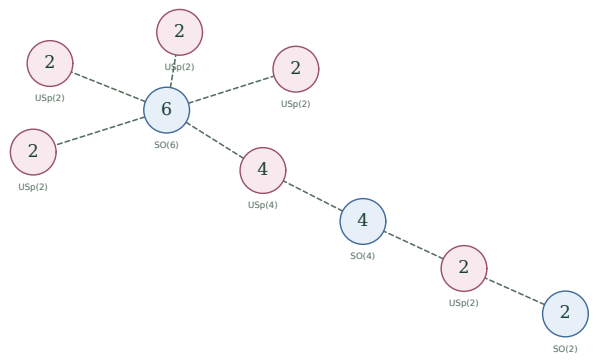

$1 + 54t^2 + 1440t^4 + 25257t^6 + 325958t^8 + O(t^9)$

**1372 2208.07270-T12-3-quot**

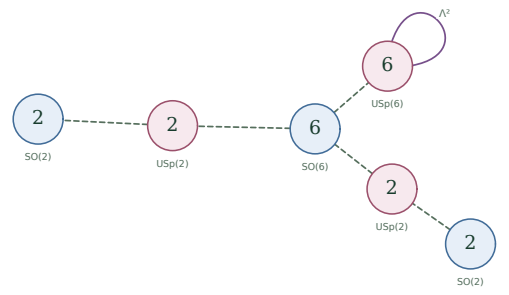

÷ $\mathbb{Z}_2$ diagonal

$1 + 8t + 45t^2 + 208t^3 + 831t^4 + 2968t^5 + 9692t^6 + 29344t^7 + 83267t^8 + 223288t^9 + 569535t^{10} + 1389120t^{11} + 3254165t^{12} + O(t^{13})$

**1373 2208.07270-T12-3-integral**

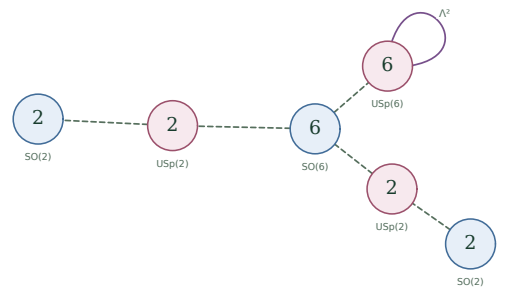

$1 + 45t^2 + 831t^4 + 9692t^6 + 83267t^8 + 569535t^{10} + 3254165t^{12} + O(t^{13})$

**1374 2208.07270-T12-1pow3-quot**

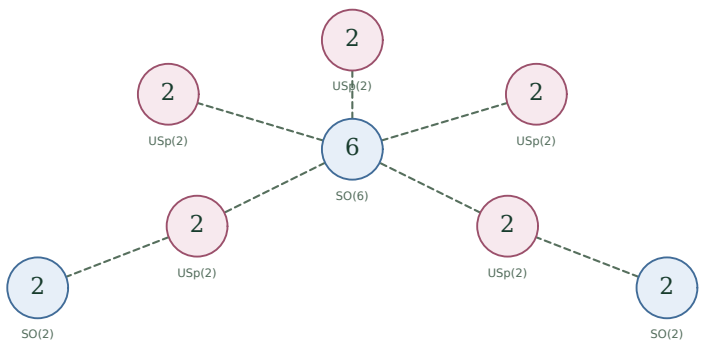


÷ $\mathbb{Z}_2$ diagonal

$1 + 8t + 51t^2 + 272t^3 + 1242t^4 + 5024t^5 + 18361t^6 + 61480t^7 + 190857t^8 + 554464t^9 + 1518870t^{10} + 3948304t^{11} + 9791797t^{12} + O(t^{13})$

**1375 2208.07270-T12-1pow3-integral**

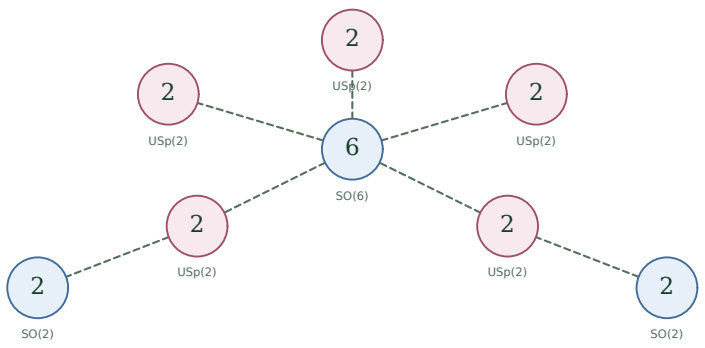


$1 + 51t^2 + 1242t^4 + 18361t^6 + 190857t^8 + 1518870t^{10} + 9791797t^{12} + O(t^{13})$

**1376 2208.07270-unitary-21**

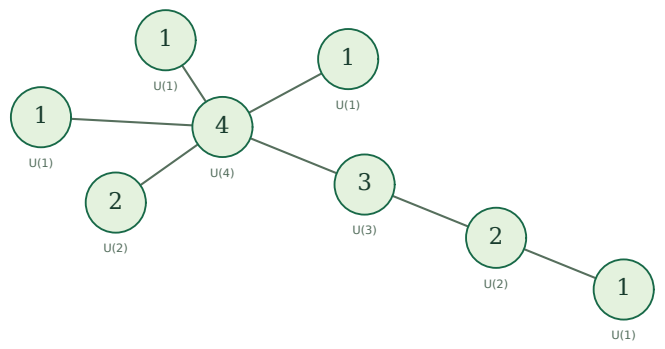


÷ U(1) diagonal

$1 + 37t^2 + 792t^4 + 12180t^6 + 145838t^8 + O(t^9)$

**1377 2208.07270-unitary-22**

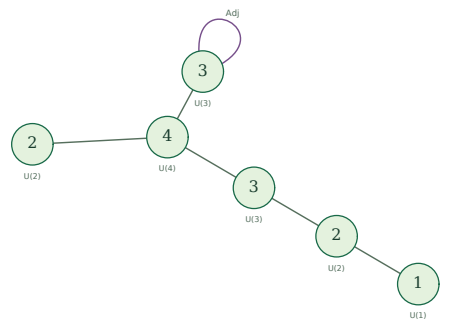

÷ U(1) diagonal

$1 + 35t^2 + 660t^4 + 8743t^6 + 90244t^8 + O(t^9)$

| QUIVER | HILBERT SERIES |
|---|---|
| **1378 2208.07270-unitary-30** 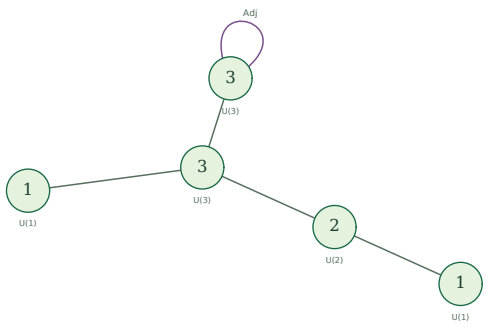 ÷ U(1) diagonal | $1 + 16t^2 + 16t^3 + 148t^4 + 248t^5 + 1093t^6 + 2168t^7 + 6818t^8 + O(t^9)$ |
| **1379 2208.07270-unitary-37** 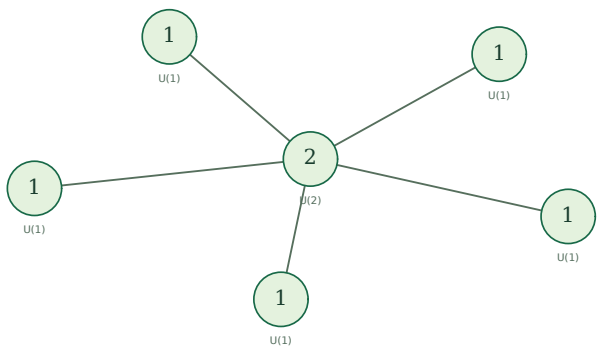 ÷ U(1) diagonal | $1 + 15t^2 + 32t^3 + 116t^4 + 352t^5 + 863t^6 + 2112t^7 + 4854t^8 + 10176t^9 + 20851t^{10} + 40736t^{11} + 76009t^{12} + O(t^{13})$ |
| **1380 2208.07270-unitary-38** 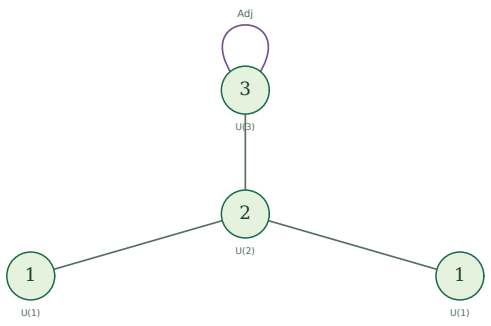 ÷ U(1) diagonal | $1 + 9t^2 + 16t^3 + 49t^4 + 128t^5 + 298t^6 + 632t^7 + 1402t^8 + 2728t^9 + 5324t^{10} + 9944t^{11} + 17946t^{12} + O(t^{13})$ |
| **1381 2208.07270-F4-curve_231-S3-quot** 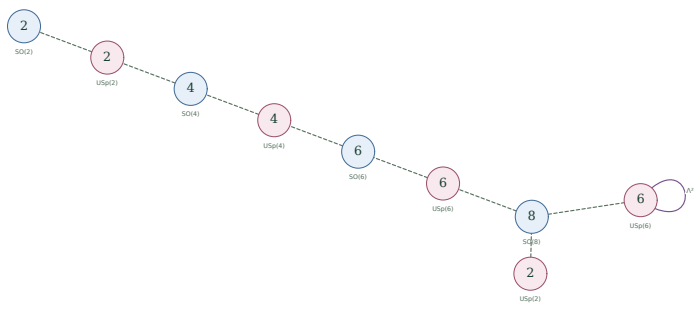 ÷ ℤ₂ diagonal | $1 + 52t^2 + 1403t^4 + 26078t^6 + O(t^7)$ |
| **1382 2208.07270-F4-curve_231-S3-integral** 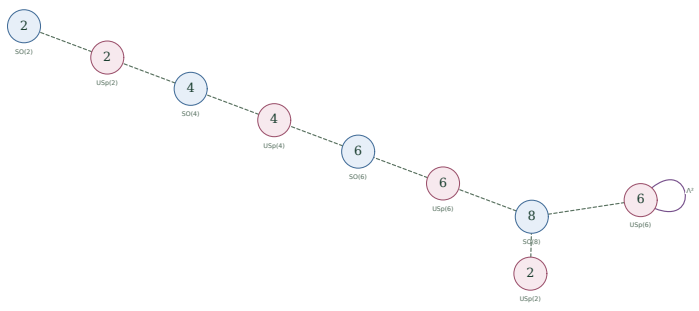 | $1 + 36t^2 + 811t^4 + 13902t^6 + O(t^7)$ |
| **1383 2208.07270-F4-curve_2231-quot** 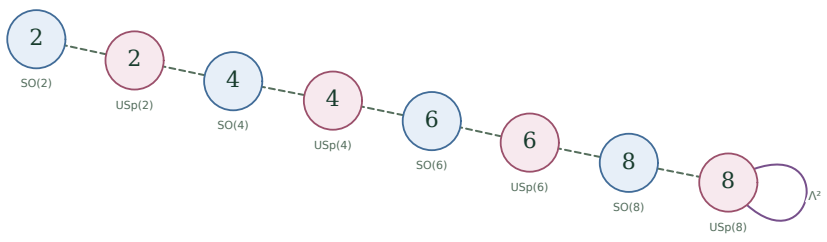  ÷ ℤ₂ diagonal | $1 + 52t^2 + 1377t^4 + 24752t^6 + 338951t^8 + O(t^9)$ **Paper typo:** Compared with its cited source, Eq. (5.17a) loses the square on $(1+t^2)^2$ and duplicates the central 10782 coefficient. The original rational expression and independent OSp sum give 339651, not 338951, at $t^8$; the PL coefficient is -26, not -726. |

| QUIVER | HILBERT SERIES |
|---|---|

**1384 2208.07270-F4-curve_2231-integral**

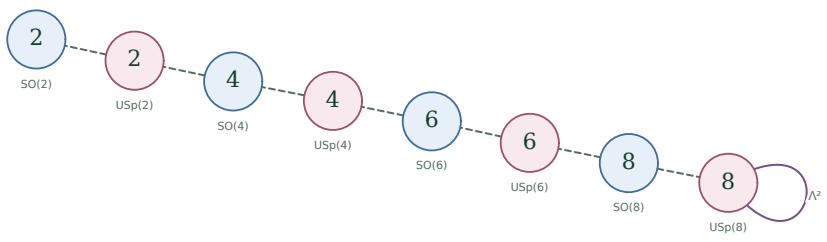

$1 + 36t^{2} + 801t^{4} + 13296t^{6} + O(t^{7})$

**1385 2208.07270-SU4-unitary-1pow2**

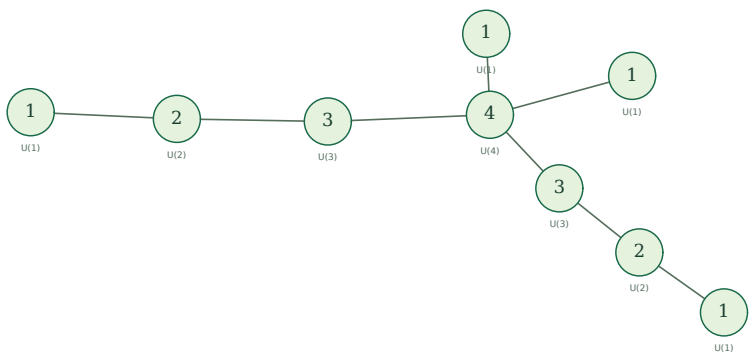

÷ U(1) diagonal

$1 + 64t^{2} + 2156t^{4} + 49035t^{6} + 832376t^{8} + 11170539t^{10} + 123240632t^{12} + 1150637073t^{14} + 9298241055t^{16} + 66216202892t^{18} + 421718414235t^{20} + 2431558547462t^{22} + 12823770808393t^{24} + 62403806228172t^{26} + 282310404939196t^{28} + 1195031191158327t^{30} + O(t^{31})$

**1386 2208.07270-SU4-unitary-2**

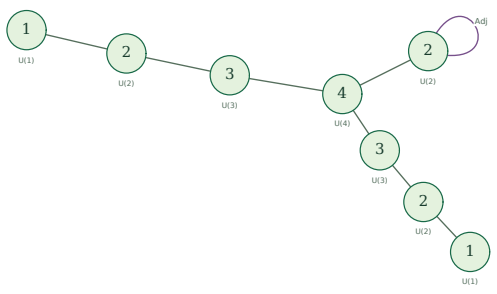

÷ U(1) diagonal

$1 + 63t^{2} + 2023t^{4} + 43428t^{6} + 696086t^{8} + 8860325t^{10} + 93249581t^{12} + 835269497t^{14} + 6509509030t^{16} + 44911956047t^{18} + 278222007609t^{20} + 1565674022207t^{22} + 8082467996707t^{24} + 38595225758861t^{26} + 171701062128429t^{28} + 716059431853538t^{30} + O(t^{31})$

**1387 2208.07270-E6-Satake**

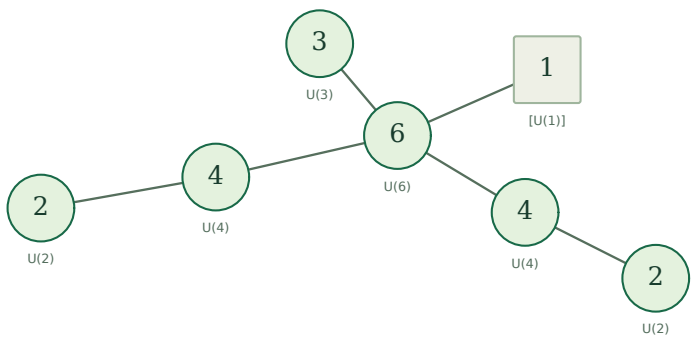


$1 + 78t^{2} + 3158t^{4} + 86787t^{6} + 1797641t^{8} + 29702895t^{10} + 406169555t^{12} + 4720049996t^{14} + 47575754291t^{16} + 422801389413t^{18} + 3357794934897t^{20} + 24101809777891t^{22} + 157869469048851t^{24} + 951454962040858t^{26} + 5314125074537538t^{28} + 27678826736346693t^{30} + O(t^{31})$

## 2209.15016 · Fibrations and Hasse diagrams for 6d SCFTs (18)

**1388 2209.15016-Yd-d4** / **d = 4**

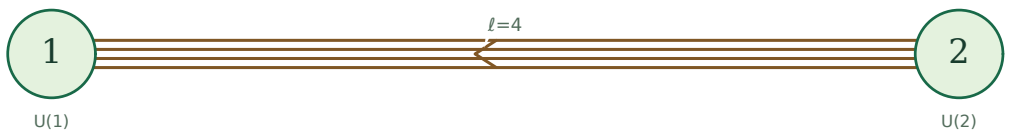


÷ U(1) diagonal · node c

$1 + 8t^{2} + 27t^{4} + 64t^{6} + 125t^{8} + 216t^{10} + 343t^{12} + 512t^{14} + 729t^{16} + 1000t^{18} + 1331t^{20} + 1728t^{22} + 2197t^{24} + 2744t^{26} + 3375t^{28} + 4096t^{30} + O(t^{31})$

**1389 2209.15016-Yd-d5** / **d = 5**

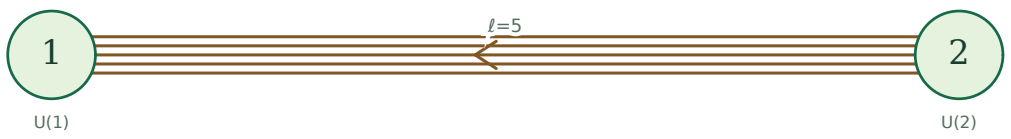


÷ U(1) diagonal · node c

$1 + 3t^{2} + 6t^{3} + 6t^{4} + 14t^{5} + 21t^{6} + 24t^{7} + 39t^{8} + 52t^{9} + 60t^{10} + 84t^{11} + 105t^{12} + 120t^{13} + 155t^{14} + 186t^{15} + 210t^{16} + 258t^{17} + 301t^{18} + 336t^{19} + 399t^{20} + 456t^{21} + 504t^{22} + 584t^{23} + 657t^{24} + 720t^{25} + 819t^{26} + 910t^{27} + 990t^{28} + 1110t^{29} + 1221t^{30} + O(t^{31})$

**QUIVER** | **HILBERT SERIES**

**1390 2209.15016-Yd-d6** / d = 6

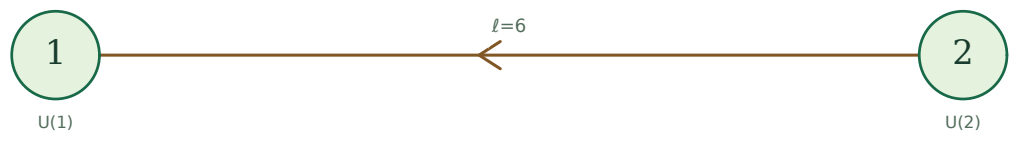


÷ U(1) diagonal · node c

$1 + 3t^{2} + 13t^{4} + 26t^{6} + 55t^{8} + 89t^{10} + 147t^{12} + 212t^{14} + 309t^{16} + 415t^{18} + 561t^{20} + 718t^{22} + 923t^{24} + 1141t^{26} + 1415t^{28} + 1704t^{30} + O(t^{31})$

**1391 2209.15016-J-family-k11-k22** / k1 = 1 · k2 = 2

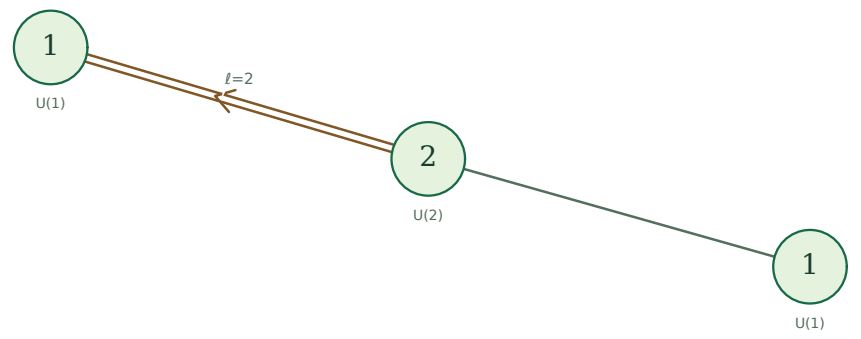


÷ U(1) diagonal · node c

$1 + 6t + 21t^{2} + 56t^{3} + 126t^{4} + 252t^{5} + 462t^{6} + 792t^{7} + 1287t^{8} + 2002t^{9} + 3003t^{10} + 4368t^{11} + 6188t^{12} + 8568t^{13} + 11628t^{14} + 15504t^{15} + 20349t^{16} + 26334t^{17} + 33649t^{18} + 42504t^{19} + 53130t^{20} + 65780t^{21} + 80730t^{22} + 98280t^{23} + 118755t^{24} + 142506t^{25} + 169911t^{26} + 201376t^{27} + 237336t^{28} + 278256t^{29} + 324632t^{30} + O(t^{31})$

**1392 2209.15016-J-family-k11-k24** / k1 = 1 · k2 = 4

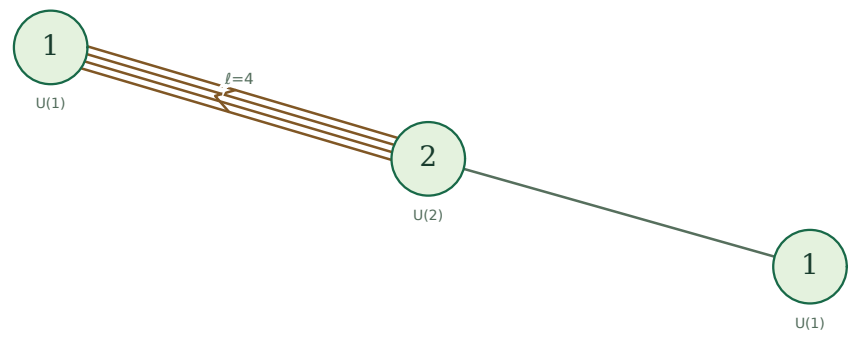


÷ U(1) diagonal · node c

$1 + 6t^{2} + 10t^{3} + 20t^{4} + 44t^{5} + 77t^{6} + 120t^{7} + 210t^{8} + 312t^{9} + 458t^{10} + 680t^{11} + 951t^{12} + 1292t^{13} + 1784t^{14} + 2350t^{15} + 3062t^{16} + 3992t^{17} + 5075t^{18} + 6368t^{19} + 8012t^{20} + 9872t^{21} + 12076t^{22} + 14752t^{23} + 17781t^{24} + 21272t^{25} + 25442t^{26} + 30082t^{27} + 35392t^{28} + 41572t^{29} + 48433t^{30} + O(t^{31})$

**1393 2209.15016-J-family-k11-k25** / k1 = 1 · k2 = 5

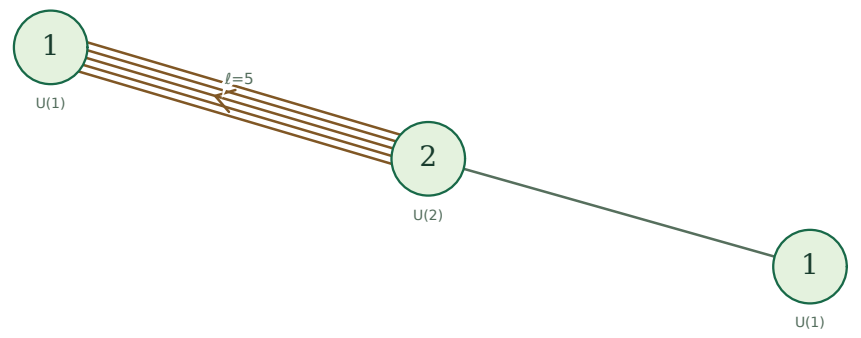


÷ U(1) diagonal · node c

$1 + 6t^{2} + 32t^{4} + 102t^{6} + 278t^{8} + 623t^{10} + 1274t^{12} + 2354t^{14} + 4105t^{16} + 6730t^{18} + 10606t^{20} + 16036t^{22} + 23562t^{24} + 33607t^{26} + 46908t^{28} + 64028t^{30} + O(t^{31})$

**1394 2209.15016-J-family-k12-k21** / k1 = 2 · k2 = 1

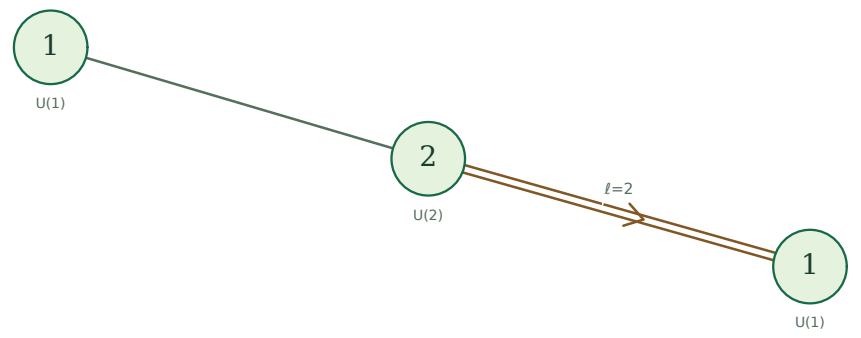


÷ U(1) diagonal · node c

$1 + 6t + 21t^{2} + 56t^{3} + 126t^{4} + 252t^{5} + 462t^{6} + 792t^{7} + 1287t^{8} + 2002t^{9} + 3003t^{10} + 4368t^{11} + 6188t^{12} + 8568t^{13} + 11628t^{14} + 15504t^{15} + 20349t^{16} + 26334t^{17} + 33649t^{18} + 42504t^{19} + 53130t^{20} + 65780t^{21} + 80730t^{22} + 98280t^{23} + 118755t^{24} + 142506t^{25} + 169911t^{26} + 201376t^{27} + 237336t^{28} + 278256t^{29} + 324632t^{30} + O(t^{31})$

**1395 2209.15016-J-family-k12-k22** / k1 = 2 · k2 = 2

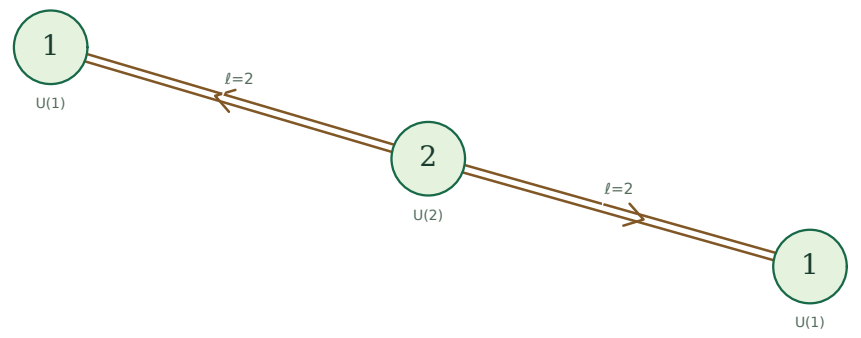


÷ U(1) diagonal · node c

$1 + 15t^{2} + 84t^{4} + 300t^{6} + 825t^{8} + 1911t^{10} + 3920t^{12} + 7344t^{14} + 12825t^{16} + 21175t^{18} + 33396t^{20} + 50700t^{22} + 74529t^{24} + 106575t^{26} + 148800t^{28} + 203456t^{30} + O(t^{31})$

| QUIVER | HILBERT SERIES |
|---|---|

**1396 2209.15016-J-family-k12-k24** / **k1 = 2 · k2 = 4**

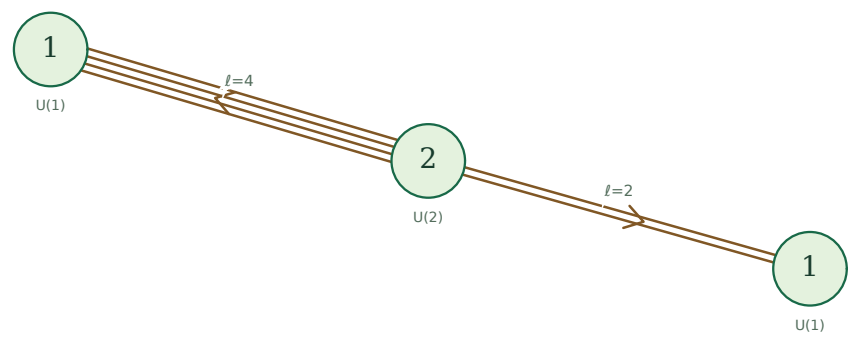


÷ U(1) diagonal · node c

$1 + 6t^{2} + 35t^{4} + 111t^{6} + 308t^{8} + 689t^{10} + 1418t^{12} + 2618t^{14} + 4579t^{16} + 7504t^{18} + 11845t^{20} + 17905t^{22} + 26334t^{24} + 37555t^{26} + 52452t^{28} + 71588t^{30} + O(t^{31})$

**1397 2209.15016-J-family-k13-k24** / **k1 = 3 · k2 = 4**

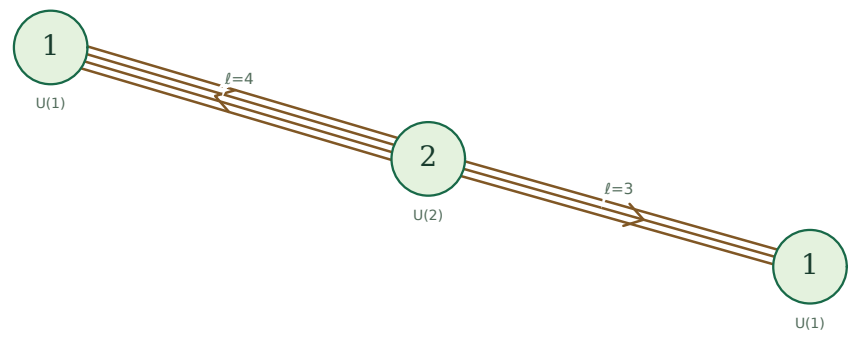


÷ U(1) diagonal · node c

$1 + 6t^{2} + 20t^{4} + 20t^{5} + 50t^{6} + 78t^{7} + 105t^{8} + 196t^{9} + 259t^{10} + 400t^{11} + 557t^{12} + 720t^{13} + 1050t^{14} + 1320t^{15} + 1795t^{16} + 2284t^{17} + 2855t^{18} + 3704t^{19} + 4520t^{20} + 5680t^{21} + 6925t^{22} + 8320t^{23} + 10215t^{24} + 12076t^{25} + 14545t^{26} + 17146t^{27} + 20080t^{28} + 23740t^{29} + 27470t^{30} + O(t^{31})$

**1398 2209.15016-J-family-k13-k25** / **k1 = 3 · k2 = 5**

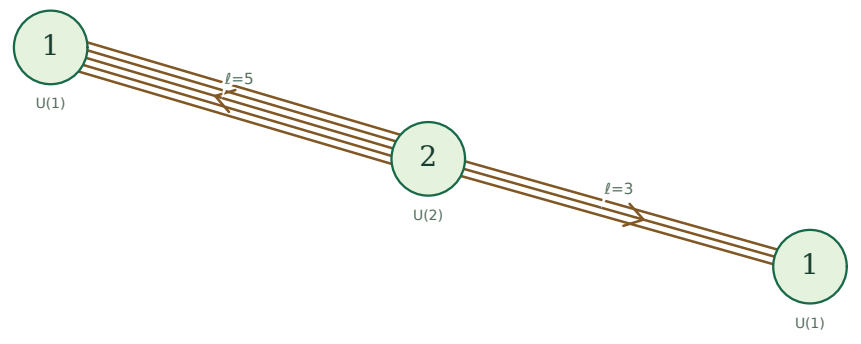


÷ U(1) diagonal · node c

$1 + 6t^{2} + 20t^{4} + 74t^{6} + 197t^{8} + 424t^{10} + 873t^{12} + 1627t^{14} + 2779t^{16} + 4592t^{18} + 7231t^{20} + 10875t^{22} + 15990t^{24} + 22851t^{26} + 31751t^{28} + 43417t^{30} + O(t^{31})$

**1399 2209.15016-J-family-k13-k26** / **k1 = 3 · k2 = 6**

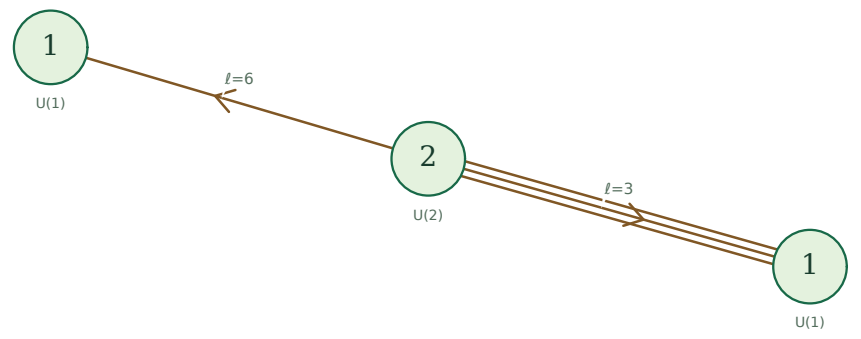


÷ U(1) diagonal · node c

$1 + 6t^{2} + 20t^{4} + 50t^{6} + 28t^{7} + 105t^{8} + 106t^{9} + 196t^{10} + 260t^{11} + 336t^{12} + 520t^{13} + 631t^{14} + 920t^{15} + 1138t^{16} + 1498t^{17} + 1920t^{18} + 2296t^{19} + 3046t^{20} + 3550t^{21} + 4591t^{22} + 5368t^{23} + 6636t^{24} + 7866t^{25} + 9268t^{26} + 11168t^{27} + 12905t^{28} + 15406t^{29} + 17722t^{30} + O(t^{31})$

**1400 2209.15016-J33**

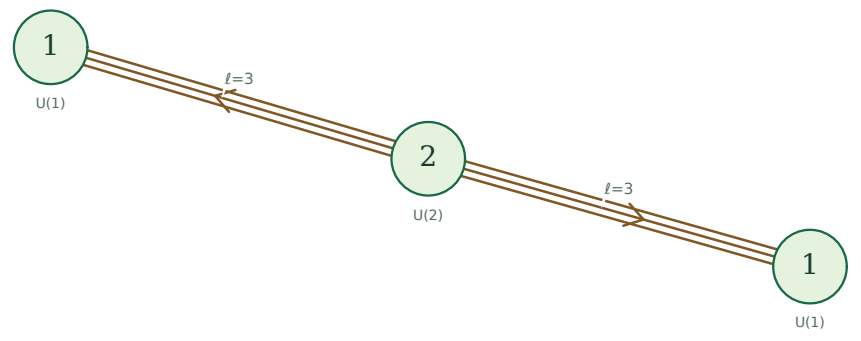


÷ U(1) diagonal · node c

$1 + 6t^{2} + 36t^{4} + 114t^{6} + 318t^{8} + 711t^{10} + 1466t^{12} + 2706t^{14} + 4737t^{16} + 7762t^{18} + 12258t^{20} + 18528t^{22} + 27258t^{24} + 38871t^{26} + 54300t^{28} + 74108t^{30} + O(t^{31})$

**1401 2209.15016-J33-Z3**

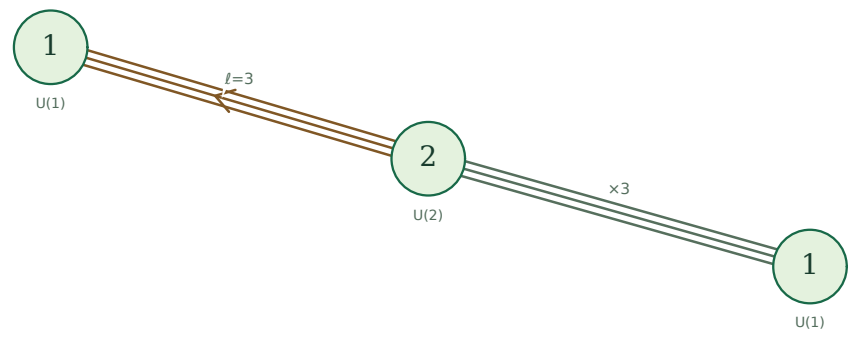


÷ U(1) diagonal · node c

$1 + 4t^{2} + 18t^{4} + 50t^{6} + 128t^{8} + 273t^{10} + 544t^{12} + 982t^{14} + 1691t^{16} + 2738t^{18} + 4284t^{20} + 6430t^{22} + 9406t^{24} + 13353t^{26} + 18584t^{28} + 25286t^{30} + O(t^{31})$

| QUIVER | HILBERT SERIES |
|---|---|
| **1402 2209.15016-J33-Z3xZ3** 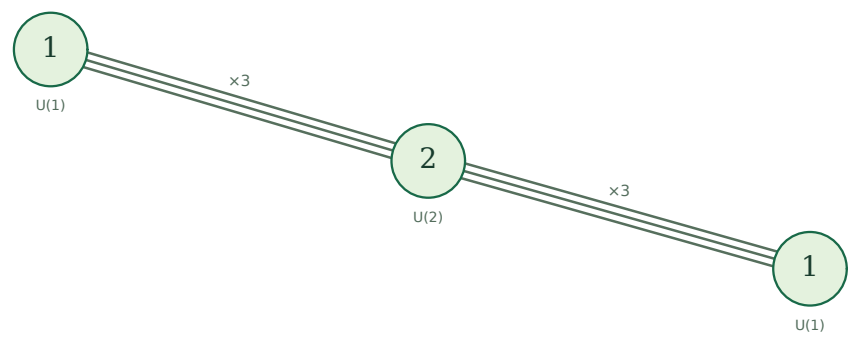  ÷ U(1) diagonal · node c | $1 + 2t^{2} + 8t^{4} + 22t^{6} + 50t^{8} + 103t^{10} + 202t^{12} + 354t^{14} + 601t^{16} + 966t^{18} + 1494t^{20} + 2228t^{22} + 3246t^{24} + 4583t^{26} + 6356t^{28} + 8628t^{30} + O(t^{31})$ |
| **1403 2209.15016-J32** 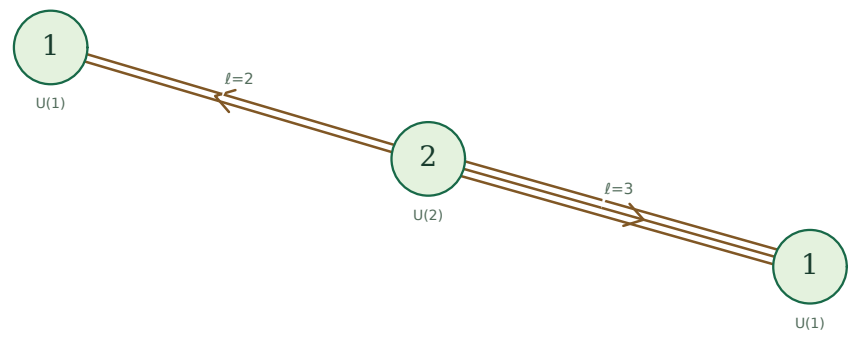  ÷ U(1) diagonal · node c | $1 + 6t^{2} + 12t^{3} + 20t^{4} + 50t^{5} + 85t^{6} + 132t^{7} + 234t^{8} + 350t^{9} + 506t^{10} + 764t^{11} + 1063t^{12} + 1442t^{13} + 2000t^{14} + 2636t^{15} + 3422t^{16} + 4484t^{17} + 5691t^{18} + 7136t^{19} + 8996t^{20} + 11084t^{21} + 13540t^{22} + 16576t^{23} + 19965t^{24} + 23876t^{25} + 28586t^{26} + 33796t^{27} + 39736t^{28} + 46726t^{29} + 54417t^{30} + O(t^{31})$ |
| **1404 2209.15016-J32-Z2** 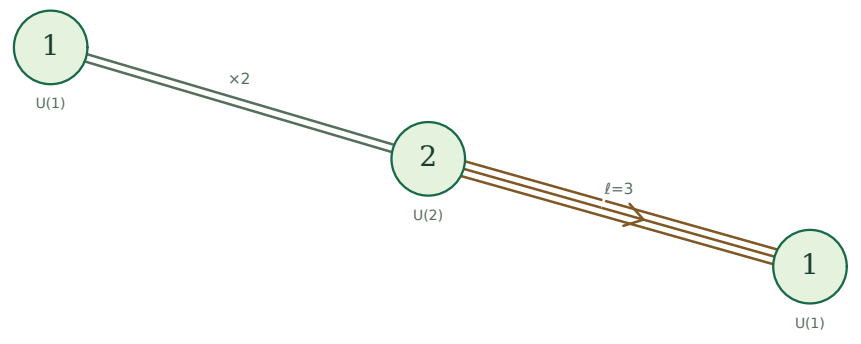  ÷ U(1) diagonal · node c | $1 + 4t^{2} + 8t^{3} + 12t^{4} + 28t^{5} + 49t^{6} + 72t^{7} + 126t^{8} + 188t^{9} + 268t^{10} + 400t^{11} + 557t^{12} + 748t^{13} + 1034t^{14} + 1360t^{15} + 1758t^{16} + 2296t^{17} + 2913t^{18} + 3640t^{19} + 4582t^{20} + 5640t^{21} + 6878t^{22} + 8408t^{23} + 10123t^{24} + 12088t^{25} + 14462t^{26} + 17088t^{27} + 20074t^{28} + 23588t^{29} + 27463t^{30} + O(t^{31})$ |
| **1405 2209.15016-J32-Z6** 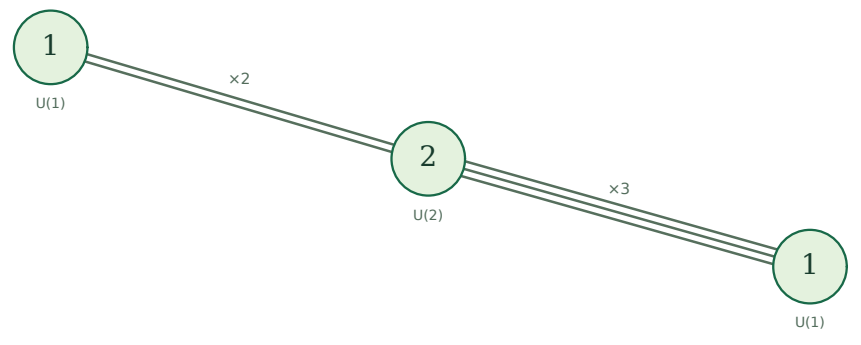  ÷ U(1) diagonal · node c | $1 + 2t^{2} + 4t^{3} + 6t^{4} + 12t^{5} + 21t^{6} + 28t^{7} + 50t^{8} + 72t^{9} + 100t^{10} + 148t^{11} + 205t^{12} + 268t^{13} + 372t^{14} + 484t^{15} + 620t^{16} + 808t^{17} + 1021t^{18} + 1264t^{19} + 1594t^{20} + 1952t^{21} + 2370t^{22} + 2896t^{23} + 3479t^{24} + 4136t^{25} + 4952t^{26} + 5836t^{27} + 6840t^{28} + 8036t^{29} + 9343t^{30} + O(t^{31})$ |

## 2210.02966 · Poisson brackets for some Coulomb branches (4)

| | |
|---|---|
| **1406 2210.02966-finite-D5** 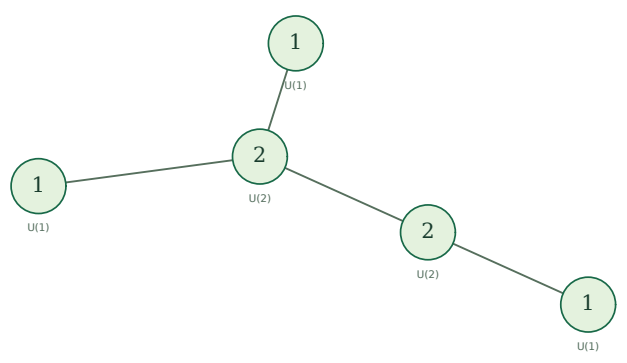 ÷ U(1) diagonal | $1 + 12t + 78t^{2} + 364t^{3} + 1365t^{4} + 4368t^{5} + 12376t^{6} + 31824t^{7} + 75582t^{8} + 167960t^{9} + 352716t^{10} + 705432t^{11} + 1352078t^{12} + 2496144t^{13} + 4457400t^{14} + 7726160t^{15} + 13037895t^{16} + 21474180t^{17} + 34597290t^{18} + 54627300t^{19} + 84672315t^{20} + 129024480t^{21} + 193536720t^{22} + 286097760t^{23} + 417225900t^{24} + 600805296t^{25} + 854992152t^{26} + 1203322288t^{27} + 1676056044t^{28} + 2311801440t^{29} + 3159461968t^{30} + O(t^{31})$ |
| **1407 2210.02966-pyramid-Nf4-n2-sigma0** / **Nf = 4 · n = 2 · sigma = 0** 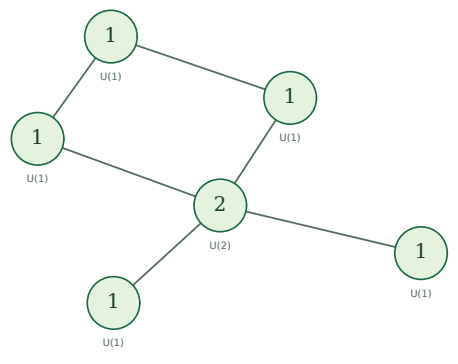 ÷ U(1) diagonal | $1 + 19t^{2} + 24t^{3} + 173t^{4} + 328t^{5} + 1162t^{6} + 2312t^{7} + 6130t^{8} + 11752t^{9} + 26192t^{10} + 47984t^{11} + 94974t^{12} + 165808t^{13} + 301944t^{14} + 503712t^{15} + 860607t^{16} + 1380096t^{17} + 2240731t^{18} + 3470304t^{19} + 5406767t^{20} + 8120016t^{21} + 12222676t^{22} + 17869536t^{23} + 26120192t^{24} + 37291056t^{25} + 53153166t^{26} + 74297496t^{27} + 103600812t^{28} + 142113704t^{29} + 194365938t^{30} + O(t^{31})$ |

| QUIVER | HILBERT SERIES |
|---|---|
| **1408 2210.02966-pyramid-Nf4-n2-sigma1** / **Nf = 4 · n = 2 · sigma = 1** 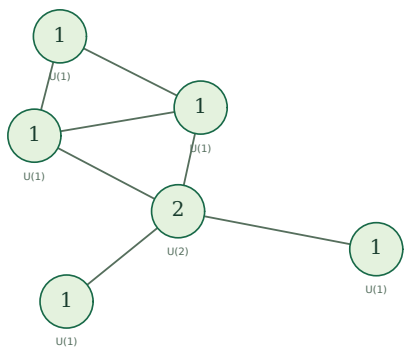 ÷ U(1) diagonal | $1 + 19t^{2} + 197t^{4} + 1370t^{6} + 7192t^{8} + 30510t^{10} + 109740t^{12} + 345914t^{14} + 979243t^{16} + 2535241t^{18} + 6087783t^{20} + 13707700t^{22} + 29197436t^{24} + 59247932t^{26} + 115207436t^{28} + 215706804t^{30} + O(t^{31})$ |
| **1409 2210.02966-pyramid-Nf4-n2-sigma2** / **Nf = 4 · n = 2 · sigma = 2** 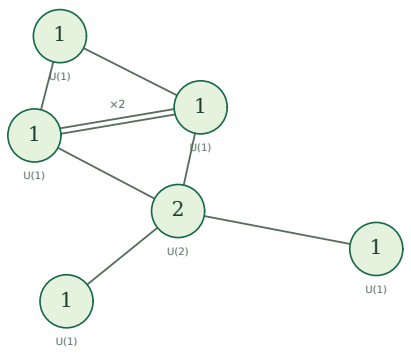 ÷ U(1) diagonal | $1 + 19t^{2} + 173t^{4} + 24t^{5} + 1042t^{6} + 328t^{7} + 4760t^{8} + 2312t^{9} + 17908t^{10} + 11352t^{11} + 58394t^{12} + 43912t^{13} + 170346t^{14} + 143400t^{15} + 453745t^{16} + 412480t^{17} + 1119263t^{18} + 1074144t^{19} + 2584763t^{20} + 2580008t^{21} + 5637702t^{22} + 5792248t^{23} + 11697782t^{24} + 12277912t^{25} + 23226840t^{26} + 24768584t^{27} + 44348442t^{28} + 47856336t^{29} + 81762932t^{30} + O(t^{31})$ |

## 2210.04921 · Exploring Seiberg-like Dualities with Eight Supercharges (5)

| | |
|---|---|
| **1410 2210.04921-U2-F6-det** 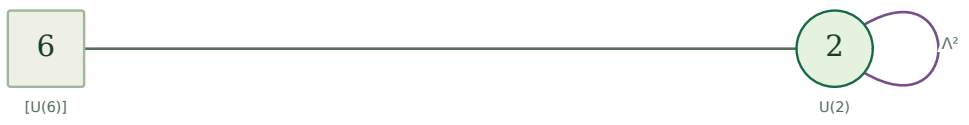 | $1 + t^{2} + 2t^{4} + 2t^{5} + 2t^{6} + 4t^{7} + 4t^{8} + O(t^{9})$ |
| **1411 2210.04921-U2-F5-det** 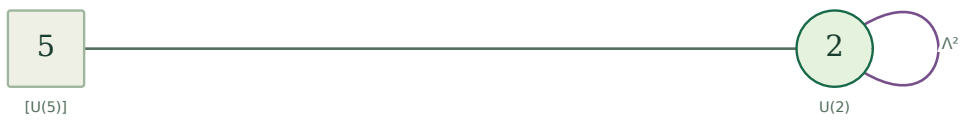 | $1 + t^{2} + 4t^{4} + 7t^{6} + 13t^{8} + 20t^{10} + 33t^{12} + 45t^{14} + 66t^{16} + 87t^{18} + 117t^{20} + O(t^{21})$ |
| **1412 2210.04921-U2-F4-det** 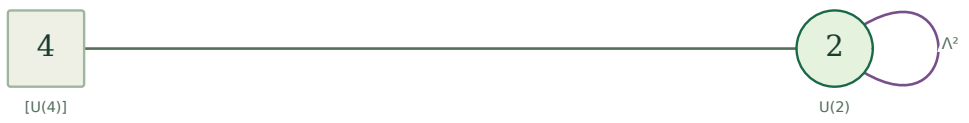 | $1 + t^{2} + 2t^{3} + 3t^{4} + 4t^{5} + 6t^{6} + 8t^{7} + O(t^{8})$ |
| **1413 2210.04921-U2-F3-det** 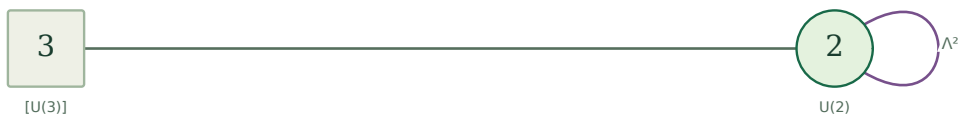 | $1 + 4t^{2} + 13t^{4} + 28t^{6} + 55t^{8} + 92t^{10} + 147t^{12} + O(t^{13})$ |

| QUIVER | HILBERT SERIES |
|---|---|

**1414 2210.04921-U3-F7-det**

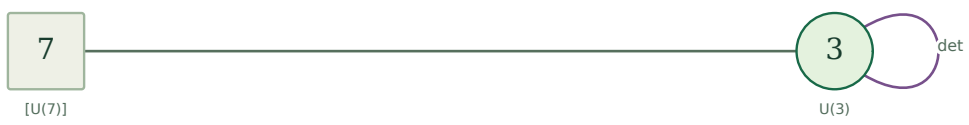


$1 + t^{2} + 4t^{4} + 8t^{6} + 17t^{8} + 29t^{10} + 54t^{12} + 86t^{14} + 141t^{16} + 213t^{18} + 322t^{20} + O(t^{21})$

## 2301.02409 · 3d N=4 mirror symmetry with 1-form symmetry (58)

**1415 2301.02409-SQED-charge-q::N1-q3** / N = 1 · q = 3

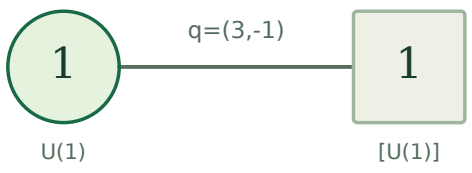


$1 + t^{2} + 2t^{3} + t^{4} + 2t^{5} + 3t^{6} + 2t^{7} + 3t^{8} + 4t^{9} + 3t^{10} + 4t^{11} + 5t^{12} + 4t^{13} + 5t^{14} + 6t^{15} + 5t^{16} + 6t^{17} + 7t^{18} + 6t^{19} + 7t^{20} + 8t^{21} + 7t^{22} + 8t^{23} + 9t^{24} + 8t^{25} + 9t^{26} + 10t^{27} + 9t^{28} + 10t^{29} + 11t^{30} + O(t^{31})$

**1416 2301.02409-SQED-charge-q::N1-q4** / N = 1 · q = 4

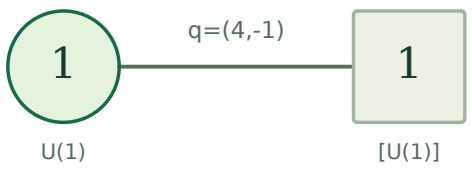


$1 + t^{2} + 3t^{4} + 3t^{6} + 5t^{8} + 5t^{10} + 7t^{12} + 7t^{14} + 9t^{16} + 9t^{18} + 11t^{20} + 11t^{22} + 13t^{24} + 13t^{26} + 15t^{28} + 15t^{30} + O(t^{31})$

**1417 2301.02409-SQED-charge-q::N1-q5** / N = 1 · q = 5

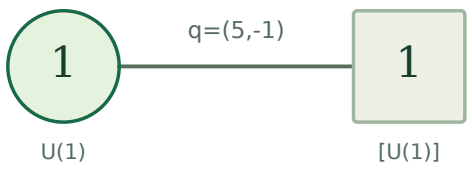


$1 + t^{2} + t^{4} + 2t^{5} + t^{6} + 2t^{7} + t^{8} + 2t^{9} + 3t^{10} + 2t^{11} + 3t^{12} + 2t^{13} + 3t^{14} + 4t^{15} + 3t^{16} + 4t^{17} + 3t^{18} + 4t^{19} + 5t^{20} + 4t^{21} + 5t^{22} + 4t^{23} + 5t^{24} + 6t^{25} + 5t^{26} + 6t^{27} + 5t^{28} + 6t^{29} + 7t^{30} + O(t^{31})$

**1418 2301.02409-SQED-charge-q::N2-q3** / N = 2 · q = 3

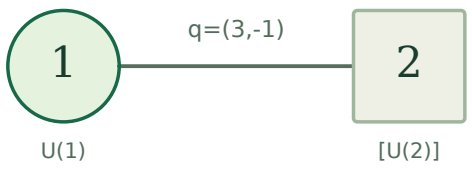


$1 + t^{2} + t^{4} + 3t^{6} + 3t^{8} + 3t^{10} + 5t^{12} + 5t^{14} + 5t^{16} + 7t^{18} + 7t^{20} + 7t^{22} + 9t^{24} + 9t^{26} + 9t^{28} + 11t^{30} + O(t^{31})$

**1419 2301.02409-SQED-charge-q::N2-q4** / N = 2 · q = 4

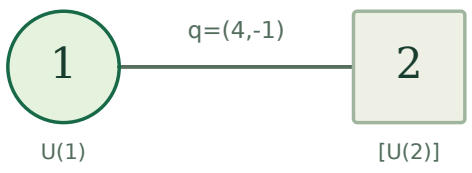


$1 + t^{2} + t^{4} + t^{6} + 3t^{8} + 3t^{10} + 3t^{12} + 3t^{14} + 5t^{16} + 5t^{18} + 5t^{20} + 5t^{22} + 7t^{24} + 7t^{26} + 7t^{28} + 7t^{30} + O(t^{31})$

| QUIVER | HILBERT SERIES |
|---|---|
| | |

**1420 2301.02409-SQED-charge-q::N2-q5** / **N = 2 · q = 5**

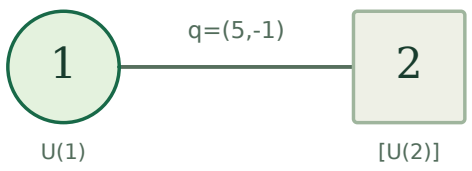


$1 + t^{2} + t^{4} + t^{6} + t^{8} + 3t^{10} + 3t^{12} + 3t^{14} + 3t^{16} + 3t^{18} + 5t^{20} + 5t^{22} + 5t^{24} + 5t^{26} + 5t^{28} + 7t^{30} + O(t^{31})$

**1421 2301.02409-SQED-charge-q::N3-q3** / **N = 3 · q = 3**

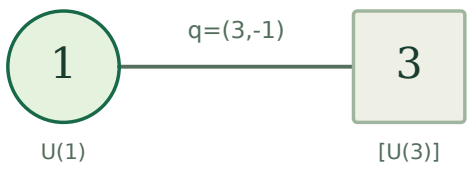


$1 + t^{2} + t^{4} + t^{6} + t^{8} + 2t^{9} + t^{10} + 2t^{11} + t^{12} + 2t^{13} + t^{14} + 2t^{15} + t^{16} + 2t^{17} + 3t^{18} + 2t^{19} + 3t^{20} + 2t^{21} + 3t^{22} + 2t^{23} + 3t^{24} + 2t^{25} + 3t^{26} + 4t^{27} + 3t^{28} + 4t^{29} + 3t^{30} + O(t^{31})$

**1422 2301.02409-SQED-charge-q::N3-q4** / **N = 3 · q = 4**

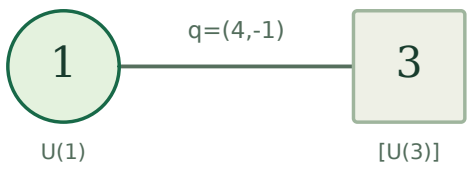


$1 + t^{2} + t^{4} + t^{6} + t^{8} + t^{10} + 3t^{12} + 3t^{14} + 3t^{16} + 3t^{18} + 3t^{20} + 3t^{22} + 5t^{24} + 5t^{26} + 5t^{28} + 5t^{30} + O(t^{31})$

**1423 2301.02409-SQED-charge-q::N3-q5** / **N = 3 · q = 5**

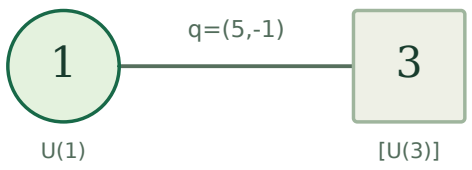


$1 + t^{2} + t^{4} + t^{6} + t^{8} + t^{10} + t^{12} + t^{14} + 2t^{15} + t^{16} + 2t^{17} + t^{18} + 2t^{19} + t^{20} + 2t^{21} + t^{22} + 2t^{23} + t^{24} + 2t^{25} + t^{26} + 2t^{27} + t^{28} + 2t^{29} + 3t^{30} + O(t^{31})$

**1424 2301.02409-abelian-explicit-character-checks::N3-k1-q2** / **N = 3 · k = 1 · q = 2**

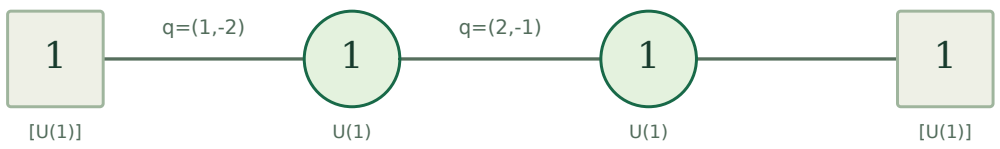


**Computed by Quiver Lab**

$1 + 4t^{2} + 15t^{4} + 32t^{6} + 65t^{8} + 108t^{10} + 175t^{12} + 256t^{14} + 369t^{16} + 500t^{18} + 671t^{20} + 864t^{22} + 1105t^{24} + 1372t^{26} + 1695t^{28} + 2048t^{30} + O(t^{31})$

Source originally Claimed · no numerical literature comparison is available.

**1425 2301.02409-abelian-explicit-character-checks::N3-k1-q3** / **N = 3 · k = 1 · q = 3**

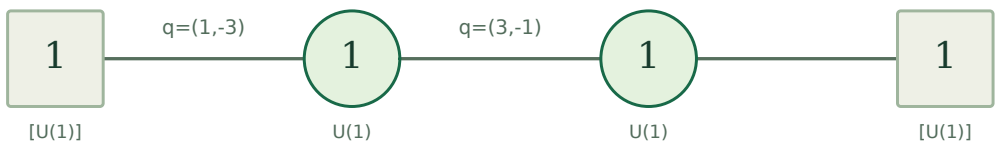


**Computed by Quiver Lab**

$1 + 4t^{2} + 9t^{4} + 24t^{6} + 45t^{8} + 72t^{10} + 119t^{12} + 176t^{14} + 243t^{16} + 340t^{18} + 451t^{20} + 576t^{22} + 741t^{24} + 924t^{26} + 1125t^{28} + 1376t^{30} + O(t^{31})$

Source originally Claimed · no numerical literature comparison is available.

**QUIVER** | **HILBERT SERIES**

**1426 2301.02409-abelian-explicit-character-checks::N3-k1-q4** / **N = 3 · k = 1 · q = 4**

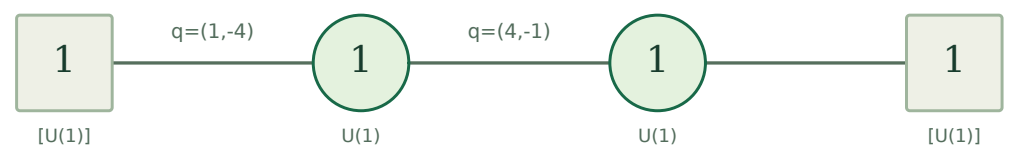


**Computed by Quiver Lab**

$1 + 4t^{2} + 9t^{4} + 16t^{6} + 35t^{8} + 60t^{10} + 91t^{12} + 128t^{14} + 189t^{16} + 260t^{18} + 341t^{20} + 432t^{22} + 559t^{24} + 700t^{26} + 855t^{28} + 1024t^{30} + O(t^{31})$

Source originally Claimed · no numerical literature comparison is available.

**1427 2301.02409-abelian-N5-k2-q2** / **N = 5 · k = 2 · q = 2**

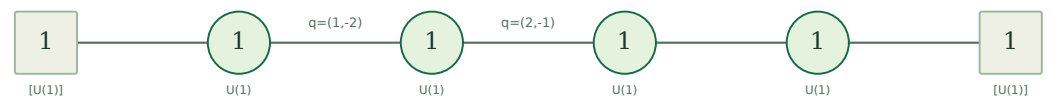


$1 + 12t^{2} + 104t^{4} + 500t^{6} + 1851t^{8} + 5488t^{10} + 14144t^{12} + 32400t^{14} + 68125t^{16} + 133100t^{18} + 245496t^{20} + 430612t^{22} + 724759t^{24} + 1176000t^{26} + 1849856t^{28} + 2829888t^{30} + O(t^{31})$

**1428 2301.02409-TSU3-SU2** / **N = 3 · k = 2**

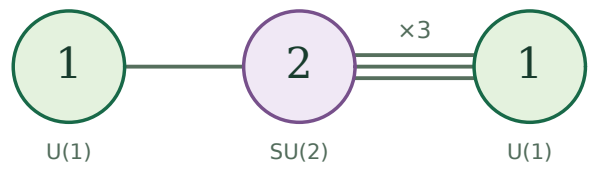


$1 + 4t^{2} + 19t^{4} + 55t^{6} + O(t^{7})$

**1429 2301.02409-TSU4-SU3** / **N = 4 · k = 3**

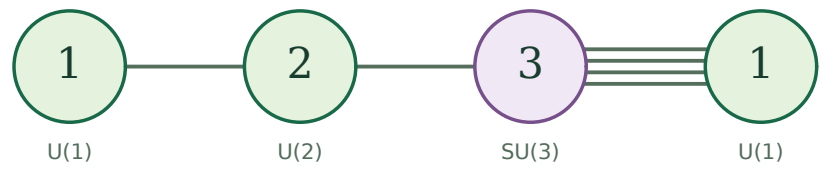


$1 + 9t^{2} + 53t^{4} + 256t^{6} + 1052t^{8} + O(t^{9})$

**1430 2301.02409-TSU4-SU2** / **N = 4 · k = 2**

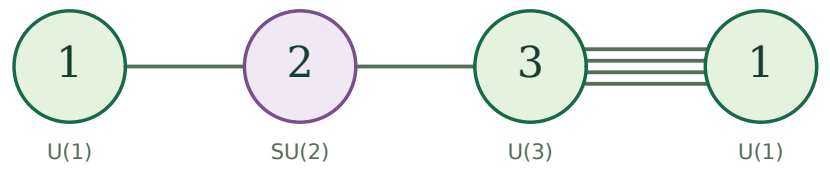


$1 + 7t^{2} + 63t^{4} + 328t^{6} + O(t^{7})$

**1431 2301.02409-TSU5-SU4** / **N = 5 · k = 4**

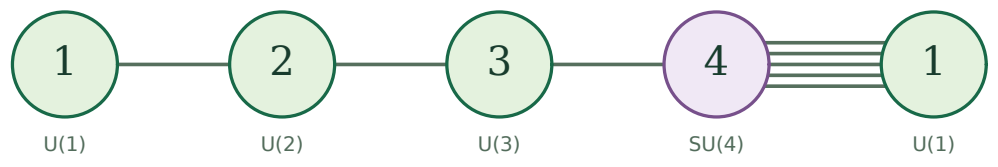


$1 + 16t^{2} + 151t^{4} + 1055t^{6} + 6053t^{8} + O(t^{9})$

**1432 2301.02409-TSU5-SU3** / **N = 5 · k = 3**

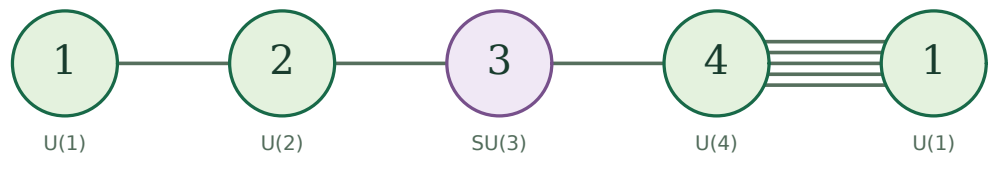


$1 + 12t^{2} + 113t^{4} + 895t^{6} + O(t^{7})$

| QUIVER | HILBERT SERIES |
|---|---|

**1433 2301.02409-TSU5-SU2** / N = 5 · k = 2

$1 + 12t^2 + 155t^4 + 1287t^6 + O(t^7)$

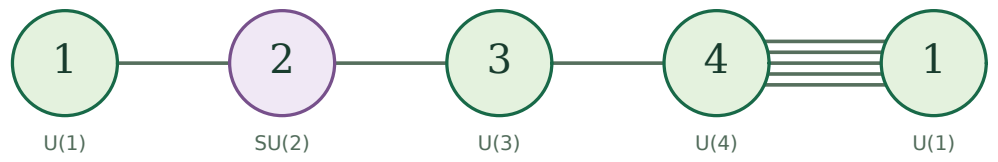


**1434 2301.02409-TSU6-SU5** / N = 6 · k = 5

$1 + 25t^2 + 349t^4 + 3524t^6 + 28449t^8 + 193832t^{10} + O(t^{11})$

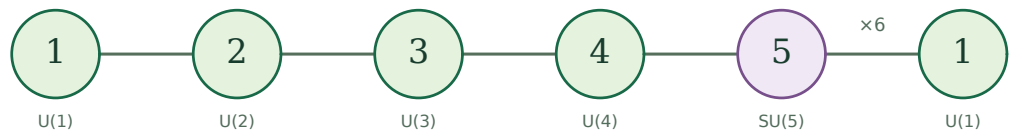


**1435 2301.02409-TSU6-SU4** / N = 6 · k = 4

$1 + 19t^2 + 253t^4 + 2526t^6 + 21157t^8 + O(t^9)$

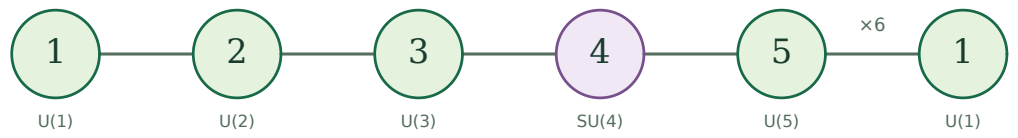


**1436 2301.02409-TSU6-SU3** / N = 6 · k = 3

$1 + 17t^2 + 233t^4 + 2658t^6 + O(t^7)$

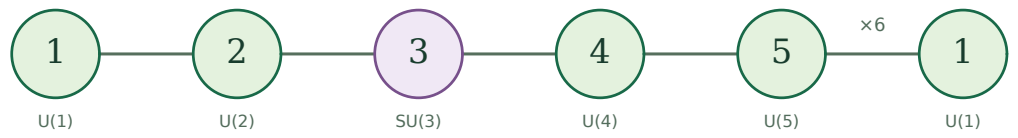


**1437 2301.02409-TSU6-SU2** / N = 6 · k = 2

$1 + 19t^2 + 325t^4 + O(t^5)$

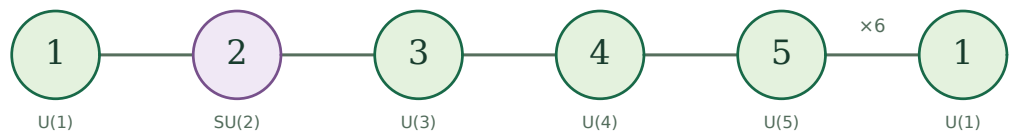


**1438 2301.02409-TSU5-SU3-charge2** / N = 5 · k = 3 · charge = 2

$1 + 12t^2 + 113t^4 + 783t^6 + 4487t^8 + 21842t^{10} + 94451t^{12} + O(t^{13})$

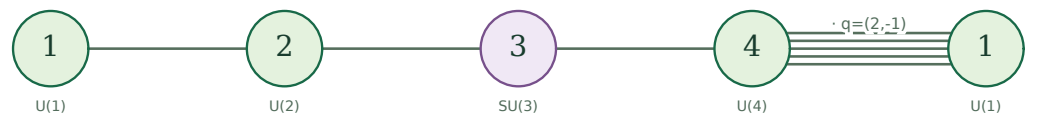


**1439 2301.02409-TSU5-SU2-charge3** / N = 5 · k = 2 · charge = 3

$1 + 12t^2 + 113t^4 + 783t^6 + 4487t^8 + 21842t^{10} + 94451t^{12} + O(t^{13})$

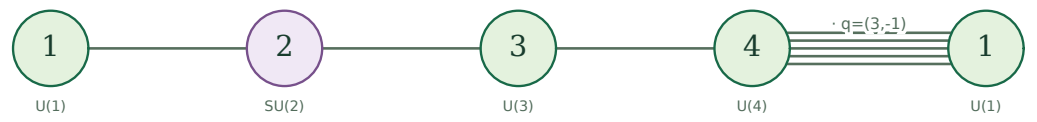

**QUIVER** | **HILBERT SERIES**

**1440 2301.02409-Trhosigma-SU-at-2**

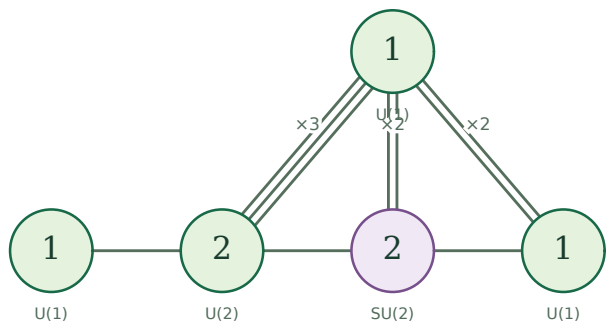


$1 + 6t^2 + 28t^4 + 101t^6 + 332t^8 + O(t^9)$

**1441 2301.02409-Trhosigma-SU-at-1**

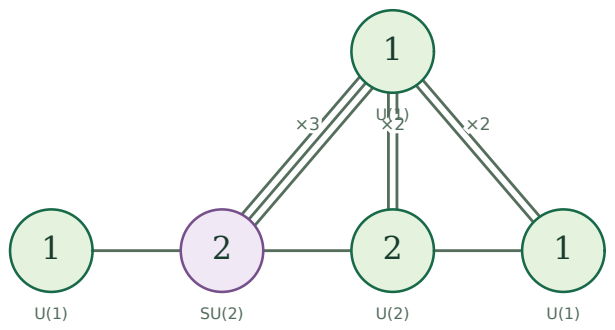


$1 + 6t^2 + 2t^3 + 24t^4 + 16t^5 + 77t^6 + 72t^7 + 228t^8 + O(t^9)$

**1442 2301.02409-Trhosigma-SU9-original**

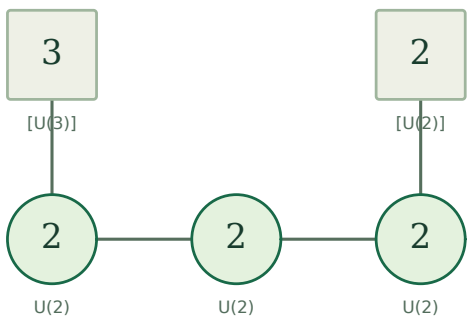


$1 + 9t^2 + 6t^3 + 54t^4 + 60t^5 + O(t^6)$

**1443 2301.02409-Trhosigma-SU9-Z2**

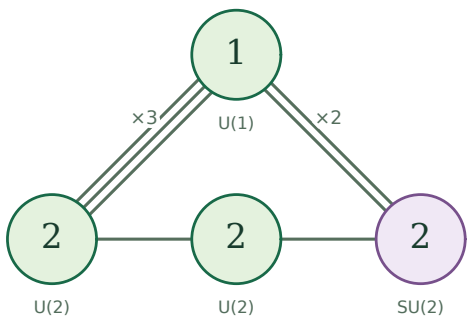


$1 + 5t^2 + 4t^3 + 30t^4 + 32t^5 + O(t^6)$

**1444 2301.02409-TSO6-Spin4**

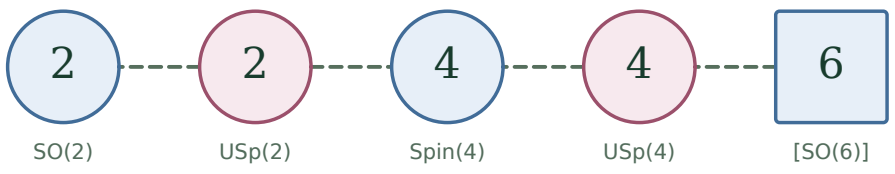


$1 + 7t^2 + 63t^4 + 328t^6 + 1476t^8 + 5390t^{10} + 17500t^{12} + O(t^{13})$

**1445 2301.02409-TSO8-Spin4**

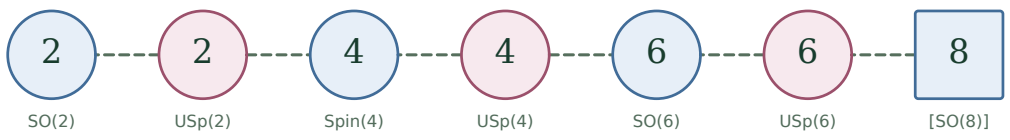


$1 + 12t^2 + 213t^4 + 1984t^6 + O(t^7)$

**1446 2301.02409-TSO8-Spin6**

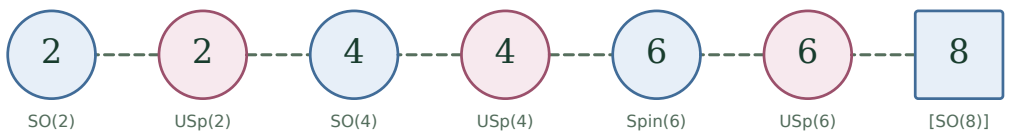


$1 + 16t^2 + 213t^4 + 2048t^6 + O(t^7)$

**QUIVER** | **HILBERT SERIES**

**1447 2301.02409-Sp2-Nf5-OSp-Spin2**

$1 + 29t^2 + 532t^4 + O(t^5)$

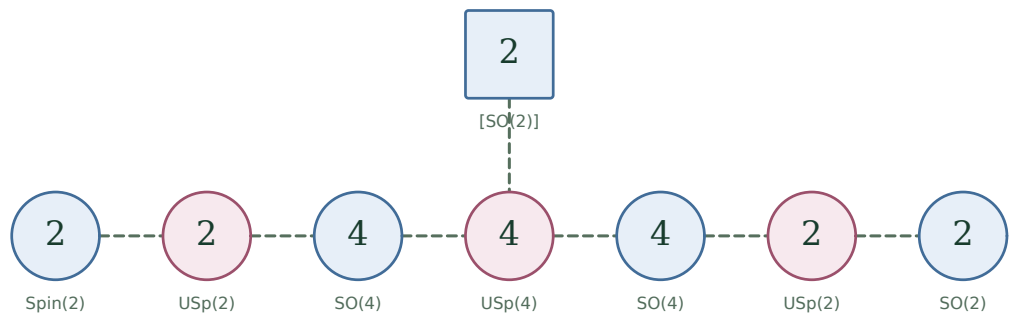


**1448 2301.02409-Sp2-Nf5-OSp-Spin4**

$1 + 21t^2 + 500t^4 + O(t^5)$

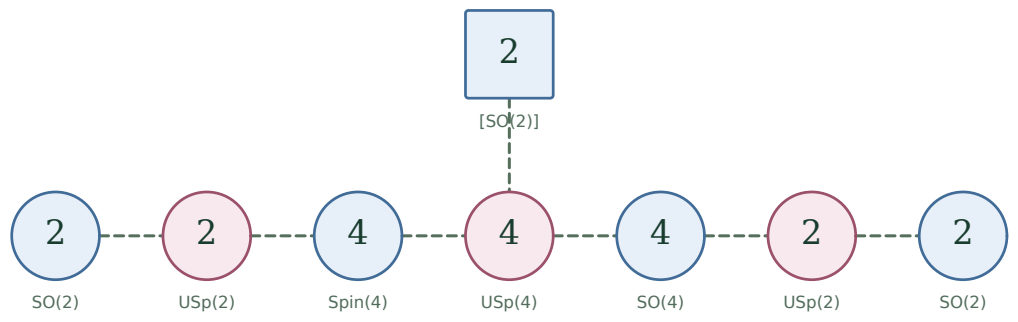


**1449 2301.02409-Sp2-Nf6-OSp-Spin2**

$1 + 46t^2 + 1233t^4 + O(t^5)$

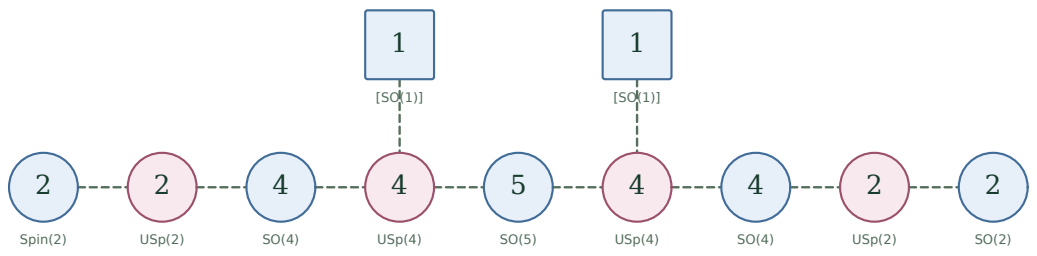


**1450 2301.02409-Sp2-Nf6-OSp-Spin4**

$1 + 34t^2 + 1077t^4 + O(t^5)$

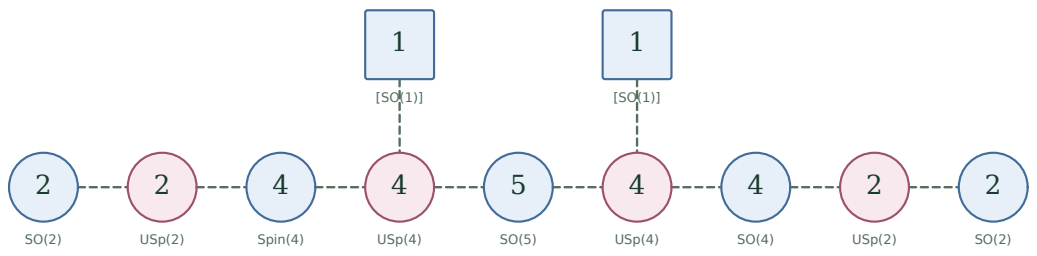


**1451 2301.02409-Sp2-Nf6-OSp-Spin5**

$1 + 30t^2 + 1089t^4 + O(t^5)$

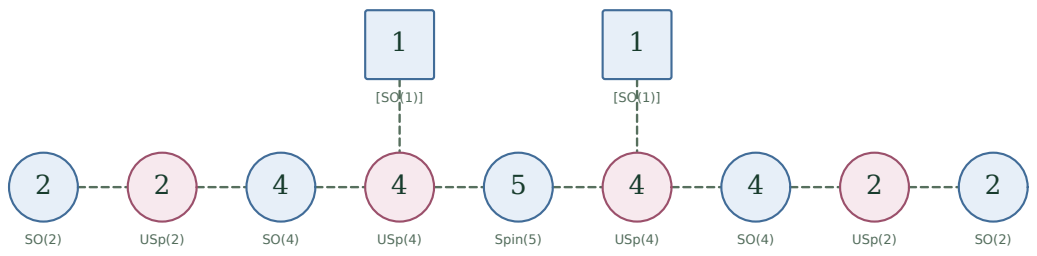


**1452 2301.02409-D7-SU-node2** / **D = 7 · SU_node = 2**

$1 + 51t^2 + 2081t^4 + O(t^5)$

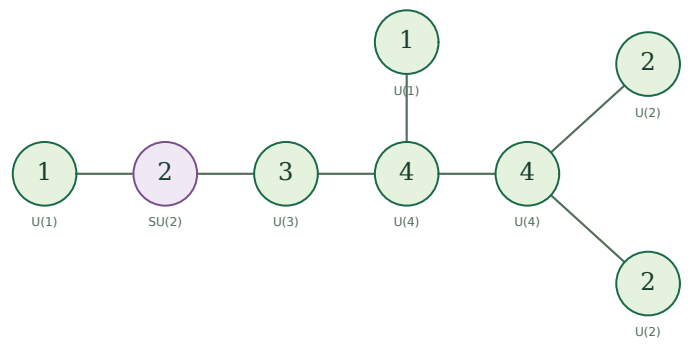


**1453 2301.02409-D7-SU-node3** / **D = 7 · SU_node = 3**

$1 + 37t^2 + 1387t^4 + O(t^5)$

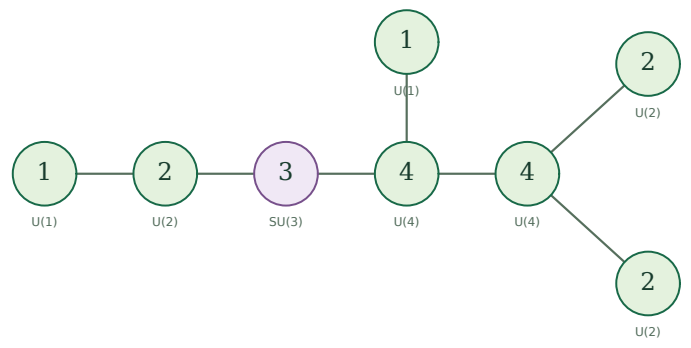

| QUIVER | HILBERT SERIES |
|---|---|

**1454 2301.02409-D7-SU-node4** / **D = 7 · SU_node = 4**

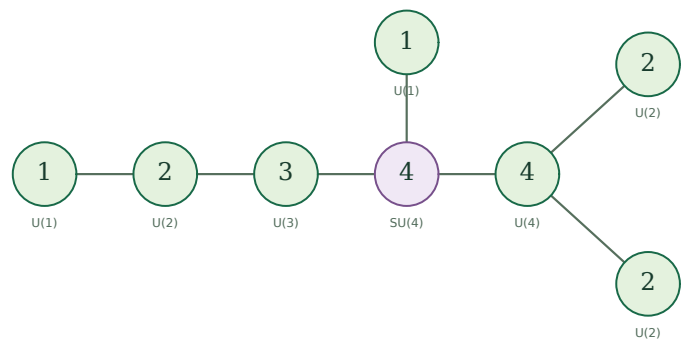


$1 + 31t^2 + 1113t^4 + O(t^5)$

**1455 2301.02409-D7-SU-node5** / **D = 7 · SU_node = 5**

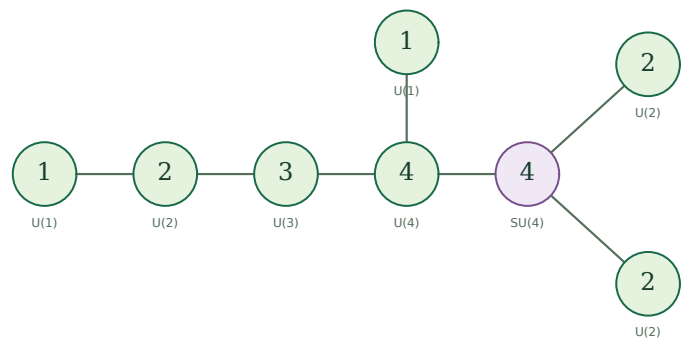


$1 + 31t^2 + 1071t^4 + O(t^5)$

**1456 2301.02409-D7-SU-node6** / **D = 7 · SU_node = 6**

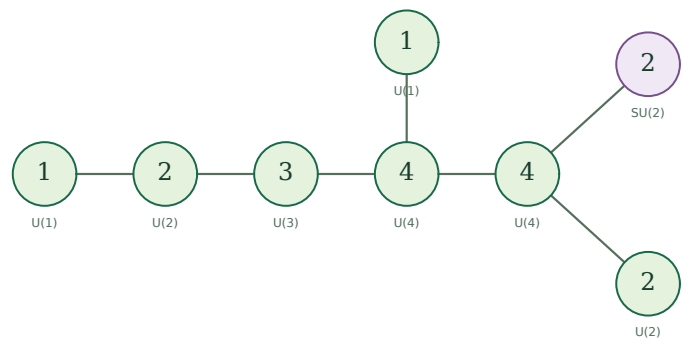


$1 + 49t^2 + 2079t^4 + O(t^5)$

**1457 2301.02409-D8-SU-node2** / **D = 8 · SU_node = 2**

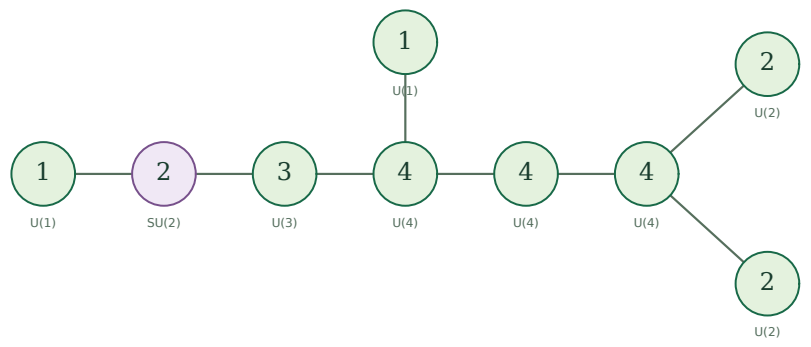


$1 + 72t^2 + 3716t^4 + O(t^5)$

**1458 2301.02409-D8-SU-node3** / **D = 8 · SU_node = 3**

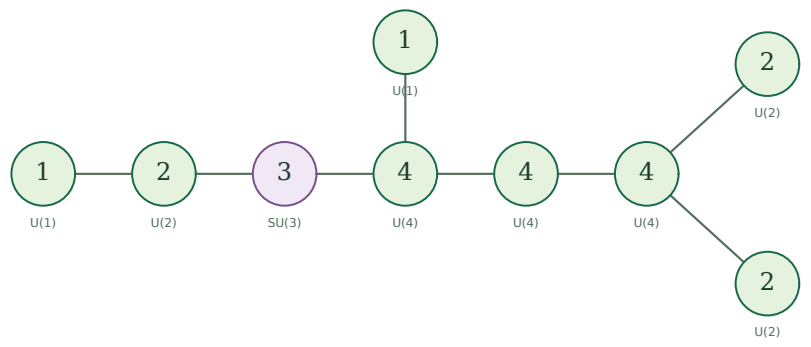


$1 + 54t^2 + 2510t^4 + O(t^5)$

**1459 2301.02409-D8-SU-node4** / **D = 8 · SU_node = 4**

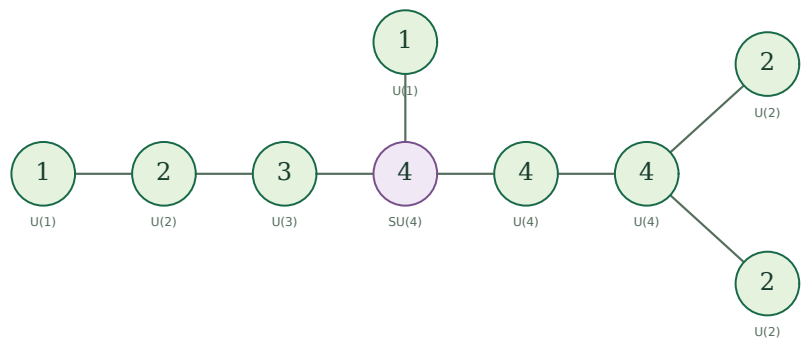


$1 + 44t^2 + 2040t^4 + O(t^5)$

**1460 2301.02409-D8-SU-node5** / **D = 8 · SU_node = 5**

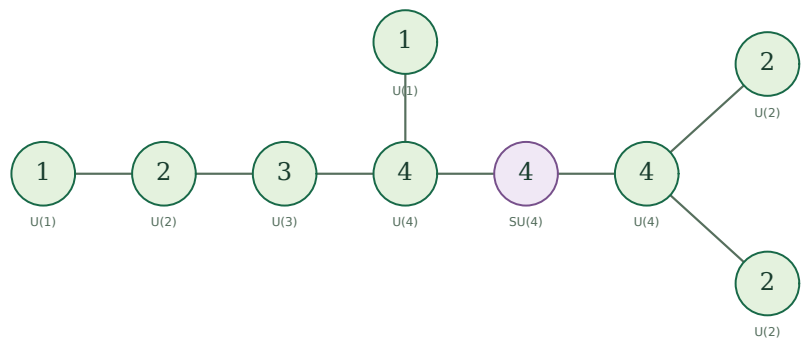


$1 + 40t^2 + 1884t^4 + O(t^5)$

| QUIVER | HILBERT SERIES |
|---|---|
| **1461 2301.02409-D8-SU-node6** / **D = 8 · SU_node = 6** | $1 + 42t^2 + 1898t^4 + O(t^5)$ |

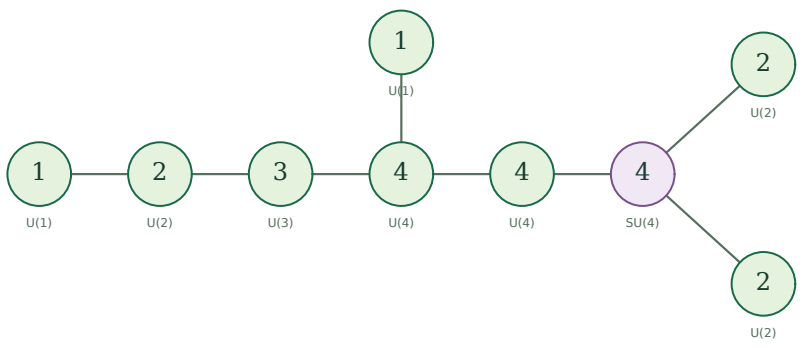


| | |
|---|---|
| **1462 2301.02409-D8-SU-node7** / **D = 8 · SU_node = 7** | $1 + 64t^2 + 3612t^4 + O(t^5)$ |

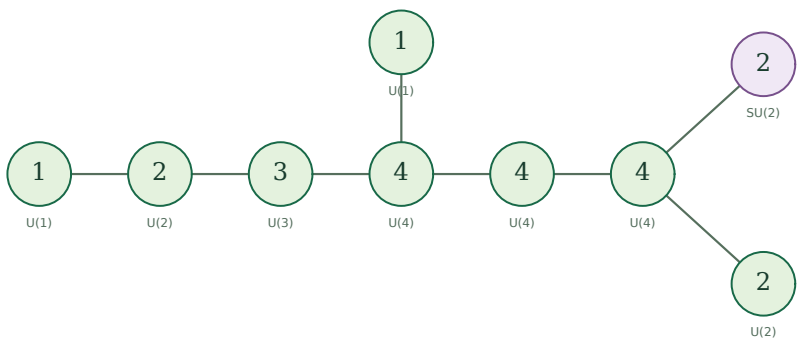


| | |
|---|---|
| **1463 2301.02409-C4-SU-node4** | $1 + 16t^2 + 330t^4 + 3290t^6 + O(t^7)$ |

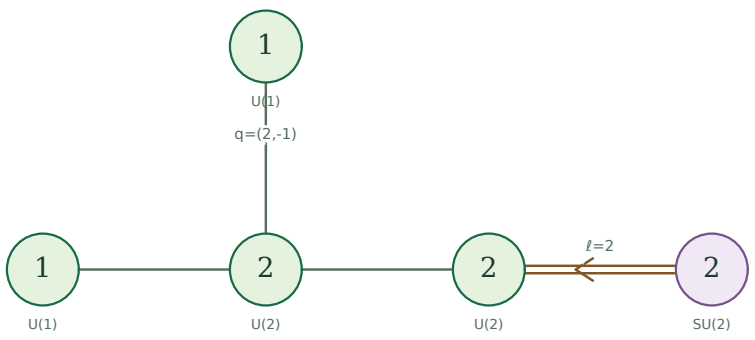


| | |
|---|---|
| **1464 2301.02409-C4-SU-node3** | $1 + 24t^2 + 362t^4 + 3514t^6 + O(t^7)$ |

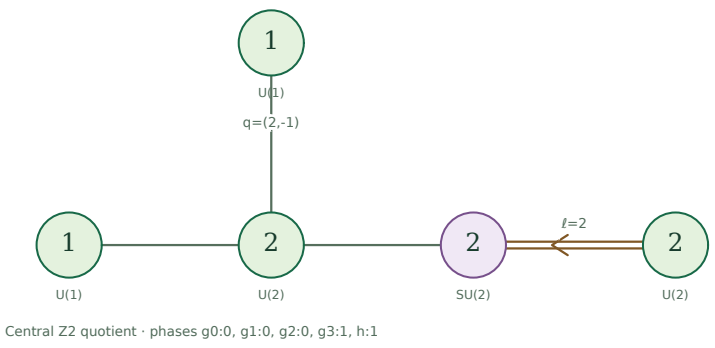


Central Z2 quotient · phases g0:0, g1:0, g2:0, g3:1, h:1

| | |
|---|---|
| **1465 2301.02409-C4-SU-node2** | $1 + 20t^2 + 334t^4 + 3350t^6 + O(t^7)$ |

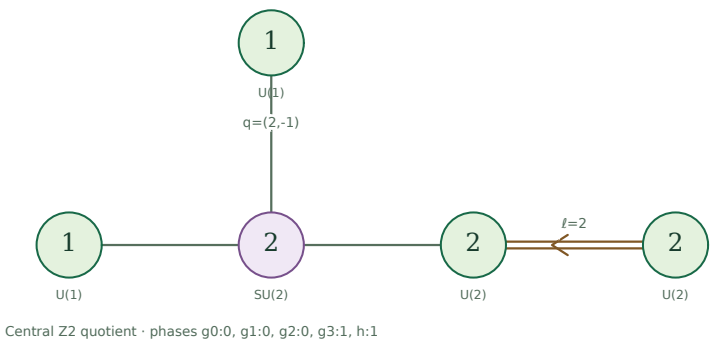


Central Z2 quotient · phases g0:0, g1:0, g2:0, g3:1, h:1

| | |
|---|---|
| **1466 2301.02409-F4-table-row2** | $1 + 24t^2 + 537t^4 + O(t^5)$ |

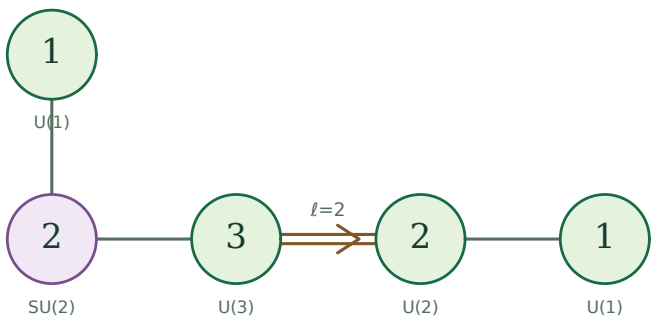


| | |
|---|---|
| **1467 2301.02409-F4-table-row3** | $1 + 16t^2 + 351t^4 + O(t^5)$ |

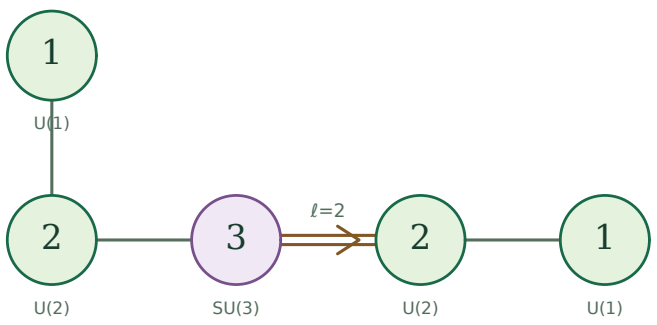

| QUIVER | HILBERT SERIES |
|---|---|

**1468 2301.02409-F4-table-row4**

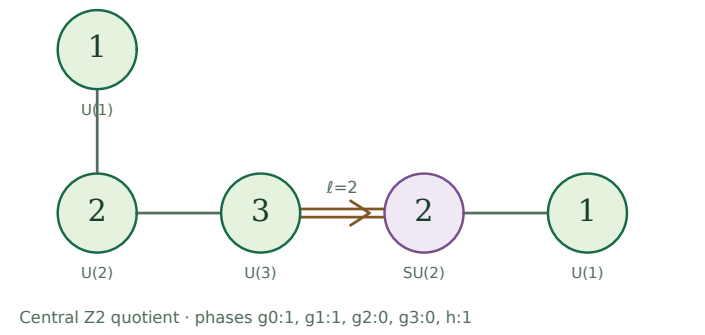




$1 + 36t^2 + 621t^4 + O(t^5)$

**1469 2301.02409-F4-table-row5**

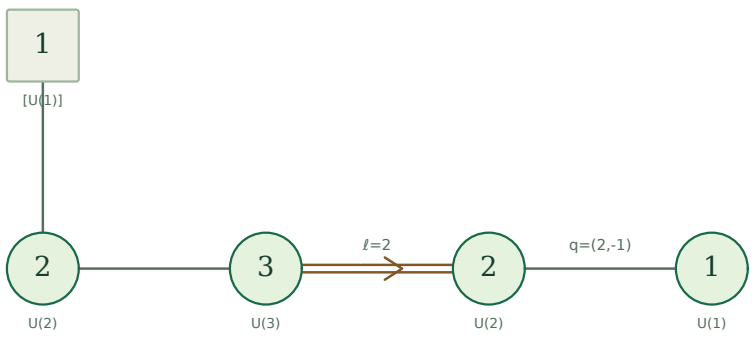


$1 + 36t^2 + 621t^4 + O(t^5)$

**1470 2301.02409-F4-table-row6**

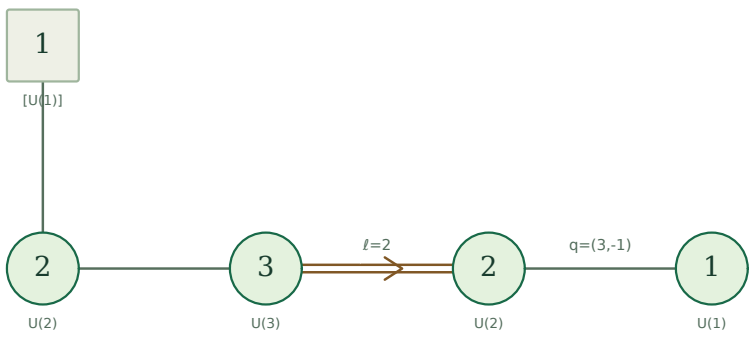


$1 + 22t^2 + 369t^4 + O(t^5)$

**1471 2301.02409-F4-table-row7**

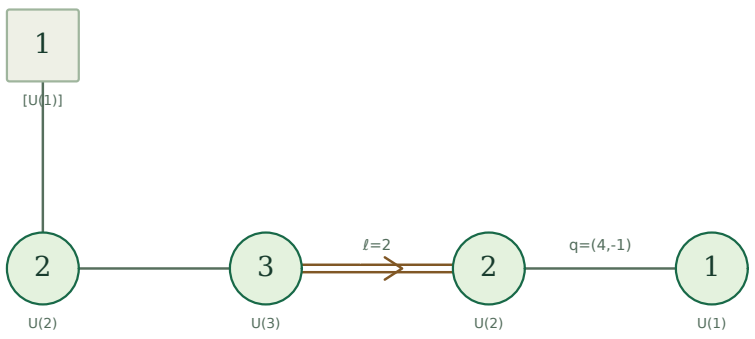


$1 + 22t^2 + 327t^4 + O(t^5)$

**1472 2301.02409-magnetic-folded-E6-integer**

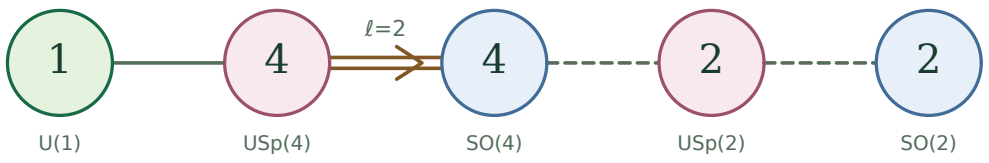


$1 + 25t^2 + 400t^4 + 3864t^6 + 26600t^8 + 141672t^{10} + 621480t^{12} + 2337280t^{14} + 7763283t^{16} + 23265515t^{18} + 63954800t^{20} + O(t^{21})$

## 2303.16939 · A tale of N cones (21)

**1473 2303.16939-D2-cone-1-2**

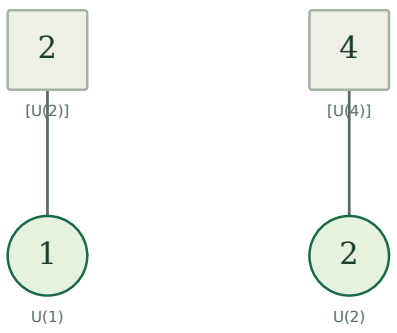


$1 + 6t^2 + 23t^4 + 67t^6 + 164t^8 + 353t^{10} + 690t^{12} + 1250t^{14} + 2131t^{16} + 3456t^{18} + 5377t^{20} + 8077t^{22} + 11774t^{24} + 16723t^{26} + 23220t^{28} + 31604t^{30} + O(t^{31})$

**1474 2303.16939-T1-n2-c1-c4**

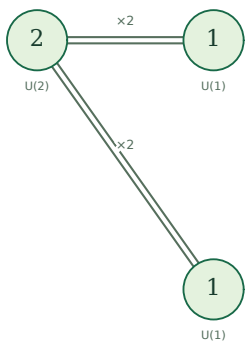


÷ U(1) diagonal

$1 + 6t^2 + 29t^4 + 89t^6 + 236t^8 + 521t^{10} + 1054t^{12} + 1934t^{14} + 3355t^{16} + 5480t^{18} + 8611t^{20} + 12991t^{22} + 19054t^{24} + 27139t^{26} + 37836t^{28} + 51596t^{30} + O(t^{31})$

| QUIVER | HILBERT SERIES |
|---|---|

**1475 2303.16939-T1-n2-c2-c3**

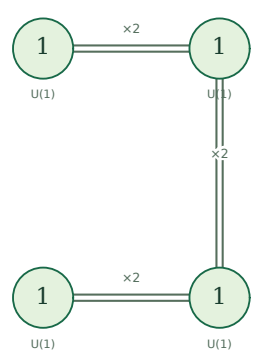

÷ U(1) diagonal

$1 + 9t^{2} + 42t^{4} + 138t^{6} + 363t^{8} + 819t^{10} + 1652t^{12} + 3060t^{14} + 5301t^{16} + 8701t^{18} + 13662t^{20} + 20670t^{22} + 30303t^{24} + 43239t^{26} + 60264t^{28} + 82280t^{30} + O(t^{31})$

**1476 2303.16939-T1-n3-c2-c3**

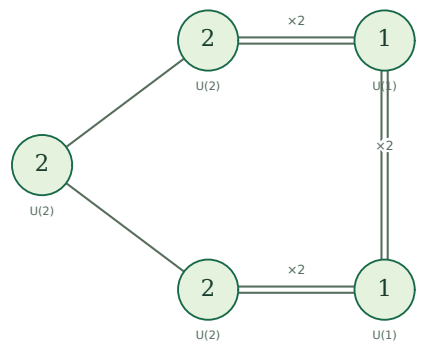

÷ U(1) diagonal

$1 + 16t^{2} + 163t^{4} + 1184t^{6} + 6686t^{8} + 31022t^{10} + 123024t^{12} + 429105t^{14} + 1345116t^{16} + 3852737t^{18} + 10215250t^{20} + 25334928t^{22} + 59272560t^{24} + 131726278t^{26} + 279699542t^{28} + 570200750t^{30} + O(t^{31})$

**1477 2303.16939-T1-n2-b1-b2-b4-b5**

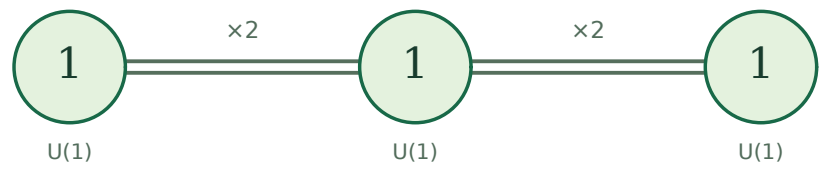


÷ U(1) diagonal

$1 + 6t^{2} + 19t^{4} + 44t^{6} + 85t^{8} + 146t^{10} + 231t^{12} + 344t^{14} + 489t^{16} + 670t^{18} + 891t^{20} + 1156t^{22} + 1469t^{24} + 1834t^{26} + 2255t^{28} + 2736t^{30} + O(t^{31})$

**1478 2303.16939-T1-n3-b1-b2-b4-b5**

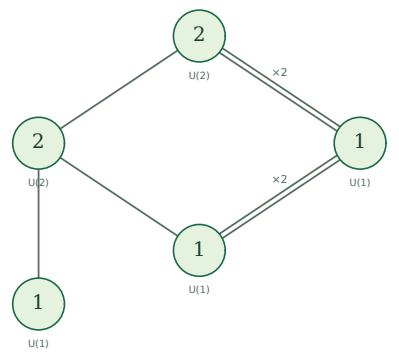

÷ U(1) diagonal

$1 + 16t^{2} + 147t^{4} + 944t^{6} + 4682t^{8} + 19072t^{10} + 66562t^{12} + 205104t^{14} + 570533t^{16} + 1456944t^{18} + 3460327t^{20} + 7722720t^{22} + 16330156t^{24} + 32938240t^{26} + 63725452t^{28} + 118806752t^{30} + O(t^{31})$

**1479 2303.16939-T1-n3-b3**

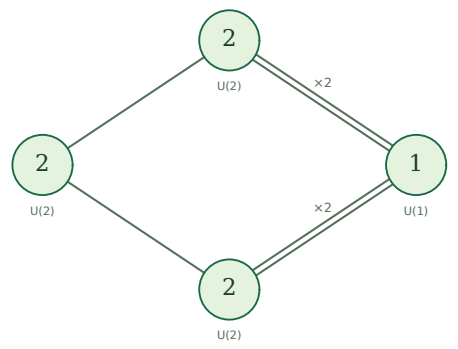

÷ U(1) diagonal

$1 + 15t^{2} + 135t^{4} + 853t^{6} + 4194t^{8} + 16974t^{10} + 58990t^{12} + 181185t^{14} + 502821t^{16} + 1281661t^{18} + 3039699t^{20} + 6776142t^{22} + 14315320t^{24} + 28852092t^{26} + 55784616t^{28} + 103946366t^{30} + O(t^{31})$

**1480 2303.16939-T1-n2-a1-a3**

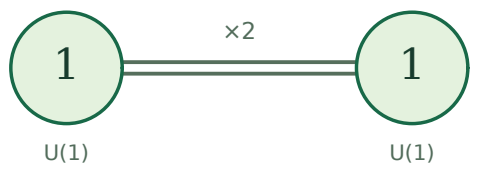


÷ U(1) diagonal

$1 + 3t^{2} + 5t^{4} + 7t^{6} + 9t^{8} + 11t^{10} + 13t^{12} + 15t^{14} + 17t^{16} + 19t^{18} + 21t^{20} + 23t^{22} + 25t^{24} + 27t^{26} + 29t^{28} + 31t^{30} + O(t^{31})$

| QUIVER | HILBERT SERIES |
|---|---|

**1481 2303.16939-T1-n3-a1-a3**

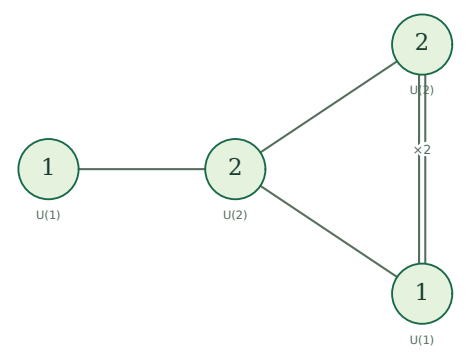

÷ U(1) diagonal

$1 + 15t^2 + 119t^4 + 649t^6 + 2730t^8 + 9478t^{10} + 28390t^{12} + 75658t^{14} + 183403t^{16} + 411125t^{18} + 863005t^{20} + 1713075t^{22} + 3240692t^{24} + 5879212t^{26} + 10281260t^{28} + 17404532t^{30} + O(t^{31})$

**1482 2303.16939-T1-n3-bottom**

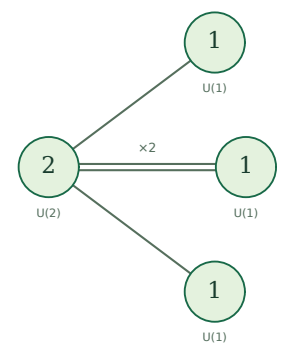

÷ U(1) diagonal

$1 + 15t^2 + 104t^4 + 475t^6 + 1659t^8 + 4802t^{10} + 12096t^{12} + 27378t^{14} + 56925t^{16} + 110473t^{18} + 202488t^{20} + 353717t^{22} + 593047t^{24} + 959700t^{26} + 1505792t^{28} + 2299284t^{30} + O(t^{31})$

**1483 2303.16939-Fig8-n1**

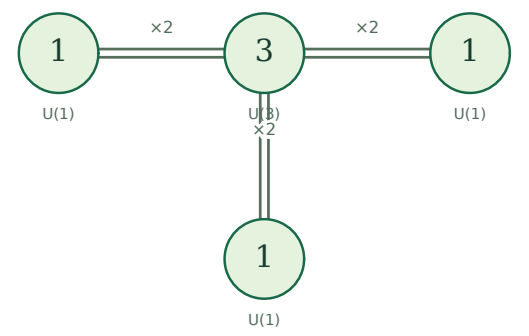

÷ U(1) diagonal

**Computed by Quiver Lab**

$1 + 5t^2 + 24t^4 + 101t^6 + 345t^8 + 1041t^{10} + 2836t^{12} + 7021t^{14} + 16098t^{16} + 34601t^{18} + 70221t^{20} + 135666t^{22} + 251127t^{24} + 447442t^{26} + 770781t^{28} + 1288523t^{30} + O(t^{31})$

Source originally Claimed · no numerical literature comparison is available.

**1484 2303.16939-Fig8-n2**

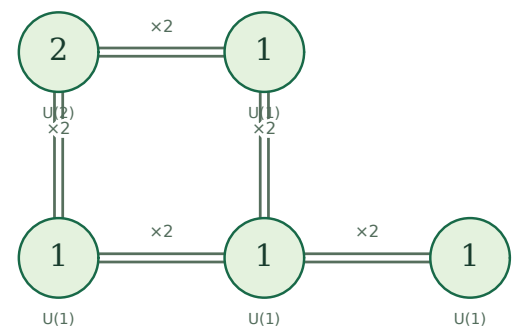

÷ U(1) diagonal

**Computed by Quiver Lab**

$1 + 8t^2 + 44t^4 + 195t^6 + 716t^8 + 2262t^{10} + 6338t^{12} + 16082t^{14} + 37563t^{16} + 81826t^{18} + 167938t^{20} + 327399t^{22} + 610352t^{24} + 1094042t^{26} + 1894134t^{28} + 3179590t^{30} + O(t^{31})$

Source originally Claimed · no numerical literature comparison is available.

**1485 2303.16939-Fig8-n5**

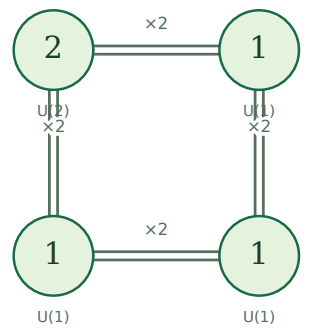

÷ U(1) diagonal

**Computed by Quiver Lab**

$1 + 5t^2 + 24t^4 + 91t^6 + 279t^8 + 746t^{10} + 1784t^{12} + 3884t^{14} + 7853t^{16} + 14929t^{18} + 26920t^{20} + 46429t^{22} + 77063t^{24} + 123674t^{26} + 192728t^{28} + 292636t^{30} + O(t^{31})$

Source originally Claimed · no numerical literature comparison is available.

**1486 2303.16939-Fig8-n6**

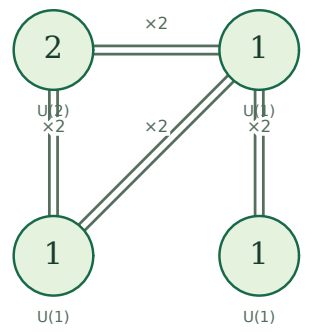

÷ U(1) diagonal

**Computed by Quiver Lab**

$1 + 7t^2 + 32t^4 + 117t^6 + 357t^8 + 946t^{10} + 2246t^{12} + 4876t^{14} + 9833t^{16} + 18651t^{18} + 33590t^{20} + 57871t^{22} + 95963t^{24} + 153910t^{26} + 239718t^{28} + 363808t^{30} + O(t^{31})$

Source originally Claimed · no numerical literature comparison is available.

| QUIVER | HILBERT SERIES |
|---|---|

**1487 2303.16939-Fig8-n7**

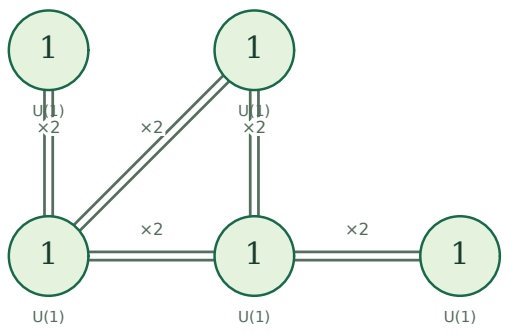


÷ U(1) diagonal

**Computed by Quiver Lab**

$1 + 8t^2 + 40t^4 + 152t^6 + 475t^8 + 1280t^{10} + 3072t^{12} + 6720t^{14} + 13629t^{16} + 25960t^{18} + 46904t^{20} + 81016t^{22} + 134615t^{24} + 216256t^{26} + 337280t^{28} + 512448t^{30} + O(t^{31})$

Source originally Claimed · no numerical literature comparison is available.

**1488 2303.16939-Fig8-n12**

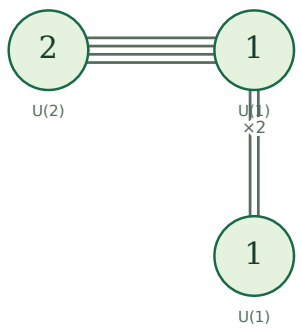


÷ U(1) diagonal

**Computed by Quiver Lab**

$1 + 6t^2 + 23t^4 + 67t^6 + 164t^8 + 353t^{10} + 690t^{12} + 1250t^{14} + 2131t^{16} + 3456t^{18} + 5377t^{20} + 8077t^{22} + 11774t^{24} + 16723t^{26} + 23220t^{28} + 31604t^{30} + O(t^{31})$

Source originally Claimed · no numerical literature comparison is available.

**1489 2303.16939-Fig8-n14**

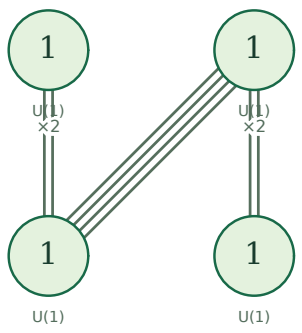


÷ U(1) diagonal

**Computed by Quiver Lab**

$1 + 7t^2 + 28t^4 + 84t^6 + 209t^8 + 455t^{10} + 896t^{12} + 1632t^{14} + 2793t^{16} + 4543t^{18} + 7084t^{20} + 10660t^{22} + 15561t^{24} + 22127t^{26} + 30752t^{28} + 41888t^{30} + O(t^{31})$

Source originally Claimed · no numerical literature comparison is available.

**1490 2303.16939-Fig8-n13**

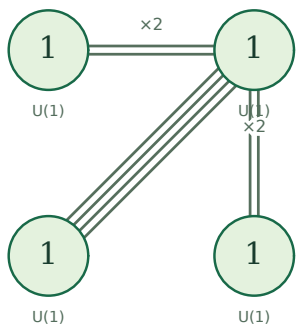


÷ U(1) diagonal

**Computed by Quiver Lab**

$1 + 7t^2 + 28t^4 + 84t^6 + 209t^8 + 455t^{10} + 896t^{12} + 1632t^{14} + 2793t^{16} + 4543t^{18} + 7084t^{20} + 10660t^{22} + 15561t^{24} + 22127t^{26} + 30752t^{28} + 41888t^{30} + O(t^{31})$

Source originally Claimed · no numerical literature comparison is available.

**1491 2303.16939-Fig8-n22**

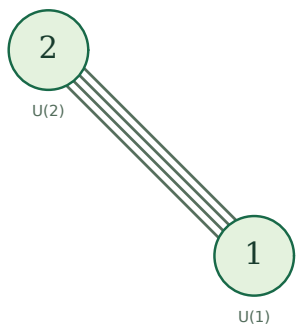


÷ U(1) diagonal

**Computed by Quiver Lab**

$1 + 3t^2 + 9t^4 + 18t^6 + 35t^8 + 57t^{10} + 91t^{12} + 132t^{14} + 189t^{16} + 255t^{18} + 341t^{20} + 438t^{22} + 559t^{24} + 693t^{26} + 855t^{28} + 1032t^{30} + O(t^{31})$

Source originally Claimed · no numerical literature comparison is available.

**1492 2303.16939-Fig8-n21**

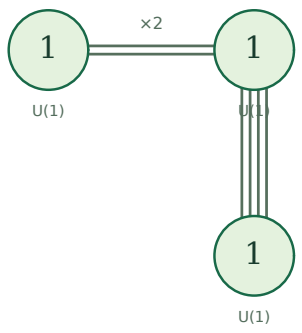


÷ U(1) diagonal

**Computed by Quiver Lab**

$1 + 4t^2 + 11t^4 + 24t^6 + 45t^8 + 76t^{10} + 119t^{12} + 176t^{14} + 249t^{16} + 340t^{18} + 451t^{20} + 584t^{22} + 741t^{24} + 924t^{26} + 1135t^{28} + 1376t^{30} + O(t^{31})$

Source originally Claimed · no numerical literature comparison is available.

| QUIVER | HILBERT SERIES |
|---|---|

**1493 2303.16939-Fig8-n27**

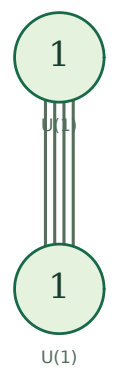


÷ U(1) diagonal

**Computed by Quiver Lab**

$1 + t^2 + 3t^4 + 3t^6 + 5t^8 + 5t^{10} + 7t^{12} + 7t^{14} + 9t^{16} + 9t^{18} + 11t^{20} + 11t^{22} + 13t^{24} + 13t^{26} + 15t^{28} + 15t^{30} + O(t^{31})$

Source originally Claimed · no numerical literature comparison is available.

## 2401.08757 · Higgs branch RG flows via decay and fission (1)

**1494 2401.08757-new-sp2-isolated**

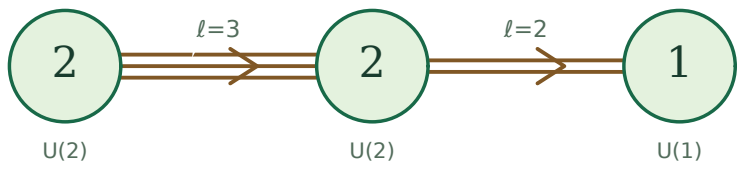


÷ U(1) diagonal · node g0

$1 + 10t^2 + 80t^4 + 359t^6 + 1295t^8 + 3751t^{10} + 9560t^{12} + 21675t^{14} + 45313t^{16} + 88055t^{18} + 161872t^{20} + 283036t^{22} + 475375t^{24} + 769802t^{26} + 1209200t^{28} + 1847322t^{30} + O(t^{31})$

## 2406.09626 · Complex symplectic contractions and 3d mirrors (6)

**1495 2406.09626-A2-contraction / N = 3**

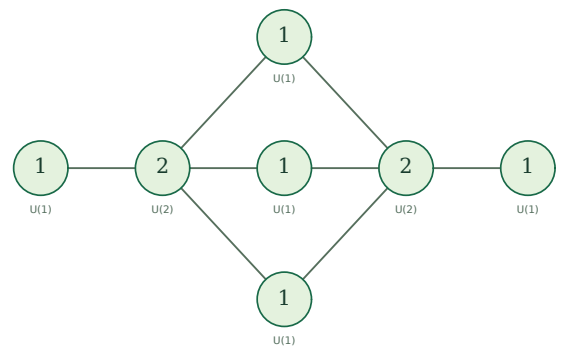

÷ U(1) diagonal

$1 + 18t^2 + 223t^4 + 1966t^6 + 13444t^8 + 74590t^{10} + O(t^{11})$

**1496 2406.09626-A2-mirror / N = 3**

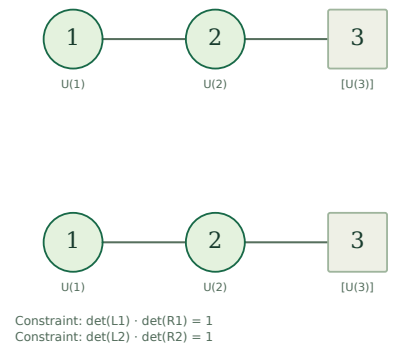

Constraint: det(L1) · det(R1) = 1 · Constraint: det(L2) · det(R2) = 1

$1 + 2t^2 + 13t^4 + 38t^6 + 115t^8 + 284t^{10} + 666t^{12} + 1392t^{14} + 2781t^{16} + 5170t^{18} + 9239t^{20} + 15730t^{22} + 25943t^{24} + 41284t^{26} + 64052t^{28} + 96696t^{30} + O(t^{31})$

**1497 2406.09626-A3-contraction / N = 4**

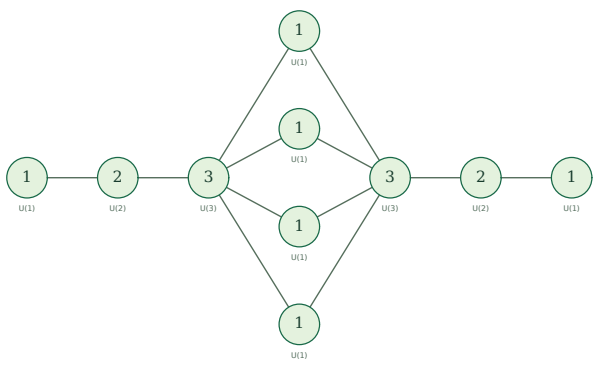

÷ U(1) diagonal

$1 + 33t^2 + 559t^4 + 6685t^6 + 61900t^8 + 491584t^{10} + O(t^{11})$

**Paper typo:** The printed 6685 at $t^6$ should be 6605. Every other independently checked coefficient through $t^8$ agrees, and the source graph and global form are unambiguous.

**1498 2406.09626-A3-mirror / N = 4**

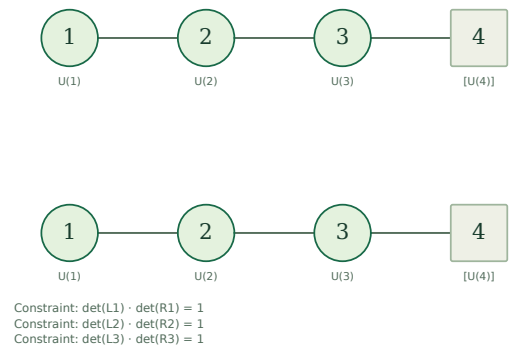

Constraint: det(L1) · det(R1) = 1 · Constraint: det(L2) · det(R2) = 1 · Constraint: det(L3) · det(R3) = 1

$1 + 3t^2 + 28t^4 + 138t^6 + 674t^8 + 2808t^{10} + 10781t^{12} + 37479t^{14} + 121152t^{16} + 364360t^{18} + 1031724t^{20} + 2761540t^{22} + 7034768t^{24} + 17121056t^{26} + 39992192t^{28} + 89945168t^{30} + O(t^{31})$

| QUIVER | HILBERT SERIES |
|---|---|

**1499 2406.09626-D3-nilcone-product / n = 3**

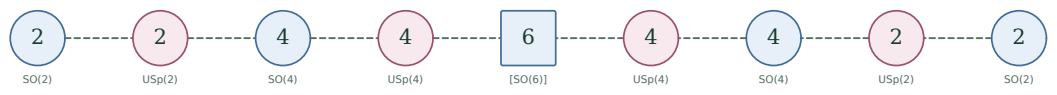


$1 + 30t^2 + 463t^4 + 4578t^6 + 30899t^8 + O(t^9)$

**Paper typo:** The printed series is not the square of the SO(6) nilpotent-cone series identified in the source. The exact Casimir product and an independent calculation of the literal quiver agree on the corrected coefficients.

**1500 2406.09626-D-contraction-source-typo / source_n = 2 · fugacities = 3**

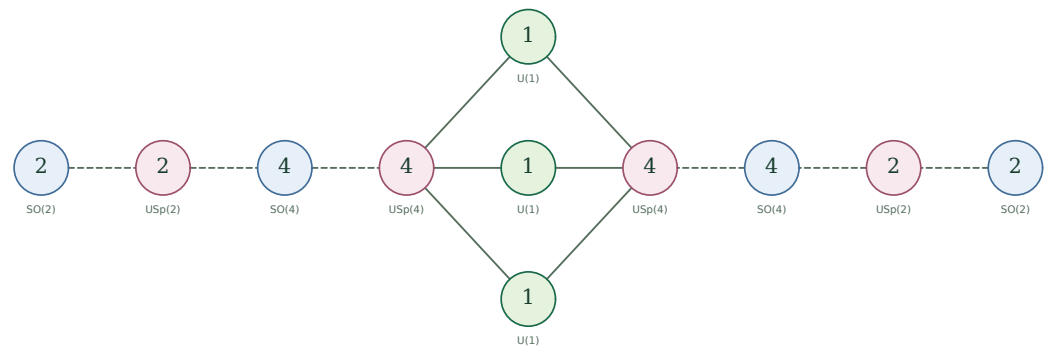


Overall diagonal $\mathbb{Z}_2$ quotient · integral and half-integral magnetic sectors

$1 + 33t^2 + 559t^4 + 6157t^6 + 47942t^8 + O(t^9)$

**Paper typo:** Under the documented, inferred n=3 interpretation, independent refined monopoles give 6605 at $t^6$ and 61900 at $t^8$, versus 6157 and 47942 printed. The last printed $t^4$ is interpreted as $t^8$; the geometric quotient check passes, but the printed coefficients do not.

## 2411.12802 · An exceptionally simple family of Orthosymplectic 3d N=4 rank-0 SCFTs (1)

**1501 2411.12802-rankzero-n4-integer-lattice / n = 4 · diagonalZ2 = False**

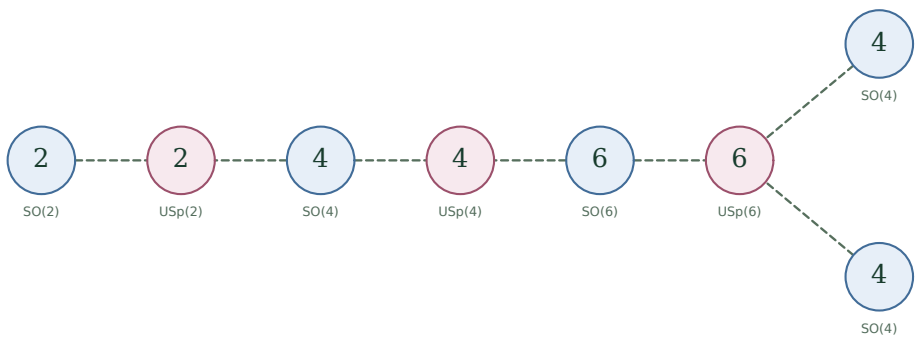


**Computed by Quiver Lab**

$1 + 72t^2 + 2794t^4 + 72088t^6 + 1345657t^8 + 19254720t^{10} + 220776050t^{12} + 2100053920t^{14} + 17030339728t^{16} + 120328614400t^{18} + 753848819738t^{20} + 4248137641696t^{22} + 21790658227949t^{24} + 102759157859672t^{26} + 449268744317810t^{28} + 1834220005095944t^{30} + O(t^{31})$

Source originally Claimed · no numerical literature comparison is available.

## 2411.14531 · A Bound on 3d Mirror Pairs (44)

**1502 2411.14531-T2-r3-electric**

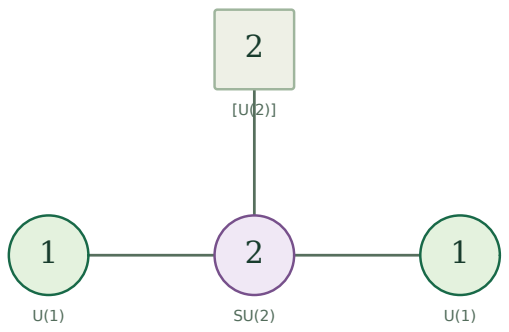


**Computed by Quiver Lab**

$1 + 6t^2 + 29t^4 + 89t^6 + 236t^8 + 521t^{10} + 1054t^{12} + 1934t^{14} + 3355t^{16} + 5480t^{18} + 8611t^{20} + 12991t^{22} + 19054t^{24} + 27139t^{26} + 37836t^{28} + 51596t^{30} + O(t^{31})$

Source originally Claimed · no numerical literature comparison is available.

**1503 2411.14531-T2-r5-electric**

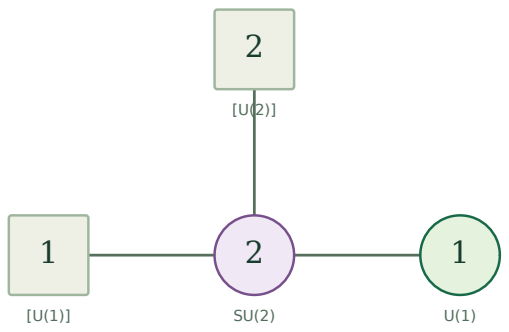


**Computed by Quiver Lab**

$1 + 3t^2 + 9t^4 + 18t^6 + 35t^8 + 57t^{10} + 91t^{12} + 132t^{14} + 189t^{16} + 255t^{18} + 341t^{20} + 438t^{22} + 559t^{24} + 693t^{26} + 855t^{28} + 1032t^{30} + O(t^{31})$

Source originally Claimed · no numerical literature comparison is available.

**1504 2411.14531-T2-r6-mirror**

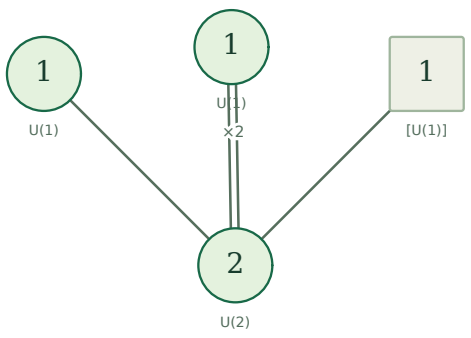


**Computed by Quiver Lab**

$1 + 15t^2 + 104t^4 + 475t^6 + 1659t^8 + 4802t^{10} + 12096t^{12} + 27378t^{14} + 56925t^{16} + 110473t^{18} + 202488t^{20} + 353717t^{22} + 593047t^{24} + 959700t^{26} + 1505792t^{28} + 2299284t^{30} + O(t^{31})$

Source originally Claimed · no numerical literature comparison is available.

| QUIVER | HILBERT SERIES |
|---|---|

**1505 2411.14531-T2-r8-mirror**

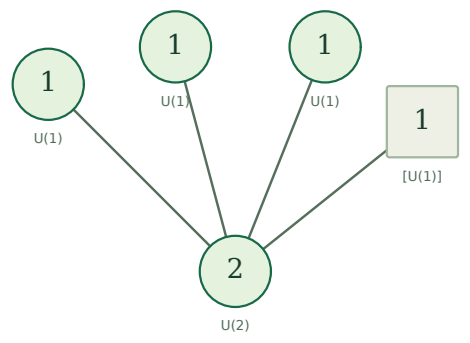


**Computed by Quiver Lab**

$1 + 28t^{2} + 300t^{4} + 1925t^{6} + 8918t^{8} + 32928t^{10} + 102816t^{12} + 282150t^{14} + 698775t^{16} + 1591876t^{18} + 3383380t^{20} + 6782139t^{22} + 12931100t^{24} + 23609600t^{26} + 41505024t^{28} + 70570332t^{30} + O(t^{31})$

Source originally Claimed · no numerical literature comparison is available.

**1506 2411.14531-T3-r1-mirror**

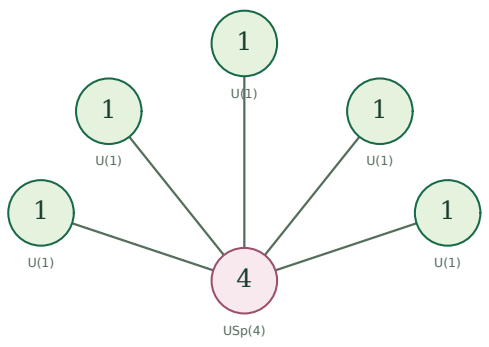


÷ $\mathbb{Z}_2$ diagonal

**Computed by Quiver Lab**

$1 + 6t^{2} + 75t^{4} + 453t^{6} + 2581t^{8} + 11457t^{10} + 45217t^{12} + 155463t^{14} + 485500t^{16} + 1382449t^{18} + 3657034t^{20} + 9044358t^{22} + 21130198t^{24} + 46888222t^{26} + 99471862t^{28} + 202604194t^{30} + O(t^{31})$

Source originally Claimed · no numerical literature comparison is available.

**1507 2411.14531-T3-r2-electric**

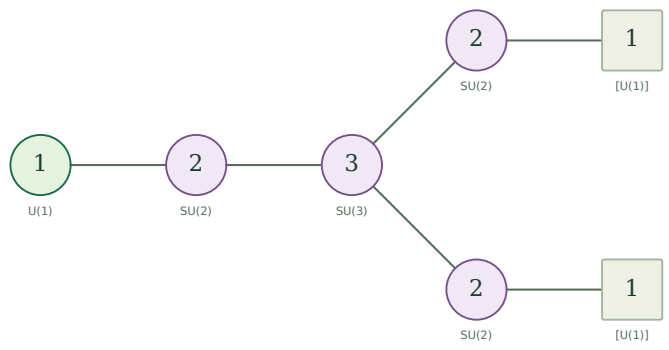


**Computed by Quiver Lab**

$1 + 3t^{2} + 30t^{4} + 134t^{6} + 623t^{8} + 2288t^{10} + 7659t^{12} + 22610t^{14} + 61400t^{16} + 153365t^{18} + 358953t^{20} + 790764t^{22} + 1656330t^{24} + 3313132t^{26} + 6368614t^{28} + 11806104t^{30} + O(t^{31})$

Source originally Claimed · no numerical literature comparison is available.

**1508 2411.14531-T3-r2-mirror**

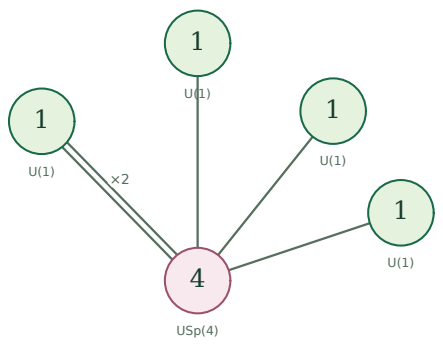


÷ $\mathbb{Z}_2$ diagonal

**Computed by Quiver Lab**

$1 + 5t^{2} + 45t^{4} + 216t^{6} + 988t^{8} + 3608t^{10} + 11954t^{12} + 34938t^{14} + 94123t^{16} + 233329t^{18} + 542933t^{20} + 1189668t^{22} + 2481366t^{24} + 4944470t^{26} + 9474992t^{28} + 17515690t^{30} + O(t^{31})$

Source originally Claimed · no numerical literature comparison is available.

**1509 2411.14531-T3-r3-electric**

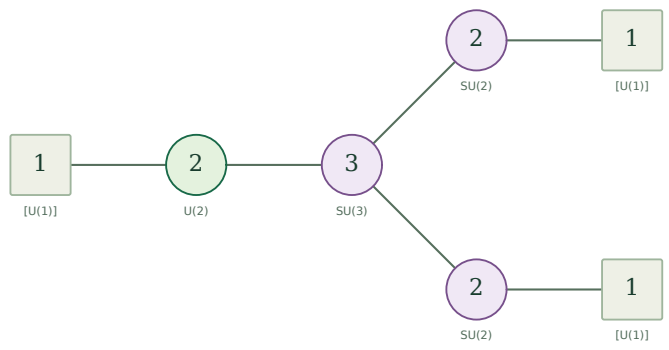


**Computed by Quiver Lab**

$1 + 3t^{2} + 30t^{4} + 134t^{6} + 623t^{8} + 2288t^{10} + 7659t^{12} + 22610t^{14} + 61400t^{16} + 153365t^{18} + 358953t^{20} + 790764t^{22} + 1656330t^{24} + 3313132t^{26} + 6368614t^{28} + 11806104t^{30} + O(t^{31})$

Source originally Claimed · no numerical literature comparison is available.

**1510 2411.14531-T3-r4-electric**

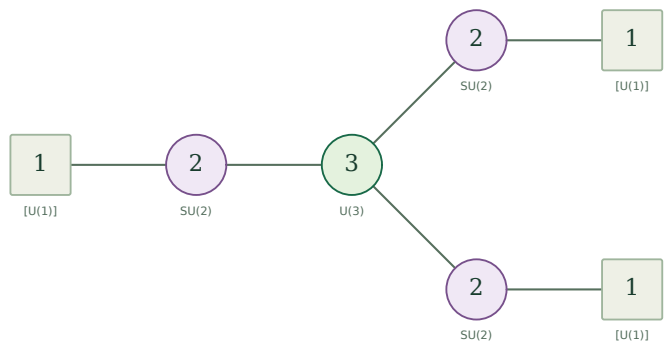


**Computed by Quiver Lab**

$1 + 3t^{2} + 30t^{4} + 134t^{6} + 623t^{8} + 2288t^{10} + 7659t^{12} + 22610t^{14} + 61400t^{16} + 153365t^{18} + 358953t^{20} + 790764t^{22} + 1656330t^{24} + 3313132t^{26} + 6368614t^{28} + 11806104t^{30} + O(t^{31})$

Source originally Claimed · no numerical literature comparison is available.

**1511 2411.14531-T3-r6-electric**

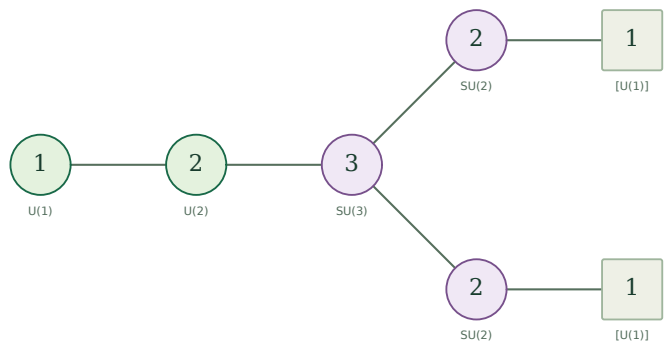


**Computed by Quiver Lab**

$1 + 8t^{2} + 72t^{4} + 458t^{6} + 2479t^{8} + 11157t^{10} + 43698t^{12} + 151156t^{14} + 472100t^{16} + 1349057t^{18} + 3573648t^{20} + O(t^{21})$

Source originally Claimed · no numerical literature comparison is available.

| QUIVER | HILBERT SERIES |
|---|---|

**1512 2411.14531-T3-r6-mirror**

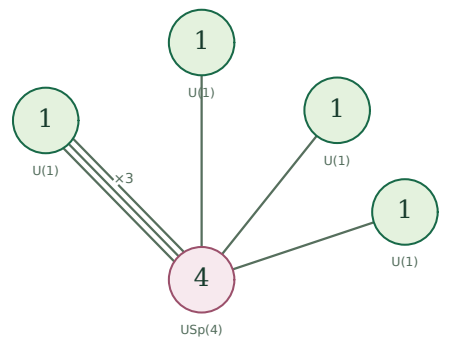


÷ $\mathbb{Z}_2$ diagonal

**Computed by Quiver Lab**

$1 + 4t^{2} + 18t^{4} + 75t^{6} + 270t^{8} + 843t^{10} + 2465t^{12} + 6576t^{14} + 16340t^{16} + 38177t^{18} + 84426t^{20} + 177532t^{22} + 357713t^{24} + 692725t^{26} + 1295006t^{28} + 2344969t^{30} + O(t^{31})$

Source originally Claimed · no numerical literature comparison is available.

**1513 2411.14531-T3-r7-electric**

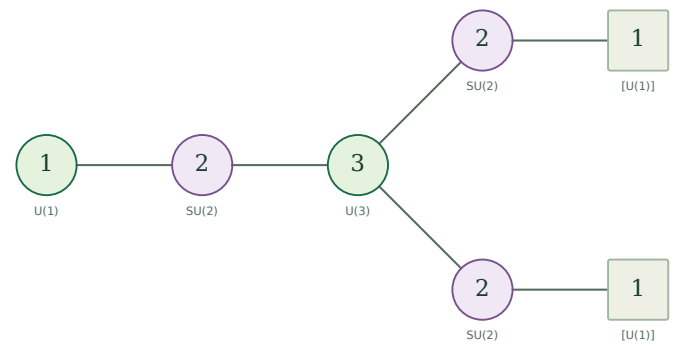


**Computed by Quiver Lab**

$1 + 6t^{2} + 62t^{4} + 384t^{6} + 2115t^{8} + 9507t^{10} + 37380t^{12} + 129354t^{14} + 404544t^{16} + 1156461t^{18} + 3065280t^{20} + 7600596t^{22} + 17787358t^{24} + 39542070t^{26} + 83999780t^{28} + 171315928t^{30} + O(t^{31})$

Source originally Claimed · no numerical literature comparison is available.

**1514 2411.14531-T3-r7-mirror**

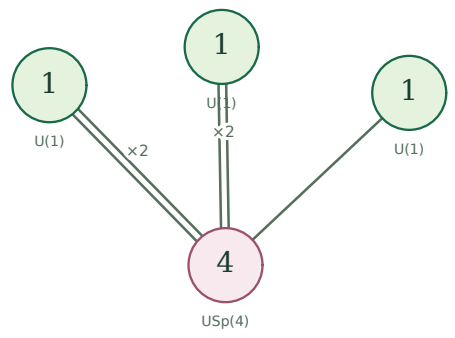


÷ $\mathbb{Z}_2$ diagonal

**Computed by Quiver Lab**

$1 + 4t^{2} + 24t^{4} + 89t^{6} + 326t^{8} + 964t^{10} + 2672t^{12} + 6582t^{14} + 15239t^{16} + 32684t^{18} + 66652t^{20} + 128767t^{22} + 238960t^{24} + 425872t^{26} + 734884t^{28} + 1228848t^{30} + O(t^{31})$

Source originally Claimed · no numerical literature comparison is available.

**1515 2411.14531-T3-r8-electric**

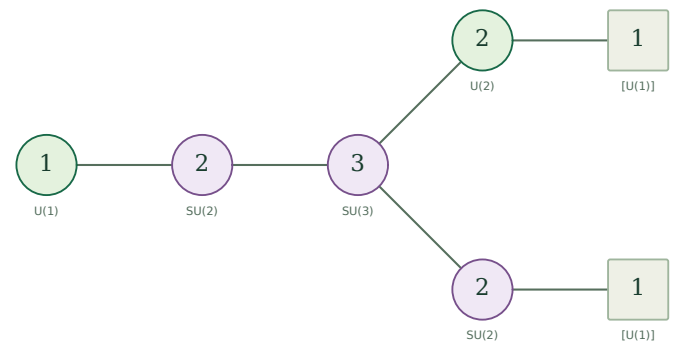


**Computed by Quiver Lab**

$1 + 6t^{2} + 62t^{4} + 384t^{6} + 2115t^{8} + 9507t^{10} + 37380t^{12} + 129354t^{14} + 404544t^{16} + 1156461t^{18} + 3065280t^{20} + 7600596t^{22} + 17787358t^{24} + 39542070t^{26} + 83999780t^{28} + 171315928t^{30} + O(t^{31})$

Source originally Claimed · no numerical literature comparison is available.

**1516 2411.14531-T3-r9-electric**

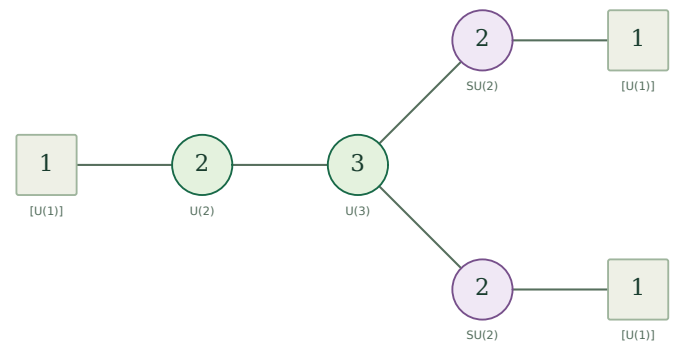


**Computed by Quiver Lab**

$1 + 8t^{2} + 72t^{4} + 458t^{6} + 2479t^{8} + 11157t^{10} + 43698t^{12} + 151156t^{14} + 472100t^{16} + 1349057t^{18} + 3573648t^{20} + 8858716t^{22} + 20725090t^{24} + 46064192t^{26} + 97835928t^{28} + 199508484t^{30} + O(t^{31})$

Source originally Claimed · no numerical literature comparison is available.

**1517 2411.14531-T3-r9-mirror**

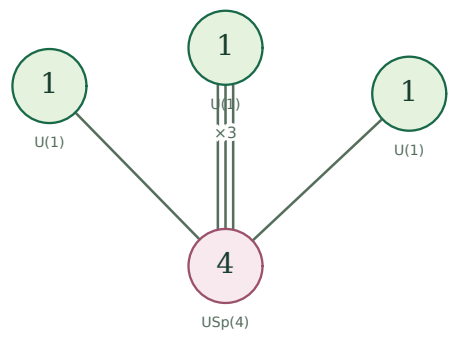


÷ $\mathbb{Z}_2$ diagonal

**Computed by Quiver Lab**

$1 + 4t^{2} + 28t^{4} + 105t^{6} + 388t^{8} + 1156t^{10} + 3200t^{12} + 7888t^{14} + 18227t^{16} + 39068t^{18} + 79536t^{20} + 153527t^{22} + 284540t^{24} + 506708t^{26} + 873528t^{28} + 1459728t^{30} + O(t^{31})$

Source originally Claimed · no numerical literature comparison is available.

**1518 2411.14531-T3-r10-electric**

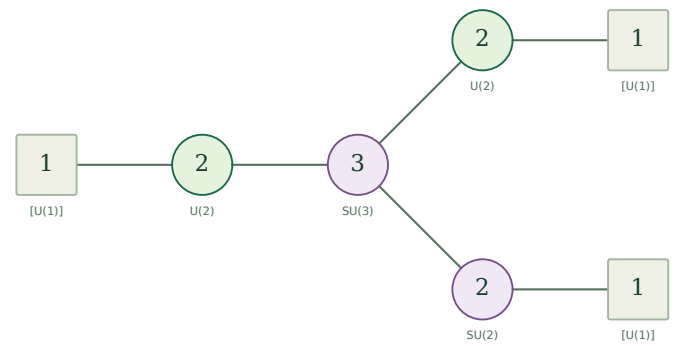


**Computed by Quiver Lab**

$1 + 6t^{2} + 62t^{4} + 384t^{6} + 2115t^{8} + 9507t^{10} + 37380t^{12} + 129354t^{14} + 404544t^{16} + 1156461t^{18} + 3065280t^{20} + O(t^{21})$

Source originally Claimed · no numerical literature comparison is available.

| QUIVER | HILBERT SERIES |
|---|---|

**1519 2411.14531-T3-r12-mirror**

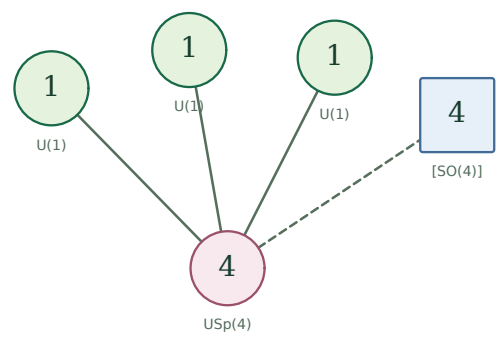


**Computed by Quiver Lab**

$1 + 4t^{2} + 24t^{4} + 89t^{6} + 326t^{8} + 964t^{10} + 2672t^{12} + 6582t^{14} + 15239t^{16} + 32684t^{18} + 66652t^{20} + 128767t^{22} + 238960t^{24} + 425872t^{26} + 734884t^{28} + 1228848t^{30} + O(t^{31})$

Source originally Claimed · no numerical literature comparison is available.

**1520 2411.14531-T4-r1-electric**

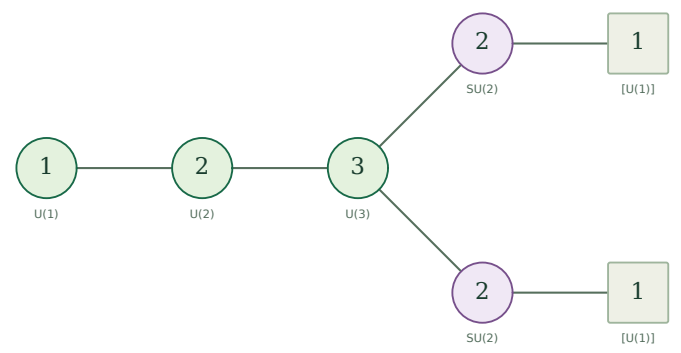


**Computed by Quiver Lab**

$1 + 15t^{2} + 171t^{4} + 1417t^{6} + 9348t^{8} + 50577t^{10} + 232895t^{12} + 935887t^{14} + 3354171t^{16} + 10900094t^{18} + 32561458t^{20} + 90396330t^{22} + 235349048t^{24} + 578937482t^{26} + 1354064898t^{28} + 3027151934t^{30} + O(t^{31})$

Source originally Claimed · no numerical literature comparison is available.

**1521 2411.14531-T4-r1-mirror**

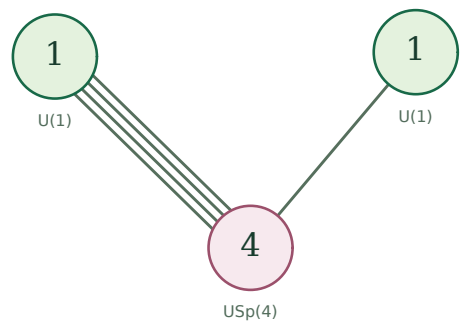


÷ $\mathbb{Z}_2$ diagonal

**Computed by Quiver Lab**

$1 + 3t^{2} + 16t^{4} + 45t^{6} + 135t^{8} + 320t^{10} + 740t^{12} + 1516t^{14} + 2997t^{16} + 5499t^{18} + 9764t^{20} + 16481t^{22} + 27055t^{24} + 42812t^{26} + 66196t^{28} + 99536t^{30} + O(t^{31})$

Source originally Claimed · no numerical literature comparison is available.

**1522 2411.14531-T4-r2-electric**

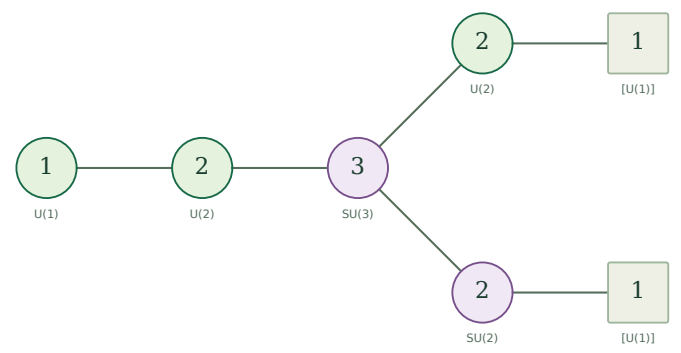


**Computed by Quiver Lab**

$1 + 11t^{2} + 135t^{4} + 1113t^{6} + 7436t^{8} + 40365t^{10} + 186577t^{12} + 751087t^{14} + 2696031t^{16} + 8769520t^{18} + 26217170t^{20} + O(t^{21})$

Source originally Claimed · no numerical literature comparison is available.

**1523 2411.14531-T4-r2-mirror**

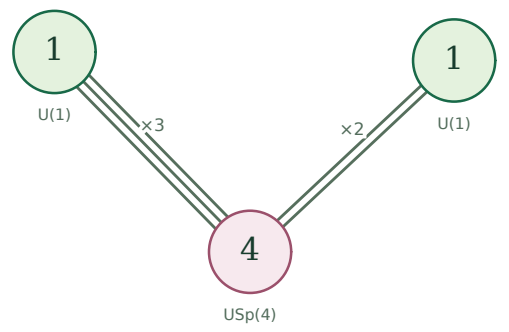


÷ $\mathbb{Z}_2$ diagonal

**Computed by Quiver Lab**

$1 + 3t^{2} + 12t^{4} + 35t^{6} + 103t^{8} + 246t^{10} + 566t^{12} + 1168t^{14} + 2307t^{16} + 4253t^{18} + 7558t^{20} + 12795t^{22} + 21023t^{24} + 33338t^{26} + 51586t^{28} + 77684t^{30} + O(t^{31})$

Source originally Claimed · no numerical literature comparison is available.

**1524 2411.14531-T4-r3-electric**

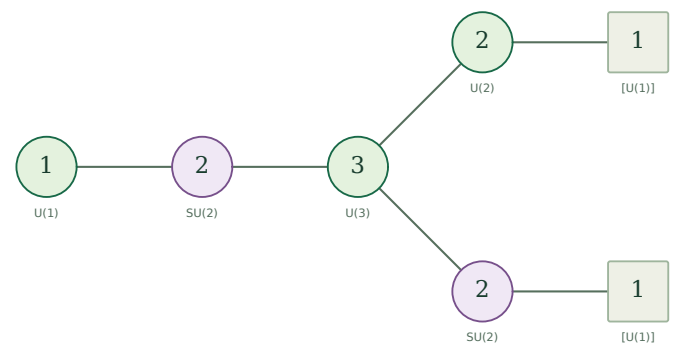


**Computed by Quiver Lab**

$1 + 11t^{2} + 135t^{4} + 1113t^{6} + 7436t^{8} + 40365t^{10} + 186577t^{12} + 751087t^{14} + 2696031t^{16} + 8769520t^{18} + 26217170t^{20} + 72822672t^{22} + 189680008t^{24} + 466752714t^{26} + 1091985516t^{28} + 2441793930t^{30} + O(t^{31})$

Source originally Claimed · no numerical literature comparison is available.

**1525 2411.14531-T4-r4-electric**

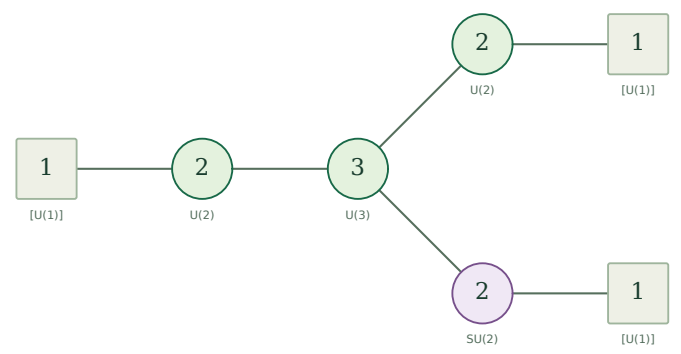


**Computed by Quiver Lab**

$1 + 15t^{2} + 171t^{4} + 1417t^{6} + 9348t^{8} + 50577t^{10} + 232895t^{12} + 935887t^{14} + 3354171t^{16} + 10900094t^{18} + 32561458t^{20} + 90396330t^{22} + 235349048t^{24} + 578937482t^{26} + 1354064898t^{28} + 3027151934t^{30} + O(t^{31})$

Source originally Claimed · no numerical literature comparison is available.

| QUIVER | HILBERT SERIES |
|---|---|

**1526 2411.14531-T4-r5-mirror**

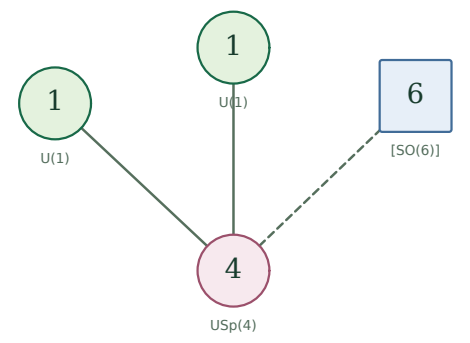


**Computed by Quiver Lab**

$1 + 3t^{2} + 16t^{4} + 45t^{6} + 135t^{8} + 320t^{10} + 740t^{12} + 1516t^{14} + 2997t^{16} + 5499t^{18} + 9764t^{20} + 16481t^{22} + 27055t^{24} + 42812t^{26} + 66196t^{28} + 99536t^{30} + O(t^{31})$

Source originally Claimed · no numerical literature comparison is available.

**1527 2411.14531-T4-r6-electric**

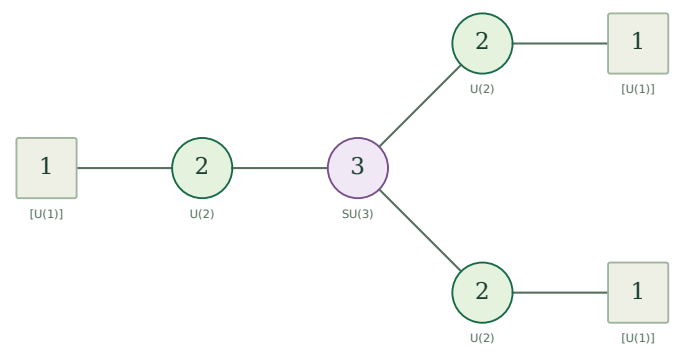


**Computed by Quiver Lab**

$1 + 9t^{2} + 119t^{4} + 963t^{6} + 6466t^{8} + 35021t^{10} + 161959t^{12} + 651667t^{14} + 2339263t^{16} + 7607852t^{18} + 22743826t^{20} + O(t^{21})$

Source originally Claimed · no numerical literature comparison is available.

**1528 2411.14531-T4-r6-mirror**

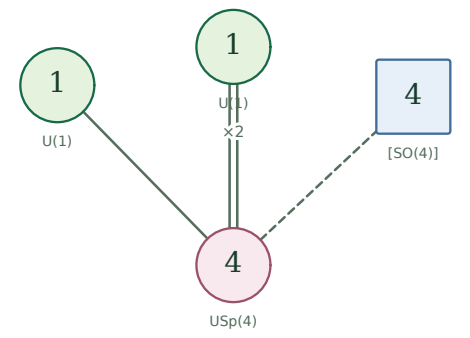


**Computed by Quiver Lab**

$1 + 3t^{2} + 12t^{4} + 31t^{6} + 93t^{8} + 216t^{10} + 498t^{12} + 1010t^{14} + 2009t^{16} + 3681t^{18} + 6554t^{20} + 11055t^{22} + 18203t^{24} + 28818t^{26} + 44638t^{28} + 67144t^{30} + O(t^{31})$

Source originally Claimed · no numerical literature comparison is available.

**1529 2411.14531-T4-r7-electric**

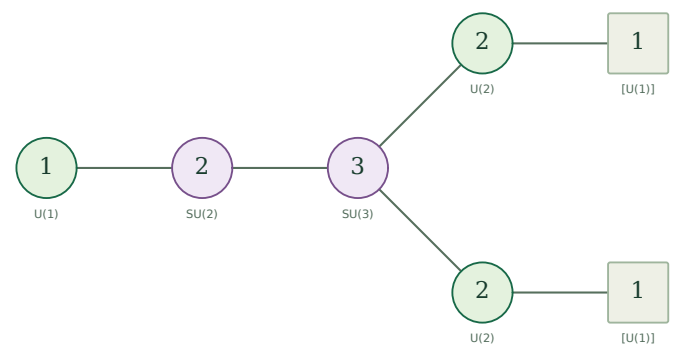


**Computed by Quiver Lab**

$1 + 9t^{2} + 119t^{4} + 963t^{6} + 6466t^{8} + 35021t^{10} + 161959t^{12} + 651667t^{14} + 2339263t^{16} + 7607852t^{18} + 22743826t^{20} + O(t^{21})$

Source originally Claimed · no numerical literature comparison is available.

**1530 2411.14531-T4-r8-mirror**

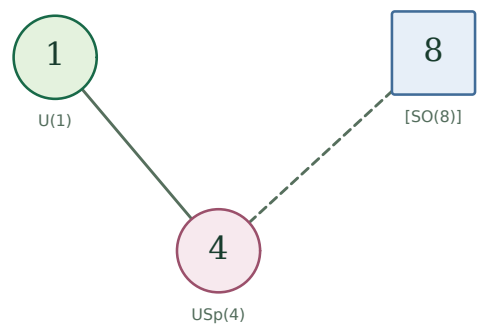


**Computed by Quiver Lab**

$1 + 2t^{2} + 9t^{4} + 17t^{6} + 44t^{8} + 77t^{10} + 154t^{12} + 250t^{14} + 431t^{16} + 656t^{18} + 1027t^{20} + 1483t^{22} + 2170t^{24} + 3003t^{26} + 4180t^{28} + 5588t^{30} + O(t^{31})$

Source originally Claimed · no numerical literature comparison is available.

**1531 2411.14531-T4-r9-electric**

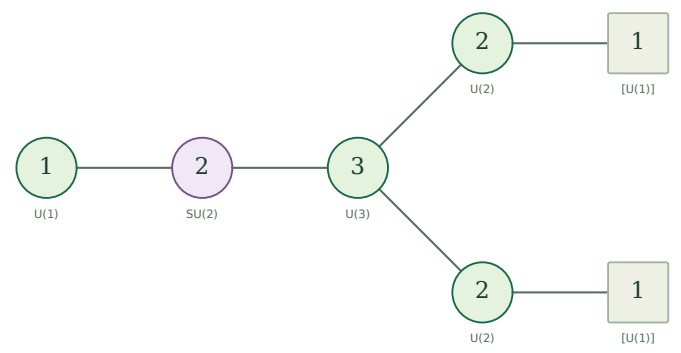


**Computed by Quiver Lab**

$1 + 18t^{2} + 291t^{4} + 3110t^{6} + 25472t^{8} + 165695t^{10} + 898514t^{12} + 4185119t^{14} + 17174684t^{16} + 63274668t^{18} + 212516498t^{20} + 658632938t^{22} + 1902503218t^{24} + O(t^{25})$

Source originally Claimed · no numerical literature comparison is available.

**1532 2411.14531-T4-r9-mirror**

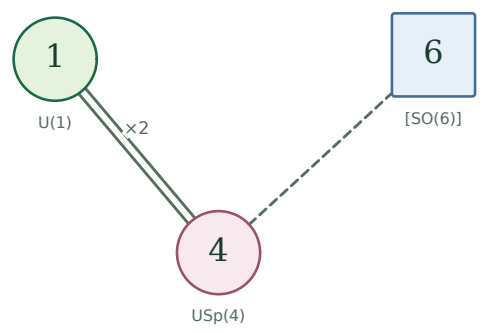


**Computed by Quiver Lab**

$1 + 2t^{2} + 5t^{4} + 11t^{6} + 28t^{8} + 51t^{10} + 96t^{12} + 162t^{14} + 277t^{16} + 430t^{18} + 663t^{20} + 971t^{22} + 1416t^{24} + 1977t^{26} + 2736t^{28} + 3682t^{30} + O(t^{31})$

Source originally Claimed · no numerical literature comparison is available.

| QUIVER | HILBERT SERIES |
| --- | --- |

**1533 2411.14531-T4-r10-electric**

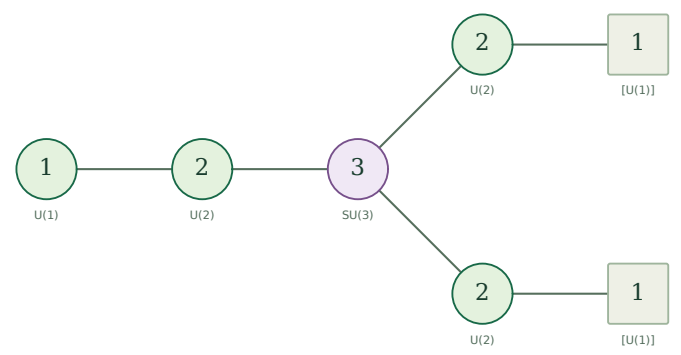


**Computed by Quiver Lab**

$1 + 14t^2 + 239t^4 + 2534t^6 + 20804t^8 + 135275t^{10} + 733796t^{12} + 3417737t^{14} + 14026586t^{16} + 51676472t^{18} + 173566322t^{20} + O(t^{21})$

Source originally Claimed · no numerical literature comparison is available.

**1534 2411.14531-T4-r10-mirror**

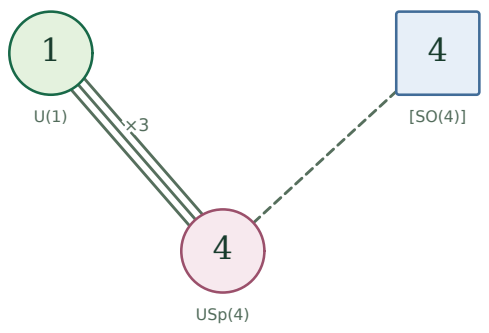


**Computed by Quiver Lab**

$1 + 2t^2 + 5t^4 + 9t^6 + 22t^8 + 41t^{10} + 80t^{12} + 132t^{14} + 223t^{16} + 346t^{18} + 539t^{20} + 787t^{22} + 1146t^{24} + 1599t^{26} + 2220t^{28} + 2984t^{30} + O(t^{31})$

Source originally Claimed · no numerical literature comparison is available.

**1535 2411.14531-T4-r11-electric**

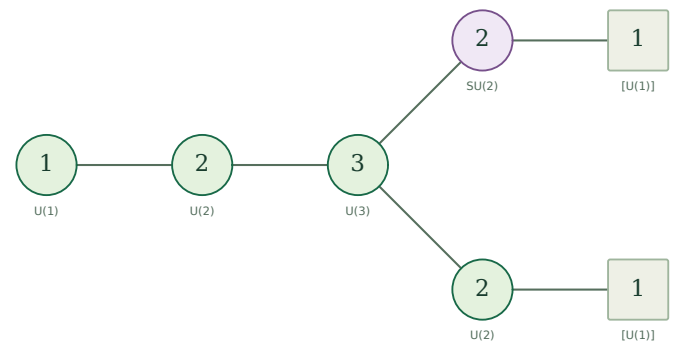


**Computed by Quiver Lab**

$1 + 24t^2 + 375t^4 + 4024t^6 + 32726t^8 + 212429t^{10} + 1149346t^{12} + 5347199t^{14} + 21922724t^{16} + 80717902t^{18} + 270970784t^{20} + O(t^{21})$

Source originally Claimed · no numerical literature comparison is available.

**1536 2411.14531-T4-r11-mirror**

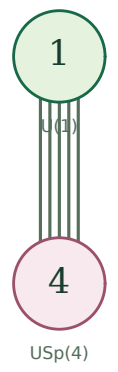


÷ $\mathbb{Z}_2$ diagonal

**Computed by Quiver Lab**

$1 + 2t^2 + 7t^4 + 15t^6 + 36t^8 + 67t^{10} + 128t^{12} + 216t^{14} + 363t^{16} + 566t^{18} + 873t^{20} + 1279t^{22} + 1854t^{24} + 2591t^{26} + 3584t^{28} + 4824t^{30} + O(t^{31})$

Source originally Claimed · no numerical literature comparison is available.

**1537 2411.14531-B4-SSSS**

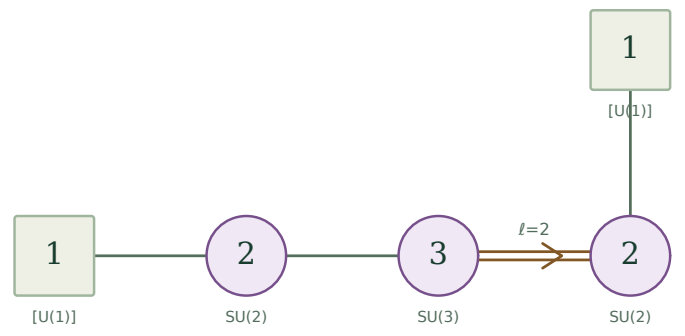


**Computed by Quiver Lab**

$1 + 11t^4 + 24t^6 + 95t^8 + 232t^{10} + 590t^{12} + 1256t^{14} + 2609t^{16} + 4920t^{18} + 8993t^{20} + 15472t^{22} + 25871t^{24} + 41488t^{26} + 64956t^{28} + 98608t^{30} + O(t^{31})$

Source originally Claimed · no numerical literature comparison is available.

**1538 2411.14531-B4-UUUS**

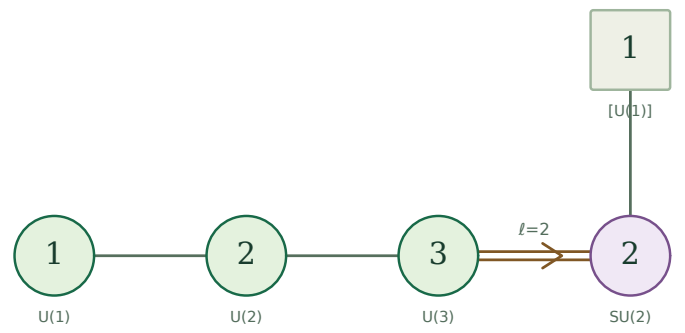


**Computed by Quiver Lab**

$1 + 15t^2 + 155t^4 + 1173t^6 + 7004t^8 + 34276t^{10} + 142465t^{12} + 516975t^{14} + 1674857t^{16} + 4930119t^{18} + 13372514t^{20} + 33801790t^{22} + 80356400t^{24} + 181018448t^{26} + 388829678t^{28} + 800574386t^{30} + O(t^{31})$

Source originally Claimed · no numerical literature comparison is available.

**1539 2411.14531-B4-USUU**

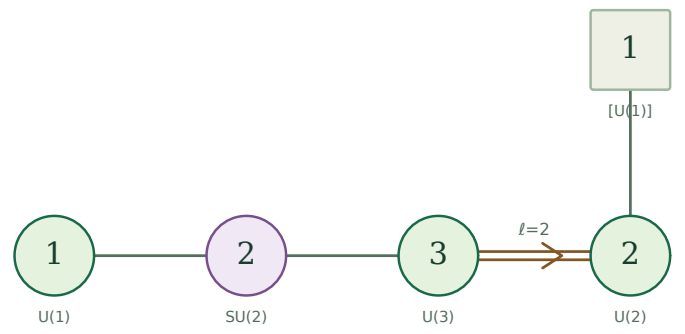


**Computed by Quiver Lab**

$1 + 13t^2 + 173t^4 + 1413t^6 + 9002t^8 + 45322t^{10} + 192185t^{12} + 704865t^{14} + 2300927t^{16} + 6804107t^{18} + 18517750t^{20} + 46911654t^{22} + 111710804t^{24} + 251950292t^{26} + 541696374t^{28} + 1116086214t^{30} + O(t^{31})$

Source originally Claimed · no numerical literature comparison is available.

| QUIVER | HILBERT SERIES |
| --- | --- |

**1540 2411.14531-eq264-part1**

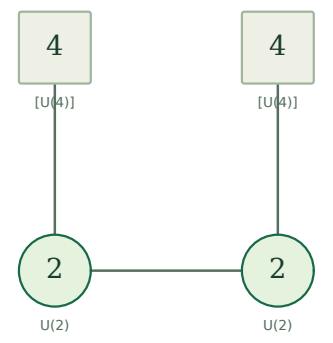


**Computed by Quiver Lab**

$1 + 2t^{2} + 9t^{4} + 22t^{6} + 58t^{8} + 126t^{10} + 275t^{12} + 532t^{14} + 1017t^{16} + 1804t^{18} + 3132t^{20} + 5174t^{22} + 8369t^{24} + 13048t^{26} + 19969t^{28} + 29718t^{30} + O(t^{31})$

Source originally Claimed · no numerical literature comparison is available.

**1541 2411.14531-eq264-part2**

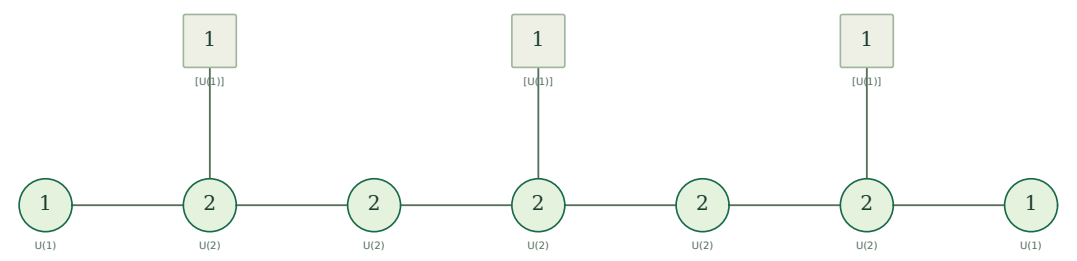

**Computed by Quiver Lab**

$1 + 31t^{2} + 32t^{3} + 495t^{4} + 960t^{5} + 5890t^{6} + 14496t^{7} + 58226t^{8} + 151712t^{9} + 486085t^{10} + 1244032t^{11} + 3460024t^{12} + 8471584t^{13} + 21324844t^{14} + 49590944t^{15} + 115652928t^{16} + O(t^{17})$

Source originally Claimed · no numerical literature comparison is available.

**1542 2411.14531-eq264-part3**

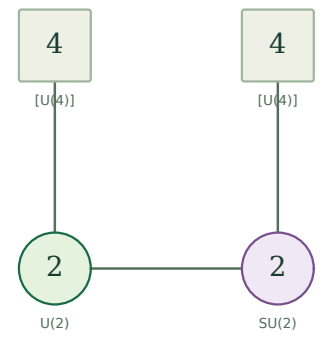


**Computed by Quiver Lab**

$1 + t^{2} + 5t^{4} + 7t^{6} + 18t^{8} + 28t^{10} + 57t^{12} + 85t^{14} + 149t^{16} + 215t^{18} + 340t^{20} + 476t^{22} + 701t^{24} + 949t^{26} + 1327t^{28} + 1747t^{30} + O(t^{31})$

Source originally Claimed · no numerical literature comparison is available.

**1543 2411.14531-eq264-part4**

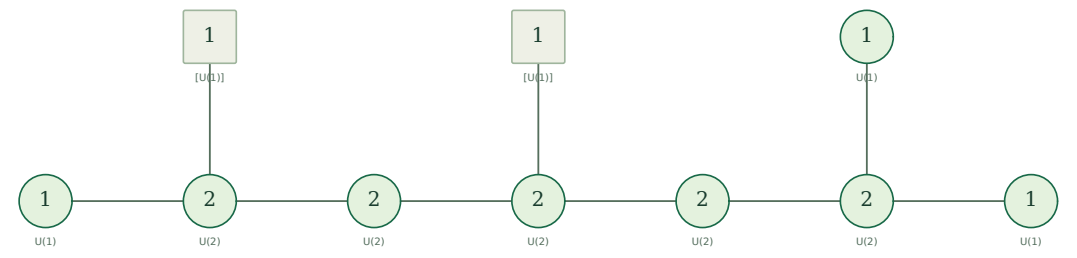

**Computed by Quiver Lab**

$1 + 44t^{2} + 64t^{3} + 931t^{4} + 2304t^{5} + 14181t^{6} + 41216t^{7} + 172451t^{8} + 504704t^{9} + 1709740t^{10} + 4779200t^{11} + 14100952t^{12} + O(t^{13})$

Source originally Claimed · no numerical literature comparison is available.

**1544 2411.14531-eq264-part5**

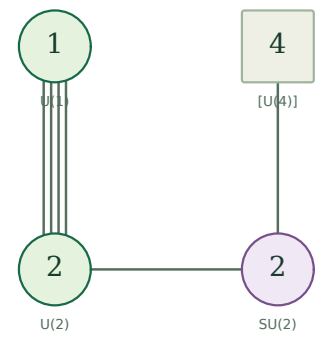


**Computed by Quiver Lab**

$1 + 2t^{2} + 11t^{4} + 24t^{6} + 72t^{8} + 148t^{10} + 339t^{12} + 638t^{14} + 1243t^{16} + 2164t^{18} + 3794t^{20} + 6192t^{22} + 10061t^{24} + 15562t^{26} + 23865t^{28} + 35316t^{30} + O(t^{31})$

Source originally Claimed · no numerical literature comparison is available.

**1545 2411.14531-eq264-part6**

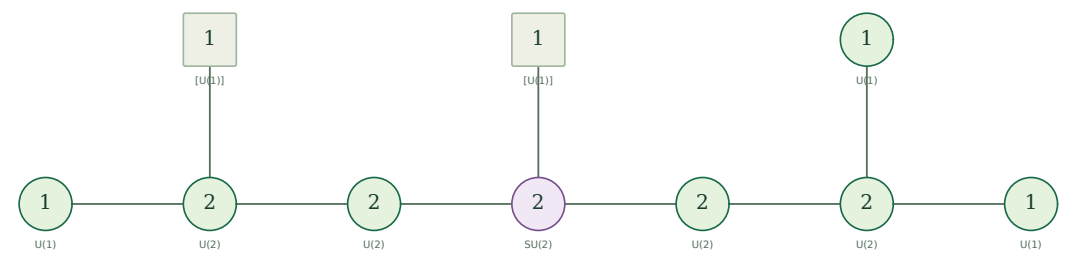

**Computed by Quiver Lab**

$1 + 43t^{2} + 875t^{4} + 12066t^{6} + 127666t^{8} + O(t^{9})$

Source originally Claimed · no numerical literature comparison is available.

## 2412.15202 · Pathway to decay and fission of orthosymplectic quiver theories (2)

**1546 2412.15202-eq2p9-star**

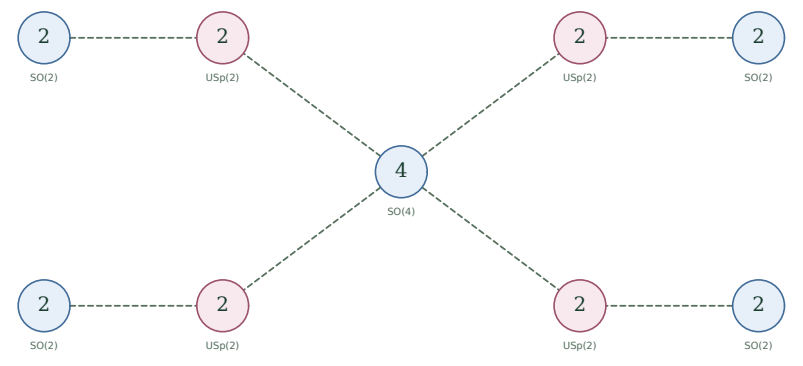

÷ $\mathbb{Z}_2$ diagonal

**Computed by Quiver Lab**

$1 + 56t^{2} + 1384t^{4} + 20650t^{6} + 215636t^{8} + 1720264t^{10} + 11106025t^{12} + 60413096t^{14} + 285191074t^{16} + 1194750560t^{18} + 4519533676t^{20} + 15651920804t^{22} + 50180163740t^{24} + 150296952736t^{26} + 423754717634t^{28} + 1131857399624t^{30} + O(t^{31})$

Source originally Claimed · no numerical literature comparison is available.

| QUIVER | HILBERT SERIES |
| --- | --- |

**1547 2412.15202-eq2p9-product**

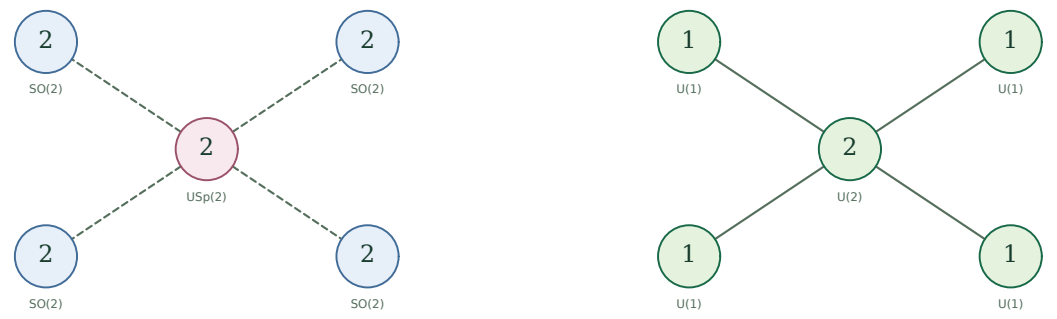


÷ U(1) diagonal on {1_c, 1_u0, 1_u1, 1_u2, 1_u3} · ÷ Z2 diagonal on {0_c, 0_o0, 0_o1, 0_o2, 0_o3}

**Computed by Quiver Lab**

$1 + 56t^2 + 1384t^4 + 20650t^6 + 215636t^8 + 1720264t^{10} + 11106025t^{12} + 60413096t^{14} + 285191074t^{16} + 1194750560t^{18} + 4519533676t^{20} + 15651920804t^{22} + 50180163740t^{24} + 150296952736t^{26} + 423754717634t^{28} + 1131857399624t^{30} + O(t^{31})$

Source originally Claimed · no numerical literature comparison is available.

## 2505.03875 · Orthosymplectic Quivers: Indices, Hilbert Series, and Generalised Symmetries (65)

**1548 2505.03875-SO4-3F-O+**

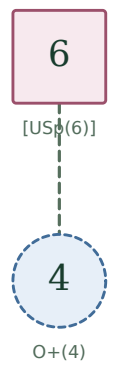


g0: θ = +

$1 + t^2 + 3t^4 + 4t^6 + 9t^8 + 11t^{10} + 19t^{12} + 24t^{14} + 37t^{16} + 45t^{18} + 63t^{20} + 76t^{22} + 101t^{24} + 119t^{26} + 151t^{28} + 176t^{30} + O(t^{31})$

**1549 2505.03875-SO4-3F-O-**

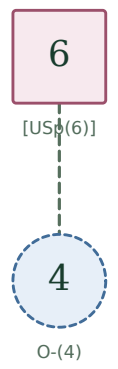


g0: θ = -

$1 + 3t^4 + 2t^6 + 9t^8 + 8t^{10} + 19t^{12} + 20t^{14} + 37t^{16} + 40t^{18} + 63t^{20} + 70t^{22} + 101t^{24} + 112t^{26} + 151t^{28} + 168t^{30} + O(t^{31})$

**1550 2505.03875-SO4-3F-Pin**

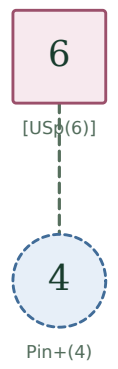


$1 + 2t^4 + t^6 + 6t^8 + 4t^{10} + 11t^{12} + 10t^{14} + 21t^{16} + 20t^{18} + 34t^{20} + 35t^{22} + 54t^{24} + 56t^{26} + 79t^{28} + 84t^{30} + O(t^{31})$

**1551 2505.03875-D2C2-3half-SO-Z2**

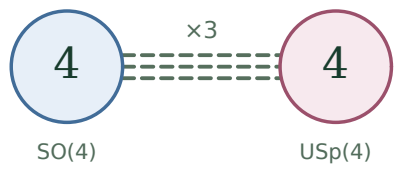


÷ ℤ₂ diagonal

$1 + 6t^4 + 6t^6 + 31t^8 + 46t^{10} + 130t^{12} + 212t^{14} + 459t^{16} + 746t^{18} + 1384t^{20} + 2178t^{22} + 3659t^{24} + 5546t^{26} + 8686t^{28} + 12700t^{30} + O(t^{31})$

**1552 2505.03875-D2C2-3half-O+-Z2**

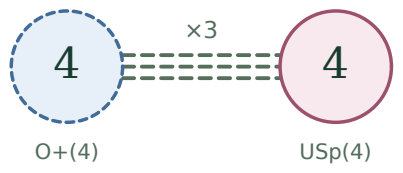


g0: θ = + · ÷ ℤ₂ diagonal

$1 + 3t^4 + 2t^6 + 14t^8 + O(t^9)$

**Paper typo:** Averaging the parent expression leaves 3ζ at $t^8$, which Eq. (4.7) drops. The unrefined coefficient is therefore 17 rather than 14, in agreement with independent magnetic sums.

| QUIVER | HILBERT SERIES |
|---|---|

**1553 2505.03875-D2C2-3half-Spin-Z2**

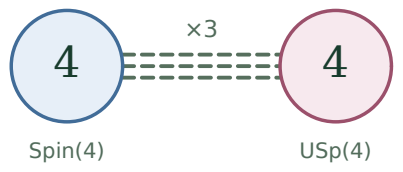


÷ $\mathbb{Z}_2$ diagonal

$1 + 5t^{4} + 3t^{6} + 20t^{8} + O(t^{9})$

**1554 2505.03875-D2C2-3half-O--Z2**

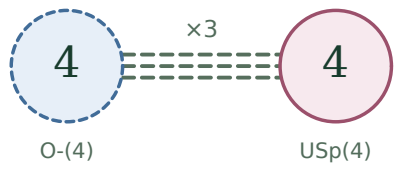


g0: θ = - · ÷ $\mathbb{Z}_2$ diagonal

$1 + 3t^{4} + 3t^{6} + 16t^{8} + 23t^{10} + 65t^{12} + 106t^{14} + 231t^{16} + 373t^{18} + 692t^{20} + 1089t^{22} + 1831t^{24} + 2773t^{26} + 4343t^{28} + 6350t^{30} + O(t^{31})$

**1555 2505.03875-D2C2-3half-SO**

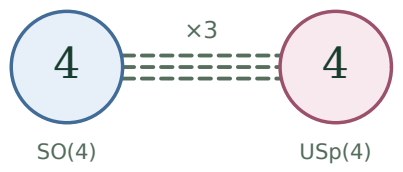


$1 + 4t^{4} + 2t^{6} + 19t^{8} + 22t^{10} + 72t^{12} + 102t^{14} + 243t^{16} + 368t^{18} + 714t^{20} + 1078t^{22} + 1865t^{24} + 2760t^{26} + 4394t^{28} + 6326t^{30} + O(t^{31})$

**1556 2505.03875-D2C2-3half-O+**

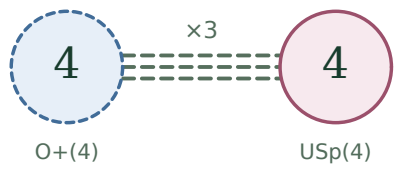


g0: θ = +

$1 + 2t^{4} + 11t^{8} + O(t^{9})$

**1557 2505.03875-D2C2-3half-Spin**

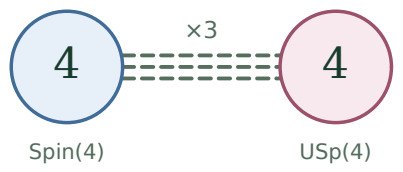


$1 + 4t^{4} + t^{6} + 14t^{8} + O(t^{9})$

**1558 2505.03875-D2C2-3half-O-**

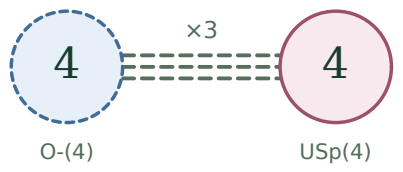


g0: θ = -

$1 + 2t^{4} + t^{6} + 10t^{8} + O(t^{9})$

**1559 2505.03875-D2C2-3half-Pin**

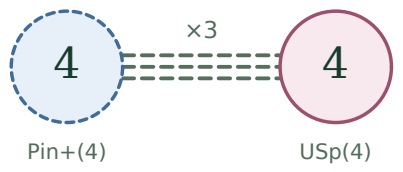


$1 + 2t^{4} + 8t^{8} + O(t^{9})$

| QUIVER | HILBERT SERIES |
|---|---|
| **1560 2505.03875-O1-USp2-SO5-delta1** 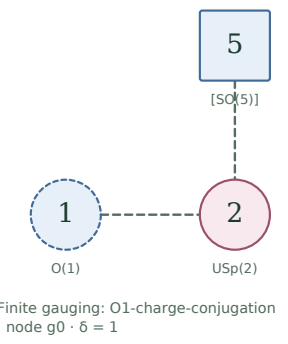 g0: δ = 1 · Finite gauging: O1-charge-conjugation · node g0 · δ = 1 | $1 + t^2 + 3t^4 + 3t^6 + 5t^8 + 5t^{10} + 7t^{12} + 7t^{14} + 9t^{16} + 9t^{18} + 11t^{20} + 11t^{22} + 13t^{24} + 13t^{26} + 15t^{28} + 15t^{30} + O(t^{31})$ |
| **1561 2505.03875-O1-double-USp2-SO4** 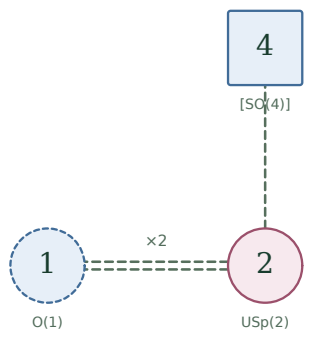 | $1 + t^2 + 3t^4 + 3t^6 + O(t^7)$ |
| **1562 2505.03875-O1-double-USp2-double-O1** 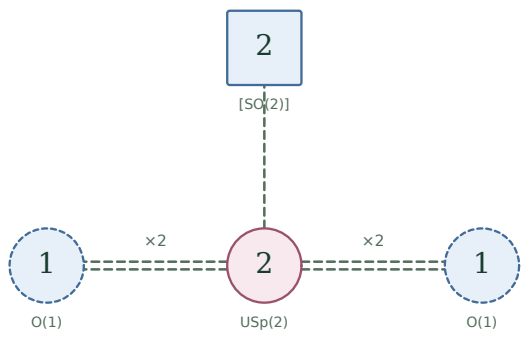 | $1 + t^2 + 3t^4 + 3t^6 + O(t^7)$ |
| **1563 2505.03875-D1C1D1-O+-O+** 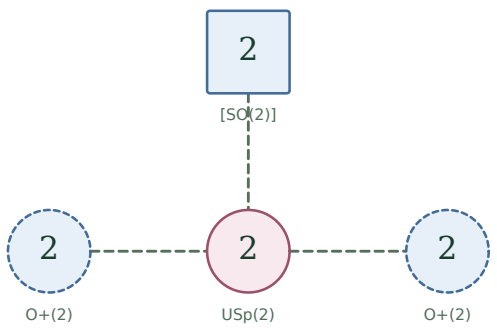 g0: θ = + · g2: θ = + | $1 + 3t^2 + 24t^4 + 72t^6 + O(t^7)$ |
| **1564 2505.03875-D1C1D1-diagC** 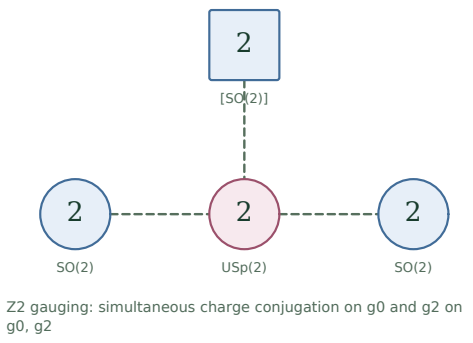 Z2 gauging: simultaneous charge conjugation on g0 and g2 on g0, g2 | $1 + 7t^2 + 44t^4 + 148t^6 + O(t^7)$ |
| **1565 2505.03875-OSpD5-Spin-Z2** 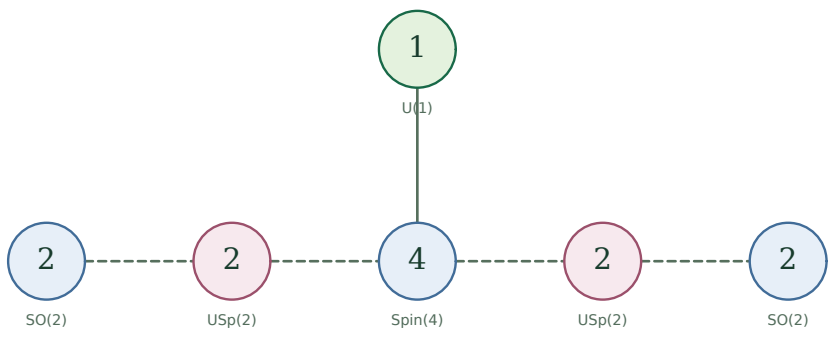 ÷ $\mathbb{Z}_2$ diagonal | $1 + 21t^2 + 394t^4 + O(t^5)$ |

| QUIVER | HILBERT SERIES |
|---|---|

**1566 2505.03875-OSpD5-O--Z2**

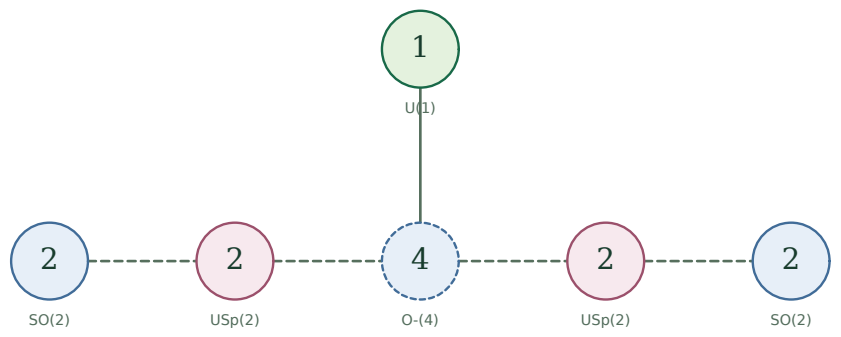


g2: θ = - · ÷ $\mathbb{Z}_2$ diagonal

$1 + 21t^2 + 394t^4 + O(t^5)$

**1567 2505.03875-OSpD5-Spin**

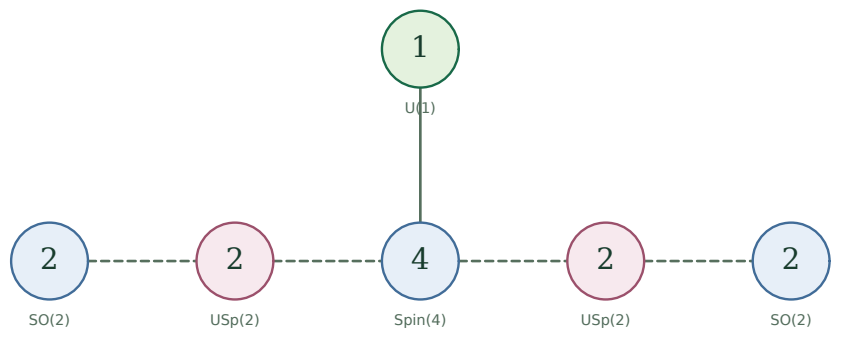


$1 + 13t^2 + 226t^4 + O(t^5)$

**1568 2505.03875-OSpD5-O-**

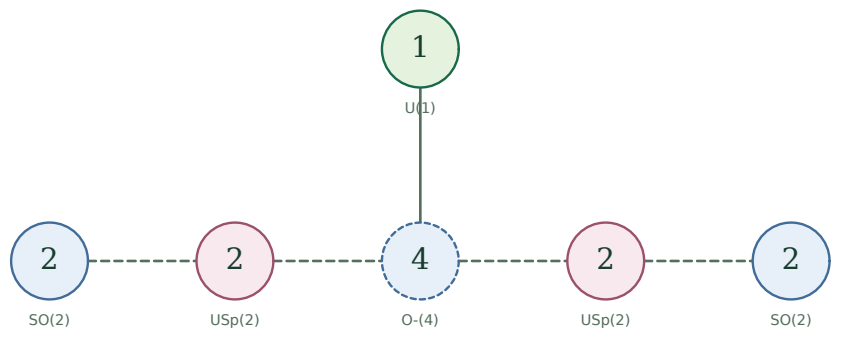


g2: θ = -

$1 + 13t^2 + 226t^4 + O(t^5)$

**1569 2505.03875-OSpD5-Pin-Z2**

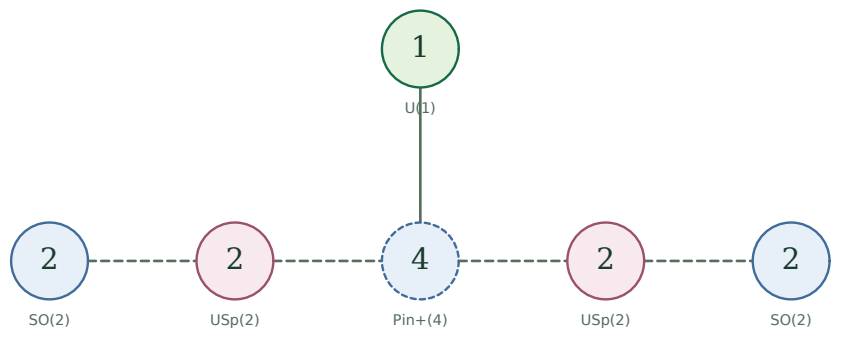


Source-prescribed Pin character average · ÷ $\mathbb{Z}_2$ diagonal

$1 + 11t^2 + 204t^4 + 1900t^6 + O(t^7)$

Conditional comparison recorded in the source database.

**1570 2505.03875-OSpD5-Pin**

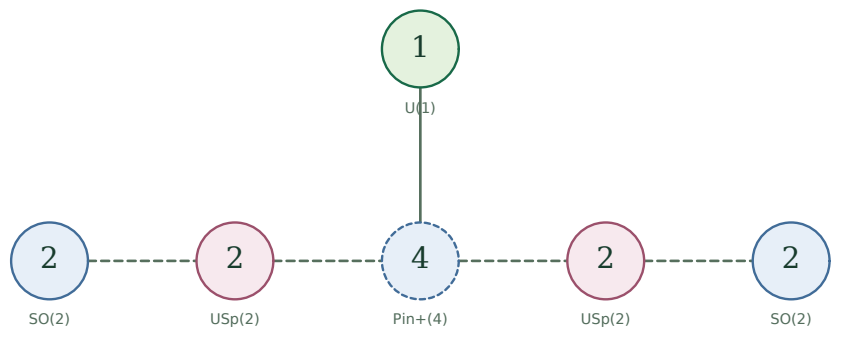


$1 + 7t^2 + 120t^4 + 1004t^6 + O(t^7)$

**1571 2505.03875-USp2-5F-discrete-1**

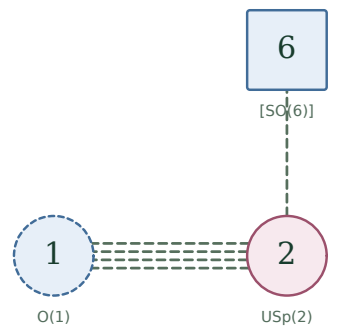


$1 + t^4 + O(t^5)$

**1572 2505.03875-USp2-5F-discrete-2**

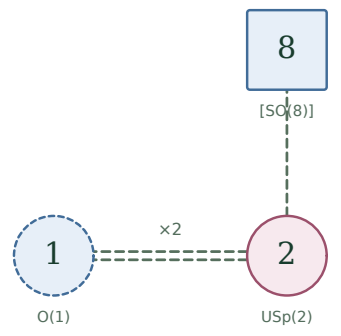


$1 + t^4 + O(t^5)$

| QUIVER | HILBERT SERIES |
|---|---|

**1573 2505.03875-USp2-5F-discrete-3**

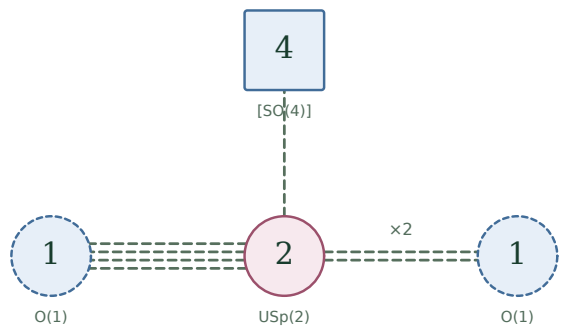


$1 + t^4 + O(t^5)$

**1574 2505.03875-SO3C2-O+**

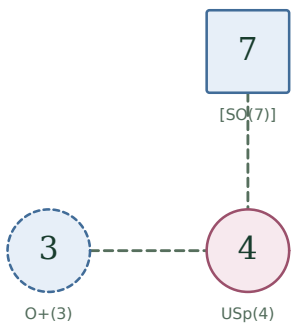


g0: θ = +

$1 + t^2 + 7t^4 + 11t^6 + 33t^8 + 53t^{10} + 114t^{12} + 178t^{14} + 319t^{16} + 475t^{18} + 761t^{20} + 1085t^{22} + 1611t^{24} + 2211t^{26} + 3108t^{28} + 4132t^{30} + O(t^{31})$

**1575 2505.03875-TSO6-k1-Spin**

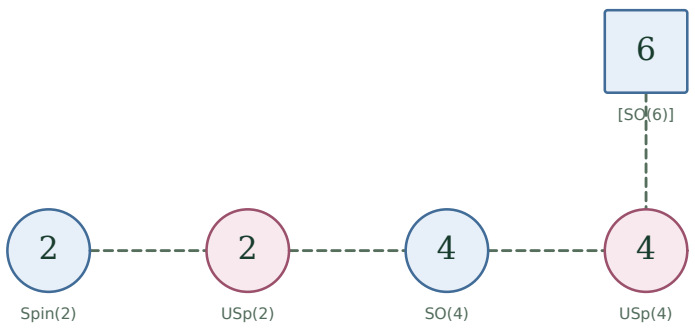


$1 + 7t^2 + 63t^4 + O(t^5)$

**1576 2505.03875-TSO6-mirror-k1-Spin**

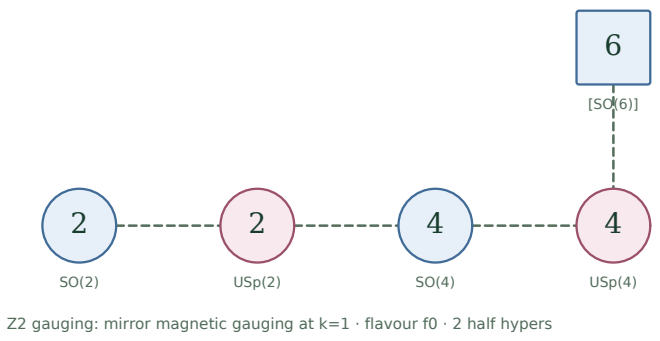


Z2 gauging: mirror magnetic gauging at k=1 · flavour f0 · 2 half hypers

$1 + 15t^2 + 119t^4 + O(t^5)$

**1577 2505.03875-TSO6-k1-O+**

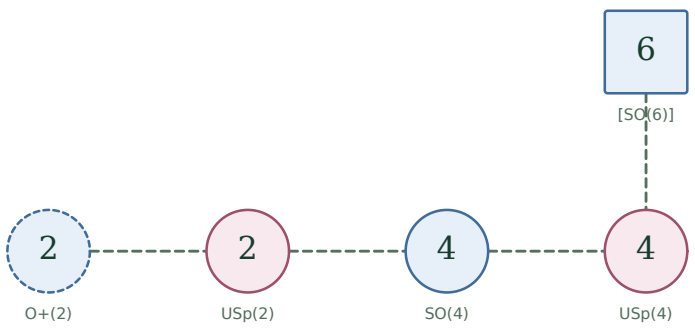


g0: θ = +

$1 + 7t^2 + 63t^4 + O(t^5)$

**1578 2505.03875-TSO6-mirror-k1-O+**

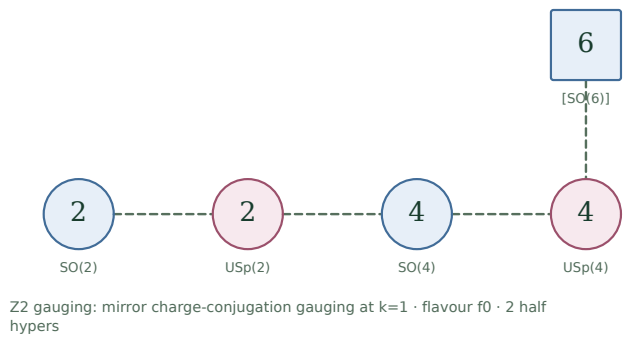


Z2 gauging: mirror charge-conjugation gauging at k=1 · flavour f0 · 2 half hypers

$1 + 15t^2 + 119t^4 + O(t^5)$

| QUIVER | HILBERT SERIES |
|---|---|

**1579 2505.03875-TSO6-mirror-k2-Spin**

$1 + 15t^2 + 119t^4 + O(t^5)$

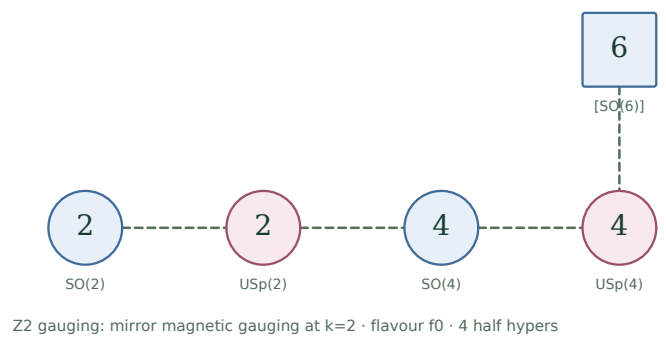


Z2 gauging: mirror magnetic gauging at k=2 · flavour f0 · 4 half hypers

**1580 2505.03875-TSO6-k2-O+**

$1 + 7t^2 + 63t^4 + O(t^5)$

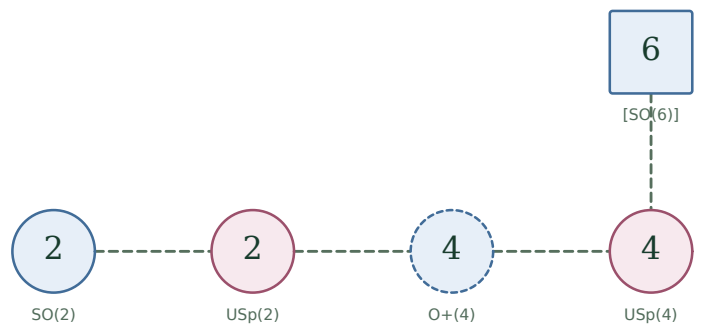


g2: θ = +

**1581 2505.03875-TSO6-mirror-k2-O+**

$1 + 15t^2 + 119t^4 + O(t^5)$

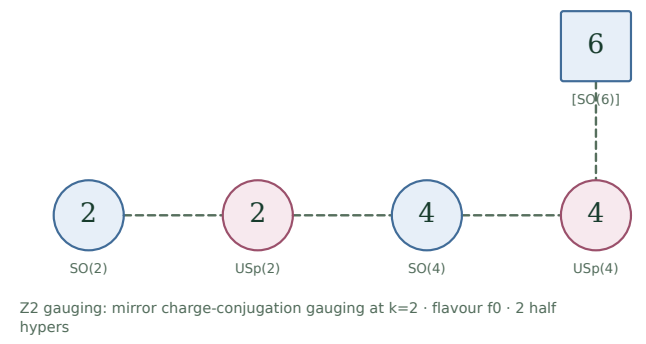


Z2 gauging: mirror charge-conjugation gauging at k=2 · flavour f0 · 2 half hypers

**1582 2505.03875-AppB-B3-5-1p2-Spin**

$1 + 15t^2 + 141t^4 + 1029t^6 + 6128t^8 + 30833t^{10} + 134417t^{12} + 518515t^{14} + 1800539t^{16} + 5708826t^{18} + 16721894t^{20} + O(t^{21})$

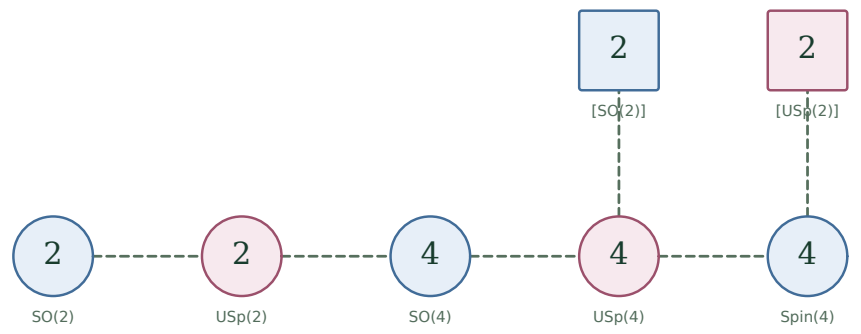


**1583 2505.03875-AppB-B3-5-1p2-O-**

$1 + 15t^2 + 141t^4 + 1029t^6 + 6128t^8 + 30833t^{10} + 134417t^{12} + 518515t^{14} + 1800539t^{16} + 5708826t^{18} + 16721894t^{20} + O(t^{21})$

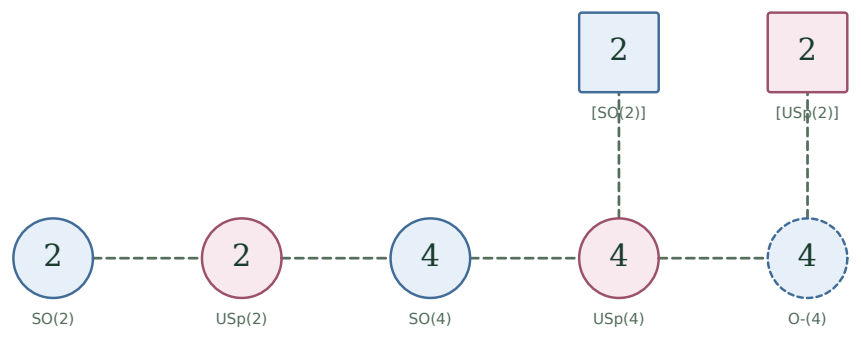


4: θ = -

**1584 2505.03875-AppB-B3-5-1p2-Pin**

$1 + 10t^2 + 85t^4 + 560t^6 + 3224t^8 + 15816t^{10} + 68281t^{12} + 261562t^{14} + 905467t^{16} + 2864436t^{18} + 8380950t^{20} + O(t^{21})$

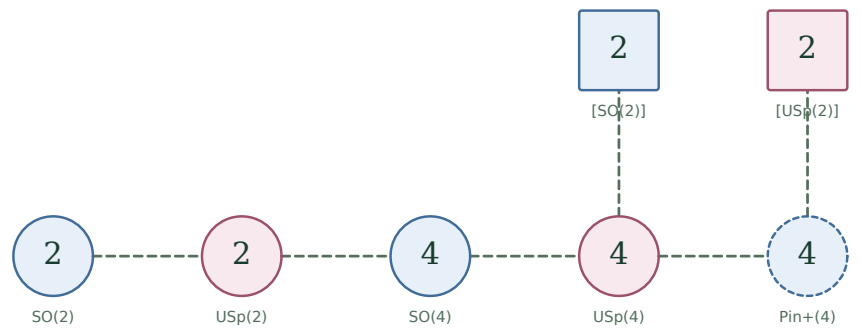

| QUIVER | HILBERT SERIES |
|---|---|

**1585 2505.03875-AppB-D4-3p2-1p2-Spin**

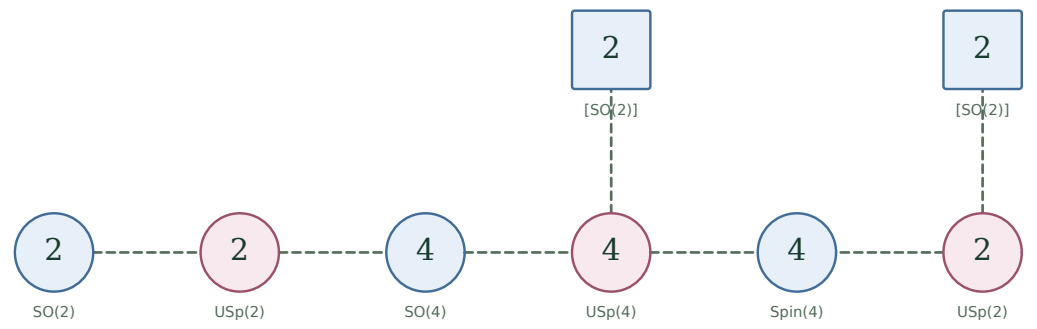


$1 + 16t^2 + 229t^4 + 2346t^6 + 18814t^8 + 121141t^{10} + 653060t^{12} + 3031145t^{14} + 12410258t^{16} + 45649332t^{18} + 153144258t^{20} + O(t^{21})$

**1586 2505.03875-AppB-D4-3p2-1p2-O-**

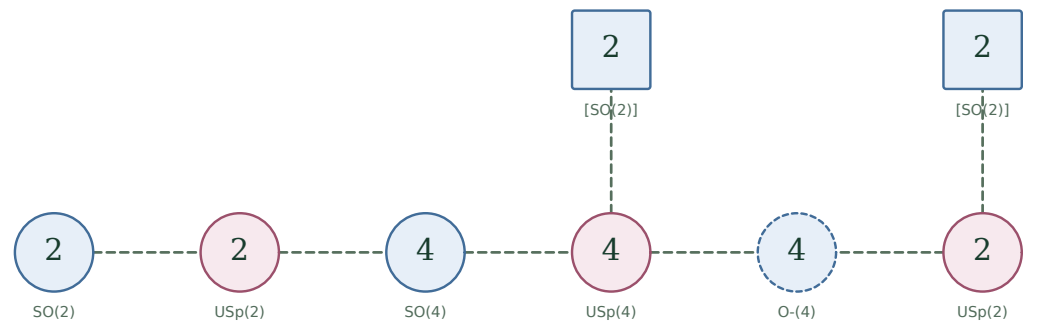


4: θ = -

$1 + 16t^2 + 229t^4 + 2346t^6 + 18814t^8 + 121141t^{10} + 653060t^{12} + 3031145t^{14} + 12410258t^{16} + 45649332t^{18} + 153144258t^{20} + O(t^{21})$

**1587 2505.03875-AppB-D4-3p2-1p2-Pin**

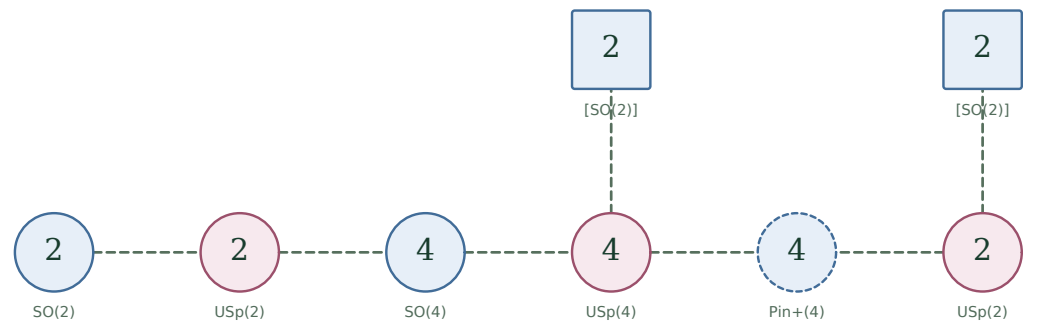


$1 + 10t^2 + 127t^4 + 1206t^6 + 9534t^8 + 60843t^{10} + 327330t^{12} + 1517077t^{14} + 6208822t^{16} + 22830996t^{18} + 76585802t^{20} + O(t^{21})$

**1588 2505.03875-AppB-B4-5-1p4-SO**

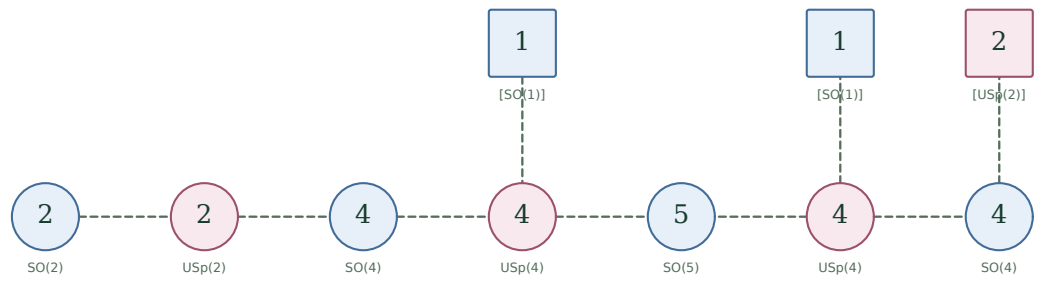


$1 + 36t^2 + 674t^4 + 8631t^6 + 84410t^8 + 669825t^{10} + 4483882t^{12} + 26024643t^{14} + 133684785t^{16} + 617619665t^{18} + 2599637184t^{20} + O(t^{21})$

**1589 2505.03875-AppB-B4-5-1p4-Spin**

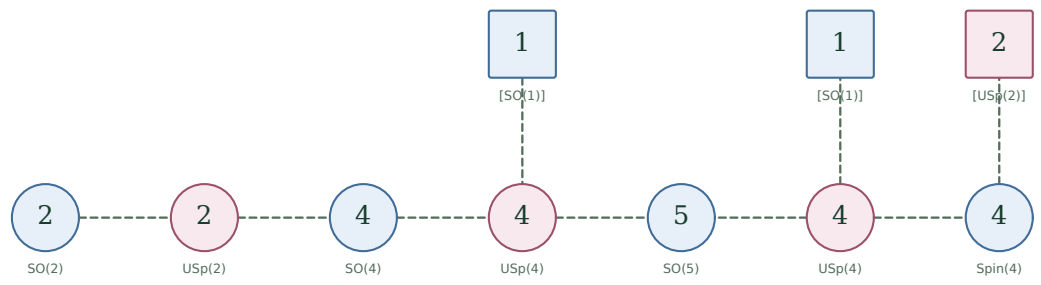


$1 + 28t^2 + 442t^4 + 5103t^6 + 46898t^8 + 358377t^{10} + 2343802t^{12} + 13405515t^{14} + 68216313t^{16} + 313216337t^{18} + 1312934992t^{20} + O(t^{21})$

**1590 2505.03875-AppB-B4-5-1p4-O-**

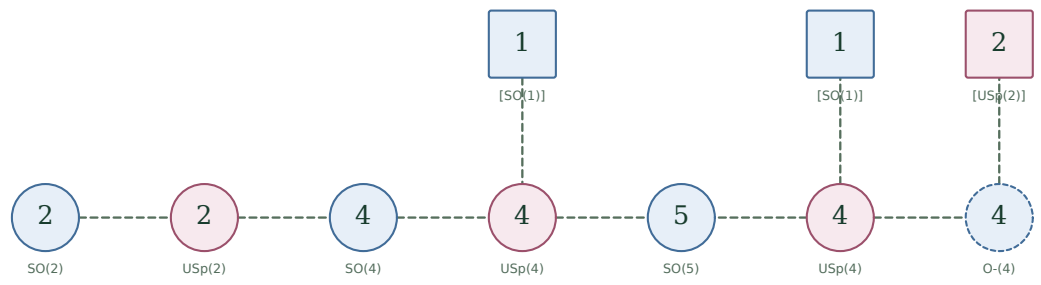


18: θ = -

$1 + 28t^2 + 442t^4 + 5103t^6 + 46898t^8 + 358377t^{10} + 2343802t^{12} + 13405515t^{14} + 68216313t^{16} + 313216337t^{18} + 1312934992t^{20} + O(t^{21})$

**1591 2505.03875-AppB-B4-5-1p4-Pin**

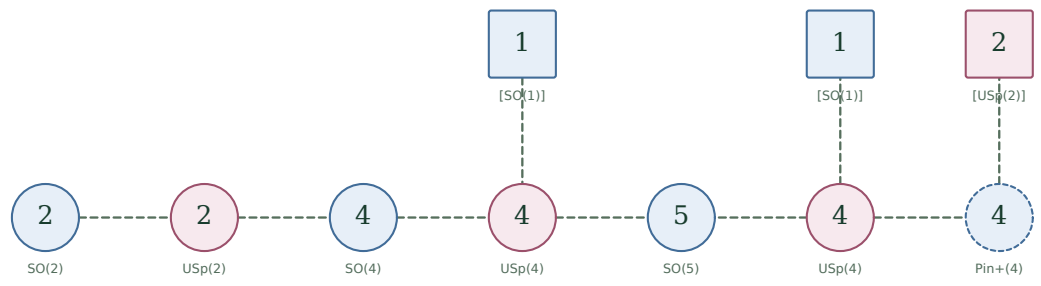


$1 + 21t^2 + 287t^4 + 3010t^6 + 26081t^8 + 191982t^{10} + 1226518t^{12} + 6910800t^{14} + 34828650t^{16} + 158901991t^{18} + 663258594t^{20} + O(t^{21})$

| QUIVER | HILBERT SERIES |
|---|---|

**1592 2505.03875-AppB-B4-5-3-1-SO**

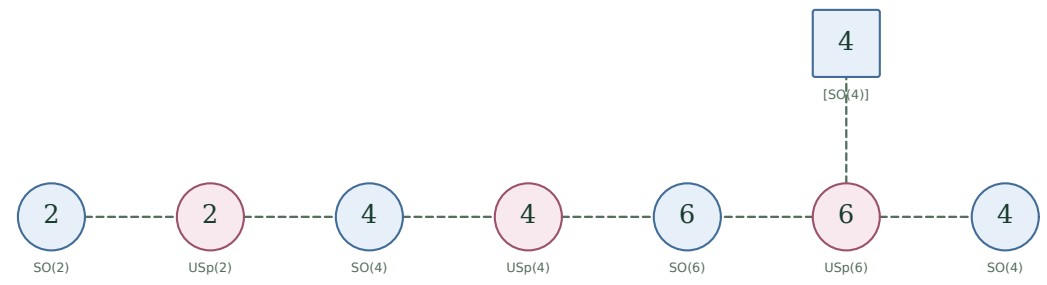


$1 + 36t^2 + 674t^4 + 8715t^6 + 87236t^8 + 718680t^{10} + 5061142t^{12} + 31260629t^{14} + 172531071t^{16} + 863004616t^{18} + 3956473971t^{20} + O(t^{21})$

**1593 2505.03875-AppB-B4-5-3-1-Spin**

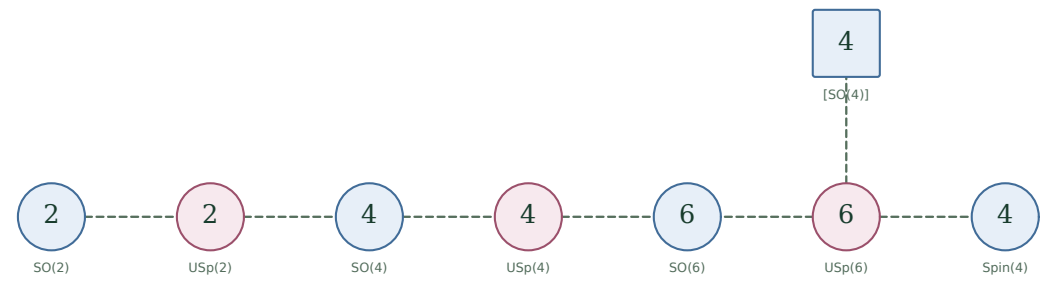


$1 + 28t^2 + 442t^4 + 5131t^6 + 48108t^8 + 381192t^{10} + 2623214t^{12} + 15981397t^{14} + 87476967t^{16} + 435361368t^{18} + 1989707963t^{20} + O(t^{21})$

**1594 2505.03875-AppB-B4-5-3-1-O-**

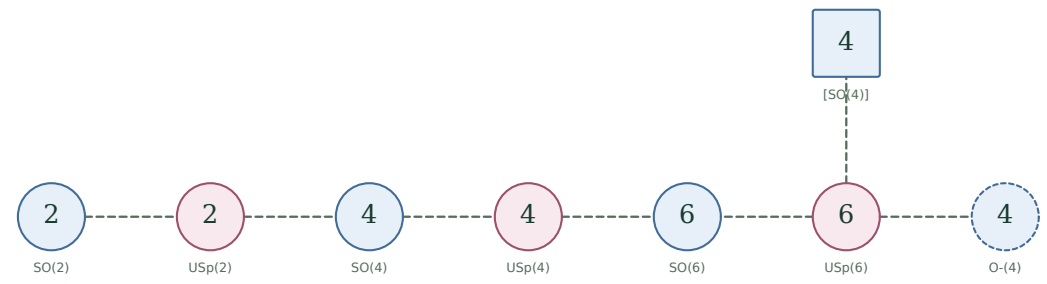


18: θ = -

$1 + 28t^2 + 442t^4 + 5131t^6 + 48108t^8 + 381192t^{10} + 2623214t^{12} + 15981397t^{14} + 87476967t^{16} + 435361368t^{18} + 1989707963t^{20} + O(t^{21})$

**1595 2505.03875-AppB-B4-5-3-1-Pin**

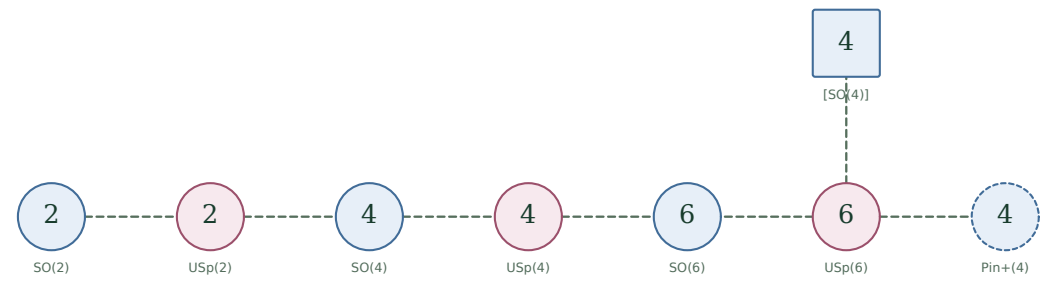


$1 + 21t^2 + 287t^4 + 3017t^6 + 26571t^8 + 202496t^{10} + 1361099t^{12} + 8175591t^{14} + 44369894t^{16} + 219676106t^{18} + 1000749470t^{20} + O(t^{21})$

**1596 2505.03875-AppB-B4-7-1p2-SO**

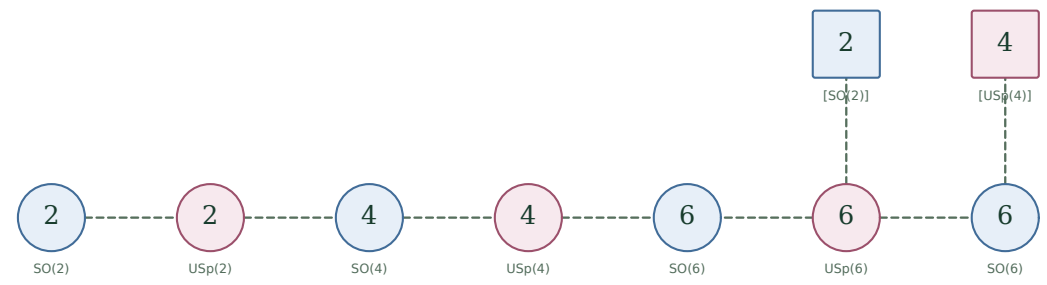


$1 + 36t^2 + 665t^4 + 8409t^6 + 81890t^8 + 654882t^{10} + 4477389t^{12} + 26899665t^{14} + 144860179t^{16} + 709780863t^{18} + 3201281653t^{20} + O(t^{21})$

**1597 2505.03875-AppB-B4-7-1p2-Spin**

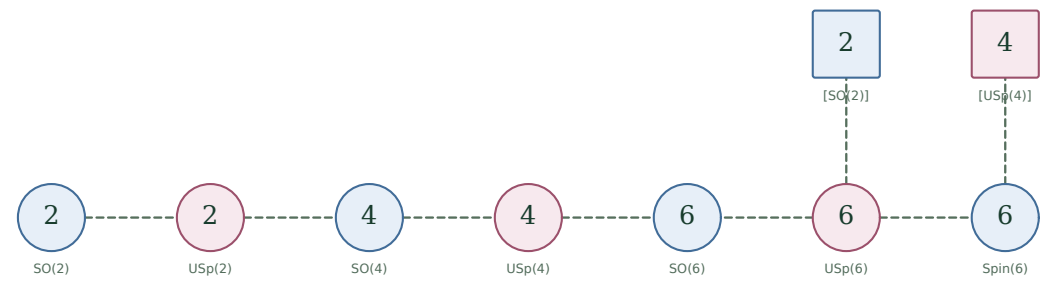


$1 + 28t^2 + 441t^4 + 5041t^6 + 46058t^8 + 353730t^{10} + 2356669t^{12} + 13923625t^{14} + 74163827t^{16} + 360749439t^{18} + 1619118093t^{20} + O(t^{21})$

**1598 2505.03875-AppB-B4-7-1p2-O-**

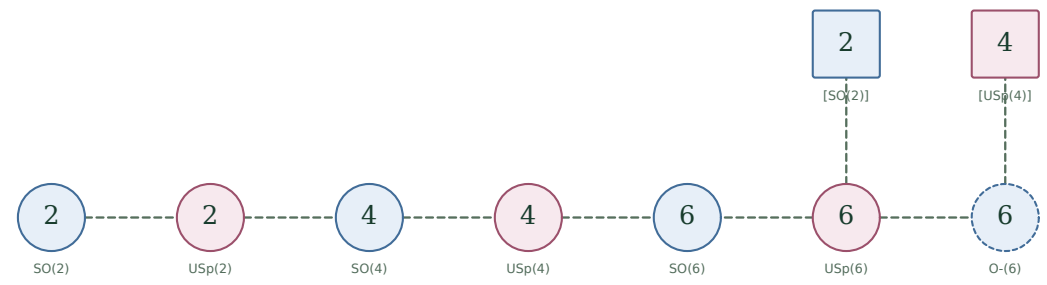


18: θ = -

$1 + 28t^2 + 441t^4 + 5041t^6 + 46058t^8 + 353730t^{10} + 2356669t^{12} + 13923625t^{14} + 74163827t^{16} + 360749439t^{18} + 1619118093t^{20} + O(t^{21})$

| QUIVER | HILBERT SERIES |
|---|---|

**1599 2505.03875-AppB-B4-7-1p2-Pin**

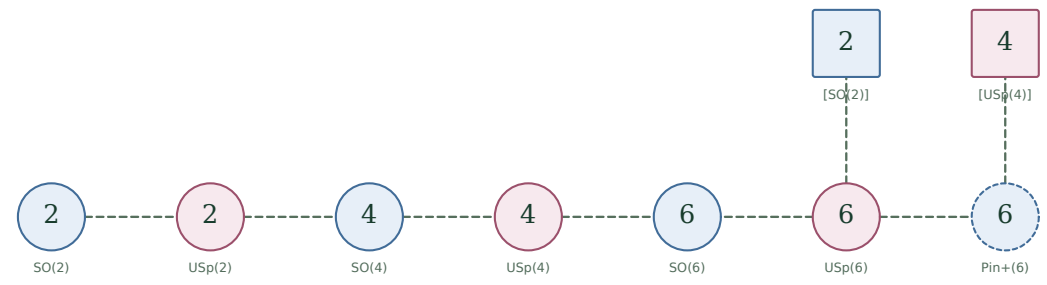


$1 + 21t^2 + 287t^4 + 2996t^6 + 25815t^8 + 190820t^{10} + 1239909t^{12} + 7206320t^{14} + 37969611t^{16} + 183357923t^{18} + 818928756t^{20} + O(t^{21})$

**1600 2505.03875-AppB-D5-3p2-1p4-SO**

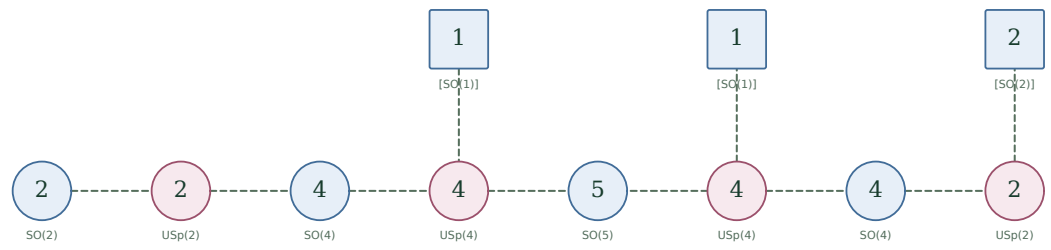


$1 + 45t^2 + 1079t^4 + 17630t^6 + 216370t^8 + 2110977t^{10} + 17028305t^{12} + 116962308t^{14} + 699968215t^{16} + 3717369030t^{18} + 17782581806t^{20} + O(t^{21})$

**1601 2505.03875-AppB-D5-3p2-1p4-Spin**

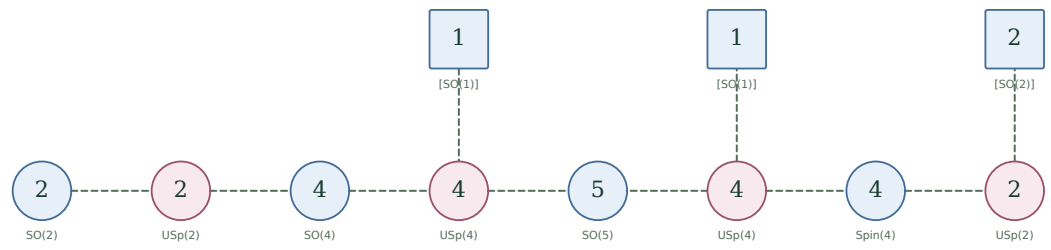


$1 + 29t^2 + 599t^4 + 9214t^6 + 110418t^8 + 1066049t^{10} + 8558209t^{12} + 58645140t^{14} + 350540615t^{16} + 1860421830t^{18} + 8896344686t^{20} + O(t^{21})$

**1602 2505.03875-AppB-D5-3p2-1p4-O-**

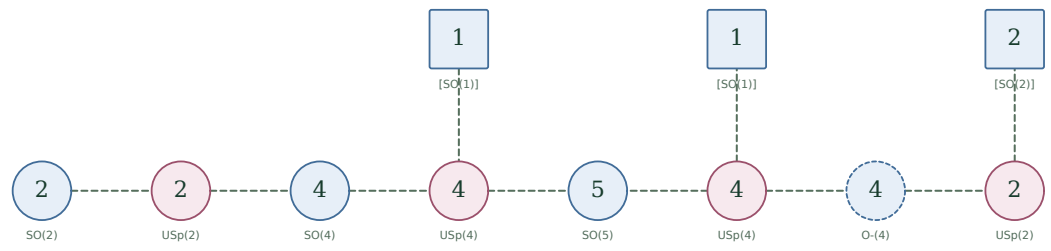


18: θ = -

$1 + 29t^2 + 599t^4 + 9214t^6 + 110418t^8 + 1066049t^{10} + 8558209t^{12} + 58645140t^{14} + 350540615t^{16} + 1860421830t^{18} + 8896344686t^{20} + O(t^{21})$

**1603 2505.03875-AppB-D5-3p2-1p4-Pin**

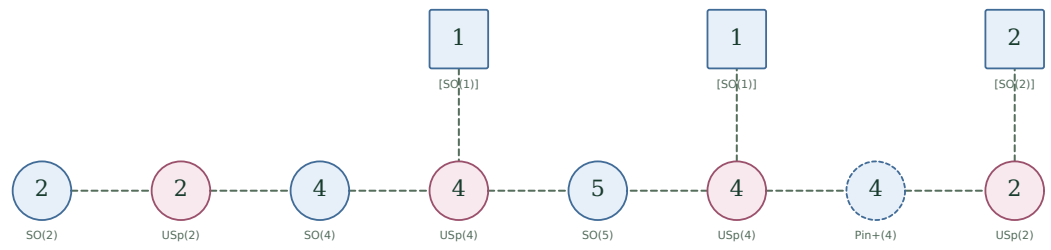


$1 + 21t^2 + 359t^4 + 5006t^6 + 57442t^8 + 543585t^{10} + 4323161t^{12} + 29486556t^{14} + 175826815t^{16} + 931948230t^{18} + 4453226126t^{20} + O(t^{21})$

**1604 2505.03875-AppB-D5-5-3-1p2-SO**

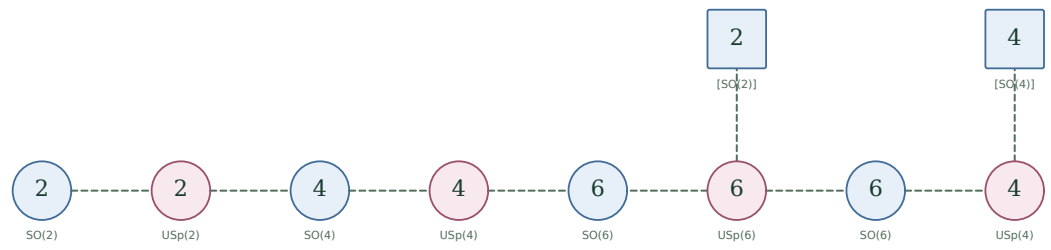


$1 + 45t^2 + 1034t^4 + 16215t^6 + 195424t^8 + 1930528t^{10} + 16268461t^{12} + 120122896t^{14} + 792104940t^{16} + 4731174753t^{18} + 25878964760t^{20} + O(t^{21})$

**1605 2505.03875-AppB-D5-5-3-1p2-Spin**

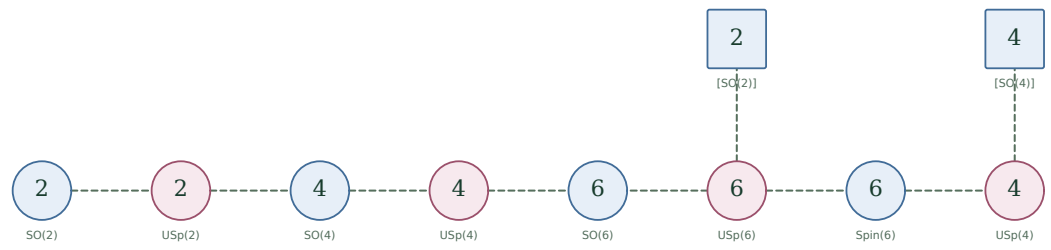


$1 + 29t^2 + 570t^4 + 8439t^6 + 99424t^8 + 972944t^{10} + 8164909t^{12} + 60172992t^{14} + 396426524t^{16} + 2366758113t^{18} + 12942930440t^{20} + O(t^{21})$

| QUIVER | HILBERT SERIES |
|---|---|
| **1606 2505.03875-AppB-D5-5-3-1p2-O-** 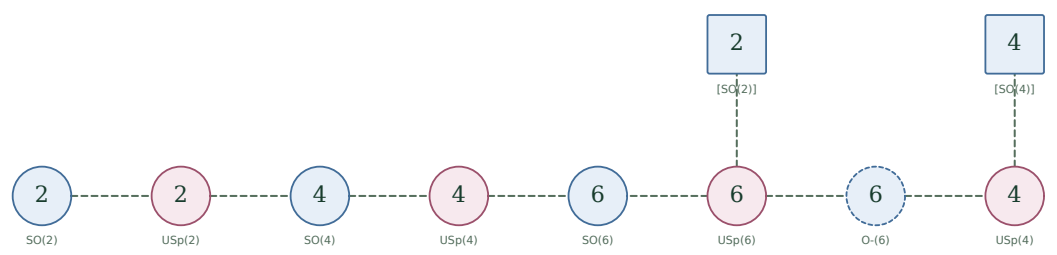 18: θ = - | $1 + 29t^2 + 570t^4 + 8439t^6 + 99424t^8 + 972944t^{10} + 8164909t^{12} + 60172992t^{14} + 396426524t^{16} + 2366758113t^{18} + 12942930440t^{20} + O(t^{21})$ |
| **1607 2505.03875-AppB-D5-5-3-1p2-Pin** 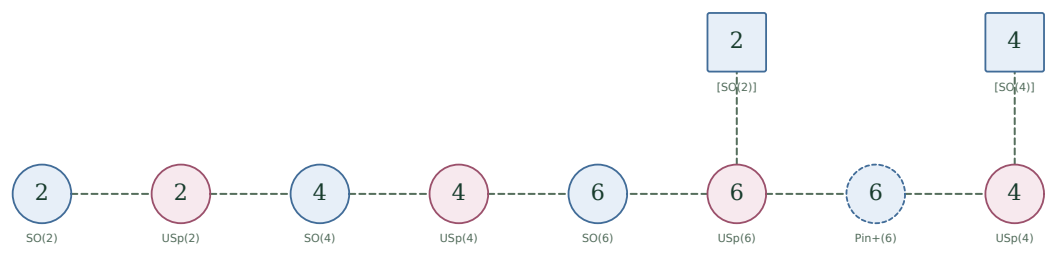 | $1 + 21t^2 + 338t^4 + 4551t^6 + 51424t^8 + 494152t^{10} + 4113133t^{12} + 30198040t^{14} + 198587316t^{16} + 1184549793t^{18} + 6474913280t^{20} + O(t^{21})$ |
| **1608 2505.03875-D5-SU2-U2** 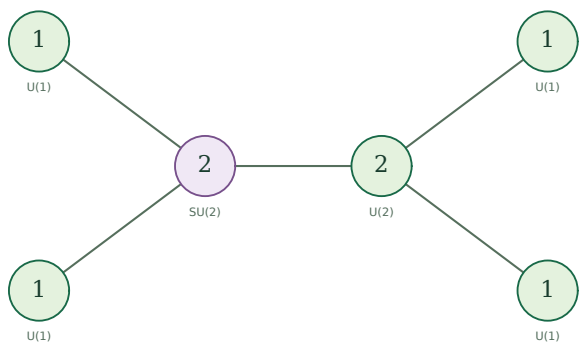 | $1 + 21t^2 + 394t^4 + O(t^5)$ |
| **1609 2505.03875-D5-SU2-SU2** 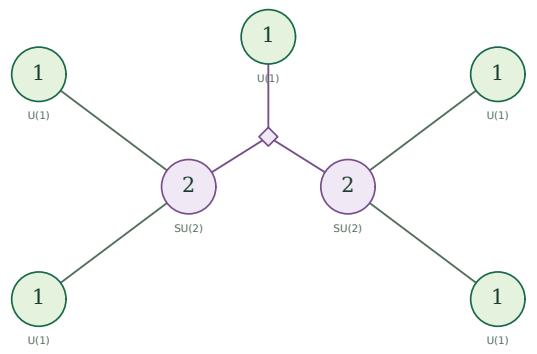 | $1 + 13t^2 + 226t^4 + O(t^5)$ Conditional comparison recorded in the source database. |
| **1610 2505.03875-D5-SU2SU2U1-Z2zeta** 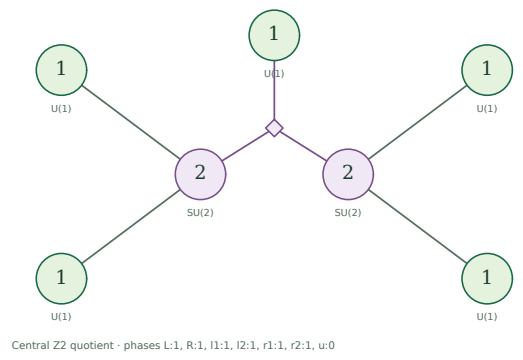 Central Z2 quotient · phases L:1, R:1, l1:1, l2:1, r1:1, r2:1, u:0 | $1 + 29t^2 + 434t^4 + O(t^5)$ |
| **1611 2505.03875-O1-USp2-SO9** 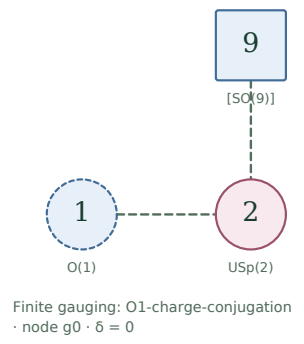 g0: δ = 0 · Finite gauging: O1-charge-conjugation · node g0 · δ = 0 | $1 + t^4 + O(t^5)$ |
| **1612 2505.03875-USp2-O1x2-SO6** 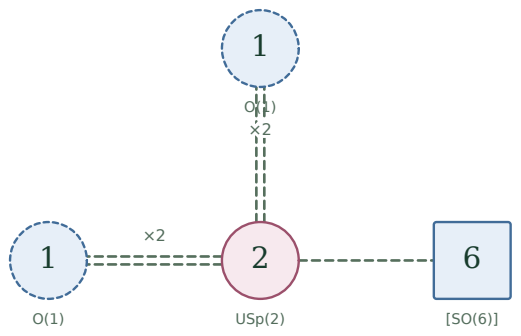 | $1 + t^4 + O(t^5)$ |

## 2510.14489 · Bootstrapping mirror pairs: The beginning of the end (27)

**1613 2510.14489-eq11__finite / n = 4**

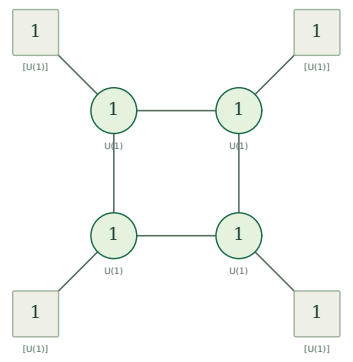

$1 + 4t^2 + 8t^3 + 20t^4 + 40t^5 + 84t^6 + 152t^7 + 285t^8 + 472t^9 + 769t^{10} + O(t^{11})$

**Paper typo:** Independent monopole and rational-function calculations give 796 at $t^{10}$, versus the printed 769; all earlier coefficients agree. A digit transposition is inferred.

**1614 2510.14489-eq12-X__finite**

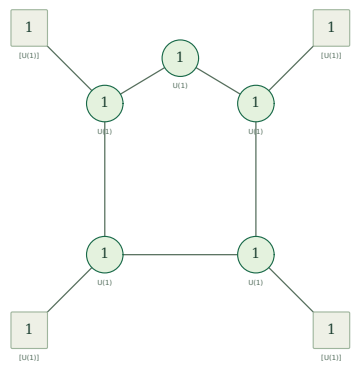

$1 + 7t^2 + 12t^3 + 41t^4 + 88t^5 + 215t^6 + 428t^7 + 902t^8 + 1656t^9 + 3114t^{10} + O(t^{11})$

**1615 2510.14489-eq12-Y__finite**

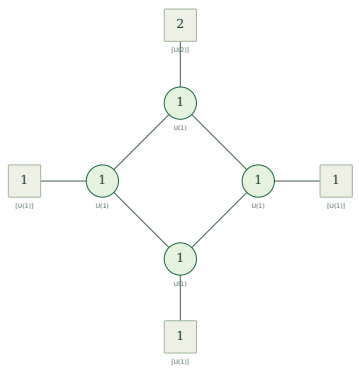

$1 + 4t^2 + 6t^3 + 16t^4 + 32t^5 + 64t^6 + 112t^7 + 209t^8 + 342t^9 + 570t^{10} + O(t^{11})$

**1616 2510.14489-eq15-X__finite**

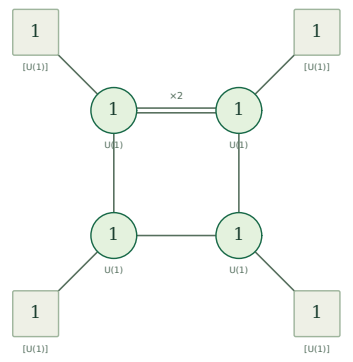

$1 + 4t^2 + 4t^3 + 20t^4 + 24t^5 + 70t^6 + 96t^7 + 221t^8 + 300t^9 + 598t^{10} + O(t^{11})$

**1617 2510.14489-eq15-Y__finite**

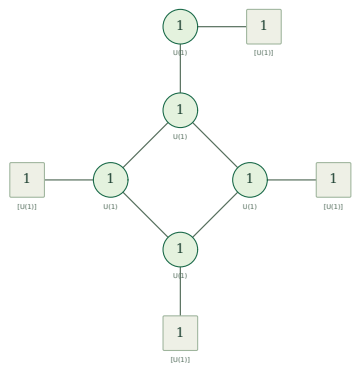

$1 + 7t^2 + 10t^3 + 43t^4 + 80t^5 + 213t^6 + 406t^7 + 884t^8 + 1584t^9 + 3042t^{10} + O(t^{11})$

**1618 2510.14489-eq19-X__finite**

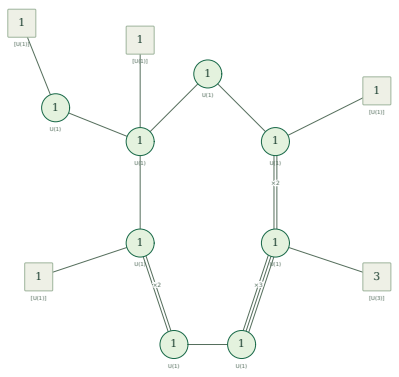

$1 + 12t^2 + 4t^3 + 92t^4 + 60t^5 + 536t^6 + 508t^7 + 2598t^8 + 3070t^9 + 10978t^{10} + O(t^{11})$

**1619 2510.14489-eq19-Y__finite**

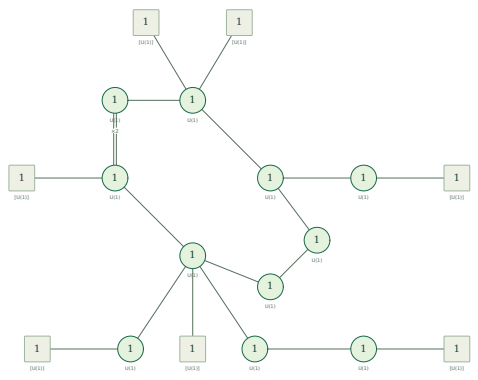

$1 + 27t^2 + 16t^3 + 362t^4 + 436t^5 + 3428t^6 + 5844t^7 + 26521t^8 + 53208t^9 + 177029t^{10} + O(t^{11})$

**QUIVER** | **HILBERT SERIES**

**1620 2510.14489-eq20-X__finite**

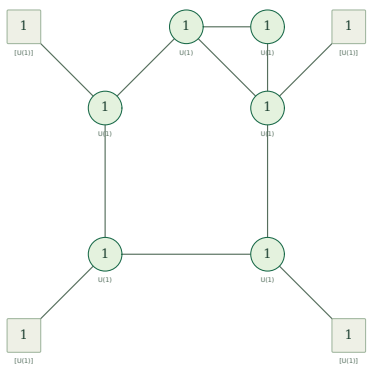

**Computed by Quiver Lab**

$1 + 8t^{2} + 12t^{3} + 53t^{4} + 108t^{5} + 310t^{6} + 632t^{7} + 1488t^{8} + 2892t^{9} + 5988t^{10} + 11016t^{11} + 20826t^{12} + 36372t^{13} + 64182t^{14} + 107016t^{15} + 178781t^{16} + 286332t^{17} + 457400t^{18} + 707288t^{19} + 1088537t^{20} + 1632324t^{21} + 2433970t^{22} + 3553184t^{23} + 5155590t^{24} + 7350276t^{25} + 10415138t^{26} + 14540076t^{27} + 20177944t^{28} + 27647240t^{29} + 37664672t^{30} + O(t^{31})$

Source originally Claimed · no numerical literature comparison is available.

**1621 2510.14489-eq20-Y__finite**

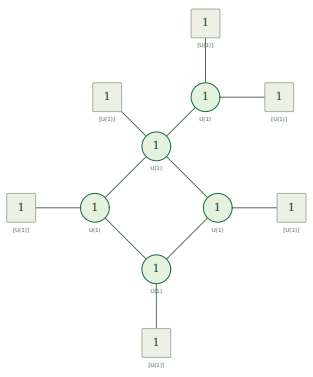

$1 + 5t^{2} + 8t^{3} + 21t^{4} + 50t^{5} + 105t^{6} + 206t^{7} + 414t^{8} + 750t^{9} + 1348t^{10} + O(t^{11})$

**1622 2510.14489-eq21-X__finite**

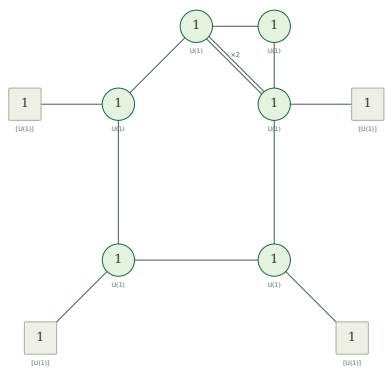

**Computed by Quiver Lab**

$1 + 8t^{2} + 8t^{3} + 49t^{4} + 80t^{5} + 252t^{6} + 472t^{7} + 1132t^{8} + 2108t^{9} + 4382t^{10} + 7856t^{11} + 14848t^{12} + 25488t^{13} + 44884t^{14} + 73980t^{15} + 123267t^{16} + 195784t^{17} + 311980t^{18} + 479436t^{19} + 736161t^{20} + 1098752t^{21} + 1634960t^{22} + 2377992t^{23} + 3444276t^{24} + 4895952t^{25} + 6926666t^{26} + 9646928t^{27} + 13369310t^{28} + 18282176t^{29} + 24877166t^{30} + O(t^{31})$

Source originally Claimed · no numerical literature comparison is available.

**1623 2510.14489-eq21-Y__finite**

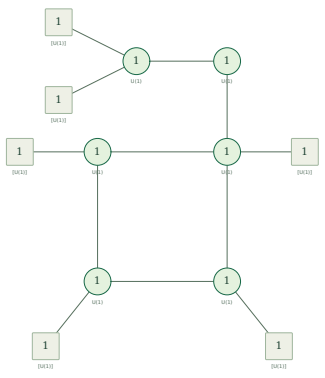

$1 + 8t^{2} + 10t^{3} + 43t^{4} + 92t^{5} + 234t^{6} + 494t^{7} + 1096t^{8} + 2150t^{9} + 4282t^{10} + O(t^{11})$

**1624 2510.14489-eq22-X__finite**

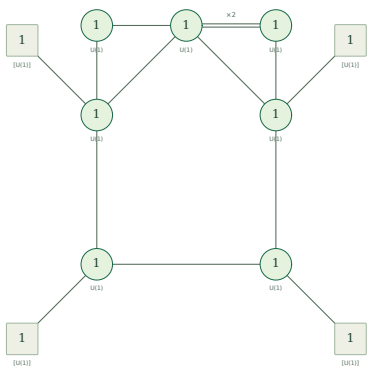

**Computed by Quiver Lab**

$1 + 9t^{2} + 6t^{3} + 66t^{4} + 78t^{5} + 374t^{6} + 572t^{7} + 1832t^{8} + 3066t^{9} + 7836t^{10} + 13360t^{11} + 29596t^{12} + 50016t^{13} + 100060t^{14} + 165852t^{15} + 307283t^{16} + 497420t^{17} + 868071t^{18} + 1370778t^{19} + 2279770t^{20} + 3513704t^{21} + 5615886t^{22} + 8459094t^{23} + 13074084t^{24} + 19276970t^{25} + 28949436t^{26} + 41850494t^{27} + 61299732t^{28} + 87023152t^{29} + 124701800t^{30} + O(t^{31})$

Source originally Claimed · no numerical literature comparison is available.

**1625 2510.14489-eq22-Y__finite**

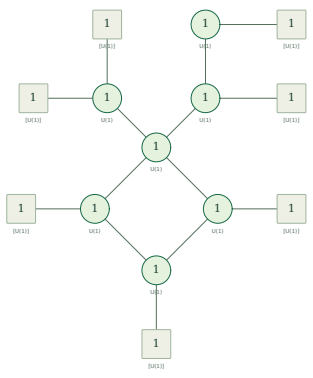

$1 + 9t^{2} + 12t^{3} + 50t^{4} + 118t^{5} + 304t^{6} + 698t^{7} + 1620t^{8} + 3424t^{9} + 7192t^{10} + O(t^{11})$

| QUIVER | HILBERT SERIES |
|---|---|

**1626 2510.14489-eq23-X__finite**

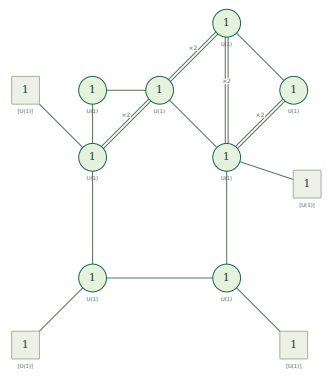

**Computed by Quiver Lab**

$1 + 10t^{2} + 6t^{3} + 60t^{4} + 78t^{5} + 308t^{6} + 544t^{7} + 1462t^{8} + 2816t^{9} + 6340t^{10} + 12198t^{11} + 24858t^{12} + 46452t^{13} + 88452t^{14} + 159458t^{15} + 288163t^{16} + 501384t^{17} + 867836t^{18} + 1460764t^{19} + 2437504t^{20} + 3979814t^{21} + 6433640t^{22} + 10217446t^{23} + 16062474t^{24} + 24876726t^{25} + 38148416t^{26} + 57752016t^{27} + 86612026t^{28} + 128435598t^{29} + 188784042t^{30} + O(t^{31})$

Source originally Claimed · no numerical literature comparison is available.

**1627 2510.14489-eq23-Y__finite**

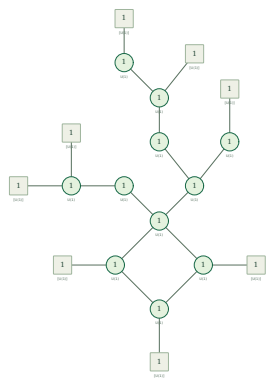

$1 + 19t^{2} + 26t^{3} + 202t^{4} + 490t^{5} + 1892t^{6} + 5176t^{7} + 15731t^{8} + 41620t^{9} + 111607t^{10} + O(t^{11})$

**1628 2510.14489-eq24__finite**

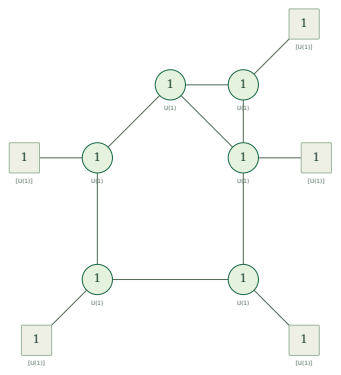

$1 + 6t^{2} + 10t^{3} + 33t^{4} + 78t^{5} + 190t^{6} + 408t^{7} + 892t^{8} + 1774t^{9} + 3504t^{10} + O(t^{11})$

**1629 2510.14489-eq25-X__finite**

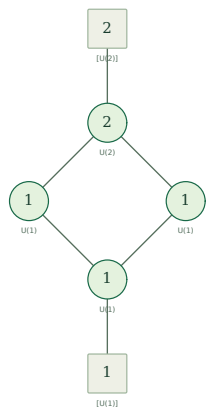

$1 + 6t^{2} + 12t^{3} + 34t^{4} + 78t^{5} + 181t^{6} + 364t^{7} + 740t^{8} + 1382t^{9} + 2524t^{10} + O(t^{11})$

**1630 2510.14489-eq25-Y__finite**

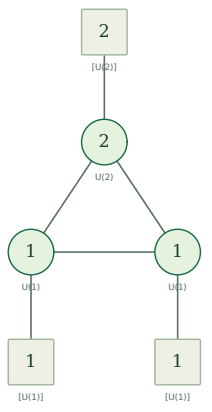

**Computed by Quiver Lab**

$1 + 5t^{2} + 28t^{4} + 105t^{6} + 339t^{8} + 920t^{10} + 2242t^{12} + 4930t^{14} + 10079t^{16} + 19267t^{18} + 34970t^{20} + 60549t^{22} + 100895t^{24} + 162354t^{26} + 253694t^{28} + 385908t^{30} + O(t^{31})$

Source originally Claimed · no numerical literature comparison is available.

**1631 2510.14489-eq27-X__finite**

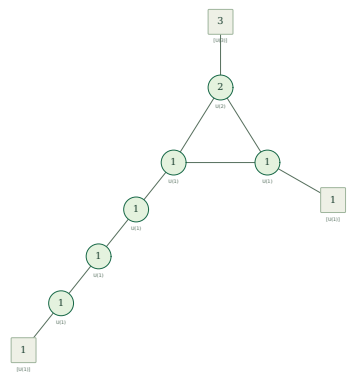

$1 + 18t^{2} + 2t^{3} + 146t^{4} + 56t^{5} + 813t^{6} + 564t^{7} + 3621t^{8} + 3516t^{9} + 13952t^{10} + O(t^{11})$

**1632 2510.14489-eq27-Y__finite**

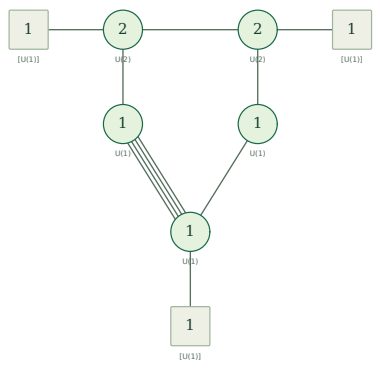

$1 + 11t^{2} + 8t^{3} + 80t^{4} + 104t^{5} + 458t^{6} + 742t^{7} + 2232t^{8} + 3872t^{9} + 9430t^{10} + O(t^{11})$

| QUIVER | HILBERT SERIES |
|---|---|
| **1633 2510.14489-eq30__finite** 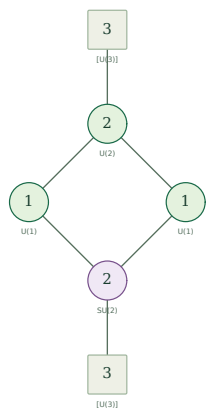 | $1 + 3t^2 + 2t^3 + 12t^4 + 12t^5 + 32t^6 + 44t^7 + 98t^8 + 136t^9 + 256t^{10} + O(t^{11})$ |
| **1634 2510.14489-eq40-X__finite** 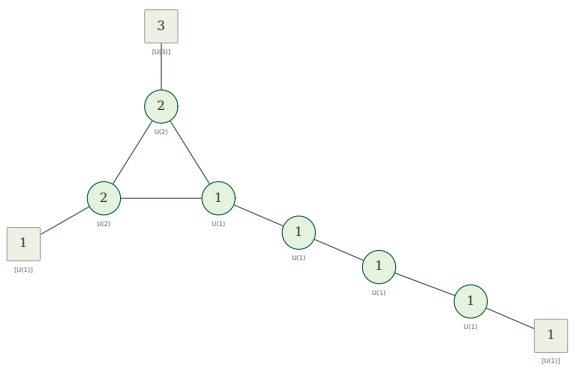 | $1 + 20t^2 + 182t^4 + 24t^5 + 1143t^6 + 368t^7 + 5663t^8 + 3000t^9 + 23769t^{10} + O(t^{11})$ |
| **1635 2510.14489-eq40-Y__finite** 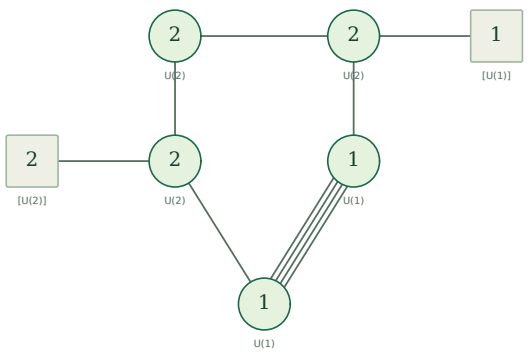 | $1 + 11t^2 + 8t^3 + 81t^4 + 112t^5 + 490t^6 + O(t^7)$ |
| **1636 2510.14489-eqB1-X__finite** 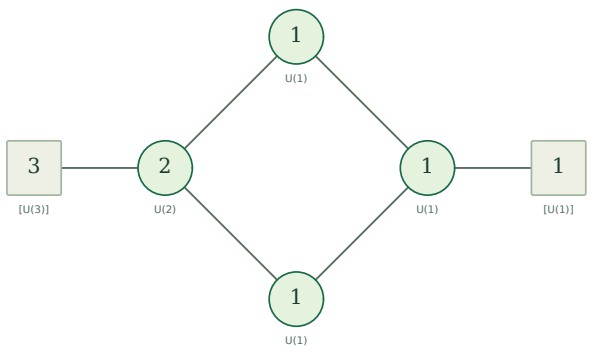 | $1 + 4t^2 + 8t^3 + 21t^4 + 42t^5 + 95t^6 + 182t^7 + 355t^8 + 636t^9 + 1137t^{10} + O(t^{11})$ |
| **1637 2510.14489-eqB1-Y__finite** 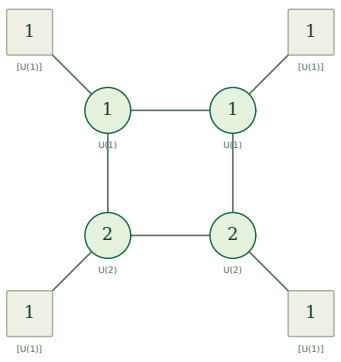 | $1 + 10t^2 + 77t^4 + 446t^6 + 2098t^8 + 8268t^{10} + O(t^{12})$ |
| **1638 2510.14489-eqB2-X__finite** 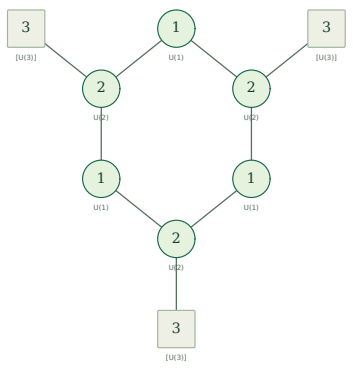 | $1 + 6t^2 + 6t^3 + 30t^4 + 54t^5 + 137t^6 + 294t^7 + 636t^8 + 1306t^9 + 2670t^{10} + O(t^{11})$ |
| **1639 2510.14489-eqB2-Y__finite** 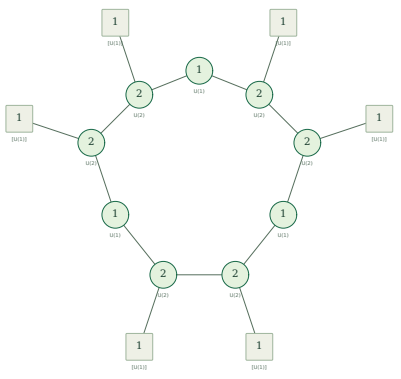 | $1 + 27t^2 + 456t^4 + 5786t^6 + 59910t^8 + 526458t^{10} + O(t^{12})$ |



\end{document}